\pdfoutput=1

\documentclass[12pt,oneside]{book}

\usepackage[greek,english]{babel}
\usepackage[T1]{fontenc}
\usepackage{amsmath,amssymb}
\usepackage{mathrsfs}
\usepackage{pgf,pgfarrows}

\title{\textbf{Aspects of locally covariant quantum field theory}\\[2cm]}
\author{\small{A thesis presented by}\\
\textbf{Jacobus Ambrosius Sanders}\\
\small{to obtain the degree of}\\
\textbf{PhD}\\
\small{from the}\\
University of York \small{(U.K.)}\\
Department of Mathematics\\
}
\date{July 2008}

\fussy

\newtheorem{definition}{Definition}[section]
\newtheorem{lemma}[definition]{Lemma}
\newtheorem{proposition}[definition]{Proposition}
\newtheorem{theorem}[definition]{Theorem}
\newtheorem{corollary}[definition]{Corollary}
\newtheorem{remark}[definition]{Remark}
\newtheorem{conjecture}[definition]{Conjecture}
\newenvironment{proof*}{\smallskip\par\noindent\emph{Proof. }
 \ignorespaces}{\hfill$\Box$\smallskip\par\ignorespaces}
\newenvironment{proofsketch*}{\smallskip\par\noindent
 \emph{Sketch of proof. }\ignorespaces}
 {\hfill$\oslash$\smallskip\par\ignorespaces}

\newcommand{\dirop}{\nabla\!\!\!\!\!\!\;/ \,}
\newcommand{\map}[3]{\ensuremath{#1\!:\!#2\!\rightarrow\!#3}}
\newcommand{\id}{\ensuremath{\mathrm{id}}}
\newcommand{\fsl}[1]{\ensuremath{#1\!\!\!\!\!\;/ \,}}

\newcommand{\N}{\ensuremath{\mathbb{N}}}

\newcommand{\R}{\ensuremath{\mathbb{R}}}
\newcommand{\C}{\ensuremath{\mathbb{C}}}
\newcommand{\Hq}{\ensuremath{\mathbb{H}}}

\newcommand{\CatAlg}{\ensuremath{\mathfrak{CAlg}}}
\newcommand{\CatTAlg}{\ensuremath{\mathfrak{TAlg}}}
\newcommand{\CatStat}{\ensuremath{\mathfrak{States}}}
\newcommand{\CatMan}{\ensuremath{\mathfrak{Man}}}
\newcommand{\CatSMan}{\ensuremath{\mathfrak{SMan}}}
\newcommand{\CatTop}{\ensuremath{\mathfrak{Top}}}
\newcommand{\CatVB}{\ensuremath{\mathfrak{VB}}}

\newcommand{\Func}[1]{\ensuremath{\mathbf{#1}}}
\newcommand{\Alg}[1]{\ensuremath{\mathcal{#1}}}
\newcommand{\Stat}[1]{\ensuremath{\mathscr{#1}}}
\newcommand{\Test}{\ensuremath{C^{\infty}}}

\begin{document}

\pagestyle{plain}
\maketitle

\frontmatter
\renewcommand{\thepage}{\arabic{page}}
\setcounter{page}{2}
\renewcommand{\thesection}{}

\section{\textbf{Abstract}}

This thesis considers various aspects of locally covariant quantum
field theory (see Brunetti et al., Commun. Math. Phys.
\textbf{237} (2003), 31--68), a mathematical framework to describe
axiomatic quantum field theories in curved spacetimes. Chapter
\ref{ch_phil} argues that the use of morphisms in this framework can
be seen as a model for modal logic. To our knowledge this is the first
interpretative description of this aspect of the framework. Chapter
\ref{ch_LCQFT} gives an exposition of locally covariant quantum field
theory which differs from the original in minor details, notably in
the new notion of nowhere-classicality and the sharpened time-slice
axiom, which puts a restriction on the state space as well as
the algebras. Chapter \ref{ch_sf} deals with the well-studied example
of the free real scalar field and includes an elegant proof of the new
general result that the commutation relations together with the Hadamard
condition on the two-point distribution of a state completely fix the
singularity structure of all $n$-point distributions. Chapter
\ref{ch_Df} describes the free Dirac field as a locally covariant
quantum field, using a new representation independent approach,
demonstrating that the physics is determined entirely by the relations
between the adjoint map, charge conjugation and Dirac operator. It also
proves the new result that the relative Cauchy evolution is related to
the stress-energy-momentum tensor in the same way as for the free scalar
field. Chapter \ref{ch_RS} studies the Reeh-Schlieder property, both in
the general setting and in specific examples. We obtain various
interesting results concerning this property in curved spacetimes, most
notably by using the idea of spacetime
deformation, but some open questions and opportunities for further
research remain. We will freely make use of smooth and
analytic wave front sets throughout. These concepts are explained in
appendix \ref{ch_ma}, using a new and elegant way to generalise results
for scalar distributions to Banach space-valued distributions, leading
to some new but expected results.

\tableofcontents
\listoffigures

\chapter{Foreword}\label{ch_foreword}

\begin{quote}
... it is not his \emph{possession} of knowledge, of
irrefutable truth, that makes the man of science, but his persistent
and recklessly critical \emph{quest} for truth.
\end{quote}
\begin{flushright}
Karl Popper, \cite{Popper} p.281
\end{flushright}

This thesis is based on research that was done at the University of
York between October 2005
and June 2008. During those three years I learned a lot about science,
about the world around us and also about myself. I feel that I have
grown a lot as a mathematician, or perhaps as a mathematical physicist,
although I don't feel that my knowledge of physics has increased much.
To a lesser extent I feel that I have grown as a philosopher of science,
especially during the preparation of chapter \ref{ch_phil} below, which
is essentially the condensation of ideas that have been in my head since
early 2004.

The most important thing that I learned about myself is exactly how
ruthless I have to be to myself from time to time in order to get things
done and to achieve the goals that I have set myself.
My working attitude is perhaps best described in
the words of my fellow PhD-student Paul Melvin, who told me time and again
that I had been working like a machine. Maybe an insult to many, but to me
these words were a compliment and they motivated me to go on and not to be
tempted too much by York's beautiful scenery, walks along the river Ouse
and the taste of lukewarm, non-sparkling, English beer.

And so I went on in the quest for truth, as Popper describes it in the
quote above.\footnote{Probably more accurate than the word ``truth'', at
least for the physical aspect of mathematical physics, would be the word
``verisimilitude'' or the phrase ``statements in which we have confidence'',
see chapter \ref{ch_phil} section \ref{sec_phback}. Of course these
alternatives are far less aesthetic.} On some occasions I managed to prove
a useful mathematical result that has consequences for the physical
theories under investigation, as this thesis will indicate. On other
occasions Mathematics denied me the proof that I was looking for and left
me in the dark as to whether my gut feeling was right or wrong. Looking
back on the results that were obtained I feel some gratification, of
course, and pride for the knowledge I now possess due to all the hard work
I have done. The dominant feeling, however, is curiosity. Curiosity aroused
by the intriguing and tantalising new questions that emerged during the
course of this research and that remain unanswered. For me too the quest
for truth still goes on, and I am grateful for every opportunity I get to
pursue it.

This thesis has been divided into six chapters and an appendix. Chapter
\ref{ch_phil} is of a more philosophical nature and the later chapters
can be read independently of the first chapter. Conversely, chapter
\ref{ch_phil} only requires a superficial understanding of the framework
of locally covariant quantum field theory, the main object of study in
this thesis. The precise mathematical formulation of this framework is
given in chapter \ref{ch_LCQFT}. Chapters \ref{ch_sf} and \ref{ch_Df}
describe in detail two examples of locally covariant quantum field
theories, namely the real free scalar field and the free Dirac field.
These chapters freely make use of the notion of wave front sets, which
is explained in appendix \ref{ch_ma}. It should be noted that the
appendix provides an elegant and new approach to generalise results for
scalar distributions to Banach space-valued ones and proves
results that are more general than those existing in the literature.
Chapter
\ref{ch_RS} studies the Reeh-Schlieder property in locally covariant
quantum field theory, both in the general axiomatic setting and in the
special examples of chapter \ref{ch_sf}. The final
chapter \ref{ch_concl} summarises the conclusions that can be drawn from
the earlier chapters and discusses some opportunities for further
research.

A remark about notations and conventions in this thesis is in order,
although most notations are either standard or defined in the text
when they are first introduced. The signatures of our spacetimes
will be $(+---)$, which agrees with most of the references, except
e.g. \cite{Hawking+,Wald,Wald2,Baer+}.
Lower case Greek letters are used to denote the components of vectors
and covectors in a coordinate basis. Lower case
Latin indices are used to indicate abstract indices of tensors (see
\cite{Wald} for a review of the abstract index notation), or to indicate
the components of vectors and covectors in a vierbein in chapter
\ref{ch_Df}. Capital Latin indices are used
to indicate the components of spinors and cospinors in a spin frame, but
for convenience these indices will often be dropped in favour of a
matrix notation, as explained in chapter \ref{ch_Df}. Einstein's
summation convention is used throughout. Retarded fundamental solutions
have their support to the future of the source function and are
indicated by a superscript $^+$. Similarly, advanced fundamental solutions
have their support in the past and are indicated by a $^-$. (A few of the
references swap the names ``retarded'' and ``advanced'', e.g.
\cite{Verch1}.) For quantisation we use the advanced-minus-retarded
fundamental solution, as in \cite{Fewster+2,Radzikowski,Wald2}.

Fourier transforms on $\R^n$ are defined by
\[
\hat{f}(k)=\int e^{-ik\cdot x} f(x)\  dx,
\]
where $\cdot$ denotes the pairing of $\R^n$ and its dual.
This is unlike e.g.
\cite{Dawson+,Fewster+,Fewster+2,Radzikowski2} who omit the minus sign
in the exponent. The Fourier inversion formula on $\R^n$ then reads:
\[
f(x)=(2\pi)^{-n}\int e^{ik\cdot x}\hat{f}(k)\ dk.
\]

For the real free scalar field in Minkowski spacetime the retarded $(+)$
and advanced $(-)$ fundamental solutions of the Klein-Gordon equation are
given by
\[
\hat{E}^{\pm}(k,l)=\lim_{\epsilon\rightarrow 0^+}
\frac{-(2\pi)^4\delta(k+l)}{(l_0\pm i\epsilon)^2-\|\mathbf{l}\|^2-m^2}
\]
where we have written $l=(l_0,\mathbf{l})$ and $\delta$ is the
four-dimensional Dirac distribution. The advanced-minus-retarded
fundamental solution is given by (using e.g. \cite{Hoermander} p.73)
\[
\hat{E}(k,l)=-(2\pi)^5i \delta(k+l)\delta(l^2-m^2)
\left(\theta(l_0)-\theta(-l_0)\right)
\]
where $\theta$ is the Heaviside distribution. The two-point distribution of
the Minkowski vacuum $\omega_0$ is given by
\[
\widehat{(\omega_0)_2}(k,l)=
\widehat{(\omega_0)_{2+}}(k,l)+\frac{i}{2}\hat{E}(k,l)=
(2\pi)^5\delta(k+l)\theta(l_0)\delta(l^2-m^2),
\]
where the symmetric part $(\omega_0)_{2+}$ has been defined implicitly.
Notice that
\[
(\omega_0)_2(\bar{f},f)=(2\pi)^{-3}\int\theta(l_0)\delta(l^2-m^2)|\hat{f}(-l)|^2
\ dl\ge 0
\]
and we have equality if $\hat{f}(l)$ is supported in the half space $l_0\ge 0$.
In other words, positive frequency functions annihilate the vacuum, because
then $\|\Phi(f)\Omega_0\|^2=(\omega_0)_2(\overline{f},f)=0$. By analogy
with Parseval's formula,
$\int\Phi(x)f(x)\ dx=(2\pi)^{-4}\int\hat{\Phi}(k)\hat{f}(-k)\ dk$ (see
\cite{Hoermander} theorem 7.1.6), we then say that the quantum field $\Phi(x)$
has positive energy in the vacuum state.
We have $WF(\omega_2)\subset\R^8\times (N^-\times N^+)$, where $N^+$
denotes the future pointing null vectors and $N^-$ the past pointing null vectors
(both including $0$) and $WF$ denotes the wave front set (see appendix
\ref{ch_ma}).

\newpage

\section{\textbf{Acknowledgements}}

I would like to thank the University of York for providing me with the
opportunity to carry out the research on which this thesis is based and
especially my supervisor, Dr. Chris Fewster, who has helped me with good
advice on many occasions and has provided useful comments on drafts of
this thesis at several stages of its development. I would also
like to thank the University of Trento for its hospitality during my
visit there in October 2007 and I am particularly grateful to Dr. Romeo
Brunetti, who corrected a misconception of mine on the relative Cauchy
evolution for the Dirac field at an early stage. Furthermore I would
like to thank Dr. Alexander Strohmaier for a very helpful discussion on
the Reeh-Schlieder property in curved spacetimes. Finally, while trying
to help me understand its meaning, Esther Sanders (MA) has spotted a
typographical error in the Greek quote on page \pageref{ch_ma}, for
which I am grateful.

\section{\textbf{Author's declaration}}

Parts of chapters \ref{ch_LCQFT} and \ref{ch_RS} are taken from a paper
that has been made available online \cite{Sanders} and was submitted to
Communications in Mathematical Physics for publication. The idea to apply
a spacetime deformation argument to the Reeh-Schlieder property is due
to Dr. Chris Fewster.

\chapter{Introduction}

Locally covariant quantum field theory was introduced in \cite{Brunetti+}
as a mathematical framework to formulate axiomatic quantum field
theories in curved spacetime and to give a precise meaning to
Einstein's general covariance principle for such theories. As such it
provides an appropriate setting for the formulation of a semi-classical
approximation to quantum gravitation. Moreover, as a matter of
principle, quantum field theories are tested in the presence of gravity,
so their formulation should not depend too much on the specific
properties of Minkowski spacetime. In particular this means that the use
of global symmetries and Fourier transformation should not be of
crucial importance.

One major advance of recent years has been the realisation that the
spectrum condition of Wightman field theories in Minkowski spacetime
can be replaced by a microlocal spectrum condition in curved spacetimes
\cite{Brunetti+1,Radzikowski}. This has allowed the formulation of
interacting quantum field theories in curved spacetime using
perturbation theory, analogous to the Minkowski spacetime case
\cite{Brunetti+2,Hollands+}. Another important idea has been the use of
spacetime deformation arguments, which use the time-slice axiom to
show that results on Minkowski spacetime can be carried over to
(diffeomorphic) curved spacetimes \cite{Fulling+}. One successful
example of this is the spin-statistics theorem proved by \cite{Verch1}.

Locally covariant quantum field theory can also serve as a reference
structure for the philosophical discussion of quantum field theories in
curved spacetime and possibly also quantum gravity. In this sense it
would be analogous to algebraic quantum field theory, which serves the
same purpose for quantum field theory in Minkowski spacetime
\cite{Haag,Clifton+,Halvorson,Clifton+2}, and indeed algebraic quantum
field theory can be recovered from locally covariant quantum field theory
\cite{Brunetti+}. Crucial aspects for the theory in this context
are its clear structure and the fact that the assumptions that are
used are believed to be weak and general enough to encompass a
sufficiently wide range of useful theories.

In this thesis we will use locally covariant quantum field theory as a
reference structure as well as for the formulation of specific
quantum field theories. In chapter \ref{ch_phil} we will study the use
of embeddings from a philosophical point of view. We believe this
aspect deserves attention, because it is essentially new. (It
differs from the setting of algebraic quantum field theory in Minkowski
spacetime, because no fixed universe is present.)
In chapters \ref{ch_LCQFT}, \ref{ch_sf} and \ref{ch_Df} we will give a
precise formulation
of locally covariant quantum field theory and describe two examples of
such theories, the real free scalar field and the free Dirac field,
including some new results concerning the microlocal spectrum
condition and Hadamard states. Chapter \ref{ch_RS} then studies the
Reeh-Schlieder property for quantum field theories in curved spacetimes.
A state with this property has many non-local correlations, which makes
this property of importance both for the physical and the philosophical
aspects of quantum field theory in curved spacetime. On top of that it
has useful and interesting mathematical implications. It is known that
many physically interesting states in many spacetimes have this property
\cite{Reeh+,Strohmaier+,Strohmaier}, but whether this covers (almost)
all physically interesting states in (almost) all interesting spacetimes
is not at all clear. We will use the locally covariant framework to
make several partial contributions towards answering this question.

\mainmatter
\renewcommand{\thesection}{\arabic{chapter}.\arabic{section}}
\setcounter{page}{14}

\chapter[Philosophical reflections]{Preliminary philosophical reflections}\label{ch_phil}

\begin{quotation}
`But how does it happen,' I said with admiration, `that you were able
to solve the mystery of the library looking at it from the outside,
and were unable to solve it when you were inside?'

`Thus God knows the world, because He conceived it in His mind, as if
from the outside, before it was created, and we do not know its rule,
because we live inside it, having found it already made.'

`So one can know things by looking at them from the outside!'
\end{quotation}
\begin{flushright}
Umberto Eco, The Name of the Rose, Third Day: Vespers
\end{flushright}

Looking at the history of science, especially the last few centuries,
it is hard to imagine what mathematics or physics would have been like
without each other. Nevertheless, the two disciplines are separate.
Indeed,
a purely mathematical argument is logically true, whether it accurately
describes the world around us or not. If a physical theory allows a
precise mathematical formulation, then the physics is
in the formulation of the model, the assumptions that are made to
arrive at it and the interpretation of the variables. Perhaps a
mathematical physicist is the most prudent of physicists, because he
checks the mathematical structure of physical theories for any
shortcomings and tries to correct these. As a result, he can lay bare
any assumptions of the theory that were previously hidden and these may
provide new insight into the physical content of the theory.

However, there is more to physics than just the mathematical structure of
its theories. There is also a philosophical side, which deals with
questions like: what is physics? and how can we hope to learn something
about the world around us in the first place? and what do our physical
theories tell us about what the world is like? The philosophy of physics
and mathematical physics are not independent of one another. Mathematical
physics can provide a clear boundary between the logical (analytic) and
the physical (synthetic) aspects of physical theories, thereby making
the job of philosophers of physics easier. On the other hand,
philosophical ideas can suggest alterations of physical theories, which
then call for a sound mathematical formulation. (As an example one may
think of Mach's principle, which influenced Einstein's thinking while
he was formulating his general theory of relativity.)

In this light, some philosophical reflections are appropriate, even
though this thesis is a work of mathematical physics. In fact, we feel
there is an even more pressing reason for such reflections, because
locally covariant quantum field theory, as described in chapter
\ref{ch_LCQFT}, is a relatively new and very general framework, whose
mathematical structure introduces some interesting new ideas.
In section \ref{sec_modal}, after an outline of some philosophical
background material, we will argue that the novel use of morphisms lends
itself excellently to make locally covariant quantum field theory a model
for modal logic. To our knowledge the current chapter provides the first
description of an interpretation of this important aspect of the theory.
The author, trained as a mathematical physicist and not as a philosopher
of physics, apologises in advance for the relatively low standard of
philosophical discussion.

\section{Philosophical background information}\label{sec_phback}

Following Kant (see e.g. \cite{Rescher}) we may divide the reality of the
world around us into two parts, namely those aspects of reality to which
we, as observers, have direct epistemic access and those parts of reality
for which this is not the case. By the phrase ``direct epistemic access''
we intend to describe all direct observations, experienced by whatever sense
of a sentient being.
Whatever we know, or believe to know, about the real world must be based on
our observations. In these observations one may discern patterns and
regularities, which lend themselves to abstraction and theoretical
description. In particular, it often happens that different senses record
certain patterns which tend to occur together. When we theorise about
these observations, we tend to construct a single theoretical
``object'', which is assumed to cause all the different perceptions.
(According to \cite{Rescher} it is these theoretical objects which
Kant calls ``things in themselves'' or ``Dinge an sich''.)

Of course there is no way of knowing anything for certain about reality beyond
the realm of our own observations, so whether our theoretical objects actually
exist will always be unknown. In fact, one may take the philosophical position
that reality consists of nothing else than ones own observations and
theories (idealism). On the other hand there is the realist position, which
postulates that there do exist things outside the realm of mere observations.
In the realist's words, the observations are \emph{appearances}, and there must
exist something that does the appearing. Note, however, that this does not
mean that the theoretical objects of a specific theory must exist (see
\cite{Rescher}).

Science is in the business of providing mathematical descriptions of
observations. As such it makes no difference for science whether one takes
an idealist or a realist position, although the meaning and importance that
an individual ascribes to science may depend on his philosophical position.
Following Popper \cite{Popper} we note that science adopts a particular
way of theorising about observations. It deals with events that are
reproducible\footnote{Popper remarks: ``It follows that any controversy
over the question whether events which are in principle unrepeatable and
unique do occur cannot be decided by science: it would be a metaphysical
controversy,'' \cite{Popper} section 8, p.24.} and describes them by
theories that are as universal in their range of application and precise
in their description of observations as possible. Whenever a theory is
falsified, i.e. whenever it has become clear that is not
in agreement with observations, the theory is discarded and science will have
to search for a better one. Another characteristic of science is, according
to Popper, the persistence in attempting to falsify theories and lay bare the
need for better ones.

The main difficulty in the characterisation of science seems to lie in
the characterisation of the way that new theories are developed. Popper,
following Hume, rejects the use of inductive logic as a characterisation
of the scientific method, because it is not clear that inductive
conclusions are justified (\cite{Popper} section 1). Although induction can
be used to formulate new theories (just like creativity or divine inspiration
for that matter), Popper holds that it is the falsifiability and testing of
such theories that is characteristic for science. The rejection of induction
is a very cautious position, which seems to fit in well with the prudent
nature of the mathematical physicist, but it does beg the question why
unfalsified theories that have withstood serious testing are useful. Indeed,
the obvious rational reason for their usefulness, namely that they will be
successful in predicting the future, is based on an inductive argument (see
\cite{NewtonSmith}) and is therefore in contradiction with Popper's position.
Another less rational explanation for Popper's choice of theory comes from
the hypothesis
that biological evolution has provided us with the inclination to choose such
theories and to have confidence in them.
This is certainly consistent with the fact that these choices have
served us well in the past. See \cite{Gibson,Smithurst} for a further
discussion along these lines.

As a final point we will comment on the social aspects of science, because
contact and discussion between different scientists is often considered to
be a crucial characteristic too. However, taking the words on direct
epistemic access at the beginning of this section seriously, an observer
should theorise on the basis of his own observations only and contact with
other observers can only be included by treating it as a form of
measurement or observation. This seems to be consistent both with relativity
theory and quantum physics. Whether the opinions of other observers are
accepted is then a question for the individual to decide and the fact that
the opinions of others carry so much weight may perhaps be explained by
another reference to evolution theory. For further comments on the social
aspects of science and the characterisation of science as a social
phenomenon we refer to \cite{NewtonSmith}.

\section[Modal logic and locally covariant QFT]{Modal logic and locally
covariant quantum field theory}\label{sec_modal}

Modal logic is the study of the truth values of statements and the validity
of arguments that involve situations that are not actually the case. It
deals with possibilities, with "it could have been that\ldots'' and
``if only\ldots'' sentences. The analysis of such sentences and arguments
is notoriously more difficult than that of proposition or predicate
logic. Nevertheless, science uses such sentences in abundance when
formulating hypothetical situations, e.g. in classical mechanics: ``if a
cylinder $C$ would roll down a slope with angle $\alpha$\ldots'', a
situation which need not actually be the case in order for us to analyse
it. Indeed, if we want to make predictions it is necessary to think of
situations that are not yet the case, but that may come about in the
future.

Modal logic decides whether an argument is valid using the idea of models
(see \cite{Forbes} for an introduction). In a specific model the validity
of an argument can be evaluated explicitly. If a statement or argument is
valid in every allowed model of a certain theory of modal logic, then the
argument is said to be valid. The most common type of model consists of
``possible worlds'', a complete alternative for how things might have been.
Objects may or may not exist at a certain possible world and propositions
and predicate statements may or may not hold. The idea of possible
worlds is well-known in quantum physics because of the ``many worlds
interpretation'' of quantum physics, in which all possible measurement
outcomes are considered to be real, but existing at different
worlds.\footnote{Van Fraassen's modal interpretation of quantum physics also
uses modal logic to describe possible measurement outcomes, although it
only considers one of these as real, see \cite{Fraassen}.}

It is sometimes said that locally covariant quantum field theory describes
quantum fields on all possible spacetimes\footnote{More precisely, one
works with globally hyperbolic spacetimes, as will be explained in chapter
\ref{ch_LCQFT}.} simultaneously. Here the word
simultaneously clearly doesn't mean ``at the same instant of time'', but
it rather means the unified, systematic way in which the quantum field
is described in all spacetimes. In fact, locally covariant quantum field
theory deals with a category whose objects can be thought of as systems,
indexed by the region of spacetime in which they live, and whose morphisms
are embeddings, each of which can be thought of as a subsystem relation.
(See
\cite{MacLane} for more information on category theory.) Moreover, the
framework assigns to each system a certain state space and it provides
a map that restricts states to subsystems, this map being the dual to
the embeddings of systems.

Taking things at face value it may be tempting to think of each
spacetime as a possible world, in the sense of modal logic, and wonder
whether there is an analogy to the many worlds interpretation of
quantum mechanics.
Actually, this idea fails at the first hurdle: a possible world in
modal logic is supposed to be a complete set of circumstances, but
when a spacetime can be embedded into a bigger one the description of
the circumstances is clearly not complete. However, we can use a less
well-known
model theory for modal logic, which uses incomplete sets of
circumstances called ``possibilities'' (see \cite{Forbes} pp.18-22). At
a possibility, not all logical sentences need to be assigned a truth
value, so they describe incomplete sets of circumstances. Moreover, a
possibility can be refined by extending the set of circumstances, i.e.
by extending the set of logical sentences which are assigned a truth
value.

To see the correspondence with locally covariant quantum field theory we
notice that we are using two types of modal operators. The first refers
to possible systems, the systems that are the subject of the theory. The
second refers to the set of possible states, for a given possible
system. The first type of operator uses incomplete worlds and
possibility semantics, whereas the second uses complete sets of
circumstances because a state should provide a complete description of
the circumstances in which a system finds itself. Putting everything
together we could identify a possibility with a pair consisting of a
system and the state it is in. A refinement necessarily corresponds to
an embedding into a supersystem together with an extension of the state.
Note that we will always identify a system with its image under a
morphism, because this seems to correspond best to the idea of
extension of circumstances and to the operational notion of
subsystem.

In general refinements of a possibility are not unique: a possible
system may have many supersystems and a state of a subsystem can be
extended to a given supersystem in more than one way. Let us now turn
to the interesting question whether a given state of a subsystem can be
extended to a given supersystem at all. First suppose that a state
cannot be extended to any supersystem. If a system is known to be in
such a state, then it must clearly be the whole universe, for otherwise
there would have to be some extension to a supersystem. Now suppose that
a state can only be extended to some supersytems, but not to others.
Such a state tells us not just something about the system under
consideration, but also about the nature of any possible supersystems.
This would be a strange situation, which would seem to indicate that we have
chosen the boundary between the system and the rest of the universe
poorly. The
assumption that every state can be extended to every supersystem is the
principle of \emph{local physical equivalence} introduced in
\cite{Fewster}.\footnote{More precisely, \cite{Fewster} definition 4.1
requires that for every supersystem every state is empirically
equivalent with a state that can be extended. This prevents us from
detecting a state that cannot be extended, but it does allow the theory
to make unphysical idealisations.}

The possibility semantics seems to fit well with our generally prudent
approach and with experimental praxis: we would like to be able to make
predictions for a certain laboratory experiment, without prescribing a
complete set of circumstances for the entire world; making assumptions
about the system in question should be enough. For practical purposes,
then, we may stick with an instrumentalist interpretation of the
framework, using a Heisenberg cut between the system and the observer
which may shift, according to which system is under consideration.
(Arguably this avoids the measurement problem by denying that the theory
deals with the universe as a whole, excepting the special case of
inextendible spacetimes.)

Let us emphasise the difference between possible worlds and possible
systems by drawing some physically relevant conclusions. The difference
between the two semantics is the idea of refinement, i.e. the embeddings
of locally covariant quantum field theory and the corresponding
extension of states. We will see in chapter \ref{ch_LCQFT} that these
embeddings have a rich structure and they are a crucial part of the
theory, because they express the idea of local covariance. Now suppose
that we may embed a subsystem $A$ into two distinct systems $B_1,B_2$,
which cannot both be embedded into a single supersystem, at
least not when we identify the images of $A$ in both $B_i$ as we have
chosen to do. (See \cite{Brunetti+} for an even more elaborate example
in their discussion of states). In other words, both $B_1$ and $B_2$ are
possible extensions of $A$, but it is not possible to have all the
circumstances of both $B_1$ and $B_2$. This implies that not all
embeddings can be actual at the same time. Furthermore, this shows that
not all possible systems can be actual at the same time.

Similar problems occur when trying to obtain a theory of quantum gravity
from locally covariant quantum field theory by allowing superpositions
over different spacetimes, each with its own classical background
gravitational field. If we would simply allow indiscriminate
superpositions, we would disregard the subsystem relation altogether.
Moreover, we would somehow jump from a theory of systems and possibilities
to a quantum gravity theory of universes and possible worlds. This
approach seems to be too naive and in fact it does not correspond to the
many worlds interpretation of quantum mechanics. Indeed, in quantum
mechanics one uses only a single system (the universe), so no
superpositions of different systems appear. The modal aspect refers to
measurement outcomes only, which can be formulated in terms of states.
When trying to quantise gravity we believe it would be better to find a
way that respects the notion of embeddings. This means we ought to allow
only superpositions of different possibilities of the same system. This
begs the question which spacetimes should be considered as the same
system, but with a different background metric. This is obviously not
the place to go into this difficult question.

\chapter[Locally covariant QFT]{Locally covariant quantum field
theory}\label{ch_LCQFT}

\begin{quote}
As before, the Peqoud steeply leaned over towards the Sperm
Whale's head, now, by the counterpoise of both heads, she regained her
even keel; though sorely strained, you may well believe. So, when on one
side you hoist in Locke's head, you go over that way; but now, on the
other side, hoist in Kant's and you come back again; but in very poor
plight. Thus, some minds for ever keep trimming boat. Oh, ye foolish!
throw all these thunderheads overboard, and then you will float light
and right.
\end{quote}
\begin{flushright}
Herman Melville, Moby Dick, Ch. 73
\end{flushright}

After the preliminary discussion in chapter \ref{ch_phil} of the meaning
of the categorical structure underlying locally covariant quantum field
theory, we now come to the actual and detailed definition of this
framework. Most of the following chapters is formulated in this
framework, so this chapter serves to establish notations as well as to
explain all the basic concepts. We will follow the original work
\cite{Brunetti+} closely, but also refer to \cite{Fewster} for a
slightly different formulation. Furthermore our definition of the time
slice axiom is slightly stronger than that of \cite{Brunetti+} and we
introduce the new notion of nowhere-classicality. The basic facts from
category theory that we will use can be found in \cite{MacLane} and we
refer to \cite{Hawking+,Wald,ONeill} for background information on general
relativity. Information on $C^*$-algebras, respectively $^*$-algebras, can
be found in \cite{Kadison+,Schmuedgen}, respectively. For the physical
applications of these algebras we refer to \cite{Bratteli+,Araki,Haag}.

\section{Operational aspects}

A quantum physical system will be described by a topological
$^*$-algebra $\Alg{A}$ with a unit $I$, whose self-adjoint elements are
the observables of the system. For technical reasons it is often
desirable to work with $C^*$-algebras, because they can be faithfully
represented as algebras of bounded operators. However, both $C^*$-algebras
and more general topological $^*$-algebras will appear in the following
chapters, so for clarity we will develop both cases alongside each other.
It will be advantageous to consider a whole class of possible systems rather
than just one.
\begin{definition}\label{def_alg}
The category $\CatTAlg$ has as its objects topological
$^*$-algebras\footnote{We recall from \cite{Schmuedgen} p.22 that a
topological $^*$-algebra is a $^*$-algebra which is also a locally convex
vector space such that the involution $^*$ is continuous and the product is
separately continuous.}
with unit $\Alg{A}$ and its morphisms are continuous, injective
$^*$-homomorphisms $\alpha$ such that $\alpha(I)=I$. The product of
morphisms is given by the composition of maps and the identity map
$\id_{\Alg{A}}$ on a given object serves as an identity morphism.
The category $\CatAlg$ is the subcategory of $\CatTAlg$ whose
objects are unital $C^*$-algebras.
\end{definition}
A morphism $\map{\alpha}{\Alg{A}_1}{\Alg{A}_2}$ in $\CatTAlg$ expresses
the fact that the system described by $\Alg{A}_1$ is a sub-system of
that described by $\Alg{A}_2$, which is called a super-system (see also
the discussion in section \ref{sec_modal}). The injectivity of the
morphisms means that, as a matter of principle, any observable of a
sub-system can always be measured, regardless of any
practical restrictions that a super-system may impose.

A state of a system $\Alg{A}$ is represented by a continuous linear
functional $\omega$ on $\Alg{A}$ which is positive, i.e.
$\omega(A^*A)\ge 0$ for all $A\in\Alg{A}$, and normalised, $\omega(I)=1$. The
set of all states on $\Alg{A}$ will be denoted by $\Alg{A}^{*+}_1$. Not all of
these states are guaranteed to be of physical interest, so it will be
convenient to have the following notion at our disposal:
\begin{definition}\label{def_states}
The category $\CatStat$ has as its objects all convex subsets
$\Stat{S}\subset\Alg{A}^{*+}_1$, for all objects $\Alg{A}$ in
$\CatTAlg$, which are closed under operations from $\Alg{A}$ (i.e.
$\frac{\omega(A^*.A)}{\omega(A^*A)}\in\Stat{S}$ if
$\omega\in\Stat{S}$ and $A\in\Alg{A}$ such that $\omega(A^*A)\not=0$) and
morphisms in $\CatStat$ are all affine maps
$\map{\sigma}{\Stat{S}_1}{\Stat{S}_2}$, i.e. maps for which
$\sigma(\lambda\omega_1+(1-\lambda)\omega_2)=\lambda\sigma(\omega_1)+
(1-\lambda)\sigma(\omega_2)$ for all $0\le \lambda\le 1$ and
$\omega_1,\omega_2\in\Stat{S}_1$. Again the product of morphisms is
given by the composition of maps and the identity map $\id_S$ on a given
object serves as an identity morphism.
\end{definition}
Each object $\Stat{S}$ is a priori a suitable candidate for a state space of a
system in $\CatTAlg$. Using the category $\CatStat$ allows us to
postpone a specific choice of state space until later.

If $\omega$ is a state on a (not necessarily topological) $^*$-algebra with unit
$\Alg{A}$, then we can perform the GNS-construction. To explain this we need the
following definitions (see \cite{Schmuedgen}):
\begin{definition}\label{def_graphtop}
A $^*$-representation of $\Alg{A}$ is called \emph{closed} if and only
if it represents $\Alg{A}$ as an algebra of closable operators on a
Hilbert space $\mathcal{H}$ which have as a common, dense and invariant
domain
$\mathscr{D}_{\pi}=\bigcap_{A\in\Alg{A}}\mathrm{dom}(\overline{\pi(A)})$.

The \emph{graph topology} of $\mathscr{D}_{\pi}$ is the locally convex
topology determined by the family of semi-norms
$\left\{\phi\mapsto\|\pi(A)\phi\|\ |\ A\in\Alg{A}\right\}$.

A closed $^*$-representation of $\Alg{A}$ on a Hilbert space
$\mathcal{H}$ is called \emph{cyclic} if and only if there is a
\emph{cyclic vector} $\phi\in\mathscr{D}_{\pi}$, i.e. a vector such that
$\pi(\Alg{A})\phi\subset\mathscr{D}_{\pi}$ is dense in the graph
topology.
\end{definition}
If $\pi$ is a closed $^*$-representation with $\mathscr{D}_{\pi}=\mathcal{H}$,
then $\pi$ represents $\Alg{A}$ as an algebra of bounded operators (by the closed
graph theorem, \cite{Kadison+} theorem 1.8.6) and hence the graph topology
coincides with the norm topology of $\mathcal{H}$. In general, a cyclic vector
for a cyclic representation $\pi$ is also \emph{weakly cyclic}, i.e.
$\pi(\Alg{A})\phi\subset\mathcal{H}$ is dense in the norm topology of
$\mathcal{H}$.

\begin{theorem}[GNS-representation]\label{GNSrep}
Let $\omega$ be a state on $\Alg{A}$. Then there
exists a closed cyclic $^*$-representation $\pi_{\omega}$ of $\Alg{A}$
on a Hilbert space $\mathcal{H}_{\omega}$ with a cyclic vector
$\Omega_{\omega}$ in the dense domain
$\mathscr{D}_{\omega}:=\mathscr{D}_{\pi_{\omega}}$ such that
$\omega(A)=\langle\Omega_{\omega},\pi_{\omega}(A)\Omega_{\omega}\rangle$
for all $A\in\Alg{A}$.

If $\pi$ is a closed cyclic $^*$-representation of $\Alg{A}$ with a
cyclic vector $\phi$ such that $\omega(A)=\langle\phi,\pi(A)\phi\rangle$
for all $A\in\Alg{A}$, then there is a unique unitary equivalence $U$
between $\pi$ and $\pi_{\omega}$ such that $U(\phi)=\Omega_{\omega}$.
\end{theorem}
This follows from theorem 8.6.4 of \cite{Schmuedgen}. The
representation $\pi_{\omega}$ is called the \emph{GNS-representation}. In
the special case that $\Alg{A}$ is a $C^*$-algebra one can show that
$\mathscr{D}_{\omega}=\mathcal{H}_{\omega}$ and the triple
$(\mathcal{H}_{\omega},\pi_{\omega},\Omega_{\omega})$ is then called the
\emph{GNS-triple} (see e.g. the GNS-construction in \cite{Kadison+} for
the $C^*$-algebraic case). In general we will call
$(\mathcal{H}_{\omega},\pi_{\omega},\Omega_{\omega},
\mathscr{D}_{\omega})$ the \emph{GNS-quadruple}.

If $\Alg{B}\subset\Alg{A}$ is a sub-$^*$-algebra and
$\omega':=\omega|_{\Alg{B}}$ then the GNS-quadruple (or GNS-triple)
associated to $\omega'$ is related to that of $\omega$ by
$\mathcal{H}_{\omega'}=\overline{\pi_{\omega}(\Alg{B})\Omega_{\omega}}$,
$\pi_{\omega'}:=P\pi_{\omega}|_{\Alg{B}}P^*$ where
$\map{P}{\mathcal{H}_{\omega}}{\mathcal{H}_{\omega'}}$ is the
orthogonal projection and $\Omega_{\omega'}:=\Omega_{\omega}$.
This follows from the uniqueness part of theorem \ref{GNSrep} and we will
often use this fact in the subsequent chapters.

\section{Spacetimes}

After these operational aspects we now turn to the physical ones. The
systems we will consider are intended to model quantum fields living in
a (region of a) spacetime which is endowed with a fixed Lorentzian metric
(a background gravitational field). The relation between sub-systems will
come about naturally by considering sub-regions of spacetime. More
precisely we consider the following:

\begin{definition}\label{def_man}
By the term \emph{globally hyperbolic spacetime} we will mean a
connected, Hausdorff, paracompact, $C^{\infty}$ Lorentzian manifold
$M=(\mathcal{M},g)$ of dimension $d=4$, which is oriented, time-oriented
and admits a Cauchy surface (i.e. a continuous hypersurface which is
intersected exactly once by every inextendible time-like curve,
see e.g. \cite{Bernal+0}).

A subset $\mathcal{O}\subset\mathcal{M}$ of a globally hyperbolic
spacetime $M$ is
called \emph{causally convex} iff for all $x,y\in\mathcal{O}$ all causal
curves from $x$ to $y$ lie entirely in $\mathcal{O}$. A non-empty open
set which is
connected and causally convex is called a causally convex region or
\emph{cc-region}. A cc-region whose closure is compact is called a
\emph{bounded cc-region}.

The category $\CatMan$ has as its objects globally hyperbolic
spacetimes $M=(\mathcal{M},g)$ and its morphisms $\Psi$ are given by all
maps $\map{\psi}{\mathcal{M}_1}{\mathcal{M}_2}$ which are smooth
isometric embeddings (i.e.
$\map{\psi}{\mathcal{M}_1}{\psi(\mathcal{M}_1)}$ is a diffeomorphism and
$\psi_*g_1=g_2|_{\psi(\mathcal{M}_1)}$) such that the orientation and
time-orientation are preserved and $\psi(\mathcal{M}_1)$ is causally
convex. Again the product of morphisms is given by the composition of
maps and the identity map $\id_M$ on a given object serves as a unit.
\end{definition}
A region $\mathcal{O}$ in a globally hyperbolic spacetime is causally
convex if and only if $\mathcal{O}$ is a globally hyperbolic region
in the sense of \cite{Hawking+} section 6.6. It then follows that
$\mathcal{O}$ is a globally hyperbolic spacetime in its own right.
However, the converse does not hold, i.e. if $\mathcal{O}$ is a globally
hyperbolic spacetime in its own right it does not follow that it is
causally convex (see e.g. the helical strip on p.177 of \cite{Kay4}).

The image of a morphism is by definition a cc-region. Notice that the
converse also holds. If $\mathcal{O}\subset\mathcal{M}$ is a cc-region
then $O=(\mathcal{O},g|_{\mathcal{O}})$ defines a globally hyperbolic
spacetime in its own right. In this case there is a canonical morphism
$I_{M,O}:O\rightarrow M$ given by the canonical embedding
$\map{\iota}{\mathcal{O}}{\mathcal{M}}$. We will often drop $I_{M,O}$
and $\iota$ from the notation and simply write $O\subset M$.

The importance of causally convex sets is that for any morphism $\Psi$
the causality structure of $M_1$ coincides with that of $\Psi(M_1)$ in
$M_2$:
\begin{equation}\label{subscript}
\psi(J_{M_1}^{\pm}(x))=J_{M_2}^{\pm}(\psi(x))
\cap\psi(\mathcal{M}_1),\quad x\in\mathcal{M}_1.
\end{equation}
If this were not the case then the behaviour of a physical
system living in $M_1$ could depend in an essential way on the
super-system, which makes it practically impossible to study the smaller
system as a sub-system in its own right. This possibility is
therefore excluded from the mathematical framework.

Equation (\ref{subscript}) allows us to drop the subscript in
$J^{\pm}_M$ if we introduce the convention that $J^{\pm}$ is always
taken in the largest spacetime under consideration. This simplifies the
notation without causing any confusion, even when
$O\subset M_1\subset M_2$ with canonical embeddings, because then we
just have $J^{\pm}(\mathcal{O}):=J^{\pm}_{M_2}(\mathcal{O})$ and
$J^{\pm}_{M_1}(\mathcal{O})=J^{\pm}(\mathcal{O})\cap\mathcal{M}_1$.
We adopt a similar convention for the domain of dependence and the
causal complement,
\begin{eqnarray}
D(\mathcal{O})&:=&D_{M_2}(\mathcal{O}),\nonumber\\
\mathcal{O}^{\perp}&:=&\mathcal{O}^{\perp_{M_2}}:=\mathcal{M}_2\setminus
\overline{J(\mathcal{O})},
\nonumber
\end{eqnarray}
and we deduce from causal convexity that
$D_{M_1}(\mathcal{O})=D(\mathcal{O})\cap\mathcal{M}_1$ and
$\mathcal{O}^{\perp_{M_1}}=\mathcal{O}^{\perp}\cap\mathcal{M}_1$.
The following lemma gives some ways of obtaining causally convex
sets in a globally hyperbolic spacetime.
\begin{lemma}\label{subspacetimes}
Let $M=(\mathcal{M},g)$ be a globally hyperbolic spacetime and
$\mathcal{O}\subset\mathcal{M}$ an open subset. Then:
\begin{enumerate}
\item the intersection of two causally convex sets is causally convex,
\item for any subset $\mathcal{Q}\subset\mathcal{M}$ the sets
$I^{\pm}(\mathcal{Q})$ are causally convex,
\item $\mathcal{O}^{\perp}$ is causally convex,
\item $\mathcal{O}$ is causally convex iff
$\mathcal{O}=J^+(\mathcal{O})\cap J^-(\mathcal{O})$,
\item  for any achronal set $\mathcal{P}\subset\mathcal{M}$ the sets
$\mathrm{int}(D(\mathcal{P}))$ and
$\mathrm{int}(D^{\pm}(\mathcal{P}))$ are causally convex,
\item if $O$ is a cc-region, then $D(O)$ is a cc-region,
\item if $\mathcal{R}\subset\mathcal{M}$ is an acausal continuous
hypersurface then $D(\mathcal{R})$, $D(\mathcal{R})\cap I^+(\mathcal{R})$
and $D(\mathcal{R})\cap I^-(\mathcal{R})$ are open and causally convex.
\end{enumerate}
\end{lemma}
\begin{proof*}
The first two items follow directly from the definitions and the fact
that a piecewise smooth, causal curve which is time-like on some
neighbourhood can be deformed to a smooth time-like curve (see e.g.
\cite{Wald} p.191 or \cite{ONeill}). The fourth follows from
$\mathcal{O}\subset J^+(\mathcal{O})\cap J^-(\mathcal{O})=
\cup_{p,q\in \mathcal{O}}(J^+(p)\cap J^-(q))$, which is contained in
$\mathcal{O}$ if and only if $\mathcal{O}$ is causally convex. The
fifth item follows from the first two and theorem 14.38 and lemma 14.6
in \cite{ONeill}.

To prove the third item, assume that $\gamma$ is a causal curve between
points in $\mathcal{O}^{\perp}$ and $p\in\overline{J(\mathcal{O})}$ lies
on $\gamma$. By perturbing one of the endpoints of $\gamma$ in
$\mathcal{O}^{\perp}$ we may ensure
that the curve is time-like (see \cite{Wald,ONeill} loc.~cit.). Then we may
perturb $p$ on $\gamma$ so that $p\in\mathrm{int}(J(\mathcal{O}))$ and
$\gamma$ is
still causal. This gives a contradiction, because there then exists a
causal curve from $O$ through $p$ to either $x$ or $y$.

For the sixth statement we note that $O$ is globally hyperbolic (see
\cite{Hawking+} section 6.6), we let $C\subset\mathcal{O}$ be a smooth Cauchy
surface for $O$ (see \cite{Bernal+1}) and note that $D(O)$ is
non-empty, connected and $D(O)=D(C)$. The causal convexity of $O$
implies that $C\subset\mathcal{M}$ is acausal, which reduces this case to
statement seven. The first part of statement seven is just lemma 14.43 and
theorem 14.38 in \cite{ONeill}. The rest of statement seven follows from
statement one and two together with the openness of $I^{\pm}(C)$.
\end{proof*}

As a matter of notation we define for any subset $S\subset T^*\mathcal{M}$
the set $-S$ by $-S:=\left\{(x,\xi)|\ (x,-\xi)\in S\right\}$ and
\begin{eqnarray}\label{defcausalN}
\mathcal{N}^+&:=&\left\{(x,\xi)\in T^*M|\ g^{\mu\nu}\xi_{\nu}
\mathrm{\ is\ a\ future\ pointing\ light-like\ vector,}\right.\nonumber\\
&&\left.\mathrm{or\ }\xi=0\right\},\nonumber\\
\mathcal{N}^-&:=&-\mathcal{N}^+,\quad\quad
\mathcal{N}:=\mathcal{N}^+\cup\mathcal{N}^-,\nonumber\\
\mathcal{V}^+&:=&\left\{(x,\xi)\in T^*M|\ g^{\mu\nu}\xi_{\nu}
\mathrm{\ is\ a\ future\ pointing\ causal\ vector,}\right.\nonumber\\
&&\left.\mathrm{or\ }\xi=0\right\},\nonumber\\
\mathcal{V}^-&:=&-\mathcal{V}^+,\quad\quad
\mathcal{V}:=\mathcal{V}^+\cup\mathcal{V}^-,\nonumber\\
\mathcal{Z}&:=&\left\{(x,0)\in T^*\mathcal{M}\right\}.\nonumber
\end{eqnarray}
Strictly speaking we should index these sets with the spacetime or manifold
on which they are defined. However, we will avoid this cumbersome notation,
because it will always be clear from the context what spacetime or manifold
is meant. In particular, when $S\subset T^*\mathcal{M}$, the expressions
$S\setminus\mathcal{Z}$ and $S\cup\mathcal{Z}$ are meant to imply that
$\mathcal{Z}$ is the zero section of $T^*\mathcal{M}$.

\section{Spacetimes with a spin structure}

In order to describe the Dirac field we need more geometric structure
than for the scalar field. This section gives the relevant definitions
to formulate a locally covariant quantum field theory in this setting.
More details on the $Spin_{1,3}$-group can be found in section
\ref{subs_spin}.

Given a globally hyperbolic spacetime $M$, the frame bundle $FM$,
which consists of all oriented, time-oriented frames of the tangent
bundle $TM$, is a principal $\mathcal{L}_+^{\uparrow}$-bundle over $M$,
where the proper orthochronous Lorentz group $\mathcal{L}_+^{\uparrow}$
acts from the right. In other words, given $e=(x,e_0,\ldots,e_3)\in FM$,
where $x\in\mathcal{M}$ and $e_a\in T_xM$ such that
$g_x(e_a,e_b)=\eta_{ab}=\mathrm{diag}(1,-1,-1,-1)$ and $e_0$ future
pointing, the action of
$\Lambda$ is defined by $R_{\Lambda}e=e'=(x,e'_0,\ldots,e'_3)$ where
$e'_a=e_b\Lambda^b_{\ a}$. The universal covering group of
$\mathcal{L}_+^{\uparrow}$ is a double covering, namely $Spin^0_{1,3}$,
the identity connected component of the Spin group.

\begin{definition}\label{def_sman}
A \emph{spin structure} on $M$ is a pair $(SM,p)$, where $SM$ is a
principal $Spin^0_{1,3}$-bundle over $M$, the \emph{spin frame bundle},
which carries a right action $R_S$, $S\in Spin^0_{1,3}$, and
$\map{p}{SM}{FM}$ is a base-point preserving bundle homomorphism such
that
\[
p\circ R_S=R_{\Lambda(S)}\circ p,
\]
where $S\mapsto\Lambda(S)$ is the canonical universal covering map
of proposition \ref{covering}.

A globally hyperbolic \emph{spin spacetime} $\hat{M}=(\mathcal{M},g,SM,p)$
is a globally hyperbolic spacetime $M=(\mathcal{M},g)$ which is endowed
with the spin structure $(SM,p)$.

The category $\CatSMan$ has as its objects globally hyperbolic
spin spacetimes $\hat{M}=(\mathcal{M},g,SM,p)$ and its morphisms
$\map{\Psi}{\hat{M}_1}{\hat{M}_2}$ are all pairs of maps $\Psi=(\psi,\chi)$
such that
\begin{enumerate}
\item $\map{\psi}{\mathcal{M}_1}{\mathcal{M}_2}$ is a morphism in
$\CatMan$ between $M_1=(\mathcal{M}_1,g_1)$ and $M_2=(\mathcal{M}_2,g_2)$,
\item $\map{\chi}{SM_1}{SM_2}$ is smooth and satisfies
$\chi\circ(R_1)_S=(R_2)_S\circ\chi$ and
$p_2\circ\chi=\widetilde{d\psi}\circ p_1$, where
$\map{\widetilde{d\psi}}{FM_1}{FM_2}$ denotes the canonical extension of
$d\psi:TM_1\rightarrow TM_2$.
\end{enumerate}
Again the product of morphisms is given by the composition of maps and
the identity map $\id_{\hat{M}}$ on a given object serves as a unit.
\end{definition}
Every globally hyperbolic spacetime admits a spin structure, which need
not be unique \cite{Dimock2}. Different spin structures on the same
spacetime define distinct spin spacetimes and are therefore to be
regarded as distinct systems. We will often drop the hat $\hat{}$ from
our notation, when it is clear from the context that we are dealing with
a spin spacetime rather than a spacetime.

To keep the framework unified it will be useful to have at our disposal a
forgetful functor $\map{\Func{F}}{\CatSMan}{\CatMan}$,
which maps the spin spacetime $(\mathcal{M},g,SM,p)$ to the spacetime
$(\mathcal{M},g)$. This functor is surjective, but not necessarily
injective. A functor $\map{\Func{A}_0}{\CatMan}{\mathfrak{C}}$ to some
category $\mathfrak{C}$ gives rise to a functor
$\map{\Func{A}}{\CatSMan}{\mathfrak{C}}$ defined by
$\Func{A}:=\Func{A}_0\circ\Func{F}$. Whenever $\Func{A}$ is of this form
we can recover $\Func{A}_0$ using the surjectivity of $\Func{F}$.

\section{Locally covariant quantum field theory}

We now come to the main set of definitions, which combine the notions
introduced above (cf. \cite{Brunetti+,Fewster}).
\begin{definition}\label{def_lcqft}
A \emph{locally covariant quantum field theory} is a covariant functor
$\map{\Func{A}}{\CatSMan}{\CatTAlg}$, written as $M\mapsto\Alg{A}_M$,
$\Psi\mapsto\alpha_{\Psi}$.

A \emph{state space} for a locally covariant quantum field theory
$\Func{A}$ is a contravariant functor
$\map{\Func{S}}{\CatSMan}{\CatStat}$, such that for all objects $M$
we have $M\mapsto\Stat{S}_M\subset(\Alg{A}_M)^{*+}_1$ and for
all morphisms $\map{\Psi}{M_1}{M_2}$ we have
$\Psi\mapsto\alpha_{\Psi}^*|_{\Stat{S}_{M_2}}$. The set $\Stat{S}_M$
is called the \emph{state space for} $M$.
\end{definition}

When it is clear that $\Psi=(\iota,\kappa)=I_{M,O}$ is a canonical
embedding $\iota:\mathcal{O}\rightarrow\mathcal{M}$,
$\kappa:SM|_{\mathcal{O}}\rightarrow SM$, of a cc-region $\mathcal{O}$
in a globally hyperbolic spacetime $\mathcal{M}$, i.e. when
$O\subset M$ as spin spacetimes, we will often simply write
$\Alg{A}_O\subset\Alg{A}_M$ instead of using $\alpha_{\Psi}$.
For a morphism $\map{\Psi}{M}{M'}$ which restricts to a morphism
$\map{\Psi|_O}{O}{O'\subset M}$ we then have
\begin{equation}\label{embedding}
\alpha_{\Psi|_O}=\alpha_{\Psi}|_{\mathcal{A}_O}
\end{equation}
rather than $\alpha_{I_{M',O'}}\circ\alpha_{\Psi|_O}=
\alpha_{\Psi}\circ\alpha_{I_{M,O}}$, as one can see from a commutative
diagram.

As a special case we may consider locally covariant quantum field
theories $\map{\Func{A}}{\CatSMan}{\CatAlg}$, which use $C^*$-algebras
only. This is a generalisation of algebraic quantum field theory (see
\cite{Brunetti+,Haag}). We will indicate it explicitly when we restrict
attention to $C^*$-algebras only.

We now proceed to define and discuss several physically
desirable properties that a locally covariant quantum field theory
and its state space may have (cf. \cite{Brunetti+}, but note that our
time-slice axiom is stronger because it places a restriction on the
state spaces as well as the algebras; see also \cite{Fewster}; the last
property is original).
\begin{definition}\label{def_LCQFTprop}
A locally covariant quantum field theory $\Func{A}$ is called
\emph{causal} iff for any two morphisms $\map{\Psi_i}{M_i}{M}$, $i=1,2$,
such that $\psi_1(\mathcal{M}_1)\subset(\psi_2(\mathcal{M}_2))^{\perp}$
in $M$ we have
$\left[\alpha_{\Psi_1}(\Alg{A}_{M_1}),\alpha_{\Psi_2}(\Alg{A}_{M_2})
\right]=\left\{0\right\}$ in $\Alg{A}_M$.

A locally covariant quantum field theory $\Func{A}$ with state space
$\Func{S}$ satisfies the \emph{time-slice axiom} iff for all morphisms
$\map{\Psi}{M_1}{M_2}$ such that $\psi(\mathcal{M}_1)$ contains a Cauchy
surface for $M_2$ we have
$\alpha_{\Psi}(\Alg{A}_{M_1})=\Alg{A}_{M_2}$ and
$\alpha_{\Psi}^*(\Stat{S}_{M_2})=\Stat{S}_{M_1}$.

A locally covariant quantum field theory $\Func{A}$ with state space
$\Func{S}$ respects \emph{local physical equivalence} iff for every
morphism $\map{\Psi}{M_1}{M_2}$ the state spaces $\Stat{S}_{M_1}$
and $\alpha_{\Psi}^*(\Stat{S}_{M_2})$ have the same weak$^*$ closures
in $\Alg{A}_{M_1}^*$.

A locally covariant quantum field theory
$\map{\Func{A}}{\CatSMan}{\CatTAlg}$ is called \emph{additive} iff
$\mathcal{A}_O=\vee_{i\in I}\mathcal{A}_{O_i}$, where the
$\left\{O_i\right\}_{i\in I}$ form a locally finite open covering of $O$
and the right-hand side denotes the smallest algebra generated by the
algebras $\mathcal{A}_{O_i}$. Similarly,
$\map{\Func{A}}{\CatSMan}{\CatAlg}$ is called \emph{additive} iff
$\mathcal{A}_{O}=\overline{\vee_{i\in I}\mathcal{A}_{O_i}}$, where we
take the completion on the right-hand side.

Given a locally covariant quantum field theory
$\map{\Func{A}}{\CatSMan}{\CatAlg}$, a state space $\Func{S}$ for
$\Func{A}$ is called \emph{locally quasi-equivalent} iff for all
$M_2$ every pair of states in $\Stat{S}_{M_2}$ is locally
quasi-equivalent, i.e. iff for every morphism $\map{\Psi}{M_1}{M_2}$
such that $\psi(\mathcal{M}_1)\subset\mathcal{M}_2$ is bounded and for
every pair of states $\omega,\omega'\in\Stat{S}_{M_2}$ the
GNS-representations $\pi_{\omega},\pi_{\omega'}$ of $\Alg{A}_{M_2}$
restricted to $\alpha_{\Psi}(\Alg{A}_{M_1})$ are quasi-equivalent
(see the discussion and definition below).
The local von Neumann algebras $\Alg{R}^{\omega}_{M_1}:=
\pi_{\omega}(\alpha_{\Psi}(\Alg{A}_{M_1}))''$ are then $^*$-isomorphic
for all $\omega\in\Stat{S}_{M_2}$.

A locally covariant quantum field theory
$\map{\Func{A}}{\CatSMan}{\CatAlg}$ with a state space
functor $\Func{S}$ is called \emph{nowhere classical} iff for every
morphism $\map{\Psi}{M_1}{M_2}$ and for every state
$\omega\in\Stat{S}_{M_2}$ the local von Neumann algebra
$\Alg{R}^{\omega}_{M_1}$ is not commutative.
\end{definition}
Note that the condition that
$\psi_1(\mathcal{M}_1)\subset(\psi_2(\mathcal{M}_2))^{\perp}$
in $M$ is
symmetric in $i=1,2$, because $\psi_i(\mathcal{M}_i)$ is open and hence:
\begin{eqnarray}
\psi_1(\mathcal{M}_1)\subset(\psi_2(\mathcal{M}_2))^{\perp}
&\Leftrightarrow&\psi_1(\mathcal{M}_1)\cap
\overline{J(\psi_2(\mathcal{M}_2))}=\emptyset\quad\Leftrightarrow
\nonumber\\
\psi_1(\mathcal{M}_1)\cap J(\psi_2(\mathcal{M}_2))=\emptyset
&\Leftrightarrow&J(\psi_1(\mathcal{M}_1))\cap
\psi_2(\mathcal{M}_2)=\emptyset.\nonumber
\end{eqnarray}
The causality condition formulates how the quantum
physical system interplays with the classical gravitational background
field, whereas the time-slice axiom expresses the existence of a causal
dynamical law. Classical theories can be described by commutative algebras,
which motivates the definition of nowhere-classicality (see also section
\ref{sec_RSintro} for comments on non-local correlations in
nowhere-classical theories). The condition of a locally quasi-equivalent
state space is more technical in nature and means that all states of a
system can be
described in the same Hilbert space representation, as long as we only
consider operations in a small (i.e. bounded) cc-region of the
spacetime. More precisely:
\begin{definition}
The \emph{folium} of a representation $\pi$ of a $C^*$-algebra
$\Alg{A}_M$ on a Hilbert space $\mathcal{H}$ is the set of all states
$\rho$ on $\Alg{A}_M$ of the form
$\omega_{\rho}(A)=Tr_{\mathcal{H}}\rho\pi(A)$
with some trace-class operator $\rho$.

Two representations are called \emph{quasi-equivalent} iff their folia
are equal.
\end{definition}

The condition that $\psi(\mathcal{M}_1)$ contains a Cauchy surface for
$M_2$ is equivalent to $D(\psi(\mathcal{M}_1))=\mathcal{M}_2$,
because a Cauchy surface $S\subset\mathcal{M}_1$ maps to a Cauchy
surface $\psi(S)$ for $D(\psi(\mathcal{M}_1))$. On the algebraic level
this yields:
\begin{lemma}\label{Cauchydevelopment}
For a locally covariant quantum field theory $\Func{A}$ with a state
space $\Func{S}$ satisfying the time-slice axiom, an object
$M=(\mathcal{M},g)\in\CatMan$ and a cc-region $O\subset M$
we have $\Alg{A}_O=\Alg{A}_{D(O)}$ and $\Stat{S}_O=\Stat{S}_{D(O)}$. If
$O$ contains a Cauchy surface of $M$ we have
$\Alg{A}_O=\mathcal{A}_M$ and $\Stat{S}_O=\Stat{S}_M$.
\end{lemma}
\begin{proof*}
Note that both $(O,g|_O)$ and $(D(O),g|_{D(O)})$ are objects of
$\CatMan$ (by lemma \ref{subspacetimes}) and that a Cauchy surface $S$
for $O$ is also a Cauchy surface for $D(O)$. (The causal convexity of
$O$ in $M$ prevents multiple intersections of $S$ by
inextendible causal curves in $D(O)$, cf. the comments below
definition \ref{def_man}.) The first statement then reduces
to the second. Leaving the canonical embedding implicit in the notation,
the result follows immediately from the time-slice axiom.
\end{proof*}

\section{Quantum fields}

The functorial dependence of an algebra $\Alg{A}_M$ on a spacetime $M$
is not specific enough for many purposes. Instead, we would like to have
certain elements in these algebras, (smeared) quantum fields, which
depend in a functorial way on the spacetime. Our formulation of such
quantum fields follows closely the treatment of
\cite{Brunetti+,Fewster,Verch1}. For simplicity we will first describe
the case of the scalar field. Here the sets of test-functions are
simply $\Test_0(M)$ in the test-function topology.\footnote{As a matter of
convention we will always identify a distribution density on a spacetime
$M$ with a distribution, using the metric volume element $d\mathrm{vol}_g$
on $M$ (see \cite{Hoermander} section 6.3). To remind the reader of this
fact we will write $\Test_0(M)$ instead of
$\Test_0(\mathcal{M})$.}\label{ftnotation}
\begin{definition}\label{def_LCfield}
The category $\CatTop$ has as objects all topological spaces and as
morphisms all continuous maps.

The functor $\map{\Func{D}}{\CatMan}{\CatTop}$ maps each object $M$
to the linear space $\Test_0(M)$ in the test-function topology and
each morphism $\Psi=(\psi)$ to the push-forward $\psi_*$, extending
functions by $0$ outside the image of $\psi$.

A \emph{locally covariant scalar quantum field} $\Phi$ is a natural
transformation\footnote{A natural transformation can only exist between
two functors with the same target category, so strictly speaking $\Phi$
should be defined as a natural transformation between $\Func{D}$ and
$\Func{F}\circ\Func{A}$, where $\map{\Func{F}}{\CatTAlg}{\CatTop}$ is
the forgetful functor.}
between the functor $\Func{D}$ and a locally covariant
quantum field theory $\Func{A}$, i.e. for each $M$ in $\CatMan$ we have
a continuous map $\map{\Phi_M}{\Test_0(M)}{\Alg{A}_M}$ such that
$\alpha_{\Psi}\circ\Phi_{M_1}=\Phi_{M_2}\circ\psi_*$ for every morphism
$\map{\Psi}{M_1}{M_2}$ in $\CatMan$.
\end{definition}

For Dirac fields we will need to use test-sections of a certain vector
bundle instead, namely the Dirac double spinor bundle $DM\oplus D^*M$,
which will be introduced in chapter \ref{ch_Df}. All we need to know
for now is that there is a functorial dependence of these vector bundles
on the spin spacetime $M$:
\begin{definition}
The category $\CatVB$ has as its objects the (finite dimensional)
vector bundles $\mathcal{X}$ on every globally hyperbolic spin spacetime
$M$ and as its morphisms the vector bundle homomorphisms
$\map{\lambda}{\mathcal{X}_1}{\mathcal{X}_2}$ such that for some
morphism $\Psi=(\psi,\chi)$ in $\CatSMan$ we have
$\pi_2\circ\lambda=\psi\circ\pi_1$, where $\pi_1,\pi_2$ are the
projections of $\mathcal{X}_1,\mathcal{X}_2$ on $M$. As usual the
products of morphisms are given by composition of maps and the
identity maps serve as units.

Given a functor $\map{\Func{X}}{\CatSMan}{\CatVB}$, written as
$M\mapsto\mathcal{X}_M$ and $\Psi\mapsto\lambda$, the functor
$\map{\Func{D}^{\Func{X}}}{\CatSMan}{\CatTop}$ maps each object $M$
to the linear space $\Test_0(\mathcal{X}_M)$ of compactly supported
smooth sections of $\mathcal{X}_M$ in the test-section
topology and each morphism $\Psi$ to the push-forward $\lambda_*$,
extending sections by $0$ outside the image of $\lambda$.
\end{definition}
In the first part of the definition above we specifically use spin
spacetimes rather than spacetimes, because the vector bundles we
have in mind, the Dirac double spinor bundles, are constructed from
the spin structure. Of course a similar definition can equally well
be made on the category of spacetimes $\CatMan$. The definition of a
locally covariant quantum field is now straightforward:
\begin{definition}
A \emph{locally covariant quantum field} $\Phi$ with test-section
functor $\Func{X}$ is a natural transformation between the functor
$\Func{D}^{\Func{X}}$ and a locally covariant quantum field theory
$\Func{A}$, i.e. for each $M$ in $\CatSMan$ we have a continuous map
$\map{\Phi_M}{\Test_0(\mathcal{X}_M)}{\Alg{A}_M}$ such that
$\alpha_{\Psi}\circ\Phi_{M_1}=\Phi_{M_2}\circ\lambda_*$ for every
morphism $\map{\Psi}{M_1}{M_2}$ in $\CatMan$, where
$\Psi\mapsto\lambda$ under $\Func{X}$.
\end{definition}
Notice that we may think of $\Phi_M$ as a generalised distributional
density, which is a section of $\mathcal{X}_M^*$, the vector bundle dual
to $\mathcal{X}_M$, and which takes values in $\Alg{A}_M$. ($\Phi_M$
need not be a distribution in the usual sense of the word, because we do
not require it to be linear.)

\chapter{The real free scalar field}\label{ch_sf}

\begin{quote}
If thou tellest thy tale in this manner, cried Don Quixote,
repeating every circumstance twice over; it will not be finished these
two days: proceed therefore, connectedly, and rehearse it, like a man
of understanding: otherwise thou hadst better hold thy tongue.
\end{quote}
\begin{flushright}
Miguel de Cervantes, Don Quixote, Vol. 1 Book 3 Ch. 6
\end{flushright}

As a first example of a locally covariant quantum field theory we will
now describe the real free scalar field in two different ways. First we
give the distributional description using the Borchers-Uhlmann algebra
in section \ref{sec_freeBU}, followed by the $C^*$-algebraic description
using the CCR-algebra (or Weyl-algebra) in section \ref{sec_freeC}.
Because the free scalar field is a well-known test ground for quantum
field theory in curved spacetime it is instructive to describe it in
some detail before we treat the more complex case of the free Dirac
field. We also give an elegant proof in proposition \ref{lem_ts2} of the
fact that the commutation relations together with the Hadamard condition
on the two-point distribution of a (not necessarily quasi-free) state
completely fix the singularity structure of all $n$-point distributions.
This result appears to be hitherto unknown in this generality.

\section{Distributional approach to the free scalar field}\label{sec_freeBU}

In this section we will make use of a topological $^*$-algebra that is
not a $C^*$-algebra, namely the Borchers-Uhlmann algebra. This algebra
naturally gives rise to unbounded field operators. After describing a
general real scalar field and the microlocal spectrum condition we will
specialise to the real free scalar field and introduce the important
class of Hadamard states. We refer to appendix \ref{ch_ma} for results
on wave front sets. Our presentation in this section is largely based on
\cite{Dimock1,Brunetti+,Radzikowski,Verch2}.

\subsection{The real scalar field}

On a spacetime $M$ in $\CatMan$ we make the following definition:
\begin{definition}\label{def_BUalg}
The Borchers-Uhlmann algebra is the direct sum
\[
\Alg{U}_M:=\oplus_{n=0}^{\infty}\Test_0(M^{\times n})
\]
(in the algebraic sense, i.e. only a finite number of terms in the sum
are non-zero), equipped with:
\begin{enumerate}
\item the product $f(x_1,\ldots,x_n)g(x_{n+1},\ldots,x_{n+m}):=
(f\otimes g)(x_1,\ldots,x_{n+m})$, extended linearly,
\item the $^*$-operation
$f(x_1,\ldots,x_n)^*:=\overline{f}(x_n,\ldots,x_1)$,
extended anti-linearly,
\item a topology such that $f_j=\oplus_n f_j^{(n)}$ converges to
$f=\oplus_n f^{(n)}$ if and only if for all $n$ we have
$f_j^{(n)}\rightarrow f^{(n)}$ in
$\Test_0(M^{\times n})$ and for some $N>0$ we have $f_j^{(n)}=0$ for all
$j$ and $n\ge N$.
\end{enumerate}
\end{definition}
More precisely, as a topological space $\Alg{U}_M$ is the strict
inductive limit $\Alg{U}_M=\cup_{N=0}^{\infty}\oplus_{n=0}^N
\Test_0(K_N^{\times n})$, where $K_N$ is an exhausting (and increasing)
sequence of compact subsets of $\mathcal{M}$ and each
$\Test_0(K_N^{\times n})$ is given the test-function topology, see
\cite{Schaefer} theorem 2.6.4.\footnote{Therefore, $\Alg{U}_M$ is an
LF-space, which is by definition the strict inductive limit of an
increasing sequence of Fr\'echet spaces.} Following our convention
for $\Test_0(M)$ we will write $\Alg{U}_M$ instead of
$\Alg{U}_{\mathcal{M}}$ (see the footnote on page \pageref{ftnotation}).
It should be noted
that the algebra $\Alg{U}_M$ restricts the field to be Hermitean by
property 2, but it does not contain any dynamical information.

\begin{lemma}\label{lem_UTalg}
The Borchers-Uhlmann algebra is a topological $^*$-algebra with unit
and a continuous linear functional $\omega$ consists of a sequence of
distributions $\omega_n$ on $M^{\times n}$, which are called the
$n$-point distributions.
\end{lemma}
\begin{proof*}
The given topology makes $\Alg{U}_M$ a locally convex topological vector
space, $^*$ is continuous and multiplication is separately continuous,
i.e. $\Alg{U}_M$ is a topological $^*$-algebra (see \cite{Schmuedgen}
p.22). The unit $I$ is $1\in\Test_0(M^{\times 0}):=\C$, i.e.
$I=1\oplus 0\oplus\ldots$. A
continuous linear functional on $\Alg{U}_M$ gives rise to continuous
linear functionals $\omega_n$ on all $\Test_0(M^{\times n})$ and vice
versa and therefore corresponds to a sequence of distributions
$\omega_n$.
\end{proof*}

The Borchers-Uhlmann algebra is not a $C^*$-algebra and it cannot be
represented faithfully as an algebra of bounded operators.
Nevertheless, most of the ideas of locally covariant quantum field
theory that apply to $C^*$-algebras also apply in the case of more general
topological $^*$-algebras.
The following proposition shows that the map $M\mapsto\Alg{U}_M$ can be
made into a covariant functor from $\CatMan$ into $\CatTAlg$.
\begin{proposition}\label{BUmorphism}
If $\map{\Psi}{M_1}{M_2}$ is a morphism in $\CatMan$ then there is a
unique injective $^*$-algebra homomorphism
$\map{\upsilon_{\Psi}}{\Alg{U}_{M_1}}{\Alg{U}_{M_2}}$ determined by
$\upsilon_{\Psi}(f):=\psi_*f=f\circ\psi^{-1}$ on $\Test_0(M_1)$, where
we extend $\psi_*f$ by $0$ outside $\psi(M_1)$.
\end{proposition}
\begin{proof*}
Using finite sums of finite tensor products of elements in
$\Test_0(M_1)$ the given relation determines $\upsilon_{\Psi}$ uniquely
on $\oplus_{n=0}^{\infty}(\Test_0(M_1))^{\otimes n}\subset\Alg{U}_{M_1}$,
where we take the algebraic direct sum and tensor product.
The map so defined is an injective $^*$-algebra homomorphism of a dense
subalgebra of $\Alg{U}_{M_1}$ into $\Alg{U}_{M_2}$ and extends by
continuity in a unique way to a $^*$-algebra homomorphism
$\upsilon_{\Psi}$ of $\Alg{U}_{M_1}$ into $\Alg{U}_{M_2}$. To prove that
$\upsilon_{\Psi}$ is injective we note that
$\upsilon_{\Psi}(f^{(n)}(x_1,\ldots,x_n))=\psi_*f^{(n)}(x_1,\ldots,x_n)
=f^{(n)}(\psi^{-1}(x_1),\ldots,\psi^{-1}(x_n))$.
\end{proof*}

\begin{definition}
The \emph{Borchers-Uhlmann functor} $\map{\Func{U}}{\CatMan}{\CatTAlg}$
assigns to each globally hyperbolic spacetime $M$ the Borchers-Uhlmann
algebra $\Alg{U}_M$ and to each morphism $\map{\Psi}{M_1}{M_2}$ the
morphism $\upsilon_{\Psi}$ of proposition \ref{BUmorphism}.
\end{definition}

\begin{proposition}\label{BUadditive}
The Borchers-Uhlmann functor $\Func{U}$ defines an additive locally
covariant quantum field theory.
\end{proposition}
\begin{proof*}
If $O=\cup_iO_i$ and $\chi_i$ is a partition of unity on $O$ such that
$\mathrm{supp}\ \chi_i\subset O_i$, then every $f\in\Test_0(O)$ can be
written as $f=\sum_if_i$ with $f_i:=f\chi_i$. The inclusion
$\Alg{U}_O\subset \vee_i\Alg{U}_{O_i}$ now follows by decomposing every
test-function in an element $A\in\Alg{U}_O$ in this way and
the converse inclusion is trivial.
\end{proof*}

A locally covariant quantum field, in the sense of definition
\ref{def_LCfield}, is given in the current setting by\footnote{
In analogy to theorem \ref{WFsum} in appendix \ref{ch_ma} we can
define the wave front set of the distribution $\Phi_M$ as
$WF(\Phi_M):=\overline{\cup_lWF(l\circ\Phi_M)}\setminus\mathcal{Z}$,
where the union is taken over all continuous linear functionals $l$ on
$\Alg{U}_M$. This makes perfect sense, provided we can generalise
lemma \ref{Banachlemma} to the case of $\Alg{U}_M$-valued
distributions. If $l=(l_n)_{n\in\N}$ is any continuous linear
functional on $\Alg{U}_M$, then $(l\circ\Phi_M)(f)=l_1(f)$ and
hence $WF(\Phi_M)=T^*M\setminus\mathcal{Z}$ by theorem 8.1.4 in
\cite{Hoermander}.}
\[
\map{\Phi_M}{\Test_0(M)}{\Alg{U}_M}:f\mapsto
0\oplus f\oplus 0\oplus\ldots\ .
\]
This takes care of the operators of the theory and the fields. Now let
us turn our attention to the states. The following class of states is
often of special interest, because they arise from the canonical
quantisation of a linear field equation.
\begin{definition}\label{def_qf}
A state $\omega$ on $\Alg{U}_M$ is called \emph{quasi-free} iff
$\omega_n=0$ for $n$ odd and for $m\ge 1$:
\[
\omega_{2m}(f_1,\dots,f_{2m})=\sum_{\pi\in\Pi_m}
\omega_2(f_{\pi(1)},f_{\pi(2)})\cdots\omega_2(f_{\pi(2m-1)},f_{\pi(2m)}),
\]
where $\Pi_m$ is the set of permutations of $\left\{1,\ldots,2m\right\}$
such that
\begin{enumerate}
\item $\pi(1)<\pi(3)<\ldots<\pi(2m-1)$,
\item $\pi(2i-1)<\pi(2i)$, $i=1,\ldots,m$.
\end{enumerate}
\end{definition}
A quasi-free state is completely determined by its two-point distribution
(note that $\omega_0=1$) and definition \ref{def_qf} tells us that the
higher $n$-point distributions can be obtained using the combinatorics that
is familiar from flat spacetime quantum field theory. Indeed, for a
$2n$-point distribution we sum over all pairings of the indices, where we
preserve the left-right ordering within each pair (we put the smaller
index of each pair on the left by the second condition on $\pi$) and we
only count every pairing once by the first condition on $\pi$.


If $\Alg{U}=\Alg{U}_M$ we may define smeared field operators
by\footnote{For these representation specific entities we drop the
subscript $M$ to ease the notation. This causes no confusion, because it
is clear that $\omega$ itself is defined on a specific spacetime.}
\begin{equation}\label{KGPhi}
\Phi^{(\omega)}(f):=\pi_{\omega}(\Phi_M(f)).
\end{equation}
These are unbounded operators on $\mathcal{H}_{\omega}$ with a common
dense and invariant domain $\mathscr{D}_{\omega}$ (see theorem \ref{GNSrep}).
We also define
$\mathcal{H}_{\omega}$-valued $n$-point distributions by
\begin{equation}\label{def_phin}
\phi^{(\omega)}_n(f_n,\ldots,f_1):=
\pi_{\omega}(f_n\otimes\ldots\otimes f_1)\Omega_{\omega}.
\end{equation}
For all $n,m$ and all $f_i,g_j\in\Test_0(M)$ we have the identity
\begin{equation}\label{phidotproduct}
\langle\phi^{(\omega)}_n(f_n,\ldots,f_1),\phi^{(\omega)}_m
(g_m\ldots g_1)\rangle=\omega_{n+m}(\overline{f}_1,\ldots,
\overline{f}_n,g_m\ldots,g_1).
\end{equation}

As our state space for $M$ we can select the class of states that satisfy
the microlocal spectrum condition of \cite{Brunetti+1}. To formulate this
condition we need to introduce some new terminology. Let $\mathcal{G}_n$
be the set of directed graphs\footnote{A directed graph is a graph in
which each edge $e$ is given a direction, so that it goes from a
source vertex to a target vertex.}
with $n$ vertices in which every edge that
appears also appears in the opposite direction. An immersion of such a
graph into $M$ assigns to every vertex $\nu_i$ a point $x_i$ and to
every edge $e_r$ from $\nu_i$ to $\nu_j$ a piecewise smooth curve
$\gamma_r$ from $x_i$ to $x_j$ and a causal covector field $k_r$ on
$\gamma_r$ which is covariantly constant ($\nabla k_r=0$) along the
curve in such a way that
\begin{enumerate}
\item if $e_{-r}$ is the edge $e_r$ in the opposite direction, then
$\gamma_{-r}$ is the curve $\gamma_r$ in the opposite direction and
$k_{-r}=-k_r$,
\item if $e_r$ is a curve from $x_i$ to $x_j$ with $i<j$ then
$k_r$ is future directed.
\end{enumerate}
Intuitively one may think of the vectors $k_r$ as ``singularities'',
``propagating'' along the curves $\gamma_r$ between points $x_i$ and
$x_j$. We now define a set of allowed singularities as follows:
\begin{eqnarray}\label{def_cone}
\Gamma_n&:=&\biggl\{(x_n,\xi_n;\ldots;x_1,\xi_1)\in T^*M^n\setminus\mathcal{Z}
|\ \exists G\in\mathcal{G}_n\mathrm{\ and\ an\ immersion\ of\ }G
\biggr.\nonumber\\
&&\left.\mathrm{\ such\ that\ }\nu_i\mapsto x_i,\mathrm{\ and\ } \xi_i=
\sum_{e_r,s(e_r)=x_i}k_r(x_i)\right\},
\end{eqnarray}
where $s(e_r)$ denotes the source of the edge $e_r$.\footnote{Note that
we have ordered the indices of $(x_n,k_n;\ldots;x_1,k_1)$ in the
opposite way to \cite{Brunetti+1}, because we want the singularities to
originate on the right-hand side in the $n$-point distributions and to
travel to the left as time progresses, cf. definition \ref{def_mSC}
below.}
\begin{definition}\label{def_mSC}
A state $\omega$ on $\Alg{U}_M$ is said to satisfy the
\emph{microlocal spectrum condition} ($\mu$SC) if and only if for
all $n\in\N$:
\[
WF(\omega_n)\subset\Gamma_n.
\]
\end{definition}
The microlocal spectrum condition restricts the set of singularities
of the $n$-point distributions to
the sets $\Gamma_n$ and the usefulness of this condition follows from
the special properties of the $\Gamma_n$:
\begin{proposition}\label{propGamma}
The sets $\Gamma_n\subset T^*M^{\times n}$ have the following
properties:
\begin{enumerate}
\item each $\Gamma_n\subset T^*M^{\times n}\setminus\mathcal{Z}$ is a
convex cone,
\item $\Gamma_n\cap-\Gamma_n=\emptyset$,
\item $\pi((\Gamma_{n_1}\cup\mathcal{Z})\times\ldots\times
(\Gamma_{n_m}\cup\mathcal{Z}))\subset\Gamma_{n_1+\ldots+n_m}\cup\mathcal{Z}$,
where $\pi$ is a permutation
acting on the indices such that
$\pi(1)<\pi(2)<\ldots<\pi(n_1)$; $\pi(n_1+1)<\ldots<\pi(n_1+n_2)$;
\ldots;$\pi(n_1+\ldots+n_{m-1}+1)<\ldots<\pi(n_1+\ldots+n_m)$.
\end{enumerate}
\end{proposition}
\begin{proof*}
We refer to \cite{Brunetti+1} lemma 4.2 for a proof of the first
property. The second property follows from the first and the third property
follows immediately from the definitions, using the unions of disjoint
graphs (cf. \cite{Brunetti+1} proposition 4.3).
\end{proof*}
It follows from the last two items that a quasi-free state satisfies
$\mu$SC if and only if $WF(\omega_2)\subset\Gamma_2$. It also seems that
the first two items are sufficient to guarantee that products of
$n$-point distributions and Wick powers can be defined
\cite{Brunetti+1, Hoermander}, even without the commutator property that
we will introduce in the next section.\footnote{Note that the difference
of two two-point distributions with the $\mu$SC does not have to be
smooth unless we also impose the commutator property \cite{Radzikowski}.
\cite{Brunetti+1}
assumes the commutator property, but it does not appear to be necessary
for their proofs.} This forms the starting point of the perturbative
treatment of interacting quantum field theories on curved spacetimes
\cite{Brunetti+2}.

\begin{proposition}\label{prop_closedsf}
One can define a state space functor $\map{\Func{Q}}{\CatMan}{\CatStat}$
for the locally covariant quantum field theory $\Func{U}$ that assigns
to each globally hyperbolic spacetime $M$ the set $\Stat{Q}_M$
of states on $\Alg{U}_M$
that satisfy the $\mu$SC.
\end{proposition}
\begin{proof*}
We first note that the set of states is convex by theorem
\ref{WFresults2}. To show that it is closed under operations from
$\Alg{U}_M$ we note that for fixed $f\in\Test_0(M^{\times m})$ and
$h\in\Test_0(M^{\times r})$ we have
\begin{eqnarray}
WF(\omega_{m+n+r}(f,x_1,\ldots,x_n,h))
&\subset&\nonumber\\
\left\{(y_1,0;\ldots;y_m,0;x_1,k_1;\ldots;x_n,k_n;z_1,0;\ldots
z_r,0)\in \Gamma_{m+n+r}\right\}&\subset&\Gamma_n,\nonumber
\end{eqnarray}
using \cite{Hoermander} theorem 8.2.12. The same holds for linear
combinations of such terms, so if $\omega(A^*A)\not=0$ then the state
$B\mapsto\frac{\omega(A^*BA)}{\omega(A^*A)}$ satisfies the $\mu$SC if
$\omega$ does.

The action of $\Func{Q}$ on morphisms is defined implicitly
by the statement that $\Func{Q}$ is a state space for $\Func{U}$. That
this action is well-defined follows from the fact that wave front sets
transform as a subset of the cotangent bundle (see appendix \ref{ch_ma})
and the cones $\Gamma_n$ are subsets of the cotangent bundle that are
constructed from the metric and hence covariant under isometric
diffeomorphisms.
\end{proof*}

The locally covariant quantum field theory $\Func{U}$ with state space
$\Func{Q}$ is not causal and does not satisfy the time-slice axiom.
These shortcomings are due to the fact that we have not put any
constraints on the dynamics or causality. This will be our next task.

\subsection{The real free scalar field}

In order to arrive at the usual description of the real free scalar field
we will put in some physically motivated restrictions. These restrictions
can be put either on the state or on the algebra and we will describe
both approaches in that order.

Classically, two operations performed in space-like separated regions
cannot influence each other. It seems reasonable to postulate that this
must remain true for the expectation values of quantum physical
operators. We therefore say that a state $\omega$ is \emph{causal} iff
(cf. definition \ref{def_LCQFTprop})
\[
\omega_n(f_1,\ldots,f_{(i},f_{i+1)},\ldots,f_n)=0,
\]
whenever $\mathrm{supp}\ f_i\subset(\mathrm{supp}\ f_{i+1})^{\perp}$.
Here $(,)$ denotes anti-symmetrisation.

A state $\omega$ on $\Alg{U}_M$ is a state of the free field iff the
dynamics is described by the Klein-Gordon equation. The classical form of
the Klein-Gordon equation is
\begin{equation}\label{KGeq}
K\phi:=(\Box+m^2+\xi R)\phi=0,
\end{equation}
where $K$ is the Klein-Gordon operator, $\Box=\nabla^a\nabla_a$ is the
d'Alembertian, $m\ge 0$ is the mass of the field $\phi\in\Test(M)$, $R$
is the Ricci scalar of $M$ and $\xi$ is a coupling parameter. Here
$\xi$ and $m$ are assumed to be independent of $M$. A state $\omega$ on
$\Alg{U}_M$ is
a \emph{state for the free field} iff for all $n\ge 1$ and all
$1\le i\le n$:
\[
\omega_n(f_1,\ldots,Kf_i ,\ldots,f_n)=0.
\]
As the Klein-Gordon operator $K$ is formally self-adjoint (or more
precisely: the dual of $K$ is an extension of $K$) these equations
can also be written as $K^{(i)}\omega_n=0$, where the upper index
indicates that $K$ acts on the $i$'th variable of the distribution
$\omega_n$.

Because we assume that the spacetime $M$ is globally hyperbolic there
are unique advanced $(-)$ and retarded $(+)$ fundamental solutions
$\map{E^{\pm}}{\Test_0(M)}{\Test(M)}$ such that $KE^{\pm}f=f$,
$E^{\pm}Kf=f$ and
$\mathrm{supp}(E^{\pm}f)\subset J^{\pm}(\mathrm{supp}\ f)$ for all
$f\in\Test_0(M)$ (see \cite{Baer+} theorem 3.3.1). Setting
$E:=E^--E^+$ we see that $Ef$ is a solution of
the Klein-Gordon equation whose intersection with each Cauchy surface of
$M$ is compact. Conversely, every solution which has compact
intersection with all Cauchy surfaces can be obtained in this way by
\cite{Dimock1} lemma A.3.

A stronger requirement than causality is the commutator property. A
state $\omega$ is said to have the \emph{commutator property} if and
only if
\begin{equation}\label{KGcomm}
\Phi^{(\omega)}(f)\Phi^{(\omega)}(h)-
\Phi^{(\omega)}(h)\Phi^{(\omega)}(f)=iE(f,h),
\end{equation}
where we view $E$ as the bidistribution
$E(f,h):=\int_M fEh\ d\mathrm{vol}_g$. This condition arises naturally
from the canonical quantisation of the classical Klein-Gordon field.

For quasi-free states the causality condition, Klein-Gordon equation and
commutator property reduce to the corresponding conditions on the
two-point distribution:
\begin{eqnarray}
\omega_{2-}(f,h)&=&0,\quad \mathrm{supp}\
f\subset(\mathrm{supp}\ h)^{\perp},\nonumber\\
K^{(1)}\omega_2&=&K^{(2)}\omega_2=0,\nonumber\\
\omega_{2-}(f,h)&:=&\omega_2(f,h)-\omega_2(h,f)=iE(f,h).\nonumber
\end{eqnarray}

Instead of putting the causality, reality, dynamics and commutator property
in the state we can incorporate this information directly in the algebra
as follows. Let $J\subset\Alg{U}_M$ be the closed $^*$-ideal generated
by all elements of the form $Kf$ or $f\otimes h-h\otimes f-iE(f,h)I$. The
quotient space
$\Alg{U}^0_M:=\Alg{U}_M/J$ is another locally convex topological vector
space (\cite{Schaefer} p.54) and the $^*$-operation, respectively
multiplication, on $\Alg{U}_M$ descends to a continuous, respectively
separately continuous, map on $\Alg{U}^0_M$. In other words, $\Alg{U}^0_M$
is another topological $^*$-algebra. The easiest way to show that the
algebra $\Alg{U}^0_M$ is not trivial is to show that it has a non-trivial
(faithful) representation.

\begin{proposition}\label{morphismBU0}
If $\map{\Psi}{M_1}{M_2}$ is a morphism in $\CatMan$ and
$\map{p_i}{\Alg{U}_{M_i}}{\Alg{U}^0_{M_i}}$, $i=1,2$, is the quotient
map, then $\upsilon_{\Psi}$ descends to an injective
$^*$-algebra homomorphism $\upsilon^0_{\Psi}$ on $\Alg{U}^0_{M_1}$.
\end{proposition}
\begin{proof*}
Let $J_i\subset\Alg{U}_{M_i}$ be the closed $^*$-ideal generated by
elements of the form $K_if$ or $f\otimes h-h\otimes f-iE_i(f,h)I$, where
$K_i$ respectively $E_i$ are the Klein-Gordon operator and its
advanced-minus-retarded fundamental solution on $M_i$. For
$f,h\in\Test_0(M_1)$ set $f':=\upsilon_{\Psi}(f)$ and
$h':=\upsilon_{\Psi}(h)$. Because of the covariance of the Klein-Gordon
operator, $K_2\circ\psi_*=\psi_*\circ K_1$, we see that
$\upsilon_{\Psi}(K_1f)=K_2f'$. Similarly we can use the uniqueness of
the advanced and retarded fundamental solutions and equation
\ref{subscript} to conclude that
$\upsilon_{\Psi}(E^{\pm}_1f)=E^{\pm}_2(f')|_{\psi(\mathcal{M}_1)}$ and
therefore $\upsilon_{\Psi}(E_1(f,h))I=E_2(f',h')I$, because on the
right-hand side we integrate over a compact region in
$\psi(\mathcal{M}_1)$. This then yields
$\upsilon_{\Psi}(f\otimes h-h\otimes f-iE_1(f,h)I)=
f'\otimes h'-h'\otimes f'-iE_2(f',h')I$. By continuity we conclude that
$\upsilon_{\Psi}(J_1)=J_2\cap\upsilon_{\Psi}(M_1)$, which means that
$\upsilon_{\Psi}$ descends to a well-defined $^*$-algebra homomorphism
$\upsilon^0_{\Psi}$ on $\Alg{U}^0_{M_1}$ which is injective.
\end{proof*}

\begin{definition}
The \emph{free field} Borchers-Uhlmann functor
$\map{\Func{U}^0}{\CatMan}{\CatTAlg}$ assigns to each globally
hyperbolic spacetime $M$ the
algebra $\Alg{U}^0_M$ and to each morphism $\map{\Psi}{M_1}{M_2}$ the
morphism $\upsilon^0_{\Psi}$ of proposition \ref{morphismBU0}.
\end{definition}

If $\map{p}{\Alg{U}_M}{\Alg{U}^0_M}$ denotes the quotient map,
then $I_0:=p(I)$ is the unit for $\Alg{U}^0_M$ and a state $\omega'$
on $\Alg{U}^0_M$ gives rise to a state $\omega:=\omega'\circ p$ on
$\Alg{U}_M$ because $p$ is continuous. By construction, $\omega$ is a
causal state for the free field with the commutator property and the
$n$-point distributions of $\omega'$ and $\omega$ are related by
$\omega_n=\omega'_n\circ p$.\footnote{Strictly speaking $\omega'_n$
is not a distribution, because it is not defined on the space of
test-functions, but rather on a quotient of that space.} The
GNS-quadruples of $\omega$ and $\omega'$ satisfy
$(\mathcal{H}_{\omega},\pi_{\omega},\Omega_{\omega},
\mathscr{D}_{\omega})=(\mathcal{H}_{\omega'},\pi_{\omega'}\circ p,
\Omega_{\omega'},\mathscr{D}_{\omega'})$ by the
uniqueness part of theorem \ref{GNSrep}.

A locally covariant quantum field $\Phi^0_M$ can be defined as
$\Phi^0_M:=p_M\circ\Phi_M$. It follows that $K\Phi^0_M=0$ in the weak
sense, i.e. $(K\Phi^0_M)(f)=\Phi^0_M(Kf)=p_M(\Phi_M(Kf))=0$. Moreover,
\[
\Phi^0_M(f)\Phi^0_M(h)-\Phi^0_M(h)\Phi^0_M(f)=p_M(f\otimes h-h\otimes f)
=iE(f,h)I_0
\]
in $\Alg{U}^0_M$, so the field $\Phi^0_M$ is an $\Alg{U}^0_M$-valued
distribution that satisfies the Klein-Gordon equation and has the
commutator property.\footnote{Again we can define the wave front
set of $\Phi^0_M$ in analogy to theorem \ref{WFsum} as
$WF(\Phi^0_M):=\overline{\cup_lWF(l\circ\Phi^0_M)}\setminus\mathcal{Z}$,
where the union is taken over all continuous linear functionals $l$ on
$\Alg{U}^0_M$. If $l=(l_n)_{n\in\N}$ is any continuous linear
functional on $\Alg{U}^0_M$, then $(l\circ\Phi^0_M)(f)=l_1(f)$ and
$(Kl\circ\Phi^0_M)(f)=l\circ\Phi^0_M(Kf)=l(\Phi^0_M(Kf))=0$, so
$WF(\Phi^0_M)\subset\mathcal{N}\setminus\mathcal{Z}$ by theorem
\ref{WFresults2} (recall the definition of $\mathcal{N}$ on page
\pageref{defcausalN}).}

A class of states that is of special importance for the free scalar
field in curved spacetime is the class of Hadamard states. The
original definition of the Hadamard condition in curved spacetimes of
Kay and Wald \cite{Kay+} is equivalent to the following
definition, due to a theorem of Radzikowski \cite{Radzikowski}.
(Recall the definition of $\mathcal{N}^{\pm}$ on page
\pageref{defcausalN}.)
\begin{definition}\label{def_hadamard}
A (not necessarily quasi-free) state $\omega$ on $\Alg{U}^0_M$ is
called a \emph{Hadamard state} iff
\[
WF(\omega_2)=\left\{(x,\xi;y,\xi')\in\mathcal{N}^-\times\mathcal{N}^+
|\ (x,-\xi)\sim (y,\xi')\right\}\setminus\mathcal{Z},
\]
where $(x,-\xi)\sim (y,\xi')$ if and only if $(x,-\xi)=(y,\xi')$ or
there is an affinely parameterised light-like geodesic between $x$ and
$y$ to which $-\xi,\xi'$ are cotangent (and hence $-\xi$ and $\xi'$ are
parallel transports of each other along the geodesic).
\end{definition}

Note that the principal symbol of $K$ is the metric $g_{\mu\nu}$, so
by theorem \ref{WFresults2} the wave front set can only contain
null-covectors. Moreover, the propagation of singularities theorem of
Duistermaat and H\"ormander (\cite{Duistermaat+} theorem 6.1.1, also
quoted in \cite{Radzikowski}) implies that these singularities
propagate under the Hamiltonian flow on $T^*M$ determined by the
principal symbol. It turns out that this means that null covectors
propagate along the null geodesics to which they are cotangent, which
gives rise to the equivalence relation $\sim$.

The two-point distributions of two Hadamard states on $\Alg{U}^0_M$ differ
by a smooth function, so the expectation value of the stress-energy-momentum
tensor of the free scalar field can be renormalised (see \cite{Radzikowski},
\cite{Wald2} section 4.6). A free field state satisfying the $\mu$SC is
Hadamard \cite{Radzikowski}. Conversely, it is known (and easy to see) that
a quasi-free Hadamard state satisfies the $\mu$SC \cite{Brunetti+1}. In
fact, we will now prove the new result that this is even true for
general (not necessarily quasi-free) Hadamard states:
\begin{proposition}\label{lem_ts2}
Let $\omega$ be a state on $\Alg{U}^0_M$ which is Hadamard on a
neighbourhood $\mathcal{W}\subset M$ of a Cauchy surface in $M$. Then
$\omega$ satisfies the $\mu$SC on $M$.
\end{proposition}
\begin{proof*}
Suppose that $(x_1,k_1;\ldots;x_n,k_n)\in WF(\omega_n)$ for $n\ge 1$.
(Note that $\omega_0=1$ is always smooth.) For each index $i$ we
have $(x_i,k_i)\in\mathcal{N}$, because of the equation of motion
(see theorem \ref{WFresults2}). Moreover, if $k_n\not=0$ then
we can apply theorem \ref{WFHvalued} first to
$\omega_n=\langle\phi^{(\omega)}_{n-1},\phi_1^{(\omega)}\rangle$
(see equation (\ref{phidotproduct})) to find
$(x_n,k_n)\in WF(\phi^{(\omega)}_1)$ and then again to
$\omega_2=\langle\phi^{(\omega)}_1,\phi_1^{(\omega)}\rangle$ to
obtain $(x_n,-k_n;x_n,k_n)\in WF(\omega_2)$.
We may then apply the propagation of singularities theorem
(\cite{Duistermaat+} theorem 6.1.1,\cite{Radzikowski}) to find
$(y,l)\in\mathcal{N}\cap T^*\mathcal{W}$ such that $(y,l)\sim (x_n,k_n)$
(see definition \ref{def_hadamard}) and $(y,-l;y,l)\in WF(\omega_2)$.
If $\omega$ is Hadamard on $\mathcal{W}$ we conclude that
$(y,l)\in\mathcal{N}^+$ and hence $(x_n,k_n)\in\mathcal{N}^+$.
Similarly, if $k_1\not=0$ then $(x_1,k_1)\in\mathcal{N}^-$. In
particular, for $n=1$ we find that
$(x_1,k_1)\in\mathcal{N}^+\cap\mathcal{N}^-=\mathcal{Z}$, so
$WF(\omega_1)=\emptyset$ and $\omega_1$ is smooth.

We now argue by contradiction. Let $n\ge 2$ be the smallest number for
which we can find a
point $(x_1,k_1;\ldots;x_n,k_n)$ in $WF(\omega_n)\setminus\Gamma_n$.
There must then be an index $i$ such that $k_i\not=0$. Assume first
that $(x_i,k_i)\in\mathcal{N}^-$. Now we interchange the points $x_i$
and $x_{i+1}$ in $\omega_n$ to find:
\begin{eqnarray}\label{npointcomm}
\omega_n(x_1,\ldots,x_n)&=&\omega_n(x_1,\ldots,x_{i+1},x_i,\ldots,x_n)\\
&&+i\omega_{n-2}(x_1,\ldots,\hat{x}_i,\hat{x}_{i+1}\ldots,x_n)
E(x_i,x_{i+1}),\nonumber
\end{eqnarray}
where the hats denote that these points are omitted. Using theorem
\ref{WFresults2} we see that $(x_1,k_1;\ldots;x_n,k_n)$ must be in the
wave front set of one of the terms on the right-hand side of equation
(\ref{npointcomm}). Suppose that it is in the wave front set of the
second term. This wave front set can be estimated by
(\cite{Hoermander} theorem 8.2.9)
\[
WF(\omega_{n-2}\otimes E)\subset (WF(\omega_{n-2})\cup\mathcal{Z})\times
(WF(E)\cup\mathcal{Z}).
\]
If $(x_1,k_1;\ldots;x_n,k_n)\in WF(\omega_{n-2}\otimes E)$, then the
assumption on $k_i$ implies
$(x_i,k_i;x_{i+1},k_{i+1})\in WF(E)\cap(\mathcal{N}^-\times T^*M)
\subset\Gamma_2$ by proposition \ref{WFE}. By the minimality of
$n$ and proposition \ref{propGamma} we find
$(x_1,k_1;\ldots;x_n,k_n)\in\Gamma_n$,
which is a contradiction. Hence, $(x_1,k_1;\ldots;x_n,k_n)$ must be in
the wave front set of the first term of equation (\ref{npointcomm})
and
\[
(x_1,k_1;\ldots;x_{i+1},k_{i+1};x_i,k_i;\ldots;x_n,k_n)\in WF(\omega_n).
\]
Proceeding in this way we can permute the point $(x_i,k_i)$ all the way
to the right. Then we have $(x_i,k_i)\in\mathcal{N}^-$ by assumption
and $(x_i,k_i)\in\mathcal{N}^+$ by the first paragraph of the proof.
Similarly, if we had started with $(x_i,k_i)\in\mathcal{N}^+$ we
could have permuted this point to the left to conclude that
$(x_i,k_i)\in\mathcal{N}^-$. In both cases we get a contradiction,
because $k_i\not=0$, but $\mathcal{N}^+\cap\mathcal{N}^-=\mathcal{Z}$.
This completes the proof.
\end{proof*}
The argument in the proof of proposition \ref{lem_ts2} can also be used to
show that the immersed graphs that occur in $WF(\omega_n)$ are disjoint
unions of pieces of light-like geodesics, to which the cotangent vectors
are parallel or anti-parallel.

For completeness we also prove a result concerning truncated $n$-point
distributions, although we will not use it in this thesis. For $n\ge 1$
we let $\mathcal{P}_n$ denote the set of all partitions of the set
$\left\{1,\ldots,n\right\}$ into pairwise disjoint ordered sets and for
each set $r$ in the partition $P\in\mathcal{P}_n$ we denote its elements
by $r(1),\ldots,r(|r|)$ where $|r|$ is the number of elements of $r$.
We then define the truncated $n$-point distributions $\omega_n^T$
implicitly through
\begin{equation}\label{def_Tnpoint}
\omega_n(x_1,\ldots,x_n)=\sum_{P\in\mathcal{P}_n}\prod_{r\in P}
\omega_{|r|}^T(x_{r(1)},\ldots,x_{r(|r|)}).
\end{equation}
Note that this equation can be solved iteratively for the $\omega^T_n$
order by order.

In their discussion of perturbative quantum field theory
\cite{Hollands+} impose the Hadamard condition together with the condition
that $\omega^T_n$ is smooth for all $n\not=2$ and
$WF(\omega^T_2)=WF(\omega_2)$. The same condition has also been considered
by Kay in \cite{Kay3}. Our result states that the
Hadamard condition already implies this condition on the truncated
$n$-point distributions, so this extra condition is superfluous.
\begin{proposition}
If $\omega$ is a (not necessarily quasi-free) Hadamard state on
$\Alg{U}^0_M$, then $\omega^T_n$ is smooth for all $n\not=2$ and
$WF(\omega^T_2)=WF(\omega_2)$.
\end{proposition}
\begin{proof*}
First note that $\omega^T_1(x_1)=\omega_1(x_1)$ and
$\omega^T_2(x_1,x_2)=\omega_2(x_1,x_2)-\omega_1(x_1)\omega_1(x_2)$ by
equation (\ref{def_Tnpoint}). Now, $\omega^T_1$ is smooth by the proof
of proposition \ref{lem_ts2} and hence $WF(\omega^T_2)=WF(\omega_2)$.
We prove the result for $n\ge 3$ by induction.

Suppose that $(x_1,k_1;\ldots;x_n,k_n)\in WF(\omega^T_n)$ and let $a$
be an index such that $k_a\not=0$. Expanding equation (\ref{def_Tnpoint})
and using the induction hypothesis it follows that
$(x_1,k_1;\ldots;x_n,k_n)$ is in the wave-front set of
\begin{eqnarray}\label{eq_unpack}
\omega_n(x_1,\ldots,x_n)&-&\sum_{i\le a-1}\omega_{n-2}(x_1,\ldots\hat{x}_i
\ldots\hat{x}_a\ldots,x_n)\omega_2(x_i,x_a)\\
&-&\sum_{i\ge a+1}\omega_{n-2}(x_1,\ldots\hat{x}_a\ldots\hat{x}_i
\ldots,x_n)\omega_2(x_a,x_i),\nonumber
\end{eqnarray}
because it cannot be in the wave front set of any of the other terms.
Notice that $(x_n,k_n)\in\mathcal{N}^+$ and
$(x_1,k_1)\in\mathcal{N}^-$, because the $\omega_n$ satisfy the
$\mu$SC by proposition \ref{lem_ts2}.

Now we note what happens when we use the commutation relations for the
indices $a$ and $a+1$ in expression (\ref{eq_unpack}). The only changes
occur in the first term and in the term $i=a+1$ under the second summation
symbol, namely:
\begin{eqnarray}
\omega_n(x_1,\ldots,x_n)-\omega_{n-2}(x_1,\ldots\hat{x}_a,\hat{x}_{a+1}
\ldots,x_n)\omega_2(x_a,x_{a+1})&=&\nonumber\\
\omega_n(x_1,\ldots,x_{a+1},x_a,\ldots,x_n)-
\omega_{n-2}(x_1,\ldots\hat{x}_{a+1},\hat{x}_a\ldots,x_n)
\omega_2(x_{a+1},x_a).&&\nonumber
\end{eqnarray}
Substituting this in expression (\ref{eq_unpack}) we see that
$(x_1,k_1;\ldots;x_n,k_n)$ is in the wave front set of
\begin{eqnarray}\label{eq_swap}
&&\omega_n(x_1,\ldots,x_{a+1},x_a,\ldots,x_n)\nonumber\\
&-&\sum_{i\le a-1\mathrm{\ or\ }i=a+1}\omega_{n-2}(x_1,\ldots\hat{x}_i\ldots
\hat{x}_a\ldots,x_n)\omega_2(x_i,x_a)\nonumber\\
&-&\sum_{i\ge a+2}\omega_{n-2}(x_1,\ldots\hat{x}_a\ldots\hat{x}_i\ldots,x_n)
\omega_2(x_a,x_i).\nonumber
\end{eqnarray}
It follows that
$(x_1,k_1;\ldots;x_{a+1},k_{a+1};x_a,k_a;\ldots;x_n,k_n)$ is
in the wave front set of expression (\ref{eq_unpack}) with $a+1$
substituted for $a$. Hence it is also
in the wave front set of $\omega^T_n$. Hence, if $k_a\not=0$ we can swap
the points
$(x_a,k_a)$ and $(x_{a+1},k_{a+1})$ in $(x_1,k_1;\ldots,x_n,k_n)$. Now move
$k_a$ to the $n$th position to see
that $(x_a,k_a)\in\mathcal{N}^+$. Then move $(x_a,k_a)$ to the first
position to find that $(x_a,k_a)\in\mathcal{N}^-$. This implies $k_a=0$,
so there can be no non-zero vector $k_a$. This proves that the
wave front set of $\omega^T_n$ is empty and hence $\omega^T_n$ is smooth
for $n\ge 3$.
\end{proof*}

\begin{definition}
The state space functor $\map{\Func{Q}^0}{\CatMan}{\CatStat}$ for the
locally covariant quantum field theory $\Func{U}^0$ assigns to every
globally hyperbolic spacetime $M$ the set of Hadamard states $\Stat{Q}^0_M$
on $\Alg{U}^0_M$.
\end{definition}
For each globally hyperbolic spacetime $M$ the set $\Stat{Q}^0_M$ is the
subset of states in
$\Stat{Q}_M$ characterised by the extra conditions that they solve the
Klein-Gordon equation and have the commutator property. This class of
states is convex and closed under operations from $\Alg{U}^0_M$, because
these extra conditions are invariant under convex linear combinations
and under operations from $\Alg{U}^0_M$. This last point uses
proposition \ref{prop_closedsf} and the fact that the Hadamard condition
implies the $\mu$SC, proposition \ref{lem_ts2}.
The action of $\Func{Q}^0$ on morphisms is
implicitly defined by the statement that $\Func{Q}^0$ is a state space
functor for $\Func{U}^0$ and this action is well-defined, because both
wave front sets and the cones $\Gamma_n$ behave covariantly under
isometric diffeomorphisms of the spacetime (see the proof of proposition
\ref{prop_closedsf}).

The following lemma contains the core of the proof of the
time-slice axiom for the free scalar field and is adapted from
\cite{Dimock1}.
\begin{lemma}\label{lem_ts1}
Let $M$ be a globally hyperbolic spacetime, $\mathcal{W}\subset M$ a
neighbourhood of a Cauchy surface and $\chi\in\Test_0(M)$ such that
$\chi\equiv 1$ on $J^+(\mathcal{W})\setminus\mathcal{W}$ and
$\chi\equiv 0$ on $J^-(\mathcal{W})\setminus\mathcal{W}$. For every
$f\in\Test_0(M)$ we have $f=f'+Kh$, where
$f':=K(\chi Ef)\in\Test_0(\mathcal{W})$ and $h:=E^-(f-f')\in\Test_0(M)$.
\end{lemma}
\begin{proof*}
Clearly $\mathrm{supp} f'\subset\mathrm{supp}(Ef)$ and
$f'\equiv 0$ on a neighbourhood of $M\setminus\mathcal{W}$, so that
$f'\in\Test_0(\mathcal{W})$. (This uses the results of \cite{Bernal+0}
and corollary A.5.4 of \cite{Baer+}.) We have
$h:=E^-(f-f')=(1-\chi)E^-f+\chi E^+f$, which is compactly supported in
$M$ and $Kh=f-f'$.
\end{proof*}

\begin{proposition}\label{U0Q0}
The locally covariant quantum field theory $\Func{U}^0$ with the state
space $\Func{Q}^0$ is causal, additive and satisfies the time-slice
axiom.
\end{proposition}
\begin{proof*}
Because $E(f,h)=0$ whenever
$\mathrm{supp}\ f\subset(\mathrm{supp}\ h)^{\perp}$ it is immediately
verified that the free field Borchers-Uhlmann functor defines a
causal locally covariant quantum field theory. Additivity follows from
proposition \ref{BUadditive} by choosing a representative in $\Alg{U}_M$
for each element of $\Alg{U}^0_M$. To prove the time-slice
axiom we suppose that $\map{\Psi}{M_1}{M_2}$ is a morphism such that
$\psi(M_1)\subset M_2$ contains a Cauchy surface $C$. For any
$f\in\Test_0(M_2)$ we use lemma \ref{lem_ts1} to find
$f'\in\Test_0(\psi(M_1))$ such that $f=f'+Kh$ for some $h\in\Test_0(M)$.
Therefore, $\Phi_{M_2}(f)=\Phi_{M_2}(f')=
\upsilon^0_{\Psi}(\Phi_{M_1}(f'\circ\psi))$. Because the elements
$\Phi_{M_2}(f)$ generate $\Alg{U}^0_{M_2}$ we conclude that
$\upsilon^0_{\Psi}$ is an isomorphism. We already noted that
$(\upsilon^0_{\Psi})^*$ maps a state satisfying the $\mu$SC on
$\Alg{U}^0_{M_2}$ to a state satisfying  the $\mu$SC on
$\Alg{U}^0_{M_1}$ (see proposition \ref{prop_closedsf}). Conversely,
every such state on $\Alg{U}^0_{M_1}$ can be obtained in this way as
follows. First such a state
gives rise to a state on $\Alg{U}^0_{\psi(M_1)}$ which satisfies the
$\mu$SC. This state in turn determines a state on $\Alg{U}^0_{M_2}$ with
the $\mu$SC by proposition \ref{lem_ts2}.
\end{proof*}

\section{A $C^*$-algebraic description of the real free scalar
field}\label{sec_freeC}

We now describe the real free scalar field as a locally covariant quantum
field theory using $C^*$-algebras, which is often convenient because
$C^*$-algebras can be represented as algebras of bounded operators
\cite{Kadison+} (in particular the GNS-representation yields an algebra
of bounded operators). We will follow the usual practice and use the
CCR-algebra or Weyl-algebra for this purpose, following
\cite{Dimock1, Wald2, Kay+, Bratteli+, Brunetti+, Verch2}. An alternative
would be to use the resolvent algebra instead \cite{Buchholz+}.

Given a globally hyperbolic spacetime $M$ we choose a smooth Cauchy
surface $C\subset M$ and consider the linear space
$\mathcal{K}_C(M):=\Test_0(C,\R)\oplus\Test_0(C,\R)$, where
$\Test_0(C,\R)$ is the space of real-valued test-functions on $C$.
An element
$(f,\dot{f})$ in $\mathcal{K}_C(M)$ specifies a unique solution $\phi$
to the Klein-Gordon equation on $M$ with initial data $\phi|_C=f$ and
$n^a\nabla_a\phi|_C=\dot{f}$, where $n^a$ is the future pointing normal
vector field on $C$. In this way $\mathcal{K}_C(M)$ can be identified
with a linear space of classical solutions to the Klein-Gordon equation.
We endow $\mathcal{K}_C(M)$ with the non-degenerate symplectic structure
\[
\sigma_C((f,\dot{f}),(h,\dot{h})):=\int_Cf\dot{h}-\dot{f}h,
\]
where we integrate with respect to the volume element associated to the
metric on $C$ that is induced by the metric $g$ of $M$. Before we
quantise the classical system that is described by the symplectic space
$(\mathcal{K}_C(M),\sigma_C)$ we show that it is independent of the
choice of Cauchy surface (see \cite{Dimock1,Verch2}).
\begin{proposition}\label{CauchyIndep}
Define the symplectic space $(\mathcal{K}(M),\sigma)$, where
$\mathcal{K}(M):=\Test_0(M,\R)/\mathrm{ker}\ E$ and
\[
\sigma(f,h):=E(f,h)=\int_MfEh\ d\mathrm{vol}_g.
\]
Then $(\mathcal{K}(M),\sigma)$ is isomorphic as a symplectic space to
$(\mathcal{K}(M)_C,\sigma_C)$ for every smooth Cauchy surface $C$.
\end{proposition}
\begin{proof*}
Note that each element $f\in\mathcal{K}(M)$ determines a unique solution
$Ef$ of the Klein-Gordon equation which has compact intersection with
each Cauchy surface of $M$, so we can define a linear map
$\map{k}{\mathcal{K}(M)}{\mathcal{K}_C(M)}$ by
$k(f):=(Ef|_C,(n^a\nabla_aEf)|_C)$. This map is surjective, because
every smooth solution of the Klein-Gordon equation which has a compact
intersection with every Cauchy surface can be obtained in this way by
\cite{Dimock1} lemma A.3. It remains to check that
$\sigma_C(k(f),k(h))=\sigma(f,h)$. Leaving the metric volume elements
on $M$ and $C$ implicit we have:
\begin{eqnarray}
\sigma(f,h)&=&\int_MfEh=\int_{J^+(C)}(KE^-f)(Eh)+
\int_{J^-(C)}(KE^+f)(Eh)\nonumber\\
&=&\int_{J^+(C)}\nabla_a((\nabla^aE^-f)(Eh))
-\nabla_a((E^-f)(\nabla^aEh))+0\nonumber\\
&&+\int_{J^-(C)}\nabla_a((\nabla^aE^+f)(Eh))
-\nabla_a((E^+f)(\nabla^aEh))+0\nonumber\\
&=&\int_C-(n_a\nabla^aE^-f)(Eh)+(E^-f)(n_a\nabla^aEh)\nonumber\\
&&+\int_C(n_a\nabla^aE^+f)(Eh)-(E^+f)(n_a\nabla^aEh)\nonumber\\
&=&\sigma_C(k(f),k(h)),\nonumber
\end{eqnarray}
where we used $KEh=0$, $E=E^--E^+$ and a partial integration (see e.g.
\cite{Wald} (B.2.26), but note the different sign convention; in this
case the sign can easily be checked by studying the example of Minkowski
spacetime).
\end{proof*}
The symplectic space $(\mathcal{K}(M),\sigma)$ gives a covariant and
Cauchy-surface independent description of the classical Klein-Gordon
field. Also note that $\sigma$ is non-degenerate by proposition
\ref{CauchyIndep}, because $\sigma_C$ is non-degenerate. To the
symplectic space $(\mathcal{K}(M),\sigma)$ we may associate the
CCR-algebra $\Alg{A}^0_M$, i.e. the (simple) $C^*$-algebra $\Alg{A}^0_M$
of canonical commutation relations \cite{Manuceau+,Bratteli+}. This
algebra is generated by the set of Weyl-operators $W(f)$,
$f\in\mathcal{K}(M)$ satisfying the Weyl-relations
\begin{equation}\label{Weylrel}
W(f)W(h)=e^{-\frac{i}{2}\sigma(f,h)}W(f+h),\quad
W(f)^*=W(-f).
\end{equation}

\begin{proposition}\label{Weyladditive}
One can define a locally covariant quantum field theory
$\map{\Func{A}^0}{\CatMan}{\CatAlg}$ which maps each $M$ to
$\Alg{A}^0_M$
and each morphism $\map{\Psi}{M}{M'}$ in $\CatMan$ to the morphism
$\map{\alpha_{\Psi}}{\Alg{A}^0_M}{\Alg{A}^0_{M'}}$ determined by
$\alpha_{\Psi}(W(f))=W'(\psi_*f)$, where $W'$ denotes the Weyl operators
that generate $\Alg{A}^0_{M'}$. This locally covariant quantum field
theory is causal and additive.
\end{proposition}
\begin{proof*}
For the proof that $\Func{A}^0$ is a causal locally covariant quantum
field theory we refer to \cite{Brunetti+}. Additivity follows from
\cite{Bratteli+} proposition 5.2.10.
\end{proof*}

Let us now explain the relation between the $C^*$-algebra $\Alg{A}^0_M$
and the Borchers-Uhlmann algebra $\Alg{U}^0_M$. If $\omega$ is a
quasi-free state on $\Alg{U}^0_M$ and $f\in\Test_0(M,\R)$, then
$\Phi^{(\omega)}(f)$ is a self-adjoint (unbounded) operator and
we can define the unitary operator $W(f):=e^{i\Phi^{(\omega)}(f)}$
(see \cite{Verch2} proposition 3.2 and \cite{Bratteli+} theorem 5.2.3 and
5.2.4). These unitary operators
satisfy the Weyl-relations (\ref{Weylrel}) and therefore generate a
$C^*$-algebra that is isomorphic to $\Alg{A}^0_M$, \cite{Bratteli+}.

In order to go in the opposite direction, i.e. to obtain the
Borchers-Uhlmann algebra from the Weyl-algebra, we need to restrict our
attention to a special class of states on $\Alg{A}^0_M$:
\begin{definition}\label{def_regular}
We call a state $\omega$ on $\Alg{A}^0_M$ \emph{regular} if and only if
for every $f\in\mathcal{K}(M)$ the unitary group
$t\mapsto\pi_{\omega}(W(tf))$ is strongly continuous with self-adjoint
(unbounded) generator $\Phi^{(\omega)}(f)$.

A regular state $\omega$ on $\Alg{A}^0_M$ is called
$C^{\infty}$-\emph{regular} if and only if the maps
\[
\omega_n(f_1,\ldots,f_n):=\partial_{t_1}\cdots\partial_{t_n}
\omega(W(t_1f_1)\cdots W(t_nf_n))|_{t_1=\ldots=t_n=0}
\]
are distributions, after extending them by linearity to $\C$-valued
test-functions.
\end{definition}
A $C^{\infty}$-regular state $\omega$ on $\Alg{A}^0_M$ also defines a
continuous state on $\Alg{U}^0_M$ via the $n$-point distributions
(see \cite{Bratteli+} or \cite{Fewster+2} section A.5).
The notation $\Phi^{(\omega)}(f)$ coincides with that of equation
(\ref{KGPhi}). This is justified, because the operators
$\Phi^{(\omega)}$ are linear in their argument and they generate an
algebra that is isomorphic to $\pi_{\omega}(\Alg{U}^0_M)$ (see
\cite{Bratteli+} lemma 5.2.12). We have for example
\begin{eqnarray}
\Phi^{(\omega)}(f)\Phi^{(\omega)}(h)-
\Phi^{(\omega)}(h)\Phi^{(\omega)}(f)&=&\nonumber\\
-\partial_s\partial_t\pi_{\omega}(W(tf)W(sh)-W(sh)W(tf))|_{s=t=0}&=&
\nonumber\\
-\partial_s\partial_t(e^{-ist\sigma(f,h)}-1)\pi_{\omega}(W(sh)W(tf))
|_{s=t=0}&=&
i\sigma(f,h)I=iE(f,h)I\nonumber
\end{eqnarray}
on a dense domain of $\mathcal{H}_{\omega}$, i.e. we recover equation
(\ref{KGcomm}). To make the correspondence with section \ref{sec_freeBU}
precise we should extend the real scalar field $\Phi^{(\omega)}$ of this
section by linearity to complex-valued test-functions.

\begin{definition}
A (not necessarily quasi-free) state $\omega$ on $\Alg{A}^0_M$ is
called \emph{Hadamard} iff $\omega$ is $C^{\infty}$-regular and
defines a Hadamard state on $\Alg{U}^0_M$.

The state space functor $\map{\Func{S}^0}{\CatMan}{\CatStat}$ for the
locally covariant quantum field theory $\Func{A}^0$ assigns to each
globally hyperbolic spacetime $M$ the set of states on $\Alg{A}^0_M$
which are locally
quasi-equivalent to a quasi-free Hadamard state.
\end{definition}
To define the state space functor we used the fact that a quasi-free
Hadamard state on $M$ restricts to a quasi-free Hadamard state on any
given sub-spacetime and the same is then true for any state locally
quasi-equivalent to a quasi-free Hadamard state. In our choice of
state space functor we have followed \cite{Brunetti+}, who also prove
some of the following properties in their theorem 3.4:
\begin{proposition}\label{prop_A0prop}
The locally covariant quantum field theory $\Func{A}^0$ with state space
$\Func{S}^0$ is causal, additive, satisfies the time-slice axiom, respects
local physical equivalence, is locally quasi-equivalent and nowhere
classical.
\end{proposition}
\begin{proof*}
We already noted causality and additivity in proposition
\ref{Weyladditive}. The condition on $\Func{A}^0$ needed for the
time-slice axiom follows from proposition
\ref{CauchyIndep}. For the condition on $\Func{S}^0$ we first note that
a state $\omega$ which is Hadamard on a neighbourhood $N$ of a Cauchy
surface $C$ is Hadamard everywhere by proposition \ref{lem_ts2} and
if $\omega$ is quasi-free on $N$ it is quasi-free everywhere
by lemma \ref{lem_ts1}. Now let $O\subset M$ be any bounded cc-region and
note that $J(\overline{O})$ has a compact intersection with $C$. This means
we can find a bounded cc-region $V\subset N$ such
that $O\subset D(V)$. If the state $\omega'$ is locally quasi-equivalent
to $\omega$ on $N$, then the map
$\pi_{\omega'}(A)\mapsto\pi_{\omega}(A)$ for all $A\in\Alg{A}^0_V$ is
well-defined and can be extended to a $^*$-isomorphism $\alpha$ of the
local von Neumann algebras $\Alg{R}^{(\omega)}_V$ and
$\Alg{R}^{(\omega')}_V$ (see \cite{Araki3} pp.212-213). It follows that
$\alpha$ restricts to a $^*$-isomorphism of the von Neumann algebras
$\Alg{R}^{(\omega)}_O$ and $\Alg{R}^{(\omega')}_O$, which proves that
the restrictions of $\pi_{\omega}$ and $\pi_{\omega'}$ to
$\Alg{A}^0_O$ are quasi-equivalent (\cite{Araki3} loc.~cit.). We can
therefore conclude that a state which is
locally quasi-equivalent to a quasi-free Hadamard state on $N$ remains
locally quasi-equivalent to a quasi-free Hadamard state. Local physical
equivalence is proved in
proposition 4.3 of \cite{Fewster}. Local quasi-equivalence follows from
\cite{Verch2} and the theory is nowhere classical because of the
Weyl-relations (\ref{Weylrel}) and the fact that the symplectic
structure $\sigma$ is not identically $0$.
\end{proof*}

If we take $\Alg{A}^0_M$ in the norm topology, then $f\mapsto W(f)$ is
not a locally covariant quantum field. Indeed, $\|W(f)-I\|=2$ for all
$f\not=0$, \cite{Bratteli+} proposition 5.2.4. However, in the strong
operator topology on $\mathcal{H}_{\omega}$ for any Hadamard state
$\omega$, $W(f_n)\rightarrow W(f)$ as $f_n\rightarrow f$ in $\Test_0(M)$
(\cite{Bratteli+} loc.~cit.). Because the locally covariant
quantum field theory $\Func{A}^0$ is additive and locally quasi-equivalent
we could define the strong topology unambiguously on a norm-dense subset
of each $\Alg{A}^0_M$, but we will not pursue this approach further. It
is worth noting, however, that $W$, as a locally covariant quantum
field, is non-linear and does not satisfy the same equation of motion as
$\Phi^0$.

\chapter{The free Dirac field}\label{ch_Df}

\begin{quote}
One's ideas must be as broad as Nature if they are to interpret Nature,
\end{quote}
\begin{flushright}
Arthur Conan Doyle, A study in scarlet, Ch. 5
\end{flushright}

After our treatment of the real free scalar field in chapter \ref{ch_sf} we
now broaden our perspective a little and describe the free Dirac field
as a locally covariant quantum field along the same lines. In section
\ref{sec_clDf} we present a construction of the classical Dirac field in
a four-dimensional globally hyperbolic spacetime, describing the necessary
algebraic, group
theoretic and geometric aspects in sufficient detail in order to point
out some pitfalls (such as the change of spacetime signature, $+---$ or
$-+++$) and to correct a few typos that appear in parts of the
literature. Our treatment differs from the existing literature by
proving that the construction is essentially independent of the chosen
representation of the Dirac algebra. More precisely, we will impose
certain relations on Dirac spinors and cospinors, concerning their
adjoints, charge conjugation and the Dirac operator. Given these
relations, different choices of representation give rise to isomorphic
Dirac spinor bundles. This shows that the physics is determined entirely
by the relations we imposed and can be described in a coherent and
unified (representation-independent) way within the locally covariant
framework. It should be noted that \cite{Dimock2} discusses a similar
idea, namely the independence of the algebras on the choice of
representation of the canonical anti-commutation relations. However,
it does not seem to consider different representations of the Dirac
algebra or to determine the theory by imposing relations between the
adjoint map, charge conjugation and the Dirac operator.

Next we will quantise the theory in section \ref{sec_LCDf}, noting that the
distributional and $C^*$-algebraic description in this case coincide. In
that section we also describe the class of Hadamard states and show that
the Hadamard condition implies the $\mu$SC, exactly as for the real free
scalar field. We discuss the causality and time-slice properties of the
free Dirac field and we indicate how Majorana spinors can be quantised
in the same, representation independent way.

In the final section of this chapter we consider the relative Cauchy
evolution of the free Dirac field. For this we use the time-slice axiom
to identify the algebra of a neighbourhood of a Cauchy surface $C_-$ in
a spin spacetime $M$ with the algebra of a neighbourhood of a Cauchy
surface
$C_+$ to the future of $C_-$. This identification is a $^*$-isomorphism,
which depends on the spin spacetime in between the two regions. A
variation of the metric and/or the spin structure in the intermediate
region can be encoded in such $^*$-isomorphisms, which is the idea
behind the relative Cauchy evolution. We will then consider the functional
derivative of the relative Cauchy evolution with respect to the metric
and prove a relation between this quantity and the
stress-energy-momentum tensor, where we describe
the latter using a point-splitting procedure.
The relation we obtain is the direct analogue of that which is already
known to hold for the free scalar field \cite{Brunetti+}.

For our presentation of the Dirac field in curved spacetime we largely
follow \cite{Dimock2,Dawson+,Fewster+}; for results on Clifford
algebras we refer to \cite{Lawson+} chapter 1.

\section{The classical free Dirac field}\label{sec_clDf}

The description of the classical Dirac field is much more involved than
that of the scalar field. Whereas the classical scalar field is a
section of a trivial vector bundle over $M$ (either $M\times\C$ or
$M\times\R$), the Dirac field is a section of a four-dimensional
complex vector bundle $DM$, the Dirac spinor bundle, that is intimately
related to the spacetime geometry.
Before we define the Dirac spinor bundle and the Dirac equation
(subsection \ref{subs_geo} and \ref{subs_Deq} respectively), we will
give a review of the Dirac algebra (subsection \ref{subs_Dalg}), i.e.
the algebra of gamma-matrices, and the Spin group (subsection
\ref{subs_spin}). This is necessary in order to prove the representation
independence of the Dirac spinor bundle in proposition \ref{equivalence}
as well as to fix our notation and to point out some confusions and typos
in the literature.

\subsection{The Dirac algebra}\label{subs_Dalg}
To add clarity to our description of the Dirac algebra we will take the
more general point of view of Clifford algebras at the beginning of this
subsection. For a detailed treatment of Clifford algebras we refer to
chapter 1 of \cite{Lawson+} (but note the difference in sign convention
in the Clifford multiplication).

Let $\R^{r,s}$ be the finite dimensional real vector space of dimension
$n=r+s$, equipped with a non-degenerate bilinear form $\Omega_{ab}$ which
has $r$ positive and $s$ negative eigenvalues. As a special case we note
that $M_0:=\R^{1,3}$ is Minkowski spacetime, where the bilinear form is
$\eta=\mathrm{diag}(1,-1,-1,-1)$ when expressed in the orthonormal basis
$g_a$, $a=0,1,2,3$, with $\|g_0\|^2=1$.
\begin{definition}
The \emph{Clifford algebra} $Cl_{r,s}$ of $\R^{r,s}$ is defined as the
real-linear associative algebra generated by a unit element $I$ and an
orthonormal basis $e_a$ of $\R^{r,s}$ subject to the Clifford
relations
\begin{equation}
e_ae_b+e_be_a=2\Omega_{ab}I.
\end{equation}

The \emph{even}, respectively \emph{odd}, \emph{subspace} of $Cl_{r,s}$
is the real-linear space spanned by monomials of even, respectively odd,
degree in the basis vectors $e_a$ and is denoted by $Cl^0_{r,s}$,
respectively $Cl^1_{r,s}$.

The \emph{Dirac algebra} $D:=Cl_{1,3}$ is the Clifford algebra of
Minkowski spacetime $M_0$ and is characterised by
\begin{equation}\label{Clifford}
g_ag_b+g_bg_a=2\eta_{ab}I.
\end{equation}
\end{definition}
The definition of Clifford algebra is independent of the choice of
basis (\cite{Lawson+} section 1.1). As a real-linear space $Cl_{r,s}$
has a basis consisting of $I$ and all elements
$e_{a_1}\cdots e_{a_m}$ with $a_1<\ldots<a_m$, $m\le r+s$, which shows
that the
dimension of $Cl_{r,s}$ is $2^{r+s}$. The even and odd subspaces are
well-defined, because the Clifford relations are purely even. Note that
the even subspace $Cl^0_{r,s}$ is a subalgebra. We will identify
$\R^{r,s}\subset Cl_{r,s}$ as the subspace of monomials of degree 1 in
the basis $e_a$. In particular we will identify $M_0\subset D$.

For convenience we define the volume element $g_5$ of the Dirac algebra
by $g_5:=g_0g_1g_2g_3$. The following lemma lends a geometric
interpretation to Clifford multiplication and will allow us to construct
the Spin group as a subset of the Dirac algebra in subsection
\ref{subs_spin}:
\begin{lemma}\label{CliffordProperties}
We have $g_5^2=-I$,
\begin{equation}\label{gamma5}
g_5vg_5^{-1}=-vg_5g_5^{-1}=-v,\quad v\in M_0.
\end{equation}
Moreover, if $u\in M_0$ has $u^2=\|u\|^2I\not=0$, with the norm
taken in $M_0$, then $u^{-1}=\frac{1}{\|u\|^2}u$ and
$v\mapsto -uvu^{-1}$ defines a reflection of $M_0$ in the hyperplane
perpendicular to $u$.
\end{lemma}
\begin{proof*}
This follows directly from the Clifford relations (\ref{Clifford}).
Indeed, we have $g_5e_a=-e_ag_5$ for each $a$, which implies equation
(\ref{gamma5}) and $g_5^2=-I$. For the last claim we compute:
\[
-uvu^{-1}=v-(uv+vu)u^{-1}=v-\frac{2\langle u,v\rangle}{\|u\|^2}u,
\quad v\in M_0.
\]
\end{proof*}

\begin{definition}
A \emph{complex representation} of the real algebra $D$ is a real-linear
representation $\map{\pi}{D}{M(n,\C)}$ for some $n\in\N$.
\end{definition}
In order to characterise the complex representations of the Dirac algebra
we first note the following. Using standard arguments with Clifford
algebras (\cite{Lawson+} theorem I.3.7, equation (I.1.7) and section
I.4) we have:
\[
D=Cl_{1,3}\simeq Cl^0_{1,4}\simeq Cl^0_{4,1},\quad
Cl_{4,1}\simeq M(4,\C).
\]
In fact, $Cl_{4,1}$ is generated by the generators $g_a$ of $D$
together with a central element $\omega$, which corresponds to the
matrix $iI\in M(4,\C)$, and hence:
\begin{equation}\label{subalgebra}
M(4,\C)\simeq\C\otimes_{\R}D.
\end{equation}
This implies that the center of $D$ is spanned by $I$ (over $\R$).
Moreover, it brings the well-known representation theory of $M(4,\C)$
into the study of complex representations of $D$. The following fundamental
theorem contains what is usually known as Pauli's theorem \cite{Pauli}.
Our method of proof is close to the approach of \cite{Waerden} and is
shorter, but less elementary, than Pauli's (loc.~cit.)
\begin{theorem}[Fundamental Theorem]\label{fundamentaltheorem}
The Dirac algebra $D$ is simple and has a unique irreducible complex
representation, up to equivalence. This is the representation
$\map{\pi_0}{D}{M(4,\C)}$ determined by $\pi_0(g_a)=\gamma_a$ with the
Dirac matrices $\gamma_a$ given by\footnote{This set of gamma-matrices
is taken from \cite{Haag} equation (I.3.45) and is the same as the
\emph{Weyl} or \emph{chiral representation} of \cite{Peskin+} equation
(3.25) up to a sign in the $\gamma_i$.}
\[
\gamma_0:=\left(\begin{array}{cc}
O&I\\
I&0
\end{array}\right),\quad
\gamma_i:=\left(\begin{array}{cc}
O&-\sigma_i\\
\sigma_i&0
\end{array}\right),
\]
where $\sigma_i$ are the Pauli matrices:
\[
\sigma_1:=\left(\begin{array}{cc}O&1\\ 1&0\end{array}\right),\quad
\sigma_2:=\left(\begin{array}{cc}O&-i\\ i&0\end{array}\right),\quad
\sigma_3:=\left(\begin{array}{cc}1&0\\ 0&-1\end{array}\right).
\]
The equivalence with another irreducible complex representation $\pi$ of
$D$ is implemented by $\pi(S)=L\pi_0(S)L^{-1}$ for all
$S\in D$, where $L\in GL(4,\C)$ is unique up to a
non-zero complex factor.

Consequently, for every set of matrices $\gamma'_a\in M(4,\C)$
satisfying equation (\ref{Clifford}) there is an
$L\in GL(4,\C)$, unique up to a non-zero complex factor, such that
\[
\gamma'_a=L\gamma_aL^{-1}.
\]
\end{theorem}
\begin{proof*}
One can show that $D\simeq M(2,\Hq)$ (\cite{Lawson+} section I.4), which
is simple, because it is a full matrix algebra. Indeed, suppose that
$J\subset D$ is an ideal which contains a non-zero element $A$. Let
$E_{ij}$ denote the matrix whose only non-zero entry is the
$(ij)$-entry, which is $1$. If the $(i_0j_0)$-entry of $A$ is $a\not=0$,
then $J$ contains
$E_{1i_0}AE_{j_01}+E_{2i_0}AE_{j_02}=aI$, $I$ being the $2\times 2$
identity matrix. As $a\in\Hq$ is invertible we have $I\in J$ and
hence $J=D$.

It can be checked by direct computation that the given matrices
$\gamma_a$ satisfy the Clifford relations (\ref{Clifford}) and therefore
extend to a representation of $D$ in $M(4,\C)$ (see \cite{Lawson+},
chapter I proposition 1.1). Any complex representation
$\map{\pi}{D}{M(n,\C)}$ extends to a complex representation
$\tilde{\pi}$ of $M(4,\C)$ by (\ref{subalgebra}), which is
irreducible if $\pi$ is irreducible. As $M(4,\C)$ has only one
irreducible representation up to equivalence (\cite{Waerden} section
16, p.75), this
determines $\pi$ up to equivalence, as stated. If $K,L\in GL(4,\C)$ are
two matrices which implement the same equivalence, then $KL^{-1}$
commutes with $D$ and hence with all of $M(4,\C)$ by (\ref{subalgebra}).
The center of $M(4,\C)$ is $\C I$, so we conclude $K=cL$ and $c\in\C$ is
non-zero because $K$ is invertible.

Note that $\pi'(g_a):=\gamma'_a$ extends to a complex representation of
$D$ in $M(4,\C)$. The last statement therefore follows from the previous
one.
\end{proof*}

For notational consistency we define $\gamma_5:=\pi_0(g_5)$. As special
cases of theorem \ref{fundamentaltheorem} we now consider the adjoint
and complex conjugate matrices, respectively, that will be used in
subsection \ref{subs_geo} to define the adjoint and charge conjugation
maps on Dirac spinors, respectively.
\begin{definition}
We say that $A,C\in GL(4,\C)$ satisfy assumption
(\ref{assumption}) w.r.t. an irreducible complex representation $\pi$ if
and only if
\begin{eqnarray}\label{assumption}
A=A^*,&\pi(g_a)^*=A\pi(g_a)A^{-1},&A\pi(n)> 0,\\
\overline{C}C=I,&-\overline{\pi(g_a)}=C\pi(g_a)C^{-1}&\nonumber
\end{eqnarray}
for all future pointing time-like vectors $n$.\footnote{On a general
representation space of complex dimension four one can define many
complex conjugations $z\mapsto\bar{z}$ and Hermitean inner products
$\langle,\rangle$. We desire to obtain certain equalities involving
adjoint and charge conjugate spinors in
section \ref{subs_geo}, which requires the complex conjugation and
Hermitean inner product to be compatible:
$\langle\overline{w},\overline{z}\rangle=\overline{\langle w,z\rangle}$.
In this case we can use the standard complex conjugation and Hermitean
inner product on $\C^4$ without loss of generality.}
\end{definition}
Here the condition $A\pi(n)>0$ means that $A\pi(n)=An^a\gamma_a$ is a
positive matrix, i.e. $\langle z,A\pi(n)z\rangle> 0$ for all non-zero
$z\in\C^4$. Note that the sets of matrices $\pi(g_a)^*$ and
$-\overline{\pi(g_a)}$, $a=0,1,2,3$, both satisfy the Clifford relations
(\ref{Clifford}), so by theorem \ref{fundamentaltheorem} the matrices
$A$ and $C$ are uniquely determined up to non-zero complex factors.

\begin{theorem}\label{ACmatrix}
For any irreducible complex representation $\pi$ of $D$ there are
$A,C\in GL(4,\C)$ which satisfy assumption (\ref{assumption})
w.r.t. $\pi$. The matrix $A$ is uniquely determined up to a positive
factor, $C$ up to a phase factor and we have $A=-C^*\overline{A}C$.
Moreover, if $A_i,C_i\in M(4,\C)$, $i=1,2$, satisfy assumption
(\ref{assumption}) w.r.t. irreducible complex representations $\pi_i$
of $D$, then there is an $L\in GL(4,\C)$, unique up to a sign,
such that $L^*A_1L=A_2$, $\overline{L}^{-1}C_1L=C_2$ and
$\pi_2=L^{-1}\pi_1L$ on $D$.
\end{theorem}
\begin{proof*}
This result is essentially already contained in \cite{Pauli}.
To prove existence in the representation $\pi_0$ we take
$A=A_0:=\gamma_0$, $C=C_0:=\gamma_2$ and check assumption
(\ref{assumption}) by direct computation, using the Clifford relations
(\ref{Clifford}). Note for example that
\[
\gamma_0n^a\gamma_a=\left(\begin{array}{cc}
n^0I+n^i\sigma_i&0\\
0&n^0I-n^i\sigma_i\end{array}\right)> 0,
\]
because $\det(n^0I\pm n^i\sigma_i)=n^an_a=1$ and
$Tr(n^0I\pm n^i\sigma_i)=2n^0>0$. Also, $-C_0^*\overline{A}_0C_0
=\gamma_2\gamma_0\gamma_2=-\gamma_0\gamma_2^2=\gamma_0=A_0$. To prove
existence in a general irreducible complex representation $\pi$ we use
theorem \ref{fundamentaltheorem} to write $\gamma_a=K\pi(g_a)K^{-1}$
for some $K\in GL(4,\C)$. One can then verify by direct computation
that $A=K^*A_0K$ and $C=\bar{K}^{-1}C_0K$ satisfy assumption
(\ref{assumption}) and $A=-C^*\overline{A}C$. This proves the existence.

The matrices $A$ and $C$ are uniquely determined up to non-zero complex
factors $a$ and $c$ by theorem \ref{fundamentaltheorem}.
Because $A=A^*$ and $\overline{C}C=I$ we see that $a\in\R$ and $|c|=1$.
Moreover, as $A\pi(n)>0$ for future pointing time-like vectors we must
have $a>0$. Now, the relation $A=-C^*\overline{A}C$ is invariant under
changes of $a$ and $c$ and we saw that for any $\pi$ there exist
matrices $A,C$ satisfying assumption (\ref{assumption}) and this
equality. Therefore any $A,C$ satisfying assumption (\ref{assumption})
w.r.t. $\pi$ necessarily satisfy this equality.

Given matrices $A_i,C_i\in GL(4,\C)$ for the representations $\pi_i$,
$i=1,2$, we can fix $K\in GL(4,\C)$ such that $\pi_1=K\pi_2K^{-1}$ on
$D$ by the fundamental theorem \ref{fundamentaltheorem}. Setting
$A'_2:=K^*A_1K$ and $C'_2:=\overline{K}^{-1}C_1K$ we can verify by
direct computation that $A'_2$ and $C'_2$ satisfy assumption
(\ref{assumption}) w.r.t. $\pi_2$, as in the first paragraph of this
proof. By the uniqueness this means that $A'_2=aA_2$ and $C'_2=cC_2$ for
some $a>0$, $|c|=1$. The desired matrix $L$ must be $L=zK$ for some
$z\not=0$, by the fundamental theorem. To get the right intertwining
relations for $A_i$ and $C_i$ we need $|z|^2=a$ and $z=c\overline{z}$,
which fixes $z$ up to a sign.
\end{proof*}

As another application of the fundamental theorem we can introduce a
determinant and trace on $D$:
\begin{definition}
The \emph{determinant} and \emph{trace} functions on $D$ are defined
by $\det S:=\det\pi(S)$ and $Tr(S):=Tr(\pi(S))$ for all $S\in D$, where
$\pi$ is any irreducible complex representation of $D$.
\end{definition}
This is well-defined by the fundamental theorem.
The following lemma will be useful in what follows:
\begin{lemma}\label{traceresults}
$Tr(g_ag_b)=4\eta_{ab}$ and
$Tr(\left[g_b,g_c\right]g_dg_a)=8(\eta_{cd}\eta_{ba}-\eta_{bd}\eta_{ca})$.
\end{lemma}
\begin{proof*}
Using the cyclicity of the trace and the Clifford relations
(\ref{Clifford}) we find:
\[
Tr(g_ag_b)=\frac{1}{2}Tr(g_ag_b+g_bg_a)=Tr(\eta_{ab}I)=4\eta_{ab}
\]
and
\begin{eqnarray}
Tr([g_b,g_c]g_dg_a)&=&Tr(g_b\left\{g_c,g_d\right\}g_a-
g_bg_d\left\{g_c,g_a\right\})\nonumber\\
&=&2Tr(\eta_{cd}g_bg_a-g_bg_d\eta_{ca})=
8(\eta_{cd}\eta_{ba}-\eta_{bd}\eta_{ca}).\nonumber
\end{eqnarray}
\end{proof*}

\subsection{The $Spin_{1,3}$ group}\label{subs_spin}

We now turn to the Spin group, which is the universal covering group
of the proper Lorentz group and which can be constructed in an
elegant way as a subset of the Dirac algebra.

\begin{definition}\label{def_spin}
The Pin and Spin groups of $Cl_{r,s}$ are defined as
\[
Pin_{r,s}:=\left\{S\in Cl_{r,s}|\ S=u_1\cdots u_k,\quad
k\in\N,\quad u_i\in\mathbb{R}^{r,s},\quad u_i^2=\pm I\right\},
\]
\[
Spin_{r,s}:=Pin_{r,s}\cap Cl^0_{r,s}.
\]

We also define the Lorentz group $\mathcal{L}:=O_{1,3}$, the proper
Lorentz group $\mathcal{L}_+:=SO_{1,3}$ and the proper orthochronous
Lorentz group $\mathcal{L}_+^{\uparrow}:=SO_{1,3}^0$, which is the
connected component of $\mathcal{L}_+$ containing the identity.
\end{definition}
The proper orthochronous Lorentz group preserves the time-orientation
as well as the orientation. Note that Pin is indeed a group and
that $I\in Pin_{r,s}$ because of the following equivalent
characterisation\footnote{The definition of the Spin group in
\cite{Choquet+} corresponds to our group $Pin_{1,3}$. In \cite{Dimock2}
and $\cite{Fewster+}$ one uses the term Spin group for the group
\[
\mathcal{S}:=\left\{S\in M(4,\mathbb{C})| \det S=1,\quad SvS^{-1}\in M_0
\mathrm{\ for\ all\ } v\in M_0 \right\}.
\]
Note that this group cannot give a double covering of the Lorentz group,
as claimed in \cite{Dimock2} (but not in \cite{Fewster+}), because for
any $S\in\mathcal{S}$ the matrices $iS,-S,-iS$ are in $\mathcal{S}$ too.
Its usefulness is based on its simple definition and the fact that
$\mathcal{S}^0=Spin^0_{1,3}$.}\label{footnote}
(cf. \cite{Choquet+} p.66 and p.334):
\begin{proposition}\label{spineq}
$Pin_{1,3}=\left\{S\in D|\ \det S=1, \forall v\in M_0\
SvS^{-1}\in M_0\right\}$.
\end{proposition}
\begin{proof*}
For $S\in Pin_{1,3}$ the map $v\mapsto SvS^{-1}$ on $M_0$ is a product
of reflections (up to a sign), by lemma \ref{CliffordProperties}, so
$SvS^{-1}\in M_0$ for all $v\in M_0$. Because $\det u=\|u\|^4I$ for all
$u\in M_0$, which can be verified by direct computation, we also have
$\det S=1$.

For the converse we suppose that $S\in D$ has $\det S=1$ and
$SvS^{-1}\in M_0$ for all $v\in M_0$. Notice that the adjoint action of
$S$ is a linear map on $M_0$ which preserves the Lorentzian inner
product, because it preserves the right-hand side of the equality
$vw+wv=2v^aw_aI$. Hence, the adjoint action of $S$ determines a
Lorentz transformation $\Lambda$, which can be written as a finite
product of reflections in non-null hyperplanes (\cite{Artin} theorem
3.20). Let $u_i$ be unit
normal vectors to these hyperplanes, where $1\le i\le k$ for some $k$.
If $k$ is even we let $T:=u_k\cdots u_1$ and otherwise we let
$T:=u_k\cdots u_1 g_5$. Notice that $T\in Pin_{1,3}$ and that in both
cases we have $SvS^{-1}=\Lambda(v)=TvT^{-1}$, by lemma
\ref{CliffordProperties}. (Each reflection is given by
$v\mapsto -u_ivu_i$ and we used $g_5$ to cancel the extra sign in case
of an odd number of reflections.) In particular,
$T^{-1}Sg_aS^{-1}T=g_a$, so by theorem \ref{fundamentaltheorem} we
have $T^{-1}S=cI$ and hence $S=cT$ for some non-zero $c\in\C$. Because
$S,T\in D$ we must have $c\in\R$ by equation (\ref{subalgebra}).
Moreover, we have $\det T=1$ as in the first paragraph and $\det S=1$ by
assumption, so $S=\pm T$. Finally, $-T=(g_1)^2T\in Pin_{1,3}$ too, so in
any case $S\in Pin_{1,3}$.
\end{proof*}
It can be seen from proposition \ref{spineq} that $Pin_{1,3}$ and
$Spin_{1,3}$ are indeed Lie groups, using the embedding of equation
(\ref{subalgebra}). We let $Spin^0_{1,3}$ denote the connected component
of $Spin_{1,3}$ which contains the identity. We now prove the following
lemma concerning the Lie algebras of these Lie groups (cf.
\cite{Lawson+} proposition I.6.1):
\begin{lemma}\label{spinspan}
The Lie algebras $spin^0_{1,3}=spin_{1,3}=pin_{1,3}$ are spanned by
$g_ag_b$, $0\le a<b\le 3$.
\end{lemma}
\begin{proof*}
We consider the curves $\map{c_i}{\left[0,1\right]}{Spin^0_{1,3}}$ for
$i=1,2,3$ and $\map{d_{ij}}{\left[0,1\right]}{Spin^0_{1,3}}$ for
$1\le i<j\le 3$ all starting at $I$ and defined by
\begin{eqnarray}
c_i(t)&:=&g_0(\cosh(t)g_0+\sinh(t)g_i)\nonumber\\
d_{ij}(t)&:=&-g_i(\cos(t)g_i-\sin(t)g_j).\nonumber
\end{eqnarray}
The derivatives of these curves at $t=0$ are the six linearly
independent elements $g_ag_b$ with $0\le a<b\le 3$. Conversely, for
every curve $S(t)\subset Pin_{1,3}$ with derivative $s$ at $t=0$ the
condition $S(t)g_aS^{-1}(t)\in M_0$ for all $t$ (see proposition
\ref{spineq}) implies
$\left[s,g_a\right]\in M_0$. If we express $s$ as a real-linear
combination of products of $g_a$'s then an elementary computation
shows that this condition implies $s=\alpha_0I + \alpha^{ab}g_ag_b$ for
some $\alpha_0,\alpha^{ab}\in\mathbb{R}$. The condition $\det S(t)=1$
implies $Tr(s)=0$ and hence $\alpha_0=0$. We conclude that all three
Lie algebras are equal and spanned by the given elements.
\end{proof*}

After these preparations we can now turn to the relation between the
Pin group and the Lorentz group. We define a mapping
$\map{\Lambda}{Pin_{1,3}}{\mathcal{L}}$ by
$S\mapsto \Lambda^a_{\ b}(S)$ such that
\begin{equation}\label{def_lambda}
Sg_bS^{-1}=g_a\Lambda^a_{\ b}(S).
\end{equation}
The matrix $\Lambda^a_{\ b}(S)$ exists by proposition \ref{spineq},
is unique and determines a Lorentz transformation because the adjoint
action of $S$ leaves the right-hand side of $vw+wv=2v^aw_aI$ invariant.

\begin{proposition}\label{covering}
The map $\Lambda$ defined in equation (\ref{def_lambda}) is a surjective
double
covering homomorphism of Lie groups, which restricts to a double covering
homomorphism $Spin^0_{1,3}\rightarrow\mathcal{L}^{\uparrow}_+$. We have:
\begin{eqnarray}
\Lambda^a_{\ b}(S)&=&\frac{1}{4}\eta^{ac}Tr(g_cSg_bS^{-1}),\nonumber\\
\Lambda^a_{\ b}(S^{-1})&=&\eta^{ac}\eta_{bd}\Lambda^d_{\ c}(S),
\nonumber\\
(d\Lambda)^{-1}(\lambda^b_{\ a})&=&\frac{1}{4}\lambda^b_{\ a}\eta^{ac}
g_bg_c,\nonumber
\end{eqnarray}
where $d\Lambda$ is the derivative
$\map{d\Lambda}{spin^0_{1,3}}{l_+^{\uparrow}}$ at $S=I$.
\end{proposition}
\begin{proof*}
(Cf. \cite{Lawson+} theorem I.2.10.) To check the homomorphism property
we note that $\Lambda^a_{\ b}(I)=\delta^a_b$ by (\ref{def_lambda}) and
for $S,T\in Pin_{1,3}$:
\begin{eqnarray}
g_a\Lambda^a_{\ c}(ST)&=&STg_cT^{-1}S^{-1}=
S(g_b\Lambda^b_{\ c}(T))S^{-1}\nonumber\\
&=&Sg_bS^{-1}\Lambda^b_{\ c}(T)=g_a\Lambda^a_{\ b}(S)
\Lambda^b_{\ c}(T)\nonumber
\end{eqnarray}
and hence $\Lambda^a_{\ c}(ST)=\Lambda^a_{\ b}(S)\Lambda^b_{\ c}(T)$.
Next we compute
\begin{eqnarray}
\Lambda^a_{\ b}(S)&=&\frac{1}{4}\eta^{ac}Tr(\eta_{cd}\Lambda^d_{\ b}(S)I)=
\frac{1}{8}\eta^{ac}Tr((g_cg_d+g_dg_c)\Lambda^d_{\ b}(S))\nonumber\\
&=&\frac{1}{4}\eta^{ac}Tr(g_cg_d\Lambda^d_{\ b}(S))=
\frac{1}{4}\eta^{ac}Tr(g_cSg_bS^{-1}),\nonumber
\end{eqnarray}
and hence also
\[
\Lambda^a_{\ b}(S^{-1})=\frac{1}{4}\eta^{ac}Tr(g_cS^{-1}g_bS)
=\frac{1}{4}\eta^{ac}\eta_{bd}\eta^{de}Tr(g_eSg_cS^{-1})=
\eta^{ac}\eta_{bd}\Lambda^d_{\ c}(S).
\]
(Of course this also follows from the fact that $\Lambda$ is a group
homomorphism and that $(\Lambda^{-1})^a_{\ b}=\eta^{ac}\eta_{bd}
\Lambda^d_{\ c}$ for $\Lambda\in\mathcal{L}$.) To find $d\Lambda$ we
expand $\Lambda(S)$ for $S=I+\epsilon s+O(\epsilon^2)$
up to second order in $\epsilon$:
\begin{eqnarray}
\Lambda^a_{\ b}(S)&=&\frac{1}{4}\eta^{ac}Tr(g_c(I+\epsilon s)
g_b(I-\epsilon s))+O(\epsilon^2)\nonumber\\
&=&\Lambda^a_{\ b}(I)+\frac{\epsilon}{4}\eta^{ac}
Tr(\left[g_b,g_c\right]s)+O(\epsilon^2),\nonumber
\end{eqnarray}
where we used the cyclicity of the trace. We can now immediately read
off $d\Lambda^a_{\ b}(s)=\frac{1}{4}\eta^{ac}Tr(\left[g_b,g_c\right]s)$.
Notice that $\dim l_+^{\uparrow}=6=\dim spin^0_{1,3}$ (real dimensions).
We will show that the map $\map{L}{l_+^{\uparrow}}{spin^0_{1,3}}$
defined by
\[
L(\lambda^a_{\ b}):=\frac{1}{4}\lambda^a_{\ b}\eta^{bc}g_ag_c
\]
is an inverse of $d\Lambda^a_{\ b}$. First note that
$\lambda^a_{\ b}\eta^{bc}+\lambda^c_{\ b}\eta^{ba}=0$ for
$\lambda^a_{\ b}\in l_+^{\uparrow}$, so $L$ is in the linear span of
$g_ag_b$ with $a<b$ and hence $L$ takes values in $spin^0_{1,3}$ by
lemma \ref{spinspan}. Now we use lemma \ref{traceresults} to compute:
\begin{eqnarray}
d\Lambda^a_{\ b}(L(\lambda^d_{\ e}))&=&\frac{1}{16}\eta^{ac}
\lambda^d_{\ e}\eta^{ef}Tr(\left[g_b,g_c\right]g_dg_f)
=\frac{1}{2}\eta^{ac}\lambda^d_{\ e}\eta^{ef}(\eta_{cd}\eta_{bf}
-\eta_{bd}\eta_{cf})\nonumber\\
&=&\frac{1}{2}(\lambda^a_{\ b}-\eta^{ae}\eta_{bd}\lambda^d_{\ e})
=\lambda^a_{\ b},\nonumber
\end{eqnarray}
where we used the symmetry properties of $\lambda^d_{\ e}$ again in the
last line.

Because $d\Lambda$ is invertible $\Lambda$ is a local diffeomorphism
(using the inverse function theorem). The surjectivity follows as in
the proof of proposition \ref{spineq} by expressing any Lorentz
transformation $\Lambda\in\mathcal{L}$ as a finite product of
reflections in non-null hyperplanes \cite{Artin}.

To find the kernel of $\Lambda$ we suppose that $S\in Pin_{1,3}$ has
$\Lambda^a_{\ b}(S)=\delta^a_{\ b}$. Then, by definition, $Sg_a=g_aS$
and $S=cI$ by the fundamental theorem \ref{fundamentaltheorem}. As
$S\in D$ we see that $c$ must be real by equation (\ref{subalgebra}).
By proposition \ref{spineq} we have $1=\det S=c^4$, so $c=\pm 1$ and
$S=\pm I$. Note that $I=g_0^2$ and $-I=g_1^2$ are both in
$Spin^0_{1,3}\subset Pin_{1,3}$, so the kernel of $\Lambda$ is
$\left\{I,-I\right\}$. It now follows that $\Lambda$ restricts to a
local diffeomorphism of $Spin^0_{1,3}$ onto
$\mathcal{L}_+^{\uparrow}$, which also is a double covering.
\end{proof*}

One can show that $\Lambda$ and its restrictions to $Pin_{1,3}$,
$Spin_{1,3}$, $Spin^0_{1,3}$ are the universal coverings of
$\mathcal{L}$, $\mathcal{L}_+$ and $\mathcal{L}_+^{\uparrow}$,
respectively and all of these are double coverings (see
\cite{Lawson+} chapter I theorem 2.10 and the remarks below).

In the next subsection we will need the following lemma, which
establishes a relationship between $Spin^0_{1,3}$ and matrices
satisfying assumption (\ref{assumption}):
\begin{lemma}\label{ACproperty}
Let $\pi$ be a complex irreducible representation of $D$ and let
$A,C\in GL(4,\mathbb{C})$ satisfy assumption (\ref{assumption}) w.r.t.
$\pi$. Then for all $S\in Spin^0_{1,3}$:
\[
\pi(S)^*A\pi(S)=A,\quad
\pi(S^{-1})C^{-1}\overline{\pi(S)}=C^{-1}.
\]
\end{lemma}
\begin{proof*}
For a unit vector $u=u^ag_a$ we have $u^2=\|u\|^2I=\pm I$ and hence
\[
\pi(u)^*A\pi(u)=u^au^b\pi(g_a)^*A\pi(g_b)=u^au^bA\pi(g_ag_b)=A\pi(u^2)
=\pm A.
\]
By definition \ref{def_spin} we must therefore have
$\pi(S)^*A\pi(S)=\pm A$ for $S\in Pin_{1,3}$. If $S=I$ the sign
is a plus, so by continuity we conclude that $\pi(S)^*A\pi(S)=A$ for all
$S\in Spin^0_{1,3}$. For $C$ we use the fact that for $u\in M_0$
\[
\pi(u^{-1})C^{-1}\overline{\pi(u)}=-\pi(u)^{-1}\pi(u)C^{-1}=-C^{-1}
\]
and hence $\pi(S^{-1})C^{-1}\overline{\pi(S)}=C^{-1}$ for all
$S\in Spin_{1,3}$, because $S$ is a product of an even number of
$u$'s.
\end{proof*}

Note that $g_5\in Spin_{1,3}\setminus Spin^0_{1,3}$. Indeed, using
$\pi_0$ and $A=A_0=\gamma_0$ in lemma \ref{ACproperty} we see that
$\gamma_5^*A_0\gamma_5=-A_0$, so $g_5$ is in $Spin_{1,3}$ by definition,
but not in $Spin^0_{1,3}$ by the lemma.

\subsection{The Dirac spinor and cospinor bundles}\label{subs_geo}

After presenting the algebraic and group theoretical background
information in the previous subsections we will now start the
formulation of the classical Dirac field in curved spacetime. Our first
task will be to construct the vector bundles in which the Dirac spinor
and cospinor fields take values. For this purpose we choose an
irreducible complex representation $\pi$ of $D$ and matrices
$A,C\in GL(4,\C)$ satisfying assumption (\ref{assumption}). Such matrices
exist by theorem \ref{ACmatrix} and we will show afterwards, in proposition
\ref{equivalence}, that different choices give rise to equivalent
constructions.

Let $M=(\mathcal{M},g,SM,p)$ be a globally hyperbolic spin spacetime.
We define the associated vector bundle
\[
DM:=SM\times_{Spin^0_{1,3}}\C^4,
\]
where $Spin^0_{1,3}$ acts on $SM$ from the right as usual and on $\C^4$ from
the left via the representation $\pi$. In other words, $DM$ is obtained from
the product bundle $SM\times\C^4$ by identifying\footnote{The claim of
\cite{Fewster+} that the map $L_S[E,z]:=[E,\pi(S)z]$ defines a left
action is to be understood as follows. If we fix the local section
$E$ of $SM$, i.e. if we choose a local gauge, then the right-hand
side is well-defined and defines a left action.}
\[
[E,z]=[R_SE,\pi(S^{-1})z],
\]
where we think of $z\in\C^4$ as a column vector. (Recall that $R_S$
denotes the right action of the group $Spin^0_{1,3}$ on the principal
vector bundle $SM$, see definition \ref{def_sman}.) We denote the dual
vector bundle by
$D^*M$ and note the equivalence relation $[E,w^*]=[R_SE,w^*\pi(S)]$, where we
used the standard anti-isomorphism $w\mapsto w^*:=\langle w,.\rangle$
between $\C^4$ and its dual $(\C^4)^*$ and we treat $w^*$ as a row vector.
There is then a canonical pairing of the fibers of $DM^*$ and $DM$ over any
point in $M$, which is given by:
\[
\langle [E,w^*],[E,z]\rangle:=w^*(z)=\langle w,z\rangle.
\]
Note that the element $E$ must be the same in both entries. This can always
be accomplished by using the equivalence relation of $DM$ or $D^*M$, because
the action of $Spin^0_{1,3}$ on each fiber of $SM$ is transitive.

\begin{definition}
The vector bundle $DM$ is the \emph{(Dirac) spinor bundle}, its elements
are \emph{(Dirac) spinors} and a section of it is a \emph{(Dirac) spinor
field}. The space of all smooth spinor fields is denoted by
$\Test(DM)$, and the space of all compactly supported smooth spinor fields
by $\Test_0(DM)$.

The dual vector bundle $D^*M$ of $DM$ is called the \emph{Dirac cospinor
bundle}, its elements are \emph{(Dirac) cospinors} and a section of it
is a \emph{(Dirac) cospinor field}. The space of all smooth cospinor
fields is denoted by $\Test(D^*M)$ and that of the compactly supported
smooth cospinor fields by $\Test_0(D^*M)$.
\end{definition}
We indicate the canonical pairing of a spinor field $u$ and a cospinor
field $v$ by writing them next to each other:
$vu(x):=\langle v(x),u(x)\rangle$. This pairing therefore defines a
sesquilinear map $\Test(D^*M)\times\Test(DM)\rightarrow\Test(M)$.
As a matter of notation we will write $-[E,z]:=[E,-z]$ and
$-[E,z^*]:=[E,-z^*]$, which is well-defined because $-I$ commutes with
the action of $\pi(D)$ on $\C^4$.

We now turn to the adjoint and charge conjugation maps. We first define
these maps for spinors and cospinors, then for spinor and cospinor
fields.

\begin{lemma}\label{+cidentities}
We can define maps $\map{^+}{DM}{D^*M}$, $\map{^+}{D^*M}{DM}$,
$\map{^c}{DM}{DM}$ and $\map{^c}{D^*M}{D^*M}$ by:
\[
[E,z]^+:=[E,z^*A]\quad [E,z^*]^+:=[E,A^{-1}z]
\]
\[
[E,z]^c:=[E,C^{-1}\overline{z}],\quad [E,z^*]^c:=[E,\overline{z}^*C].
\]
These maps are
base-point preserving vector bundle anti-isomorphisms. For
$q=[E,z]\in DM$ and $p=[E,w^*]\in D^*M$ we have:
\begin{eqnarray}\label{ac_rel}
q^{++}=q=q^{cc}&&p^{++}=p=p^{cc}\nonumber\\
q^{+c}=-q^{c+}&&p^{+c}=-p^{c+}\\
\langle q^+,p^+\rangle=&\overline{\langle p,q\rangle}&
=\langle p^c,q^c\rangle.\nonumber
\end{eqnarray}
\end{lemma}
\begin{proof*}
It follows from lemma \ref{ACproperty} that the maps $^+$ and $^c$ are
well-defined. As an example we compute for all $S\in Spin^0_{1,3}$:
\begin{eqnarray}
[R_SE,\pi(S^{-1})z]^+&=&[R_SE,z^*\pi(S^{-1})^*A]=
[R_SE,z^*A\pi(S)]\nonumber\\
&=&[E,z^*A]=[E,z]^+.\nonumber
\end{eqnarray}
The proof for the other maps is similar. By their definition the maps
$^+$ and $^c$ are seen to be anti-linear and to preserve the base-point.
That they are isomorphisms follows from the relations (\ref{ac_rel}),
which we will prove next.

From assumption (\ref{assumption}) we see that $A=A^*$,
$C^{-1}\overline{C^{-1}}=(\overline{C}C)^{-1}=I$ and $\overline{C}C=I$,
so for any $q=[E,z]$ in $DM$ and $p=[E,w^*]$ in $D^*M$ we find:
\begin{eqnarray}
q^{++}&=&[E,z^*A]^+=[E,A^{-1}A^*z]=[E,z]=q\nonumber\\
p^{++}&=&[E,A^{-1}w]^+=[E,w^*(A^{-1})^*A]=[E,w^*]=p\nonumber\\
q^{cc}&=&[E,C^{-1}\overline{z}]^c=[E,C^{-1}\overline{C^{-1}}z]
=[E,z]=q\nonumber\\
p^{cc}&=&[E,\overline{w}^*C]^c=[E,w^*\overline{C}C]=[E,w^*]=p.\nonumber
\end{eqnarray}
From theorem \ref{ACmatrix} we find that $A=-C^*\overline{A}C$, and
hence
\begin{eqnarray}
q^{+c}&=&[E,z^*A]^c=[E,\overline{z}^*\overline{A}C]=
-[E,\overline{z}^*(C^*)^{-1}A]\nonumber\\
&=&-[E,C^{-1}\overline{z}]^+=-[E,z]^{c+}=-q^{c+}.\nonumber
\end{eqnarray}
The result $p^{+c}=-p^{c+}$ now follows, because $p=q^+$ for some $q$
and hence $p^{+c}=q^c$ whereas $p^{c+}=q^{+c+}=-q^{c++}=-q^c$. Finally,
$\langle p,q\rangle=w^*(z)$ and hence:
\begin{eqnarray}
\langle q^+,p^+\rangle&=&
\langle [E,z^*A],[E,A^{-1}w]\rangle=z^*(w)=\overline{w^*(z)}=
\overline{\langle p,q\rangle}\nonumber\\
\langle p^c,q^c\rangle&=&
\langle [E,\overline{w}^*C],[E,C^{-1}\overline{z}]\rangle=
\overline{w}^*(\overline{z})=\overline{w^*(z)}=
\overline{\langle p,q\rangle}.\nonumber
\end{eqnarray}
\end{proof*}

\begin{definition}\label{def_ac}
The maps $^+$ are called the \emph{(Dirac) adjoint} maps and the maps
$^c$ are called the \emph{charge conjugation} maps.

For spinor and cospinor fields we define the adjoint maps and the charge
conjugation maps pointwise.
\end{definition}
This means that for $u\in \Test(DM)$ we have e.g. $u^+(x):=u(x)^+$. Notice
that
the adjoint and charge conjugation maps preserve the support. The
identities (\ref{ac_rel}) of lemma \ref{+cidentities} translate as:
\begin{eqnarray}\label{ac_rel2}
u^{++}=u=u^{cc}&&v^{++}=v=v^{cc}\nonumber\\
u^{+c}=-u^{c+}&&v^{+c}=-v^{c+}\\
u^+v^+=&\overline{vu}&=v^cu^c.\nonumber
\end{eqnarray}

Taking tensor products of $DM,D^*M,TM,T^*M$ we can form a mixed
spinor-tensor algebra in a natural way. In order to perform
computations in this mixed spinor-tensor algebra it will be useful
to work in suitable local frames, which we will now describe.

Given a local section
$E$ of $SM$ and an orthonormal basis $b_A$ of $\C^4$ such that
$\overline{b_A}=b_A$ we obtain local frames $E_A:=[E,b_A]$ of $DM$
and $p\circ E=e=\left\{e_a\right\}_{a=0,\ldots 3}$ of $TM$, where
$\map{p}{SM}{FM}$ is the projection of the spin structure.
We denote the dual frames of $e_a$ and
$E_A$ by $e^b$ and $E^B$, respectively, so that
$e^b(e_a)=\delta^b_a$ and $E^BE_A=\delta^B_A$, where the Kronecker
$\delta$'s are regarded as constant functions on $M$. Together these
local frames give rise to local frames for the spinor-tensor algebra.

A different local section $E'$ of $SM$ over the same region
$O\subset\mathcal{M}$ can always be expressed as $E'=R_{S^{-1}}E$, where
we allow $S$ to depend on $x\in O$, i.e. $\map{S}{O}{Spin^0_{1,3}}$. To
find the corresponding change of frames we compute:
\[
E'_A=[E',b_A]=[R_{S^{-1}}E,b_A]=[E,b_B\pi(S^{-1})^B_{\ A}]=
E_B\pi(S^{-1})^B_{\ A}.
\]
It then follows that
\[
E'_A=E_B\pi(S^{-1})^B_{\ A},\quad (E')^A=\pi(S)^A_{\ B}E^B
\]
\[
e'_a=e_b\Lambda(S^{-1})^b_{\ a},\quad (e')^a=\Lambda(S)^a_{\ b}e^b.
\]
The components of a spinor $u=E_Au^A=E'_A(u')^A$ transform under a
change of section as $(u')^A=\pi(S)^A_{\ B}u^B$ and for general
spinor-tensors we get similar expressions, e.g.
\[
(T')^{Aa}_{Bb}=\pi(S)^A_{\ C}\Lambda(S)^a_{\ c}
\pi(S^{-1})^{D}_{\ B}\Lambda(S^{-1})^d_{\ b}T^{Cc}_{Dd}.
\]

For the frame $e_a$ of $TM$ we can use
$g_{\mu\nu}e_a^{\mu}e_b^{\nu}=\eta_{ab}$ to derive
\begin{equation}\label{coframe}
e^a_{\mu}=g_{\mu\nu}\eta^{ab}e_b^{\nu},
\end{equation}
because both sides have the same action on basis vectors. It
follows that $\eta_{ab}e^a_{\mu}e^b_{\nu}=g_{\mu\nu}$,
$g^{\mu\nu}e^a_{\mu}e^b_{\nu}=\eta^{ab}$ and
$\eta^{ab}e_a^{\mu}e_b^{\nu}=g^{\mu\nu}$. A vector $x$ can be
expressed as $x=x^ae_a$ where $x^a:=e^a_{\mu}x^{\mu}$ and
similarly for covectors. We see that we can raise and lower indices
in the frame $e_a$ with $\eta^{ab}$ and $\eta_{ab}$.

Using the matrix expressions of $A,A^{-1},C^{-1},C$ in the bases
$b_A$, $b^A$ we can also express the adjoint and charge conjugation
maps in components. Because of $E_A^+=\delta_{AB}A^B_{\ C}E^C$ and
$E_A^c=E_B(C^{-1})^B_{\ A}$ we find:
\[
\begin{array}{lll}
u^+_A=\overline{u^C}\delta_{CB}A^B_{\ A}&\ &
(u^c)^A=(C^{-1})^A_{\ B}\overline{u^B},\\
(v^+)^A=(A^{-1})^A_{\ B}\delta^{BC}\overline{v_C}&\ &
v^c_A=\overline{v_B}C^B_{\ A}.
\end{array}
\]

\begin{lemma}\label{lemdefgamma}
There is a smooth section
$\gamma\in C^{\infty}(T^*M\otimes DM\otimes D^*M)$ such that:
\[
\gamma=\gamma_{a\ \ A}^{\ B}e^a\otimes E_B\otimes E^A,
\]
for every local section $E$ of $SM$, where $\gamma_{a\ \ A}^{\ B}$
denotes the entries of $\pi(g_a)$ in the basis $b^A$ and its dual basis
$b_B$.
\end{lemma}
\begin{proof*}
Using a different local section $E'=R_{S^{-1}}E$ and, using the
definitions of $E^A$, $E_B$, $e^a$ and that of $\Lambda$ in equation
(\ref{def_lambda}), we see that we can define $\gamma$ locally by the
given expression, independent of the choice of $E$. Covering the
manifold by suitable regions (e.g. contractible ones) we can then extend
$\gamma$ to a global section, which is given by the formula in any local
frame constructed from a local section $E$ of $SM$.
\end{proof*}

When we perform computations in components we will ease the notation
considerably by dropping the spinorial (capital) indices and using a
matrix notation instead, whenever this is possible. In this notation we
think of spinors as column vectors and cospinors as row vectors, so
$\gamma^A_{a\ B}u^B$ becomes $\gamma_au$ and $v_A\gamma^A_{a\ B}$
becomes $v\gamma_a$. This should cause no confusion, as long as we
remember which objects carry spinor indices and we are careful with the
non-commutative matrix products. This only works as long as no object
carries more than one upper or lower spinorial index, but this will
usually be the case in what follows. For vector fields $v$ and covector
fields $k$ we also introduce the Feynman slash notation:
$\fsl{v}:=v^a\gamma_a$, $\fsl{k}:=k_a\gamma^a$.

\subsection{The spin connection, Dirac operator and Dirac equation}\label{subs_Deq}

The dynamics of the free Dirac field is described by a partial differential
equation which contains a covariant derivative for sections of the Dirac
spinor bundle $DM$. There is a natural choice of a connection for this bundle,
which is related to the Levi-Civita connection on the tangent bundle $TM$.
The latter is therefore the  starting point of this subsection.

Let $\nabla$ be the Levi-Civita connection, i.e. the unique connection on
$TM$ which is torsion free and compatible with the metric. In local
coordinates we can define the Christoffel symbols
$\Gamma^{\rho}_{\ \mu\nu}$ in terms of the coordinate derivatives through
\[
\nabla v=(\nabla_{\mu}v^{\rho})dx^{\mu}\otimes
\frac{\partial}{\partial x^{\rho}}=
(\partial_{\mu}v^{\rho}+\Gamma^{\rho}_{\ \mu\nu}v^{\nu})dx^{\mu}
\otimes\frac{\partial}{\partial x^{\rho}}.
\]
These Christoffel symbols are given by the expression (\cite{Wald}
equation (3.1.30))
\begin{equation}\label{Christoffel}
\Gamma^{\rho}_{\ \mu\nu}=\frac{1}{2}g^{\rho\sigma}
(\partial_{\mu}g_{\nu\sigma}+\partial_{\nu}g_{\mu\sigma}-
\partial_{\sigma}g_{\mu\nu}).
\end{equation}
We may define connection coefficients $\Gamma^a_{\ bc}$ for any local
frame $\left\{e_a\right\}_{a=0,\ldots, 3}$ through
\begin{equation}\label{connectioncoeffs}
\nabla v=(\nabla_bv^a)e^b\otimes e_a=(\partial_b v^a+\Gamma^a_{\ bc}v^c)
e^b\otimes e_a,
\end{equation}
where $\partial_b=e_b^{\mu}\partial_{\mu}$ denotes the action of the
vector field $e_b$ as a derivative. The connection coefficients
$\Gamma^a_{\ bc}$ may be compared to the Christoffel symbols in a
coordinate basis\footnote{It is important to note that our indices
are not abstract indices, that only indicate the type of a tensor, but
actually number the specific vector fields of a frame.}, which yields:
\[
\partial_bv^a+\Gamma^a_{\ bc}v^c=(\partial_{\mu}v^{\rho}+
\Gamma^{\rho}_{\ \mu\nu}v^{\nu})e_b^{\mu}e^a_{\rho}
=\partial_b(v^{\rho}e^a_{\rho})-v^{\rho}\partial_be^a_{\rho}
+\Gamma^{\rho}_{\ \mu\nu}v^{\nu}e_b^{\mu}e^a_{\rho}
\]
and hence, using
$\partial_b(e^a_{\rho}e_c^{\rho})=\partial_b\delta^a_c=0$,
\begin{equation}\label{linkconnections}
\Gamma^a_{\ bc}=-e_c^{\rho}\partial_be^a_{\rho}
+e^a_{\rho}e_b^{\mu}e_c^{\nu}\Gamma^{\rho}_{\ \mu\nu}
=e^a_{\rho}\partial_be_c^{\rho}
+e^a_{\rho}e_b^{\mu}e_c^{\nu}\Gamma^{\rho}_{\ \mu\nu}.
\end{equation}
Equivalently the Levi-Civita connection can be described by the
connection one-forms $\mathbf{\omega}^a_{\ c}:=\Gamma^a_{\ bc}e^b$.
These one-forms can be regarded as a single one-form taking values
in $l_+^{\uparrow}$, because equation (\ref{Christoffel}) implies:
\begin{eqnarray}\label{antisymmetry}
g_{\tau\rho}\Gamma^{\rho}_{\ \mu\nu}+g_{\nu\rho}\Gamma^{\rho}_{\ \mu\tau}
&=&\partial_{\mu}g_{\tau\nu}\quad \Rightarrow\nonumber\\
\eta_{da}\Gamma^a_{\ bc}+\eta_{ca}\Gamma^a_{\ bd}&=&
-e_c^{\rho}\partial_b(e_d^{\sigma}g_{\rho\sigma})
+e_c^{\sigma}g_{\sigma\rho}\partial_be_d^{\rho}
+e_d^{\tau}e_b^{\mu}e_c^{\nu}(g_{\tau\rho}\Gamma^{\rho}_{\ \mu\nu}
+g_{\nu\rho}\Gamma^{\rho}_{\ \mu\tau})\nonumber\\
&=&-e_c^{\rho}e_d^{\sigma}\partial_bg_{\rho\sigma}+
e_d^{\tau}e_c^{\nu}\partial_bg_{\tau\nu}=0.
\end{eqnarray}

To find the spin connection on $DM$ we use a third equivalent
description of the Levi-Civita connection in terms of the principal
$\mathcal{L}_+^{\uparrow}$-bundle $FM$. Indeed, there is an
$l_+^{\uparrow}$-valued one-form $\mathbf{\Omega}^a_{\ c}$ on $FM$
such that for every local section $e$ of $FM$ the pull-back satisfies
$e^*\mathbf{\Omega}^a_{\ c}=\mathbf{\omega}^a_{\ c}$ and which behaves
in a particular way under the action of $\mathcal{L}_+^{\uparrow}$.
We refer to \cite{Kobayashi+} chapter 2 proposition 1.1 for the detailed
description of this behaviour, because it is not essential for our
discussion.
The one-form $\mathbf{\Omega}^a_{\ c}$ can be pulled back by
$\map{p}{SM}{FM}$ and lifted from $l_+^{\uparrow}$ to $spin^0_{1,3}$,
which yields a $spin^0_{1,3}$-valued one-form $\mathbf{\Sigma}$ on $SM$:
\[
\mathbf{\Sigma}:=(d\Lambda)^{-1}p^*(\mathbf{\Omega}^a_{\ c})
=\frac{1}{4}p^*(\mathbf{\Omega}^a_{\ c})\gamma_a\gamma^c
\]
(see proposition \ref{covering}), where we used
$\gamma^c=\eta^{cb}\gamma_b$. Because $p$ intertwines the actions of
the structure groups appropriately $\mathbf{\Sigma}$ defines a
connection on $DM$ (see \cite{Kobayashi+} loc.~cit.), which we call
the spin connection. In
a local section $E$ of $SM$ the spin connection one-forms
$\sigma^A_{b\ C}$ are given by the pull-back of $\mathbf{\Sigma}$ by
$E$. Because of $E^*p^*=(p\circ E)^*=e^*$ we obtain\footnote{Note the
mistaken sign in the expression for the spin connection in
\cite{Dimock2,Fewster+}. With the wrong sign we do not obtain a
connection on $SM$, because one of the properties of \cite{Kobayashi+}
proposition 1.1 is not satisfied, and lemma \ref{lem_covgamma}
would no longer hold.}:
\begin{equation}\label{spinconnection}
\sigma_b=\frac{1}{4}\Gamma^a_{\ bc}\gamma_a\gamma^c.
\end{equation}
We define the covariant derivative of spinor fields $u$ by\footnote{
Here again it is important that our indices are not abstract indices,
but denote the components in specific frames.}
\[
\nabla u=(\nabla_bu^A)e^b\otimes E_A=
(\partial_bu^A+\sigma^{\ A}_{b\ C}u^C)e^b\otimes E_C,
\]
and for cospinor fields $v$ via
$\partial_a(vu)=\nabla_avu+v\nabla_au$, i.e.
\[
\nabla v=(\nabla_bv_C)e^b\otimes E^C=(\partial_bv_C-
v_A\sigma^{\ A}_{b\ C})e^b\otimes E^C,
\]
In components, using the shorthand matrix notation, these definitions
read:
\begin{equation}\label{spinnabla}
\nabla_bu=\partial_bu+\sigma_bu,\quad
\nabla_bv=\partial_bv-v\sigma_b.
\end{equation}

We can now define a covariant derivative $\nabla$ for mixed
spinor-tensors as follows. For spinorial indices we use the spin
connection as in equation (\ref{spinnabla}). For tensor indices we
use the Levi-Civita connection as usual. The following lemma gives a
typical illustration:
\begin{lemma}\label{lem_covgamma}
The section $\gamma$ is covariantly constant.
\end{lemma}
\begin{proof*}
Because the coefficients $\gamma^{\ A}_{a\ B}$ in a local frame are
constant we have
\[
\nabla_b\gamma^{\ A}_{a\ B}=\sigma^{\ A}_{b\ D}\gamma^{\ D}_{a\ B}
-\Gamma^c_{\ ba}\gamma^{\ A}_{c\ B}-\sigma^{\ D}_{b\ B}\gamma^A_{a\ D},
\]
or, dropping the spinor indices and using equation
(\ref{spinconnection}):
\begin{eqnarray}
\nabla_b\gamma_a&=&\sigma_b\gamma_a-\gamma_a\sigma_b
-\Gamma^c_{\ ba}\gamma_c=\frac{1}{4}\Gamma^c_{\ bd}
(\gamma_c\gamma^d\gamma_a-\gamma_a\gamma_c\gamma^d)-
\Gamma^c_{\ ba}\gamma_c\nonumber\\
&=&\frac{1}{4}\Gamma^c_{\ bd}(\gamma_c\left\{\gamma^d,\gamma_a\right\}
-\left\{\gamma_a,\gamma_c\right\}\gamma^d-4\delta^d_a\gamma_c)\nonumber\\
&=&-\frac{1}{2}\Gamma^c_{\ bd}(\delta^d_a\gamma_c+\eta_{ac}\gamma^d)
=0.\nonumber
\end{eqnarray}
Here we used the Clifford relations (\ref{Clifford}), which also hold
pointwise for the section $\gamma_a$, and the
anti-symmetry of the $l_+^{\uparrow}$-valued connection one-form
(\ref{antisymmetry}).
\end{proof*}

\begin{remark}
We warn the reader for the following. When applying $\nabla$ to a
spinor-tensor $T$ we find the usual expression, involving the
coordinate derivative $\partial T$ and a term for each index. Because we
often leave the spinorial indices implicit to ease our notation, it is
tempting to forget the corresponding terms in the expression for
$\nabla T$.
\end{remark}

We now define the following operators:
\begin{definition}
The \emph{Dirac operator}
$\,\nabla\!\!\!\!\!\!\;/ :\!\Test(DM)\!\rightarrow\!\Test(DM)$
is the first order partial differential operator defined by
$\dirop:=\gamma^a\nabla_a$, where we view $\gamma^a$
as a map from $DM$ to itself, acting on the left.

The \emph{Dirac operator}
$\,\nabla\!\!\!\!\!\!\;/ :\!\Test(D^*M)\!\rightarrow\!\Test(D^*M)$
is defined by the same expression, $\dirop:=\gamma^a\nabla_a$, where we
now view $\gamma^a$ as a map from $D^*M$ to itself, acting on the
right.
\end{definition}
When expressed in a local frame the Dirac operator on spinor and
cospinor fields is given by:
\begin{eqnarray}
\dirop u&=&E_A(\dirop u)^A=E_A\gamma^{aA}_{\ \ \ B}\nabla_au^B
=E_A\gamma^{aA}_{\ \ \ B}(\partial_au^B+\sigma^{\ B}_{a\ C}u^C),
\nonumber\\
\dirop v&=&(\dirop v)_AE^A=(\nabla_av_B)\gamma^{aB}_{\ \ A}E^A=
(\partial_av_B-v_C\sigma^{\ C}_{a\ B})\gamma^{aB}_{\ \ A}E^A,\nonumber
\end{eqnarray}
or dropping the spinorial indices:
\begin{eqnarray}
\dirop u&=&\gamma^a\nabla_au=\gamma^a(\partial_au+\sigma_au),\nonumber\\
\dirop v&=&\nabla_av\gamma^a=(\partial_av-v\sigma_a)\gamma^a.\nonumber
\end{eqnarray}

\begin{definition}
The \emph{Dirac equation} for $u\in\Test(DM)$, respectively for
$v\in\Test(D^*M)$, is
\begin{equation}\label{Diraceqn}
(-i\dirop+m)u=0,\quad (i\dirop +m)v=0,
\end{equation}
for a constant mass $m\ge 0$.
\end{definition}
Note that the spinor field $u$ is a solution to the Dirac equation if
and only if the cospinor field $u^+$ is a solution, because of the
following lemma:
\begin{lemma}\label{swapop}
For all spinor fields $u$ and cospinor fields $v$ we have
\begin{eqnarray}
(\dirop u)^+=\dirop u^+,&&(\dirop v)^+=\dirop v^+,\nonumber\\
(\dirop u)^c=-\dirop u^c,&&(\dirop v)^c=-\dirop v^c\nonumber
\end{eqnarray}
and $u^+\fsl{n}u\ge 0$ everywhere on $M$, for any future pointing time-like
vector field $n$.
\end{lemma}
\begin{proof*}
Using assumption (\ref{assumption}) and the fact that the entries of
$A$ and $C$ are constant we can compute in a local frame $E$:
\begin{eqnarray}
(\dirop v)^c&=&((\partial_av-v\sigma_a)\gamma^a)^c
=(\partial_a\overline{v}-\overline{v\sigma_a})\overline{\gamma^a}C
\nonumber\\
&=&-(\partial(\overline{v}C)-\overline{v}C\sigma_a)\gamma^a
=-\dirop(\overline{v}C)=-\dirop v^c,\nonumber\\
(\dirop u)^+&=&(\gamma^a(\partial_au+\sigma_au))^+
=(\partial_au^*+u^*\sigma_a^*)(\gamma^a)^*A\nonumber\\
&=&(\partial_a(u^*A)-u^*A\sigma_a)\gamma^a=
\dirop(u^*A)=\dirop u^+,\nonumber
\end{eqnarray}
where the minus sign in the last line appears because the order of the
two factors of $\gamma$ in the expression (\ref{spinconnection})
for $\sigma_a$ needs to be reversed. It follows that
\begin{eqnarray}
(\dirop v)^+&=&(\dirop v^{++})^+=(\dirop v^+)^{++}=\dirop v^+\nonumber\\
(\dirop u)^c&=&(\dirop u^+)^{+c}=-(\dirop u^+)^{c+}=(\dirop u^{+c})^+
=-(\dirop u^{c+})^+=-\dirop u^c.\nonumber
\end{eqnarray}
Finally, if $u(x)=[E,z]$ at $x\in M$, then
$u^+(x)\fsl{n}(x)u(x)=\langle z,A\fsl{n}(x) z\rangle\ge 0$, because
$\fsl{n}(x)$ is just $n(x)$ considered as an element of the Dirac
algebra.
\end{proof*}

We have now used all parts of assumption (\ref{assumption}) to define
the adjoint and charge conjugation maps and to establish their
interrelations with each other (equation (\ref{ac_rel2})), with the
Dirac equation and with the time-orientation of the spacetime (lemma
\ref{swapop}). The next proposition shows that these relations
completely characterise the Dirac field, independent of the choice
of representation and of the matrices $A,C\in GL(4,\C)$ satisfying
assumption (\ref{assumption}).

\begin{proposition}\label{equivalence}
Consider the Dirac spinor bundle $DM_0$ and cospinor bundle $D^*M_0$,
defined analogously to $DM$ and $D^*M$ but using $\pi_0$ instead of
$\pi$. Let $^{\dagger}$ and $^-$ be defined
analogously to $^+$ and $^c$, using the matrices $A_0=\gamma_0$ and
$C_0=\gamma_2$, and let $\dirop_0$ be the Dirac operator defined through
the representation $\pi_0$. Then there exists a base-point preserving,
vector bundle isomorphism $\map{\lambda}{DM}{DM_0}$ with induced
isomorphism $\map{\lambda^*}{D^*M}{D^*M_0}$ such that
$\lambda\circ^+=^{\dagger}\circ\lambda^*$,
$\lambda\circ^c=^-\circ\lambda$ and
$\lambda\circ\dirop=\dirop_0\circ\lambda$. This isomorphism is unique,
up to an overall sign.
\end{proposition}
\begin{proof*}
On each fiber the bundle isomorphism $\lambda$ must be given by the
formula $\lambda:[E,z]\mapsto [E,Lz]_0$ for some $L\in GL(4,\C)$ by the
fundamental theorem \ref{fundamentaltheorem}. The
induced morphism $\lambda^*$ is then
$\lambda^*:[E,z^*]\mapsto [E,z^*L^{-1}]_0$. To make $\lambda$
well-defined and to obtain the correct intertwining with adjoint
operation, charge conjugation and the Dirac operator we need:
\[
LA^{-1}=A_0(L^*)^{-1},\quad LC^{-1}=C_0^{-1}\overline{L},\quad
L\pi=\pi_0L.
\]
Using the fact that $A_0=A_0^{-1}$ we can apply theorem \ref{ACmatrix}
to conclude that $L$ exists and is unique up to a sign. By continuity
$L$ must be locally constant on $M$, but then the map
$\lambda:[E,z]\mapsto[E,Lz]_0$ is globally well-defined and is unique up
to a global sign (by connectedness of $M$).
\end{proof*}
The bundle isomorphisms which intertwine the relations between the
adjoint operation, charge conjugation and the Dirac operator form a
group. Because this group preserves these relations it will also
leave the theory's predictions invariant, so we can think of this group
as a gauge group for the Dirac field. Choosing a specific representation
is then like fixing a gauge, leaving as a residual gauge freedom only
the involutive bundle isomorphism $u\mapsto -u$. We will divide out
this residual gauge freedom after quantisation of the Dirac field.

A change in the signature convention, $\tilde{\eta}:=-\eta$, should not
have any physical consequences either and indeed it can be seen that
this is easily compensated for by changing the sign in equation
(\ref{Clifford}). This does not change the Dirac algebra and any other
constructions that follow from it, although we do get signs for all
covectors when raising or lowering indices with
$\tilde{\eta}$.\footnote{Alternatively we could set
$\tilde{\gamma}_a:=i\gamma_a$, in which case the Dirac algebra would
become $D\simeq Cl_{3,1}$. Because $Cl^0_{3,1}=Cl^0_{1,3}$ we have
$Spin^0_{1,3}=Spin^0_{3,1}$, so nothing changes in the representation of
the Spin group. We can also keep the same matrices $A,C$, which now must
satisfy the relations $-\tilde{\gamma}_a^*=A\tilde{\gamma}_aA^{-1}$ and
$\overline{\tilde{\gamma}}_a=C\tilde{\gamma}_aC^{-1}$.
The spinor and cospinor bundle and the adjoint and charge conjugation
maps remain the same as before. All relations between these operations
and the Dirac equation remain valid if we drop the factor $i$ in front
of the Dirac operator in the Dirac equation.

Notice that a complex
irreducible representation of $Cl_{1,3}$ extends to an irreducible
representation of $M(4,\C)$ and therefore also gives a complex
irreducible representation of $Cl_{3,1}$ and vice versa. The standard
Clifford algebra isomorphism $Cl_{3,1}\simeq M(4,\R)$ appears if and
only if the representation of $Cl_{1,3}$ is a Majorana representation,
i.e. iff $\overline{\gamma}_a=-\gamma_a$. In that case we also find
\[
Pin_{3,1}\simeq \left\{S\in M(4,\R)| \det S=1,
\forall v\in M_0 SvS^{-1}\in M_0\right\}\not=Pin_{1,3}.
\]
The group $Pin_{3,1}$ is also sometimes called Spin group (see e.g.
\cite{Choquet+} p.334).}

\section{The free Dirac field as a LCQFT}\label{sec_LCDf}

Now that we have described spinor and cospinor fields and their equation
of motion, the Dirac equation, it is time to quantise the theory. In
subsection \ref{quantisation} we will describe a suitable space of
classical solutions and quantise the theory on a single spin spacetime,
largely following \cite{Dimock2,Dawson+,Fewster+}.
Then we consider Hadamard states and states with the microlocal spectrum
condition in subsection \ref{subs_DHad}, obtaining the new result that
every Hadamard state satisfies the microlocal spectrum condition, just
as in the scalar field case. Finally we show that that we indeed
obtain a locally covariant quantum field theory in subsection
\ref{subs_DHad} and we establish some of its properties. There we also
note that different choices of representation give rise to equivalent
theories.

\subsection{Quantisation}\label{quantisation}

It will be convenient to deal with spinor and cospinor fields
simultaneously, for which purpose we introduce the following:
\begin{definition}\label{def_double}
The \emph{double spinor bundle} is defined as the vector bundle
$DM\oplus D^*M$.
A \emph{double spinor} (field) is a smooth section of this vector
bundle. The space of double spinors will be denoted by
$\mathcal{D}(M):=\Test(DM\oplus D^*M)$.

The space of \emph{double test-spinors} is the space of
compactly supported smooth double spinors
$\mathcal{D}_0(M):=\Test_0(DM\oplus D^*M)$ in the topology of
uniform convergence on a fixed compact set.

We define the adjoint map and charge conjugation map as anti-linear
isomorphisms from $DM\oplus D^*M$ to itself by
$[E,z,w^*]^+:=[E,A^{-1}w,z^*A]$ and
$[E,z,w^*]^c:=[E,C^{-1}\overline{z},\overline{w}^*C]$, respectively.
For double spinors we define the adjoint and charge conjugation maps
pointwise, i.e. $(u\oplus v)^+:=v^+\oplus u^+$ and
$(u\oplus v)^c:=u^c\oplus v^c$. We also
introduce the operators $D:=(-i\dirop+m)\oplus(i\dirop+m)$ and
$\tilde{D}=(i\dirop+m)\oplus(-i\dirop+m)$.
\end{definition}
Note that the Dirac equation (\ref{Diraceqn}) translates as $Df=0$ for
$f\in\mathcal{D}(M)$.
Every $u_1\oplus v_1\in\mathcal{D}(M)$ determines a distribution on the
double test-spinors $\mathcal{D}_0(M)$ by the sesquilinear pairing
\begin{equation}\label{eq_pairing}
\langle u_1\oplus v_1,u_2\oplus v_2\rangle:=
\int_M u_1^+u_2-v_2v_1^+ d\mathrm{vol}_g.
\end{equation}
This pairing is non-degenerate, but not positive (because $A$ itself is
not a positive matrix). The following relations hold between the adjoint
and charge conjugate maps, $D$ and the pairing of equation
(\ref{eq_pairing}):
\begin{lemma}\label{lem_pairprop}
We have for all $f\in\mathcal{D}_0(M)$ and $h\in\mathcal{D}(M)$:
\begin{enumerate}
\item $Dh^+=(Dh)^+$, $Dh^c=(Dh)^c$ and $h^{c+}=-h^{+c}$,
\item $\langle f^+,h^+\rangle=\langle f^c,h^c\rangle=
-\overline{\langle f,h\rangle}=-\langle h,f\rangle$,
\item $f^+,f^c\in\mathcal{D}_0(M)$ and
$\langle f,Dh\rangle=\langle Df,h\rangle$.
\end{enumerate}
\end{lemma}
\begin{proof*}
The first set of equations follows from lemma \ref{swapop}. E.g.
\[
((-i\dirop +m)u)^c=i(\dirop u)^c+(mu)^c=-i\dirop u^c+mu^c=
(-i\dirop +m)u^c
\]
and similarly for cospinors $v$, which implies $Df^c=(Df)^c$. The second
item follows directly from the equations (\ref{ac_rel2}) and
(\ref{eq_pairing}). For the last item we note that
$\mathrm{supp}\ h^c=\mathrm{supp}\ h^+=\mathrm{supp}\ h$ for every
$h\in\mathcal{D}(M)$ by definition \ref{def_ac} and that we can perform
a partial integration as follows \cite{Dimock2}: note that for all
$u\oplus v\in\mathcal{D}(M)$ we have
$\nabla_a(v\gamma^au)=(\dirop v)u+v\dirop u$, because $\gamma$ is
covariantly constant. If either $u$ or $v$ is compactly supported we
can integrate this equation over $M$ to get
$\int_M(\dirop v)u=-\int_Mv\dirop u$. Together with equation
(\ref{eq_pairing}) this implies the result.
\end{proof*}

The second order operator $\tilde{D}D=D\tilde{D}$ has as its principal
part the wave operator $\Box=g^{\mu\nu}\nabla_{\mu}\nabla_{\nu}$, which
is diagonal in the spinorial indices. Due to global hyperbolicity of $M$
there exist\footnote{As a slight abuse of notation we use the same
symbols $E^{\pm}$ here as in chapter \ref{ch_sf}, although the operators
are not the same. (They do not even have the same domain.)}
unique advanced (-) and retarded (+) fundamental solutions
$\map{E^{\pm}}{\mathcal{D}_0(M)}{\mathcal{D}(M)}$ for the operator
$D\tilde{D}$, i.e. for all $f\in\mathcal{D}_0(M)$ we have
$D\tilde{D}E^{\pm}f=f=E^{\pm}\tilde{D}Df$ and
$\mathrm{supp}(E^{\pm}f)\subset J^{\pm}(\mathrm{supp}\ f)$
(see \cite{Baer+} theorem 3.3.1).
The fundamental solutions $E^{\pm}$ help us to find fundamental
solutions for the operator $D$ as follows (see \cite{Dimock2}):
\begin{proposition}\label{prop_fundS}
The maps $\map{S^{\pm}}{\mathcal{D}_0(M)}{\mathcal{D}(M)}$ defined by
$S^{\pm}:=\tilde{D}E^{\pm}$ are the unique advanced ($-$) and retarded
($+$) fundamental (left and right) solutions for $D$ such that
$\mathrm{supp}\ S^{\pm}f\subset J^{\pm}(\mathrm{supp}\ f)$ for all
$f\in\mathcal{D}_0(M)$. Moreover, $S^{\pm}f^+=(S^{\pm}f)^+$,
$S^{\pm}f^c=(S^{\pm}f)^c$ and
$\langle f,S^{\pm}h\rangle=\langle S^{\mp}f,h\rangle$ for all
$f,h\in\mathcal{D}_0(M)$.
\end{proposition}
\begin{proof*}
For $f\in\mathcal{D}_0(M)$ we see that $S^{\pm}f$ has the correct
support property and $DS^{\pm}f=f$, so $S^{\pm}$ is a right fundamental
solution. For the fact they are left fundamental solutions and their
uniqueness we refer to \cite{Dimock2} theorem 2.1. Given
$f,h\in\mathcal{D}_0(M)$ we choose $\chi\in\Test_0(M)$ with
$\chi\equiv 1$ on the compact sets
$J^{\pm}(\mathrm{supp}\ f)\cap J^{\mp}(\mathrm{supp}\ h)$. Using lemma
\ref{lem_pairprop} and the support properties we then compute:
\begin{eqnarray}
\langle f,S^{\pm}h\rangle&=&\langle DS^{\mp}f,S^{\pm}h\rangle
=\langle D(\chi S^{\mp}f),\chi S^{\pm}h\rangle\nonumber\\
&=&\langle \chi S^{\mp}f,D(\chi S^{\pm}h)\rangle=
\langle S^{\mp}f,D S^{\pm}h\rangle=\langle S^{\mp}f,h\rangle.\nonumber
\end{eqnarray}
Finally we compute for all $f,h\in\mathcal{D}_0(M)$:
\begin{eqnarray}
\langle S^{\pm}f^+,h\rangle&=&\langle f^+,S^{\mp}h\rangle
=\langle (DS^{\pm}f)^+,S^{\mp}h\rangle=
\langle D(S^{\pm}f)^+,S^{\mp}h\rangle\nonumber\\
&=&\langle (S^{\pm}f)^+,DS^{\mp}h\rangle=
\langle (S^{\pm}f)^+,h\rangle,\nonumber
\end{eqnarray}
from which it follows that $S^{\pm}f^+=(S^{\pm}f)^+$. The proof for charge
conjugation is similar.
\end{proof*}

Analogous to the scalar field case we define $S:=S^--S^+$. The proof of
proposition \ref{prop_fundS} also works for the spinor and cospinor
cases separately, yielding fundamental solutions $S^{\pm}_{sp}$,
$S_{sp}$, $S^{\pm}_{cosp}$ and $S_{cosp}$ using the obvious notation. By
the uniqueness part of the proposition we find
$S=S_{sp}\oplus S_{cosp}$, $(S_{sp}u)^+=S_{cosp}u^+$ and
$\int_Mv(S_{sp}u)=-\int_M(S_{cosp}v)u$
(see \cite{Dimock2,Fewster+}).

\begin{lemma}\label{positivity}
$\mathrm{ker}\ S=D(\mathcal{D}_0(M))$ and the bilinear map
$(f,h):=i\langle f,Sh\rangle$ defines an inner product
on $\mathcal{D}_0(M)/\mathrm{ker}\ S$. The adjoint and charge conjugation
maps descend to this quotient space too and
$(f^+,h^+)=(f^c,h^c)=\overline{(f,h)}=(h,f)$.
\end{lemma}
\begin{proof*}
If $f=Dh$ for $h\in\mathcal{D}_0(M)$ then $Sf=S^-Dh-S^+Dh=0$, so
$D(\mathcal{D}_0(M))\subset\mathrm{ker}\ S$.
Conversely, if $Sf=0$ with $f\in\mathcal{D}_0(M)$ then
$h:=S^-f=S^+f$ has its support in the compact set
$J^+(\mathrm{supp}\ f)\cap J^-(\mathrm{supp}\ f)$ and $f=Dh$, so
$D(\mathcal{D}_0(M))=\mathrm{ker}\ S$
Now $(f,h)=-i\langle Sf,h\rangle=\overline{(h,f)}$ by proposition
\ref{prop_fundS} and lemma \ref{lem_pairprop}, so $(,)$ is a
well-defined sesquilinear map on $\mathcal{D}_0(M)/\mathrm{ker}\ S$.
By proposition \ref{prop_fundS} again the adjoint and
charge conjugation maps descend to the quotient space
$\mathcal{D}_0(M)/\mathrm{ker}\ S$
and with $^*$ denoting either $^+$ or $^c$ we compute:
\[
(f^*,h^*)=i\langle f^*,Sh^*\rangle=i\langle f^*,(Sh)^*\rangle
=-i\overline{\langle f,Sh\rangle}=\overline{(f,h)}=(h,f).
\]
It remains to show that $(f,f)\ge 0$ for all $f\in\mathcal{D}_0(M)$,
with equality only if $Sf=0$. Leaving the metric induced volume
element implicit in the notation we have,
\[
(u\oplus v,u\oplus v)=i\int_M u^+S_{sp}u-(S_{cosp}v)v^+
=i\int_Mu^+S_{sp}u+vS_{sp}v^+
\]
and for any Cauchy surface $C\subset M$ and $D_{sp}:=-i\dirop+m$:
\begin{eqnarray}
i\int_M u^+S_{sp}u&=&i\int_{J^+(C)}(D_{sp}S_{sp}^-u)^+S_{sp}u+
i\int_{J^-(C)}(D_{sp}S_{sp}^+u)^+S_{sp}u\nonumber\\
&=&i\int_{J^+(C)}(S_{sp}^-u)^+(D_{sp}S_{sp}u)+i\nabla_a((S_{sp}^-u)^+
\gamma^aS_{sp}u)\nonumber\\
&&+i\int_{J^-(C)}(S_{sp}^+u)^+(D_{sp}S_{sp}u)+i\nabla_a((S_{sp}^+u)^+
\gamma^aS_{sp}u)\nonumber\\
&=&-\int_{J^+(C)} \nabla_a((S_{sp}^-u)^+\gamma^aS_{sp}u)
-\int_{J^-(C)} \nabla_a((S_{sp}^+u)^+\gamma^aS_{sp}u)\nonumber\\
&=&\int_C n_a(S_{sp}^-u-S_{sp}^+u)^+\gamma^aS_{sp}u
=\int_C (S_{sp}u)^+\fsl{n}S_{sp}u,\nonumber
\end{eqnarray}
where $n^a$ is the future pointing normal vector field to $C$ and we
used \cite{Wald} equation (B.2.26) for the final partial integration.
The integrand is smooth and pointwise positive by lemma \ref{swapop},
so the result follows.
\end{proof*}
We define $\mathcal{L}(M):=\overline{\mathcal{D}_0(M)/\mathrm{ker}\ S}$
to be the Hilbert space completion in the inner product $(,)$. The
continuous extensions of $^+$ and $^c$ to $\mathcal{L}(M)$ will be
denoted by the same symbol.

In order to quantise the free Dirac field we first define the following.
\begin{definition}\label{boxtimes}
The \emph{exterior tensor product} $\mathcal{V}_1\boxtimes\mathcal{V}_2$
of two vector bundles $\mathcal{V}_i$ over $\mathcal{M}_i$ with fiber
$V_i$, $i=1,2$, is the vector bundle over $\mathcal{M}_1\times\mathcal{M}_2$
whose
fiber is $V_1\otimes V_2$ and whose local trivialisations are determined
by local trivialisations $O_i\times V_i$ of $\mathcal{V}_i$ as
$(O_1\times O_2)\times (V_1\otimes V_2)$.
\end{definition}
We can extend the adjoint and charge conjugation maps from
$DM\oplus D^*M$ to its exterior powers $(DM\oplus D^*M)^{\boxtimes m}$
by anti-linear extension of
\[
(p_1\otimes\ldots\otimes p_m)^+:=(p_1^+\otimes\ldots\otimes p_m^+),
\]
\[
(p_1\otimes\ldots\otimes p_m)^c:=(p_1^c\otimes\ldots\otimes p_m^c),
\]
for all $p_1,\ldots,p_m\in DM\oplus D^*M$.

\begin{definition}
The \emph{Dirac Borchers-Uhlmann algebra} $\Alg{F}_M$ is the direct sum
\[
\Alg{F}_M:=\oplus_{n=0}^{\infty}\Test_0((DM\oplus D^*M)^{\boxtimes n})
\]
(in the algebraic sense), equipped with:
\begin{enumerate}
\item the product $f(x_1,\ldots,x_n)h(x_{n+1},\ldots,x_{n+m}):=
(f\otimes h)(x_1,\ldots,x_{n+m})$, extended linearly,
\item the $^*$-operation
$f(x_1,\ldots,x_n)^*:=f^+(x_n,\ldots,x_1)=(f(x_n,\ldots,x_1))^+$,
extended anti-linearly,
\item a topology such that $f_j=(\oplus_nf_j^{(n)})$ converges to $f=(f^{(n)})$
if and only if for all $n$ we have $f_j^{(n)}\rightarrow f^{(n)}$ in
$\Test_0((DM\oplus D^*M)^{\boxtimes n})$ and for some $N>0$ we have
$f_j^{(n)}=0$ for all $j$ and $n\ge N$.
\end{enumerate}
\end{definition}
Much like the Borchers-Uhlmann algebra for the real scalar field,
$\Alg{F}_M$ is the strict inductive limit
$\Alg{F}_M=\cup_{N=0}^{\infty}\oplus_{n=0}^N
\Test_0((DM\oplus D^*M)|_{K_N}^{\boxtimes n})$, where $K_N$ is an
exhausting (and increasing) sequence of compact subsets of
$\mathcal{M}$, the vector bundle $(DM\oplus D^*M)|_{K_N}^{\boxtimes n}$
is the restriction of $(DM\oplus D^*M)^{\boxtimes n}$ to
$(K_N)^{\times n}$ and each
$\Test_0((DM\oplus D^*M)|_{K_N}^{\boxtimes n})$ is given the
test-function topology, see \cite{Schaefer} theorem 2.6.4.

As in lemma \ref{lem_UTalg} one can show that $\Alg{F}_M$ is a
topological $^*$-algebra and that a continuous state $\omega$ on
$\Alg{F}_M$ consists of a sequence of $n$-point distributions
$(\omega_n)$ acting on the smooth, compactly supported sections of
$(DM\oplus D^*M)^{\boxtimes n}$. Another analogy with the scalar field
case is that $\Alg{F}_M$ does not carry any dynamical information or
anti-commutation relation. As in chapter \ref{ch_sf} this can be
remedied by dividing out a certain ideal:
\begin{definition}
We define the \emph{free Dirac Borchers-Uhlmann algebra} as the
topological $^*$-algebra $\Alg{F}^0_M:=\Alg{F}_m/J$,
where $J\subset\Alg{F}_M$ is the closed $^*$-ideal generated by all
elements of the form $Df$ or $f\otimes h+h\otimes f-(f,h)I$, where
$f,h\in\mathcal{D}_0(M)$.
\end{definition}

We have presented the Dirac Borchers-Uhlmann algebras $\Alg{F}_M$ and
$\Alg{F}^0_M$ to indicate the analogy with the Borchers-Uhlmann algebras
$\Alg{U}_M$ and $\Alg{U}^0_M$ of the real scalar field. However, there is a
more direct way to obtain $\Alg{F}^0_M$, which is analogous to the
algebra $\Alg{A}^0_M$ for the free scalar field. This approach also shows
that $\Alg{F}^0_M$ can be completed to a $C^*$-algebra.

\begin{proposition}\label{prop_FC*}
The algebra $\Alg{F}^0_M$ can be completed to a $C^*$-algebra
$\overline{\Alg{F}}^0_M$, which is the unique $C^*$-algebra generated by
elements $B_M(f)$, $f\in\mathcal{L}(M)$, such that
\begin{enumerate}
\item $f\mapsto B_M(f)$ is $\C$-linear,
\item $B_M(f^+)=B_M(f)^*$,
\item $\left\{B_M(f)^*,B_M(g)\right\}=(f,g)I$.
\end{enumerate}
\end{proposition}

Recall that $f\in\mathcal{L}(M)=\overline{\mathcal{D}_0(M)/\mathrm{ker}\ S}$,
so the equation of motion is implicit.

\begin{proof*}
To prove that $\Alg{F}^0_M$ has a $C^*$-norm we note that it is an
infinite-dimensional Clifford algebra, by the anti-commutation
relations. Each finite dimensional Clifford algebra has a $C^*$-norm,
because it can be represented faithfully as an algebra of bounded
operators \cite{Lawson+}. The algebra $\overline{\Alg{F}}^0_M$ is the
$C^*$-algebraic inductive limit of its finite dimensional subalgebras.
We refer to \cite{Bratteli+} theorem 5.2.5 (or
\cite{Araki} lemma 4.1 and 3.3) for a proof that
$\overline{\Alg{F}}^0_M$ is the unique $C^*$-algebra generated by
elements $B_M(f)$ with $f\in\mathcal{L}(M)$ and satisfying the stated
properties.
\end{proof*}

We formulate a charge conjugation map on the algebra
$\overline{\Alg{F}}^0_M$ as follows:
\begin{proposition}\label{AlphaC}
The map $f\mapsto f^{c+}$ gives rise to a $*$-isomorphism $\alpha_C$ of
$\overline{\Alg{F}}^0_M$ determined by $\alpha_C(B_M(f)):=B_M(f^{c+})$
and we have $\alpha_C^2B_M(f)=-B_M(f)$.
\end{proposition}
\begin{proof*}
Note that $f\mapsto f^{c+}$ is a linear isomorphism of $\mathcal{L}(M)$,
because of $(f^{c+},g^{c+})=(g^c,f^c)=(f,g)$ by lemma \ref{positivity}.
The result then follows (see \cite{Bratteli+} section 5.2.2.1, p.18).
\end{proof*}

We have now quantised the Dirac field on a single spin spacetime. In
subsection \ref{subs_LCDF} we will investigate the properties of this
construction as a locally covariant quantum field theory. The map
$\map{B_M}{\mathcal{D}_0(M)}{\overline{\Alg{F}}^0_M}$ of proposition
\ref{prop_FC*} will then be a candidate for a locally covariant quantum
field. For now we will only introduce the following notation.
\begin{definition}\label{def_psi}
Leaving the quotient map $\mathcal{D}_0(M)\mapsto\mathcal{L}(M)$
implicit, we define the maps
$\map{\psi_M}{\Test_0(D^*M)}{\overline{\Alg{F}}^0_M}$ and
$\map{\psi_M^+}{\Test_0(DM)}{\overline{\Alg{F}}^0_M}$ by
\[
\psi_M(v):=B_M(0\oplus v),\quad \psi_M^+(u):=B_M(u\oplus 0).
\]
We also define
\[
\psi_M^c(v^c):=\alpha_C(B_M(0\oplus v^c))=\psi_M(v)^*,
\]
\[
\psi_M^{+c}(u^c):=\alpha_C(B_M(u^c\oplus 0))=\psi_M^+(u)^*.
\]
\end{definition}

\begin{proposition}\label{prop_psi}
The maps $B_M$, $\psi_M$ and $\psi_M^+$ are distributions valued in
the $C^*$-algebra $\overline{\Alg{F}}^0_M$ and:
\begin{enumerate}
\item $\psi_M^+(u)=\psi_M(u^+)^*$,
\item $\left\{\psi_M^+(u),\psi_M(v)\right\}=(v^+\oplus 0,u\oplus 0)I
=i\int_M v(S_{sp}u) I$ and all the other anti-commutators vanish,
\item $(-i\dirop+m)\psi_M=0$ and $(i\dirop+m)\psi_M^+=0$, where
we have set $(\gamma_aU)(v):=U(v\gamma_a)$ and
$(\nabla_aU)(v):=-U(\nabla_av)$ for any distribution $U$ on smooth
sections of $D^*M$.
\end{enumerate}
\end{proposition}
\begin{proof*}
The first two items follow straightforwardly from the definitions of
$\psi_M$, $\psi_M^+$, $B_M$ and the inner product $(,)$. For the third
we have:
\[
((-i\dirop+m)\psi_M)(v)=\psi_M((i\dirop+m)v)=B_M(D(0\oplus v))=0,
\]
because $SD(0\oplus v)=0$, and similarly for $(i\dirop+m)\psi_M^+=0$.

It remains to show that $B_M$, $\psi_M$ and $\psi_M^+$ are
$C^*$-algebra-valued distributions. The $C^*$-sub-algebra of
$\overline{\Alg{F}}^0_M$
generated by $I,\psi_M(v),\psi_M(v)^*$ is a Clifford algebra which is
isomorphic to
$M(2,\mathbb{C})$ and an explicit isomorphism is given by
$\psi_M(v)\mapsto\left(\begin{array}{cc}0&\sqrt{c}\\ 0&0\end{array}
\right)$, where $c=(0\oplus v,0\oplus v)=i\int_Mv(S_{sp}v^+)>0$. It
follows that $\|\psi_M(v)\|=\sqrt{c}$ is the operator norm of the
corresponding matrix, i.e.\footnote{The factor 2 in \cite{Fewster+}
remark 2, p.340 seems to be erroneous. The sign is due to the fact that
\cite{Fewster+} uses for $S$ the retarded-minus-advanced rather than
the advanced-minus-retarded fundamental solution.}
\[
\|\psi_M(v)\|^2=i\int_Mv(S_{sp}v^+)d\mathrm{vol}_g.
\]
In the test-spinor topology we then have continuous maps
$v\mapsto v\oplus v^+\mapsto i\int_Mv(S_{sp}v^+)$, from which
it follows that $v\mapsto\psi_M(v)$ is norm continuous, i.e. it is
a $C^*$-algebra-valued distribution. The proof for $\psi_M^+$
is analogous and the result for $B_M$ then follows.
\end{proof*}

\begin{remark}
The quantisation of Majorana spinors proceeds in a largely analogous
way. The space of test-spinors
$C_0^{\infty}(DM)/\mathrm{ker}\ S_{sp}$ can be given the inner product
$(u_1,u_2)':=(u_1\oplus 0,u_2\oplus 0)$ and then completed
to a Hilbert space $\mathcal{L}'(M)$. The map $^c$ provides a conjugation
map on this space,
so we can quantise to obtain a $C^*$-algebra $\overline{\Alg{F}}'_M$
generated by elements $B_M'(u)$, $u\in\mathcal{L}'(M)$, satisfying
\begin{enumerate}
\item $u\mapsto B_M'(u)$ is $\mathbb{C}$-linear,
\item $B_M'(u^c)=B_M'(u)^*$ and
\item $\left\{B_M'(u_1)^*,B_M'(u_2)\right\}=(u_1,u_2)'I$.
\end{enumerate}
For cospinors $v$ we can then define $B_M'(v):=B_M'(v^+)^*$, which adds
nothing new to the algebra. However,
$\left\{B_M'(v)^*,B_M'(u)\right\}=(v^{+c},u)'I$, but on the other hand
$\left\{\psi_M(v)^*,\psi_M^+(u)\right\}=\left\{B_M(0\oplus v)^*,
B_M(u\oplus 0)\right\}=0$. Furthermore, the charge conjugation map
$\alpha_C$ used in propositions \ref{AlphaC} and \ref{InvariantAlgebra}
reduces to the identity map in the case of Majorana spinors.
\end{remark}

To conclude this subsection we deal with the residual gauge freedom:
\begin{proposition}\label{InvariantAlgebra}
The bundle isomorphism $\map{\lambda}{DM}{DM}$ defined by
$\lambda(u):=-u$ gives rise to an involutive $^*-isomorphism$
$\tau$ of $\overline{\Alg{F}}^0_M$. The set
$\Alg{B}_M\subset\overline{\Alg{F}}^0_M$ of $\tau$-invariant
elements is a $C^*$-algebra and is generated by elements of
the form $B_M(f)B_M(g)$ (i.e. it is the even subalgebra of
$\overline{\Alg{F}}^0_M$). The $^*$-isomorphism $\alpha_C$ restricts to
an involutive $^*$-isomorphism of $\Alg{B}_M$.
\end{proposition}
\begin{proof*}
Note that $\lambda$ extends to $\mathcal{D}_0$ as $\lambda(f)=-f$ and
then descends to $\mathcal{L}(M)$, where it is the linear map $-I$. It
then gives rise to the map $\tau$ on $\overline{\Alg{F}}^0_M$ defined by
$\tau(B_M(f)):=B_M(\lambda(f))=-B_M(f)$ extended as an algebra
homomorphism. As $(-I)^2=I$ we see that $\tau^2=\id$ and we can define
the linear space $\Alg{B}_M$ of $\tau$-invariant elements. As
$\tau$ is a $^*$-isomorphism, $\Alg{B}_M$ is a closed $^*$-subalgebra
of $\overline{\Alg{F}}^0_M$ and hence a $C^*$-algebra in its own right.
Clearly, $\Alg{B}(M)$ contains all even powers of $B_M$, i.e.
$B_M(f)B_M(g)$. Conversely, any $\tau$-invariant
$A\in\overline{\Alg{F}}^0_M$ can be approximated by a sequence of
polynomials $A_n$, which we can choose to be $\tau$-invariant. As the
$\tau$-invariant polynomials $A_n$ only contain even powers of the
$B_M(f)$'s we see that these even polynomials
generate $\Alg{B}_M$. Finally, $\alpha_C$ maps the even polynomials
onto themselves, so $\alpha_C(\Alg{B}_M)=\Alg{B}_M$, and
$\alpha_C^2B_M(f)=-B_M(f)=\tau(B_M(f))$ implies that $\alpha_C^2$ is the
identity on $\Alg{B}_M$.
\end{proof*}

Although the physical information should be contained entirely in the
gauge-invariant algebra $\Alg{B}_M$, it will be convenient to have
$\overline{\Alg{F}}^0_M$ at our disposal too, because the putative
locally covariant quantum fields $B_M$, $\psi_M$ and $\psi_M^+$ take
values in $\overline{\Alg{F}}^0_M$ rather than $\Alg{B}_M$.

\subsection{States of the Dirac field}\label{subs_DHad}

If $\omega$ is a state on $\overline{\Alg{F}}^0_M$ and
$(\mathcal{H}_{\omega},\pi_{\omega},\Omega_{\omega})$
its GNS-triple, then we may consider for each
$n\in\N$ the $\mathcal{H}_{\omega}$-valued distribution on
$DM\oplus D^*M$ defined by:
\[
\varphi_n(f_n,\ldots,f_1):=\pi_{\omega}(B_M(f_n)\cdots B_M(f_1))
\Omega_{\omega}
\]
and $B^{(\omega)}(f):=\pi_{\omega}(B_M(f))$.
\begin{definition}\label{def_DmSC}
A state $\omega$ on $\overline{\Alg{F}}^0_M$ is called \emph{Hadamard}
if and only if
\[
WF(\varphi_1)\subset\mathcal{N}^+.
\]

A state $\omega$ on $\overline{\Alg{F}}^0_M$ satisfies the
\emph{microlocal spectrum condition} ($\mu$SC) if and only if
$WF(\omega_n)\subset\Gamma_n$ for all $n\in\N$.

A state $\omega$ on $\Alg{B}_M$ or $\Alg{F}^0_M$ is called \emph{Hadamard},
respectively satisfies the \emph{microlocal spectrum condition}, if and
only if it can be extended to
a state on $\overline{\Alg{F}}^0_M$ which is Hadamard, respectively
satisfies the $\mu$SC.
\end{definition}
Note that every state on $\Alg{B}_M$ can be
extended to a state on $\overline{\Alg{F}}^0_M$ by the Hahn-Banach
theorem \cite{Kadison+} and
every state on $\Alg{F}^0_M$ has an extension to
$\overline{\Alg{F}}^0_M$ only if it is continuous in the $C^*$-norm,
in which case the extension is unique by continuity. The
Hadamard condition on $\Alg{B}_M$ is independent of the choice of
extension, because it depends solely on the two-point distribution, as
the following proposition shows, and the same is true for the $\mu$SC
by proposition \ref{lem_Dts2} below. The following proposition also shows
that the definition of Hadamard states on $\overline{\Alg{F}}^0_M$ is
analogous to definition \ref{def_hadamard} for the free scalar field.
\begin{proposition}
For a state $\omega$ on $\overline{\Alg{F}}^0_M$ the following three
conditions are equivalent:
\begin{enumerate}
\item $\omega$ is Hadamard,
\item the two-point distribution
$\omega_2(f_1,f_2):=\omega(B_M(f_1)B_M(f_2))$ has
\[
WF(\omega_2)\subset\mathcal{C}:=\left\{(x,\xi;y,\xi')\in
\mathcal{N}^-\times\mathcal{N}^+|\ (x,-\xi)\sim (y,\xi')\right\},
\]
where again $(x,-\xi)\sim (y,\xi')$ if and only if $(x,-\xi)=(y,\xi')$
or there is an affinely parameterised light-like geodesic between $x$ and
$y$ to which $-\xi,\xi'$ are cotangent (and hence $-\xi$ and $\xi'$ are
parallel transports of each other along the geodesic),
\item there is a two-point distribution $w$ such that
$\omega_2(f_1,f_2)=iw(Df_1,f_2)$ and $WF(w)\subset\mathcal{C}$.
\end{enumerate}
\end{proposition}
\begin{proof*}
First note that $\omega_2$ is a bidistribution on $DM\oplus D^*M$,
because $B_M$ is an $\overline{\Alg{F}}^0_M$-valued distribution and
multiplication in $\overline{\Alg{F}}^0_M$ and $\omega$ are continuous.
For the equivalence of
the first two statements we adapt the argument in \cite{Strohmaier+},
proposition 6.1. If $WF(\omega_2)\subset\mathcal{C}$, then
$WF(\varphi_1)\subset\mathcal{N}^+$ by theorem \ref{WFHvalued}, so
$\omega$ is Hadamard. For the
converse we suppose that $\omega$ is Hadamard. Again by theorem
\ref{WFHvalued} we see that
$WF(\omega_2)\subset\mathcal{N}^-\times\mathcal{N}^+$. Defining
$\tilde{\omega}_2(f_1,f_2):=\omega_2(f_2,f_1)$ we find
$WF(\tilde{\omega}_2)\cap WF(\omega_2)=\emptyset$. Now,
$(\omega_2+\tilde{\omega}_2)(f_1,f_2)=2i\langle f_1^+,Sf_2\rangle$, so
$WF(\omega_2)\subset WF(S)\cup WF(\tilde{\omega}_2)$ and hence
$WF(\omega_2)\subset WF(S)\subset WF(E)$, because $S=\tilde{D}E$.
By proposition \ref{WFE} and $E=E^--E^+$ we find
$WF(E)\cap(\mathcal{N}^-\times\mathcal{N}^+)\subset\mathcal{C}$ and
therefore $WF(\omega_2)\subset\mathcal{C}$.

To prove the equivalence of the second and third statement we now assume
that $\omega_2(f_1,f_2)=iw(Df_1,f_2)$ and
$WF(w)\subset\mathcal{C}$. If $D^*$ is the formal adjoint of $D$,
then $WF(\omega_2)= WF((D^*\otimes I)w)\subset WF(w)\subset\mathcal{C}$.
For the converse we suppose that $\omega_2$ satisfies condition $2$
and we choose a smooth real-valued function $\chi^+$ on $M$ such that
$\chi^+\equiv 0$ to the past of some Cauchy surface $C_-$ and such that
$\chi^-:=1-\chi^+\equiv 0$ to the future of another Cauchy surface
$C_+$. We then define
$w(f_1,f_2):=-i\omega_2(\chi^+S^-f_1+\chi^-S^+f_1,f_2)$. Note that
$w$ is a bidistribution which is well-defined, because $\chi^+S^-f_1$
and $\chi^-S^+f_1$ are compactly supported. It is easy to verify that
$iw(Df_1,f_2)=\omega_2(f_1,f_2)$. We now estimate the wave front set
of $w$ as follows. The wave front set of $S^{\pm}=\tilde{D}E^{\pm}$ is
contained in $WF(E^{\pm})$, which we collected in proposition \ref{WFE}.
Then we may apply theorem 8.2.9 and 8.2.13 in \cite{Hoermander} to
estimate the wave front sets of the tensor product
$\chi^{\pm}(y)S^{\mp}(y,x)\delta(y',x')$ and the compositions in
$iw(x,x')=\sum_{\pm}\int\omega_2(y,y')(\chi^{\pm}(y)S^{\mp}(y,x)
\delta(y',x'))$ respectively and, using
$WF(\omega_2)\subset\mathcal{C}$, we find:
\begin{eqnarray}
WF(iw)&\subset&\cup_{\pm}\left\{(x,k;x',k')|\ \exists
(y,l;y',l')\in WF(\omega_2) \mathrm{\ such\ that\ }\right.\nonumber\\
&&\left.(y,-l;x,k;y',-l';x',k')\in WF(S^{\pm}\otimes\delta)
\right\}\nonumber\\
&\subset& WF(\omega_2).\nonumber
\end{eqnarray}
\end{proof*}
Using scaling limits one can even show that $WF(\omega_2)=\mathcal{C}$
if $WF(\omega_2)\subset\mathcal{C}$. In this form the equivalence of
statements $2$ and $3$ was already known (\cite{Sahlmann+} definitions
5.1 and 5.3 and theorem 5.8). The first characterisation appears to
be new, but is analogous to the result for the free scalar field
\cite{Strohmaier+}.

The following definition of quasi-free states is analogous to the free
field case, definition \ref{def_qf}.
\begin{definition}\label{def_qfD}
A state $\omega$ on either of the algebras $\Alg{F}^0_M$ or
$\overline{\Alg{F}}^0_M$ is called \emph{quasi-free}
if and only if $\omega_n=0$ for $n$ odd and for $m\ge 1$:
\[
\omega_{2m}(f_1,\dots,f_{2m})=\sum_{\pi\in\Pi_m}
\omega_2(f_{\pi(1)},f_{\pi(2)})\cdots\omega_2(f_{\pi(2m-1)},f_{\pi(2m)}),
\]
where $\Pi_m$ is the set of permutations of $\left\{1,\ldots,2m\right\}$
such that
\begin{enumerate}
\item $\pi(1)<\pi(3),\ldots<\pi(2m-1)$,
\item $\pi(2i-1)<\pi(2i)$, $i=1,\ldots,m$.
\end{enumerate}

The set of all states on $\Alg{F}^0_M$ satisfying the $\mu$SC will be
denoted by $\Stat{R}^0_M$.

The set of all states on $\overline{\Alg{F}}^0_M$ which are locally
quasi-equivalent to a quasi-free Hadamard state is denoted by
$\Stat{T}^0_M$.

The set of all states on $\Alg{B}_M$ which are locally quasi-equivalent
to the restriction of a state on $\overline{\Alg{F}}^0_M$ which is in
$\Stat{T}^0_M$ is denoted by $\Stat{T}_M$.
\end{definition}
Adapting the proof of proposition \ref{lem_ts2} we now show that every
Hadamard state on $\overline{\Alg{F}}^0_M$ satisfies the $\mu$SC.
(The quasi-free Hadamard case was essentially known, because the proof
is the same as for the scalar field algebra $\Alg{U}^0_M$, see
\cite{Brunetti+1} proposition 4.3).

\begin{proposition}\label{lem_Dts2}
Let $\omega$ be a state on $\Alg{F}^0_M$ which is Hadamard on a
neighbourhood $\mathcal{W}\subset M$ of a Cauchy surface in $M$. Then
$\omega$ satisfies the $\mu$SC on $M$.
\end{proposition}
\begin{proof*}
The proof is completely analogous to that of proposition \ref{lem_ts2},
with the following modifications. We use the parts of theorems
\ref{WFresults2} and \ref{WFHvalued} that pertain to distributions on
vector bundle-valued sections and the operator $P=\tilde{D}D$. The
propagation of the wave front set in this case follows from the work of
\cite{Dencker} concerning polarisation sets. We use the fact that
$WF(S)=WF(\tilde{D}E)\subset WF(E)$ and finally we need to use the
anti-commutation relations instead of the commutation relations:
\begin{eqnarray}\label{npointanticomm}
\omega_n(x_1,\ldots,x_n)&=&-\omega_n(x_1,\ldots,x_{i+1},x_i,\ldots,x_n)\\
&&i\omega_{n-2}(x_1,\ldots\hat{x}_i,\hat{x}_{i+1}\ldots,x_n)
S(x_i,x_{i+1}).\nonumber
\end{eqnarray}
Here we view $S$ as the bidistribution $S(f,g):=\langle f,Sg\rangle=-i(f,g)$,
cf. lemma \ref{positivity}. Finally we need to replace $\phi_1$ and
$\Phi^{(\omega)}$ by $\varphi_1$ and $B^{(\omega)}$ respectively.
\end{proof*}

We now check that
$\Stat{R}^0_M$, $\Stat{T}^0_M$ and $\Stat{T}_M$ are suitable candidates
to construct a state space functor, in the sense of definition
\ref{def_states} (cf. proposition \ref{prop_closedsf} and \ref{U0Q0} for
analogous results for the scalar field):
\begin{proposition}\label{prop_closedD}
The set $\Stat{R}^0_M$, $\Stat{T}^0_M$, respectively $\Stat{T}_M$, is
convex and closed under operations from $\Alg{F}^0_M$,
$\overline{\Alg{F}}^0_M$, respectively $\Alg{B}_M$.
\end{proposition}
\begin{proof*}
$\Stat{R}^0_M$ is convex by theorem \ref{WFresults2}. To show that it is
closed under operations from $\Alg{F}^0_M$ we note that for fixed
$f\in\Test_0((DM\oplus D^*M)^{\boxtimes m})$ and
$h\in\Test_0((DM\oplus D^*M)^{\boxtimes r})$ we have
\begin{eqnarray}
WF(\omega_{m+n+r}(f,x_1,\ldots,x_n,h))&\subset&\nonumber\\
\left\{(y_1,0;\ldots;y_m,0;x_1,k_1;\ldots;x_n,k_n;z_1,0;\ldots
;z_r,0)\in \Gamma_{m+n+r}\right\}&\subset&\Gamma_n,\nonumber
\end{eqnarray}
using \cite{Hoermander} theorem 8.2.12, which can also be applied to
vector bundle-valued sections by using local sections and expressing
$\omega_n$ as a sum of components (cf. appendix \ref{ch_ma}). The same
inclusion holds for linear combinations of such
terms, so if $\omega(A^*A)\not=0$ then the state
$B\mapsto\frac{\omega(A^*BA)}{\omega(A^*A)}$ satisfies the
$\mu$SC if $\omega$ does.

$\Stat{T}^0_M$ is convex, because all quasi-free Hadamard states of
$\overline{\Alg{F}}^0_M$ are locally quasi-equivalent \cite{DAntoni+}.
From the definition of $\Stat{T}^0_M$ we see that it is closed under
operations from $\overline{\Alg{F}}^0_M$.

To see that $\Stat{T}_M$ is convex we note that the restrictions of any
two states $\omega_1,\omega_2$ in $\Stat{T}^0_M$ to $\Alg{B}_M$ are
locally quasi-equivalent. Indeed, for any bounded cc-region a
$^*$-isomorphism between the von Neumann algebras
$\pi_{\omega_i}(\overline{\Alg{F}}^0_M)''$ restricts to a $^*$-isomorphism
of the von Neumann algebras $\pi_{\omega_i}(\Alg{B}_M)''$ and the claim
then follows from \cite{Araki3} pp.212-213. That $\Stat{T}_M$ is closed
under operations from $\Alg{B}_M$ follows directly from definition
\ref{def_qfD} again.
\end{proof*}

The following lemma is analogous to the free field case, lemma
\ref{lem_ts1}:
\begin{lemma}\label{lem_Dts1}
Let $M$ be a globally hyperbolic spin spacetime, $\mathcal{W}\subset M$ a
neighbourhood of a Cauchy surface and $\chi\in\Test_0(M)$ such that
$\chi\equiv 1$ on $J^+(\mathcal{W})\setminus\mathcal{W}$ and $\chi\equiv 0$
on $J^-(\mathcal{W})\setminus\mathcal{W}$. For every $f\in\mathcal{D}_0(M)$
we have $f=f'+Dh$, where $f':=D(\chi Sf)\in\mathcal{D}_0(\mathcal{W})$ and
$h:=S^-(f-f')\in\mathcal{D}_0(M)$.
\end{lemma}
\begin{proof*}
Notice that $\mathrm{supp}\ f'\subset\mathrm{supp}(Sf)$ and $f'\equiv 0$
on a neighbourhood of $M\setminus\mathcal{W}$, so
$f'\in\mathcal{D}_0(\mathcal{W})$. We
have $h=S^-(f-f')=(1-\chi)S^-f+\chi S^+f$ which is compactly supported
in $M$ and $Dh=f-f'$.
\end{proof*}
If, in the situation of lemma \ref{lem_Dts1}, we set $\chi^+:=\chi$ and
$\chi^-:=1-\chi$ we note that $\chi^+Sf+S^+f$ and $\chi^-Sf-S^-f$ have
compact support in $M$ and hence
\begin{equation}\label{timeslice1}
SD(\chi^{\pm}Sf)=\mp Sf.
\end{equation}

\subsection{The free Dirac field as a LCQFT}\label{subs_LCDF}

In the previous subsections we have quantised the Dirac field and
discussed interesting classes of states on a
single spin spacetime. In this section we will show how the free Dirac
field can be described as a locally covariant quantum field theory.
For that purpose we will need to investigate how our quantisation and
our classes of states behave under morphisms, paying special attention
to the residual gauge freedom that arises from the choice of
representation (see proposition \ref{equivalence}).

\begin{proposition}\label{prop_2lambda}
Given a morphism $\map{\Psi}{M_1}{M_2}$ in $\CatSMan$ such that
$\Psi=(\psi,\chi)$ (see definition \ref{def_sman}), there exist
exactly two bundle isomorphisms
$\map{\lambda_{\pm}}{DM_1}{DM_2|_{\psi(M)}}$ which intertwine the
adjoint, charge conjugation and Dirac operator and we have
$\lambda_+(u)=-\lambda_-(u)$. Each of these bundle isomorphisms gives
rise to a morphism $\map{\beta_{\pm}}{\overline{\Alg{F}}^0_{M_1}}
{\overline{\Alg{F}}^0_{M_2}}$ in $\CatAlg$ and we have
$\beta_+=\tau\circ\beta_-$ where $\tau$ is the $^*$-isomorphism of
$\overline{\Alg{F}}^0_{M_2}$ of proposition
\ref{InvariantAlgebra}.
\end{proposition}
\begin{proof*}
First we note that $\chi(SM_1)=SM_2|_{\psi(M)}$, because $\chi$ maps the
fiber of $SM_1$ over $x\in M_1$ to the fiber of $SM_2$ over $\psi(x)$
and it intertwines the action of the structure group and the bundle
projection appropriately. Note in particular that $\chi$ is a
diffeomorphism of each fiber, because the action of the structure group
is transitive. Because the Dirac spinor bundle is constructed from the
spin frame bundle we can apply proposition \ref{equivalence} to conclude
that there are only two bundle isomorphisms
$\map{\lambda_{\pm}}{DM_1}{DM_2|_{\psi(M)}}$ which intertwine the
adjoint, charge conjugation and Dirac operator in the required way
and that
$\lambda_+(u)=-\lambda_-(u)$.

The bundle isomorphisms $\lambda_{\pm}$ extend in the canonical way to
the cospinor bundle and the double spinor bundle and we denote these
extensions by the same symbol. Therefore the $\lambda_{\pm}$ give rise
to two linear maps
$\map{(\lambda_{\pm})_*}{\mathcal{D}(M_1)}{\mathcal{D}(M_2)}$. Now let
$S^{\pm}_i$ be the fundamental advanced $(-)$ and retarded $(+)$ solution
to the
Dirac equation on the globally hyperbolic spin spacetimes $M_i$. Let
$f\in\mathcal{D}_0(\psi(M_1))$, let $D_i$ denote the operator $D$ on
the spin spacetime $M_i$ and fix a sign $s=\pm$. Then we have
(cf. the proof of proposition \ref{morphismBU0}):
\[
f=(\lambda_s)_*\lambda_s^*f=(\lambda_s)_*(D_1S^{\pm}_1\lambda_s^*f)
=D_2((\lambda_s)_*S^{\pm}_1\lambda_s^*f)
\]
and
\[
\mathrm{supp}((\lambda_s)_*S^{\pm}_1\lambda_s^*f)\subset
\psi(J^{\pm}(\mathrm{supp}(\lambda_s^*f)))=
J^{\pm}(\mathrm{supp}\ f)\cap \psi(M_1)
\]
by causal convexity. The uniqueness part of proposition \ref{prop_fundS}
now shows that
$S_2^{\pm}|_{\psi(M_1)}=(\lambda_s)_*S_1^{\pm}\lambda_s^*$ and hence
$S_2|_{\psi(M_1)}=(\lambda_s)_*S_1\lambda_s^*$.

If $f\in\mathcal{D}_0(M_1)$ has $S_1f=0$, then $f=D_1h$ for some
$h\in\mathcal{D}_0(M_1)$ by lemma \ref{positivity} and hence
$S_2(\lambda_s)_*f=S_2(\lambda_s)_*D_1h=S_2D_2(\lambda_s)_*h=0$.
Therefore
$\map{(\lambda_s)_*}{\mathcal{D}_0(M_1)}{\mathcal{D}_0(M_2)}$
descends to a map
$\map{\kappa_s}{\mathcal{L}(M_1)}{\mathcal{L}(M_2)}$ (see lemma
\ref{positivity}) which is injective and isometric. Indeed, leaving the
metric induced volume elements implicit:
\[
(f,h)_{\mathcal{L}(M_1)}=i\int_{M_1} fS_1h=i\int_{M_2}\kappa_sf
S_2\kappa_sh=(\kappa_sf,\kappa_sh)_{\mathcal{L}(M_2)},
\]
because $\kappa_sf=(\lambda_s)_*f\equiv 0$ outside $\psi(M_1)$. It now
follows that there are morphisms $\map{\beta_{\pm}}
{\overline{\Alg{F}}^0_{M_1}}{\overline{\Alg{F}}^0_{M_2}}$ in $\CatAlg$
defined by $\beta_{\pm}B_{M_1}(f):=B_{M_2}(\kappa_{\pm}f)$ (see
\cite{Bratteli+}). Finally, as $\lambda_+(f)=-\lambda_-(f)$ we have
$\beta_+B_{M_1}(f)=-\beta_-B_{M_1}(f)$ and hence
$\beta_+=\tau\circ\beta_-$.
\end{proof*}

To describe the $\lambda_{\pm}$ explicitly we fix a choice of complex
irreducible representation $\pi$ of the Dirac algebra $D$ and of
matrices $A,C\in GL(4,\C)$ satisfying assumption (\ref{assumption}).
We use this choice to construct the Dirac spinor bundle on every spin
spacetime in $\CatSMan$. For the morphism $\map{\Psi}{M_1}{M_2}$ with
$\Psi=(\psi,\chi)$
we can then define
\begin{equation}\label{gaugechoice}
\lambda_{\pm}([E,z]_1):=[\chi(E),\pm z]_2
\end{equation}
and note that this is a linear, base-point preserving bundle
homomorphism, which is well-defined because $\chi$ intertwines the right
action of $Spin^0_{1,3}$ on both spin frame bundles. Moreover,
$\lambda_{\pm}$ intertwines the adjoint and charge conjugation maps,
because $\pm I$ commutes with $A$ and $C$. Hence, $\lambda_{\pm}$ are
the maps of proposition \ref{prop_2lambda}.

Choosing the representation $\pi$ and the matrices $A,C$ does not fix
the gauge completely. There are still bundle-automorphisms of the Dirac
spinor bundle that leave all physical equations invariant. There are two
ways to proceed in order to deal with this residual gauge freedom. The
first is to fix it by hand in a locally covariant way. The second is to
divide out the gauge freedom. We will present both approaches in that
order.

To fix the residual gauge freedom we need to choose a sign for the
Dirac spinor bundle on each spin spacetime in a locally covariant way.
The following proposition shows that this can be done.
\begin{proposition}\label{prop_gaugefunc}
Fix a choice of $\pi$, $A$ and $C$. We can define a locally covariant
quantum field theory $\map{\overline{\Func{F}}^0}{\CatSMan}{\CatAlg}$
which assigns to a globally hyperbolic spin spacetime $M$ the algebra
$\overline{\Alg{F}}^0_M$ and to every morphism $\map{\Psi}{M_1}{M_2}$
the morphism $\phi_{\Psi}:=\beta_+$ of proposition
\ref{prop_2lambda} associated to the bundle isomorphism $\lambda_+$
defined in equation (\ref{gaugechoice}).
\end{proposition}
\begin{proof*}
The maps are well-defined, so we only need to check that they define a
covariant functor. The identity morphism gets mapped to the identity
morphism and for a composition of morphisms $\Psi=\Psi_1\circ\Psi_2$ we
have in the obvious notation,
$\lambda_+=(\lambda_1)_+\circ(\lambda_2)_+$ and hence
$\beta_+=(\beta_1)_+\circ(\beta_2)_+$, which proves the proposition.
\end{proof*}

The functor $\overline{\Func{F}}^0$ of proposition \ref{prop_gaugefunc}
seems to depend on the choice of $\pi$, $A$ and $C$, however we will now
prove that all choices give rise to equivalent functors. Recall that two
functors $\map{\Func{F}_i}{\CatSMan}{\CatTAlg}$, $i=1,2$, are equivalent
iff there is a natural transformation between $\Func{F}_1$ and
$\Func{F}_2$, given by maps
$\map{T_M}{\Alg{F}^1_M}{\Alg{F}^2_M}$, such that each $T_M$ is an
isomorphism in the category $\CatTAlg$ (see e.g.
\cite{MacLane,Brunetti+}).\footnote{The fact underlying the
proof of proposition \ref{prop_2natural} is the following. Given a
choice of representation $\pi$ and matrices $A,C$ we can define a
functor from the category $\CatSMan$ to the category $\CatVB$ of vector
bundles over spin manifolds, which maps each spin spacetime to the
associated Dirac bundle and which uses $\lambda_+$ to describe
embeddings. Different choices of representation then give rise to
equivalent functors and there is an equivalence from such a functor to
itself which is given by the bundle isomorphism $[E,z]\mapsto[E,-z]$.}
\begin{proposition}\label{prop_2natural}
Let $\pi$ and $\pi'$ be two complex irreducible representations of the
Dirac algebra $D$ and let $A,C$ and $A',C'$ be matrices in
$GL(4,\C)$ satisfying assumption (\ref{assumption}) w.r.t. $\pi$ and
$\pi'$, respectively. We let $\overline{\Func{F}}^0$ and
$(\overline{\Func{F}}')^0$ be the corresponding functors of proposition
\ref{prop_gaugefunc}. Then there are two equivalences between the
functors $\overline{\Func{F}}^0$ and $(\overline{\Func{F}}')^0$.
\end{proposition}
\begin{proof*}
Let $L_{\pm}$ be the matrices of theorem \ref{ACmatrix} which
intertwine $\pi,A,C$ and $\pi',A',C'$. For each globally hyperbolic
spin spacetime $M$ we then define bundle isomorphisms of the Dirac
spinor bundle by
\[
\eta^{\pm}_M([E,z]'):=[E,L_{\pm}z]
\]
and we let $\beta^{\pm}_M$ be the associated $^*$-isomorphisms of
$\overline{\Alg{F}}^0_M$ as in proposition \ref{prop_2lambda}.

Now let $\map{\Psi}{M_1}{M_2}$ be a morphism in $\CatSMan$ and let
$\lambda_+$ and $\lambda'_+$ be the bundle homomorphisms of equation
(\ref{gaugechoice}) using the representation $\pi$ and $\pi'$
respectively. Notice that
$\eta^{\pm}_{M_2}\circ\lambda_+=\lambda'_+\circ\eta^{\pm}_{M_1}$. The
corresponding equation for the $^*$-isomorphisms of
$\overline{\Alg{F}}^0_M$ is then $\beta^{\pm}_{M_2}\circ\phi_{\Psi}
=\phi'_{\Psi}\circ\beta^{\pm}_{M_1}$, which shows that $\beta^{\pm}_M$
defines an equivalence for each choice of the sign.
\end{proof*}

\begin{corollary}
We can define a locally covariant quantum field theory
$\map{\Func{F}^0}{\CatSMan}{\CatTAlg}$ which assigns to every
globally hyperbolic spin
spacetime $M$ the algebra $\Alg{F}^0_M$ and to every morphism
$\map{\Psi}{M_1}{M_2}$ the restriction of the morphism $\phi_{\Psi}$ in
$\CatAlg$ to $\Alg{F}^0_M$. If we define the functor $(\Func{F}')^0$ in
the same way, but for a different choice of representation $\pi'$ and
matrices $A'$, $C'$, then there are two equivalences between
$\Func{F}^0$ and $(\Func{F}')^0$.
\end{corollary}
\begin{proof*}
This follows from propositions \ref{prop_gaugefunc} and
\ref{prop_2natural}. We only need to check that for each morphism the
image $\phi_{\Psi}(\Alg{F}^0_{M_1})$ is contained in $\Alg{F}^0_{M_2}$,
which follows from the definition of $\phi_{\Psi}$ in the proof of
proposition \ref{prop_2lambda}.
\end{proof*}

The following corollary describes the result of dividing out the
residual gauge symmetry:
\begin{corollary}
We can define a locally covariant quantum field theory
$\map{\Func{B}}{\CatSMan}{\CatAlg}$ which assigns to every
globally hyperbolic spin
spacetime $M$ the algebra $\Alg{B}_M$ and to every morphism
$\map{\Psi}{M_1}{M_2}$ the restriction of the morphism $\phi_{\Psi}$ in
$\CatAlg$ to $\Alg{B}_M$. If we define the functor $\Func{B}'$ in the
same way, but for a different choice of representation $\pi'$ and
matrices $A'$, $C'$, then there is an equivalence between
$\Func{B}$ and $\Func{B}'$.
\end{corollary}
\begin{proof*}
Again this follows from propositions \ref{prop_gaugefunc} and
\ref{prop_2natural} if we check that for each morphism the image
$\phi_{\Psi}(\Alg{B}_{M_1})$ is contained in $\Alg{B}_{M_2}$, which
follows from the definition of $\phi_{\Psi}$ in the proof of proposition
\ref{prop_2lambda}. Note that the two natural transformations of
proposition \ref{prop_2natural} coincide on the even algebras
$\Alg{B}_M$, so now we only find one equivalence.
\end{proof*}

Finally we prove the properties of the locally covariant quantum field
theories and their associated state spaces:
\begin{proposition}\label{F0R0}
We can define a state space functor $\Func{R}^0$ for the locally
covariant quantum field theory $\map{\Func{F}^0}{\CatSMan}{\CatAlg}$,
which assigns to every globally hyperbolic spin spacetime $M$ the set
$\Stat{R}^0_M$.
Together these functors satisfy the time-slice axiom, additivity and
nowhere-classicality.
\end{proposition}
\begin{proof*}
We know that each $\Stat{R}^0_M$ is convex and closed under operations
from $\Alg{F}^0_M$ by proposition \ref{prop_closedD}. To see that the
functor $\Func{R}^0$ is well-defined we need to show that
$\phi_{\Psi}^*(\Stat{R}^0_{M_2})\subset\Stat{R}^0_{M_1}$ for every
morphism $\map{\Psi}{M_1}{M_2}$. This holds, because a state $\omega$
on $\Alg{F}^0_{M_2}$ which satisfies the $\mu$SC restricts to a state
on $\Alg{F}^0_{\psi(M_1)}$ satisfying the $\mu$SC and hence maps to a state
on $\Alg{F}^0_{M_1}$ satisfying the $\mu$SC, as in the proof of
proposition \ref{prop_closedsf}.

Additivity follows as in proposition \ref{U0Q0}, by using a
partition of unity after choosing representatives in $\Alg{F}_M$.
The time-slice axiom follows from lemma \ref{lem_Dts1} and proposition
\ref{lem_Dts2} and covariance of the functors $\Func{F}^0$ and
$\Func{R}^0$. To prove nowhere-classicality we use that fact that for
each globally hyperbolic spin spacetime $M$ the $C^*$-algebra
$\overline{F}^0_M$ is simple,
\cite{Bratteli+} theorem 5.2.5. It is also non-commutative, because any
subspace of $\mathcal{L}(M)$ of dimension at least $2$ generates a
non-commutative Clifford-algebra, which is a sub-algebra of
$\overline{\Alg{F}}^0_M$. By our definition of the $\mu$SC, definition
\ref{def_DmSC}, the state $\omega$ of $\Alg{F}^0_M$ extends to a state
on $\overline{\Alg{F}}^0_M$, which is necessarily faithful, because
$\overline{F}^0_M$ is simple. It follows
that $\pi_{\omega}(\overline{\Alg{F}}^0_M)$ is not commutative and
hence the dense sub-algebra $\pi_{\omega}(\Alg{F}^0_M)$ cannot be
commutative.
\end{proof*}

\begin{proposition}\label{barF0}
We can define a state space functor $\Func{T}^0$ for the locally
covariant quantum field theory
$\map{\overline{\Func{F}}^0}{\CatSMan}{\CatAlg}$, which assigns to every
globally hyperbolic
spin spacetime $M$ the set $\Stat{T}^0_M$. Together these functors
satisfy the time-slice axiom, local physical equivalence,
local quasi-equivalence, additivity and nowhere-classicality.
\end{proposition}
\begin{proof*}
We know that each $\Stat{T}^0_M$ is convex and closed under operations
from $\overline{\Alg{F}}^0_M$ by proposition \ref{prop_closedD}. To see
that the functor $\Func{T}^0$ is well-defined we need to show that
$\phi_{\Psi}^*(\Stat{T}^0_{M_2})\subset\Stat{T}^0_{M_1}$ for every
morphism $\map{\Psi}{M_1}{M_2}$. For this we first notice that a
quasi-free Hadamard state $\omega_2$ on $\overline{\Alg{F}}^0_{M_2}$
restricts to a quasi-free Hadamard state $\omega'_1$ on
$\overline{\Alg{F}}^0_{\psi(M_1)}$ which maps to a quasi-free Hadamard
state $\omega_1$ on $\overline{\Alg{F}}^0_{M_1}$. Next we note that a
state which is locally quasi-equivalent to $\omega_2$ restricts to a
state which is locally quasi-equivalent to $\omega'_1$ and then maps
to a state which is locally quasi-equivalent to $\omega_1$, by
covariance.

The additivity and nowhere-classicality of $\overline{\Func{F}}^0$ and
$\Func{T}^0$ follow from proposition \ref{F0R0} by taking the norm
closure. The same is true for the time-slice axiom, if we notice in
addition that a state which is locally quasi-equivalent to a quasi-free
Hadamard state on a neighbourhood of a Cauchy surface extends to a state
with the same property, just like in proposition \ref{prop_A0prop}.
Because
$\overline{\Alg{F}}^0_M$ is simple (theorem 5.2.5 in \cite{Bratteli+})
the local physical equivalence follows from proposition 4.3 in
\cite{Fewster}. Local quasi-equivalence of $\Func{T}^0$ was already
shown in the proof of proposition \ref{prop_closedD}.
\end{proof*}

\begin{proposition}\label{BTprop}
We can define a state space functor $\Func{T}$ for the locally
covariant quantum field theory $\map{\Func{B}}{\CatSMan}{\CatAlg}$,
which assigns to every globally hyperbolic
spin spacetime $M$ the set $\Stat{T}_M$. Together
these functors satisfy causality, the time-slice axiom, local
quasi-equivalence and additivity.
\end{proposition}
\begin{proof*}
We know that each $\Stat{T}_M$ is convex and closed under operations
from $\Alg{B}_M$ by proposition \ref{prop_closedD}. The fact that
the functor $\Func{T}$ is well-defined follows from proposition
\ref{barF0}.

To prove causality we choose $f_1,f_2,g_1,g_2\in\mathcal{D}_0(M)$
such that $\mathrm{supp}\ f_i\subset O$ and
$\mathrm{supp}\ g_i\subset O^{\perp}$ for some cc-region
$O\subset M$. Then
\begin{eqnarray}
\left[B(f_1)B(f_2),B(g_1)B(g_2)\right]&=&\nonumber\\
B(f_1)\left[B(f_2),B(g_1)B(g_2)\right]+
\left[B(f_1),B(g_1)B(g_2)\right]B(f_2)&=&\nonumber\\
B(f_1)(\left\{B(f_2),B(g_1)\right\}B(g_2)-
B(g_1)\left\{B(f_2),B(g_2)\right\})&&\nonumber\\
+(\left\{B(f_1),B(g_1)\right\}B(g_2)-
B(g_1)\left\{B(f_1),B(g_2)\right\})B(f_2)&=&0,\nonumber
\end{eqnarray}
because all the anti-commutators vanish in $\Alg{F}^0_M$
by the support properties of the $f_i$ and $g_i$.

To prove the time-slice axiom we let $\map{\Psi}{M_1}{M_2}$ be any
morphism in $\CatSMan$ for which $\psi(M_1)$ contains a Cauchy surface
of $M_2$. We then use the time-slice axiom of proposition \ref{barF0}
together with the fact that the isomorphism $\phi_{\Psi}$ preserves the
even and odd subspaces of the $\overline{\Alg{F}}^0_{M_i}$.

That all states in $\Stat{T}_M$ are locally quasi-equivalent follows
from the fact that the restrictions of all states in $\Stat{T}^0_M$
to $\Alg{B}_M$ are locally quasi-equivalent, which was shown in the
proof of proposition \ref{prop_closedD}.

Additivity follows as in proposition \ref{U0Q0}, by using a
partition of unity after choosing representatives in $\Alg{F}_M$.
\end{proof*}

Note that $B_M$, $\psi_M$ and $\psi^+_M$ are linear locally covariant
quantum fields with values in $\Alg{F}_M$, but not in $\Alg{B}_M$. To
find a locally covariant quantum field for the theory $\Func{B}$ we
could choose e.g. the non-linear field $\Theta_M(f):=B_M(f)B_M(f)$,
but we have little need for this field in what follows, because the
linear field $B_M$ is much easier to work with.

\section[Relative Cauchy evolution and the energy-momentum tensor]
{Relative Cauchy evolution and the stress-energy-momentum tensor for
the free Dirac field}

Using lemma \ref{lem_Dts1}, which is an explicit expression for the
time-slice axiom, we can now consider the relative Cauchy evolution
of the free Dirac field (cf. \cite{Brunetti+}). This means that we
will consider the $^*$-isomorphism $\beta$ between the algebras of
two cc-regions $N_{\pm}$ in a spin spacetime $M$, $N_+$ being to the
future of $N_-$ and each containing a Cauchy surface for $M$.
We study how $\beta$ varies when we vary the metric and/or the spin
structure in a compact set in the region between $N_-$ and $N_+$. We
will show that we obtain commutators with the
stress-energy-momentum tensor, in complete analogy with the case of
the free scalar field (\cite{Brunetti+} theorem 4.3).

As a preparation we will first discuss the stress-energy-momentum
tensor in subsection \ref{sec_semt}, where we use a point-splitting
procedure to obtain an expression for its commutator with a smeared
field operator.

\subsection{The stress-energy-momentum tensor}\label{sec_semt}

In a local frame $e_a$ the stress-energy-momentum tensor for the
classical free Dirac field $\psi$ on a spin spacetime $M$ has the form
\begin{equation}\label{SEMT1}
T_{ab}=\frac{i}{2}\left(\psi^+\gamma_{(a}\nabla_{b)}\psi
-\nabla_{(a}\psi^+\gamma_{b)}\psi\right),
\end{equation}
where the brackets around indices denote symmetrisation as an
idempotent operation. (In the following, indices between $||$
are not to be excluded from the symmetrisation over.)
Following \cite{Fewster+} we want to find a point-split
bidistribution which acts on scalar test-functions and which is
analogous to $T_{ab}$.
For this purpose we use the components $\gamma^{\ A}_{a\ B}$ of
$\gamma_a$ in a spin frame $E_A$. Recall that these components are
constant and note that
\begin{eqnarray}\label{SEMT2}
T^s_{ab}(x,y)&:=&\frac{i}{2}\left((\psi^+E_A)(x)\gamma^{\ A}_{(a\ |B|}
(E^Be_{b)}^{\mu}\nabla_{\mu}\psi)(y)\right.\nonumber\\
&&\left.
-(e_{(a}^{\mu}\nabla_{|\mu}\psi^+E_{A|})(x)\gamma^{\ A}_{b)\ B}
(E^B\psi)(y)\right)
\end{eqnarray}
reduces to $T_{ab}$ in the limit $y\rightarrow x$. We write $T^s_{ab}$
as a bidistribution of scalar test-functions $f,h$ after
performing a partial integration, $\int\nabla_{\mu}(e_a^{\mu}vu)=0$:
\begin{eqnarray}\label{SEMT3}
T^s_{ab}(f,h)&=&\frac{i}{2}\left(-\psi^+(E_Af)\gamma^{\ A}_{(a\ |B}
\psi(\nabla_{\mu|}(E^Be_{b)}^{\mu}h))\right.\nonumber\\
&&\left.
+\psi^+(\nabla_{\mu}(e_{(a}^{\mu}E_{|A|}f))\gamma^{\ A}_{b)\ B}
\psi(E^Bh)\right).
\end{eqnarray}
Equation (\ref{SEMT3}) can be promoted to the quantised case by
replacing $\psi$ and $\psi^+$ with the operator-valued distributions
$\psi_M$ and $\psi^+_M$ of definition \ref{def_psi}. The expression
(\ref{SEMT2}) can be viewed as a formal expression for the same
bidistribution when we substitute the quantised field operators.

\begin{proposition}\label{SEMTresult}
Writing $\gamma_a(u\oplus v):=(\gamma_au)\oplus(v\gamma_a)$ and
$R(u\oplus v):=u\oplus -v$ we have for all $f\in\mathcal{D}_0$ and
$h\in C_0^{\infty}(M)$:
\begin{eqnarray}
\int_M \left[B_M(f),T^s_{ab}(x,x)\right]h(x)d\mathrm{vol}_g(x)&=&
\nonumber\\
\frac{1}{2}\left\{(\nabla_{(a}B_M)(\gamma_{b)}(SRf) h)-
B_M(\gamma_{(b}\nabla_{a)}(SRf) h)\right\}.\nonumber
\end{eqnarray}
\end{proposition}
\begin{proof*}
For $f=u\oplus v$ we use proposition \ref{prop_psi} to obtain:
\begin{eqnarray}
\left\{B_M(f),(\psi^+_ME_A)(h)\right\}&=&(v^+\oplus 0,E_Ah\oplus 0)I=
-i\langle(S_{cosp}v)(E_A),h\rangle I\nonumber\\
\left\{B_M(f),(E^Be_b^{\mu}\nabla_{\mu}\psi_M)(h)\right\}&=&
-(0\oplus u^+,0\oplus\nabla_{\mu}(e_b^{\mu}E^Bh))I\nonumber\\
&=&-i\langle E^B(\nabla_b S_{sp}u),h\rangle I\nonumber\\
\left\{B_M(f),(e_a^{\mu}\nabla_{\mu}\psi^+_ME_A)(h)\right\}&=&
-(v^+\oplus 0,\nabla_{\mu}(e_a^{\mu}E_Ah)\oplus 0)I\nonumber\\
&=&i\langle (\nabla_a S_{cosp}v)(E_A),h\rangle I\nonumber\\
\left\{B_M(f),(E^B\psi_M)(h)\right\}&=&(0\oplus u^+,0\oplus E^Bh)I
=i\langle E^B(S_{sp}u),h\rangle I\nonumber
\end{eqnarray}
where the pairing $\langle,\rangle$ on the right-hand side denotes
the action of a scalar
distribution on $h$. Together with equation (\ref{SEMT2}), the
anti-commutation relations and
$\left[A,BC\right]=\left\{A,B\right\}C-B\left\{A,C\right\}$ this implies
\begin{eqnarray}
\left[B_M(f),T^s_{ab}(x,y)\right]&=&
\frac{1}{2}\left\{((S_{cosp}v)(E_A))(x)\gamma^{\ A}_{(a\ |B|}
(E^Be_{b)}^{\mu}\nabla_{\mu}\psi_M)(y)\right.\nonumber\\
&&-(\psi^+_ME_{A})(x)\gamma^{\ A}_{(a\ |B|}
(E^B(\nabla_{b)}S_{sp}u))(y)\nonumber\\
&&
+((\nabla_{(a}S_{cosp}v)(E_A))(x)\gamma^{\ A}_{b)\ B}
(E^B\psi_M)(y)\nonumber\\
&&
-\left.
(e_{(a}^{\mu}\nabla_{|\mu}\psi^+_ME_{A|})(x)\gamma^{\ A}_{b)\ B}
(E^B(S_{sp}u))(y)\right\}.\nonumber
\end{eqnarray}
In this expression we may take the coincidence limit, which yields:
\begin{eqnarray}
\left[B_M(f),T^s_{ab}(x,x)\right]&=&\frac{1}{2}\left\{
\nabla_{(b}\psi_M((S_{cosp}v)\gamma_{a)})(x)
-\psi^+_M(\gamma_{(a}\nabla_{b)}(S_{sp}u))(x)\right.\nonumber\\
&&\left.
-\nabla_{(a}\psi^+_M(\gamma_{b)}S_{sp}u)(x)
+\psi_M(\nabla_{(a}(S_{cosp}v)\gamma_{b)})(x)\right\}\nonumber\\
&=&\frac{1}{2}\left\{\nabla_{(a}B_M(\gamma_{b)}SRf)(x)
-B_M(\gamma_{(b}\nabla_{a)}(SRf))(x)\right\},\nonumber
\end{eqnarray}
from which the result follows.
\end{proof*}
Note that the point-split stress-energy-momentum tensor of
equation (\ref{SEMT2}) can also be used to renormalise the
expectation value of the stress-energy-momentum tensor by
subtracting a term that cancels out the divergence. In the proof
of proposition
\ref{SEMTresult}, however, we used the tensor in a commutator,
so any (divergent) multiple of the unit operator $I$ cancels out.
For that reason we did not need to subtract any divergent part.
For more details on the stress-energy-momentum tensor, its
renormalisation and its conservedness we refer to
\cite{Moretti,Hollands+2}, which deal with the real scalar field.

The result of proposition \ref{SEMTresult}
can be written for spinor and cospinor fields separately as:
\begin{eqnarray}\label{Tcommutator}
\int_M \left[\psi_M(v),T^s_{ab}(x,x)\right]h(x)d\mathrm{vol}_g(x)&=&
\nonumber\\
-\frac{1}{2}\left\{\nabla_{(a}\psi_M(\gamma_{b)}S_{cosp}vh)
-\psi_M(\gamma_{(b}\nabla_{a)}(S_{cosp}vh))\right\}\\
\int_M \left[\psi^+_M(u),T^s_{ab}(x,x)\right]h(x)d\mathrm{vol}_g(x)&=&
\nonumber\\
\frac{1}{2}\left\{\nabla_{(a}\psi^+_M(\gamma_{b)}S_{sp}uh)
-\psi^+_M(\gamma_{(b}\nabla_{a)}(S_{sp}uh))\right\}.\nonumber
\end{eqnarray}

\subsection{Relative Cauchy evolution}

In this subsection we will prove that we can obtain the expressions on
the right-hand side of equation (\ref{Tcommutator}) also via a
relative Cauchy evolution. We will first have to explain what such a
relative Cauchy evolution means in the case of the free Dirac field
theory $\overline{\Func{F}}^0$ (cf.\cite{Brunetti+}).

Suppose that we have two objects in $\CatSMan$,
$M=(\mathcal{M},g,SM,p)$ and $M'=(\mathcal{M},g',SM',p')$ ,
where the manifold $\mathcal{M}$ is the same in both cases and such that
both spin spacetimes are the same outside a compact subset
$K\subset\mathcal{M}$, i.e.
$g'|_{\mathcal{M}\setminus K}=g|_{\mathcal{M}\setminus K}$ and
$(SM',p')|_{\mathcal{M}\setminus K}=(SM,p)|_{\mathcal{M}\setminus K}$. Now
let $N^{\pm}\subset M$ be cc-regions, each containing
a Cauchy surface for $M$ and such that $K$ lies to the future of $N^-$
(i.e. $K\subset J^+(N^-)\setminus N^-$ in $M$ and hence also in $M'$)
and to the past of $N^+$. We view $N^{\pm}$ as objects in
$\CatSMan$ and consider the canonical embedding morphisms
$\map{\iota^{\pm}}{N^{\pm}}{M}$ and $\map{(\iota')^{\pm}}{N^{\pm}}{M'}$.
By the time-slice axiom, proposition \ref{barF0}, these give rise
to $^*$-isomorphisms
$\map{\beta^{\pm}}{\overline{\Alg{F}}^0_{N^{\pm}}}
{\overline{\Alg{F}}^0_M}$ and
$\map{(\beta')^{\pm}}{\overline{\Alg{F}}^0_{N^{\pm}}}
{\overline{\Alg{F}}^0_{M'}}$. We then define the
$^*$-automorphism $\beta_{g'}$ of $\overline{\Alg{F}}^0_M$ by
\begin{equation}\label{defrelCE}
\beta_{g'}:=\beta^+\circ((\beta')^+)^{-1}\circ(\beta')^-\circ(\beta^-)^{-1}.
\end{equation}
This $^*$-automorphism can easily be characterised in terms of its
action on the generators $B_M(f)$ of $\overline{\Alg{F}}^0_M$ as
follows:
\begin{proposition}\label{RCE}
If $f\in\mathcal{D}_0(M)$ with $\mathrm{supp}\ f\subset N^+$, then
$\beta_g B_M(f)=B_M(T_gf)$, where
\[
T_{g'}f=D'\chi_+S'D\chi_-Sf.
\]
Here the superscripts on $D$ and $S$ indicate whether they are the
objects defined on $M$ or $M'$ and the smooth functions
$\chi_{\pm}$ are such that $\chi_{\pm}\equiv 1$ to the past of some
Cauchy surface in $N^{\pm}$ and $\chi_{\pm}\equiv 0$ to the future of
some other Cauchy surface in $N^{\pm}$.
\end{proposition}
\begin{proof*}
Note that $(\beta')^-\circ(\beta^-)^{-1}B_M(\tilde{f})=
B_{M'}(\tilde{f})$ for any $\tilde{f}\in\mathcal{D}_0(N^-)$. Similarly,
for $f'\in\mathcal{D}_0(N^+)$ we have
$\beta^+\circ((\beta')^+)^{-1}B_{M'}(f')=B_M(f')$.
The functions $\chi_{\pm},1-\chi_{\pm}$ have been chosen appropriately
in order to apply equation (\ref{timeslice1}). We then have
$S\tilde{f}=Sf$ and hence
$B_M(\tilde{f})=B_M(f)$, where $\tilde{f}:=D\chi_-Sf$.
Notice that $\tilde{f}$ indeed has a compact support in $N^-$.
Similarly we have $B_{M'}(\tilde{f})=B_{M'}(f')$, where
$f':=D'\chi_+S'\tilde{f}$ has support in $N^+$. Putting
everything together yields for $f'=T_{g'}f$:
\begin{eqnarray}
\beta_{g'} B_M(f)&=&\beta_{g'} B_M(\tilde{f})=
\beta^+\circ((\beta')^+)^{-1}B_{M'}(\tilde{f})\nonumber\\
&=&\beta^+\circ((\beta')^+)^{-1}B_{M'}(f')=B_M(f').\nonumber
\end{eqnarray}
\end{proof*}

We will want to compute the variation of the $^*$-isomorphism
$\beta_{g'}$ with respect to the metric $g'$. For this purpose we suppose
that the
compact set $K\subset\mathcal{M}$ has a contractible neighbourhood $O$
which doesn't intersect either $N^{\pm}$. Now let
$\epsilon\mapsto g_{\epsilon}$ be a smooth curve from $[0,1]$ into the
space of Lorentzian metrics on $\mathcal{M}$ starting at $g$ and such
that $g_{\epsilon}=g$ outside $K$ for every $\epsilon$.

The spin bundle $SM_{\epsilon}$ must be trivial over the contractible
region $O$. If we assume it to be diffeomorphic to $SM$ outside $K$
we can simply take $SM_{\epsilon}:=SM$ as a manifold and, choosing a
fixed complex irreducible representation $\pi$ and matrices $A,C$
satisfying assumption (\ref{assumption}) to construct the Dirac spinor
bundle, we obtain $DM_{\epsilon}=DM$. The deformation of the spin
structure is contained entirely in the $\epsilon$-dependence of
the projection $\map{p_{\epsilon}}{SM}{FM_{\epsilon}}$. Now let $E$ be
a section of $SM$ over $O$ and set
$(e_{\epsilon})_a:=p_{\epsilon}(E)$.
We require that $e_{\epsilon}$ varies smoothly with $\epsilon$ and that
$(e_{\epsilon})_a=(e)_a=p(E)$ outside $K$. To show that projections
$p_{\epsilon}$ with these properties exist we can apply the Gram-Schmidt
orthonormalisation procedure for all $\epsilon$ simultaneously, starting
with the frame $(e)_a$, which yields a smooth family of frames
$(e_{\epsilon})_a$. The assignment $p_{\epsilon}:E\mapsto e_{\epsilon}$
then determines $p_{\epsilon}$ completely, because of the intertwining
properties of $p_{\epsilon}$ and the transitive action of
$Spin^0_{1,3}$ on the spin frame bundle. The family of frames
$e_{\epsilon}$ determines principal fiber bundle isomorphisms
$\map{f_{\epsilon}}{FM_{\epsilon}}{FM}$ between the frame
bundles by
\[
f_{\epsilon}:\left\{(e_\epsilon)_a\right\}\mapsto\left\{(e)_a\right\}
\]
on $K$ and extending it by the identity on the rest of $\mathcal{M}$.
By definition $f_{\epsilon}$ intertwines the action of
$\mathcal{L}_+^{\uparrow}$ on the frame bundles.

There may be many deformations of the spin structure, i.e. many families
of projections $p_{\epsilon}$ which satisfy our requirements. However,
the variation of $D_{\epsilon} f$ will not depend on this choice.
Indeed, if $p'_{\epsilon}$ is a different deformation of the spin
structure, then $e'_{\epsilon}:=p'_{\epsilon}(E)=
R_{\Lambda_{\epsilon}}e_{\epsilon}=p_{\epsilon}(R_{S_{\epsilon}}E)$
for some smooth curve $S_{\epsilon}$ in $Spin^0_{1,3}$. However,
$v\in DM_{\epsilon}=DM$ and $D_{\epsilon}v$ are invariant under the action
of the gauge group $Spin^0_{1,3}$ and therefore the variation will be too.

On each spin spacetime
$M_{\epsilon}=(\mathcal{M},g_{\epsilon},SM,p_{\epsilon})$ we can now
quantise the Dirac field and obtain relative Cauchy evolutions
$\beta_{\epsilon}:=\beta_{g_{\epsilon}}$ on
$\overline{\Alg{F}}^0_M$ as in equation \ref{defrelCE}.
\begin{proposition}\label{varRCE}
Writing $\delta:=\partial_{\epsilon}|_{\epsilon=0}$ we have for all
$f\in\mathcal{D}_0(M)$ with $\mathrm{supp}\ f\subset N^+$:
\[
\delta(\beta_{\epsilon}B_M(f))=B_M((\delta D_{\epsilon})Sf).
\]
\end{proposition}
\begin{proof*}
Using the fact that $B_M$ is a $C^*$-algebra-valued distribution and
proposition \ref{RCE} we find:
\begin{eqnarray}
\delta(\beta_{\epsilon}B_M(f))&=&\delta(B_M(D_{\epsilon}\chi_+
S_{\epsilon}D\chi_-Sf))
=B_M(\delta(D_{\epsilon}\chi_+S_{\epsilon})D\chi_-Sf)\nonumber\\
&=&B_M(\delta(D_{\epsilon})\chi_+SD\chi_-Sf)
+B_M(D\chi_+\delta(S_{\epsilon})D\chi_-Sf).\nonumber
\end{eqnarray}
Now, because $D\chi_-Sf\in\mathcal{D}_0(N^-)$ and $N^-$ is to the
past of $K$ we see that
$\delta(S_{\epsilon})D\chi_-Sf$ vanishes on $J^-(N^-)$ and that
$\chi_+\delta(S_{\epsilon})D\chi_-Sf$ has compact support.
Because $B_M$ solves the Dirac equation we conclude that the second
term vanishes. The first term can be rewritten using equation
(\ref{timeslice1}), which yields:
\[
\delta(\beta_{\epsilon}B_M(f))=
B_M(\delta(D_{\epsilon})\chi_+Sf)=
B_M(\delta(D_{\epsilon})Sf).
\]
For the last equality we used the fact that $\delta(D_{\epsilon})$ is
supported in $K$, where $\chi_+\equiv 1$.
\end{proof*}

To compute the variation of $D_{\epsilon}$ we may work in a local frame
on the contractible region $O$, because that is where
$\delta(D_{\epsilon})$ is supported.
Recall that $D=(-i \dirop +m)\oplus(i\dirop+m)$, so essentially we just
need to find the variation of $\dirop_{\epsilon}$ on spinor and
cospinor fields. In fact, we will next show that it is sufficient to know
the variation of this operator on cospinor fields, because we can then
derive the case of spinor fields using the adjoint map. This uses the fact
that the Dirac adjoint map is independent of $\epsilon$. Also note that
the components $\gamma^{\ B}_{a\ A}$, in a local frame determined by the
section $E$ of $SM$ over $O$, are constant and independent of $\epsilon$.
This follows immediately from the definition of $\gamma$, in lemma
\ref{lemdefgamma}.
\begin{lemma}\label{adjointvar}
For $v\in C_0^{\infty}(D^*M)$ we have
$\delta(\dirop)v=(\delta(\dirop)v^+)^+$.
\end{lemma}
\begin{proof*}
Because the adjoint operation between spinor and cospinor fields
is continuous we have:
\[
\delta(\dirop) v=\partial_{\epsilon}\dirop_{\epsilon}v|_{\epsilon=0}
=\partial_{\epsilon}(\dirop_{\epsilon}v^+)^+|_{\epsilon=0}
=(\partial_{\epsilon}\dirop_{\epsilon}v^+|_{\epsilon=0})^+
=(\delta(\dirop)v^+)^+.
\]
\end{proof*}

We now start the computation of the variation of the Dirac operator
on a cospinor field. For this purpose we will work in components and
in local coordinates on the contractible neighbourhood $O$. To
ease the notation we will drop the subscript $\epsilon$
on the local frame $e_a^{\mu}$. As $\gamma^a$ is independent of
$\epsilon$ we may use equations (\ref{spinconnection}) and
(\ref{linkconnections}) to vary the following equation for
$v\in\Test_0(D^*M_0)$:
\begin{eqnarray}\label{var0}
\dirop v&=&\left(\partial_av-\frac{1}{4}\Gamma^c_{\ ab}v\gamma_c\gamma^b
\right)\gamma^a\nonumber\\
&=&e_a^{\alpha}\left(\partial_{\alpha}v
+\frac{1}{4}e_b^{\beta}\left\{\partial_{\alpha}e^c_{\beta}
-e^c_{\gamma}\Gamma^{\gamma}_{\ \alpha\beta}\right\}
v\gamma_c\gamma^b\right)\gamma^a,
\end{eqnarray}
which yields:
\begin{eqnarray}\label{var1}
\delta\dirop v&=&\delta e_a^{\alpha}e^d_{\alpha}\nabla_dv\gamma^a
-\frac{1}{4}\delta e_b^{\beta}e^d_{\beta}\Gamma^c_{\ ad}v
\gamma_c\gamma^b\gamma^a
+\frac{1}{4}\partial_a\delta e^c_{\beta}e_b^{\beta}v
\gamma_c\gamma^b\gamma^a\nonumber\\
&&
-\frac{1}{4}\delta e^c_{\gamma}e_a^{\alpha}e_b^{\beta}
\Gamma^{\gamma}_{\ \alpha\beta}v\gamma_c\gamma^b\gamma^a
-\frac{1}{4}\delta\Gamma^{\gamma}_{\ \alpha\beta}
e_a^{\alpha}e_b^{\beta}e^c_{\gamma}v\gamma_c\gamma^b\gamma^a,
\end{eqnarray}
where we inserted a factor
$\delta^{\gamma}_{\beta}=e_{\beta}^de_d^{\gamma}$ twice to
simplify the first two terms.

We now define $D_c:=i\dirop+m$ acting on cospinor fields and we
try to get terms with this operator acting on $v$ or on the
whole expression. These are harmless when we compute
$B_M(\delta\dirop Sf)$, because $B_M$ and $v=Sf$ solve the Dirac
equation. We start by performing what is essentially an integration
by parts as follows:
\begin{eqnarray}\label{var2}
\frac{1}{4}\partial_a\delta e^c_{\beta}e_b^{\beta}v
\gamma_c\gamma^b\gamma^a&=&
\frac{-i}{4}D_c(\delta e^c_{\beta}e_b^{\beta}v\gamma_c\gamma^b)
+\frac{i}{4}\delta e^c_{\beta}e_b^{\beta}D_c(v\gamma_c\gamma^b)
\nonumber\\
&&-\frac{1}{4}\delta e^c_{\beta}\partial_ae_b^{\beta}v\gamma_c
\gamma^b\gamma^a
-\frac{1}{4}\delta e^c_{\beta}e_b^{\beta}\Gamma^d_{\ ac}
v\gamma_d\gamma^b\gamma^a\nonumber\\
&&
+\frac{1}{4}\delta e^c_{\beta}e_b^{\beta}\Gamma^b_{\ ad}
v\gamma_c\gamma^d\gamma^a\nonumber\\
&=&\frac{-i}{4}D_c(\delta e^c_{\beta}e_b^{\beta}v\gamma_c\gamma^b)
+\frac{i}{4}\delta e^c_{\beta}e_b^{\beta}(D_cv)\gamma_c\gamma^b\\
&&-\frac{1}{4}\delta e^c_{\beta}e_b^{\beta}\nabla_av
\left[\gamma_c\gamma^b,\gamma^a\right]
-\frac{1}{4}\delta e^c_{\beta}\partial_ae_b^{\beta}v\gamma_c
\gamma^b\gamma^a\nonumber\\
&&+\frac{1}{4}\delta e_b^{\beta}e^d_{\beta}\Gamma^c_{\ ad}
v\gamma_c\gamma^b\gamma^a
+\frac{1}{4}\delta e^c_{\gamma}e_d^{\gamma}\Gamma^d_{\ ab}
v\gamma_c\gamma^b\gamma^a.\nonumber
\end{eqnarray}
Because $\left[\gamma_c\gamma^b,\gamma^a\right]
=\gamma_c\left\{\gamma^b,\gamma^a\right\}
-\left\{\gamma_c,\gamma^a\right\}\gamma^b
=2\eta^{ab}\gamma_c-2\delta^a_c\gamma^b$
we can write:
\begin{eqnarray}\label{var3}
-\frac{1}{4}\delta e^c_{\beta}e_b^{\beta}\nabla_av
\left[\gamma_c\gamma^b,\gamma^a\right]&=&
-\frac{1}{2}\delta(g_{\mu\beta}\eta^{cd} e_d^{\mu})e_b^{\beta}
\eta^{ab}\nabla_av\gamma_c
+\frac{1}{2}\delta e^c_{\beta}e_b^{\beta}\nabla_cv\gamma^b
\nonumber\\
&=&
-\frac{1}{2}\delta g_{\mu\beta}\eta^{cd} e_d^{\mu}e_b^{\beta}
\eta^{ab}\nabla_av\gamma_c
-\delta e_d^{\mu}e^a_{\mu}\nabla_av\gamma^d\nonumber\\
&=&
\frac{1}{2}\delta g^{\alpha\beta}e^a_{\alpha}e^b_{\beta}
\nabla_av\gamma_b
-\delta e_a^{\alpha}e^d_{\alpha}\nabla_dv\gamma^a.
\end{eqnarray}
When substituting equations (\ref{var2}) and (\ref{var3}) in
(\ref{var1}) we can recombine the terms
\[
\frac{-1}{4}\delta e^c_{\beta}\partial_ae_b^{\beta}
v\gamma_c\gamma^b\gamma^a-\frac{1}{4}\delta e^c_{\gamma}e_a^{\alpha}
e_b^{\beta}\Gamma^{\gamma}_{\ \alpha\beta}v\gamma_c\gamma^b\gamma^a
=\frac{-1}{4}\delta e^c_{\gamma}e_d^{\gamma}\Gamma^d_{\ ab}v
\gamma_c\gamma^b\gamma^a
\]
to obtain
\begin{eqnarray}\label{var4}
\delta\dirop v&=&
\frac{-i}{4}D_c(\delta e^c_{\beta}e_b^{\beta}v\gamma_c\gamma^b)
+\frac{i}{4}\delta e^c_{\beta}e_b^{\beta}(D_cv)\gamma_c\gamma^b\\
&&+\frac{1}{2}\delta g^{\alpha\beta}e^a_{\alpha}e^b_{\beta}
\nabla_av\gamma_b
-\frac{1}{4}\delta\Gamma^{\gamma}_{\ \alpha\beta}
e_a^{\alpha}e_b^{\beta}e^c_{\gamma}v\gamma_c\gamma^b\gamma^a.\nonumber
\end{eqnarray}
Note that the variations of the frame $\delta e_a^{\alpha}$ cancel out,
except in the terms with $D_c$. Therefore, the final answer will not
depend on variations of the frame, as desired.

In the last term of equation (\ref{var4}) we can use the
symmetry of the Christoffel symbol in the lower indices:
\begin{eqnarray}\label{var5}
-\frac{1}{4}\delta\Gamma^{\gamma}_{\ (\alpha\beta)}
e_a^{\alpha}e_b^{\beta}e^c_{\gamma}v\gamma_c\gamma^b\gamma^a&=&
-\frac{1}{4}\delta\Gamma^{\gamma}_{\ \alpha\beta}
e_a^{\alpha}e_b^{\beta}e^c_{\gamma}v\gamma_c\eta^{ab}
=-\frac{1}{4}\delta\Gamma^{\gamma}_{\ \alpha\beta}g^{\alpha\beta}
e^c_{\gamma}v\gamma_c\nonumber\\
&=&
-\frac{1}{4}\delta g^{\gamma\mu}g_{\mu\nu}\Gamma^{\nu}_{\ \alpha\beta}
g^{\alpha\beta}e^c_{\gamma}v\gamma_c
-\frac{1}{4}\partial_{\alpha}\delta g_{\beta\mu}e_a^{\mu}
g^{\alpha\beta}v\gamma^a\nonumber\\
&&+\frac{1}{8}\partial_{\mu}\delta g_{\alpha\beta}e_a^{\mu}
g^{\alpha\beta}v\gamma^a
\end{eqnarray}
We handle the last term as before:
\begin{eqnarray}\label{var6}
\frac{1}{8}\partial_a\delta g_{\alpha\beta}g^{\alpha\beta}v\gamma^a
&=&\frac{-i}{8}D_c(\delta g_{\alpha\beta}g^{\alpha\beta}v)
+\frac{i}{8}\delta g_{\alpha\beta}g^{\alpha\beta}D_cv\nonumber\\
&&-\frac{1}{8}\delta g_{\alpha\beta}\partial_ag^{\alpha\beta}v\gamma^a
\nonumber\\
&=&\frac{-i}{8}D_c(\delta g_{\alpha\beta}g^{\alpha\beta}v)
+\frac{i}{8}\delta g_{\alpha\beta}g^{\alpha\beta}D_cv\\
&&-\frac{1}{8}\delta g^{\alpha\beta}\partial_ag_{\alpha\beta}v
\gamma^a,\nonumber
\end{eqnarray}
where we used $\delta g_{\alpha\beta}\partial_ag^{\alpha\beta}=
-\delta g^{\alpha\beta}g_{\alpha\mu}g_{\beta\nu}\partial_ag^{\mu\nu}
=\delta g^{\alpha\beta}\partial_ag_{\alpha\beta}$.
The second term in equation (\ref{var5}) is:
\begin{eqnarray}\label{var7}
-\frac{1}{4}\partial_{\alpha}\delta g_{\beta\mu}e_a^{\mu}
g^{\alpha\beta}v\gamma^a&=&
\frac{1}{4}\partial_b(\delta g^{\alpha\beta}g_{\alpha\mu}g_{\beta\nu})
e_a^{\mu}e^b_{\rho}g^{\rho\nu}v\gamma^a\nonumber\\
&=&\frac{1}{4}\partial_b(\delta g^{\alpha\beta}e^a_{\alpha}e^b_{\beta})
v\gamma_a-\frac{1}{4}\delta g^{\alpha\beta}g_{\alpha\mu}g_{\beta\nu}
\partial_b(e_a^{\mu}e^b_{\rho}g^{\nu\rho})v\gamma^a\nonumber\\
&=&
\frac{1}{4}\nabla_b(\delta g^{\alpha\beta}e^a_{\alpha}e^b_{\beta})
v\gamma_a\nonumber\\
&&-\frac{1}{4}\delta g^{\alpha\beta}\left(\Gamma^a_{\ bc}e^c_{\alpha}
e^b_{\beta}+\Gamma^b_{\ bc}e^a_{\alpha}e^c_{\beta}\right)v\gamma_a
\nonumber\\
&&
-\frac{1}{4}\delta g^{\alpha\beta}g_{\alpha\mu}g_{\beta\nu}
\partial_b(e_a^{\mu}e^b_{\rho}g^{\rho\nu})v\gamma^a.
\end{eqnarray}
The first term of equation (\ref{var7}) is
\begin{equation}\label{var8}
\frac{1}{4}\nabla_b(\delta g^{\alpha\beta}e^a_{\alpha}e^b_{\beta})
v\gamma_a=
\frac{1}{4}\nabla_b(\delta g^{\alpha\beta}e^a_{\alpha}e^b_{\beta}
v\gamma_a)-
\frac{1}{4}\delta g^{\alpha\beta}e^a_{\alpha}e^b_{\beta}
\nabla_bv\gamma_a.
\end{equation}
The other terms can be simplified using equation
(\ref{linkconnections}) and some computation, which yields:
\begin{eqnarray}\label{var9}
-\frac{1}{4}\delta g^{\alpha\beta}\left(\Gamma^a_{\ bc}e^c_{\alpha}
e^b_{\beta}+\Gamma^b_{\ bc}e^a_{\alpha}e^c_{\beta}+
g_{\alpha\mu}g_{\beta\nu}\eta^{ac}
\partial_b(e_c^{\mu}e^b_{\rho}g^{\rho\nu})\right)v\gamma_a&=&
\nonumber\\
-\frac{1}{4}\delta g^{\alpha\beta}\left(
e^a_{\gamma}\partial_{\beta}e_c^{\gamma}e^c_{\alpha}
+e^a_{\gamma}\Gamma^{\gamma}_{\ \beta\alpha}
+e^a_{\alpha}\partial_{\mu}e_c^{\mu}e^c_{\beta}
+e^a_{\alpha}\Gamma^{\mu}_{\ \mu\beta}
\right.\quad\ \ &&\nonumber\\
\left.
+e^a_{\alpha}g_{\beta\nu}\partial_{\rho}g^{\rho\nu}
+e^a_{\alpha}\partial_be^b_{\beta}
+g_{\alpha\mu}\eta^{ac}\partial_{\beta}e_c^{\mu}
\right)v\gamma_a&=&\nonumber\\
-\frac{1}{4}\delta g^{\alpha\beta}\left(
-\partial_{\beta}e^a_{\alpha}
+e^a_{\gamma}\Gamma^{\gamma}_{\ \beta\alpha}
-e^a_{\alpha}\partial_ce^c_{\beta}
+e^a_{\alpha}\Gamma^{\mu}_{\ \mu\beta}
\right.\quad\ \ &&\nonumber\\
\left.
-e^a_{\alpha}g^{\rho\nu}\partial_{\rho}g_{\beta\nu}
+e^a_{\alpha}\partial_be^b_{\beta}
+g_{\alpha\mu}\eta^{ac}\partial_{\beta}e_c^{\mu}
\right)v\gamma_a&=&\nonumber\\
-\frac{1}{4}\delta g^{\alpha\beta}\left(
-\eta^{ac}e_c^{\mu}\partial_{\beta}g_{\alpha\mu}
+e^a_{\gamma}\Gamma^{\gamma}_{\ \beta\alpha}
+e^a_{\alpha}\Gamma^{\mu}_{\ \mu\beta}
-e^a_{\alpha}g^{\rho\nu}\partial_{\rho}g_{\beta\nu}
\right)v\gamma_a&=&\nonumber\\
-\frac{1}{8}\delta g^{\alpha\beta}\left(
-2e^a_{\gamma}g^{\gamma\mu}\partial_{\beta}g_{\alpha\mu}
+e^a_{\gamma}g^{\gamma\mu}(2\partial_{\beta}g_{\alpha\mu}
-\partial_{\mu}g_{\alpha\beta})
\right.\quad\ \ &&\nonumber\\
\left.
+e^a_{\alpha}g^{\mu\gamma}\partial_{\beta}g_{\mu\gamma}
-2e^a_{\alpha}g^{\rho\nu}\partial_{\rho}g_{\beta\nu}
\right)v\gamma_a&=&\nonumber\\
\frac{1}{8}\delta g^{\alpha\beta}\left(
e^a_{\gamma}g^{\gamma\mu}
\partial_{\mu}g_{\alpha\beta}
+2e^a_{\alpha}g_{\beta\mu}g^{\rho\nu}\Gamma^{\mu}_{\ \rho\nu}
\right)v\gamma_a.&&
\end{eqnarray}
Substituting equations (\ref{var5}-\ref{var9}) into (\ref{var4})
yields:
\begin{eqnarray}\label{var10}
\delta\dirop v&=&
\frac{-i}{4}D_c(\delta e^c_{\beta}e_b^{\beta}v\gamma_c\gamma^b)
+\frac{i}{4}\delta e^c_{\beta}e_b^{\beta}(D_cv)\gamma_c\gamma^b
\nonumber\\
&&-\frac{i}{8}D_c(\delta g_{\alpha\beta}g^{\alpha\beta}v)
+\frac{i}{8}\delta g_{\alpha\beta}g^{\alpha\beta}D_cv\nonumber\\
&&+\frac{1}{4}\delta g^{\alpha\beta}e^a_{\alpha}e^b_{\beta}
\nabla_av\gamma_b
+\frac{1}{4}\nabla_b(\delta g^{\alpha\beta}e^a_{\alpha}e^b_{\beta}
v\gamma_a).
\end{eqnarray}
Using lemma \ref{adjointvar} and introducing $D_s:=-i\dirop+m$ we find
for a spinor field $u\in C^{\infty}(DM)$:
\begin{eqnarray}\label{var11}
\delta\dirop u&=&
\frac{i}{4}D_s(\delta e^c_{\beta}e_b^{\beta}\gamma^b\gamma_cu)
-\frac{i}{4}\delta e^c_{\beta}e_b^{\beta}\gamma^b\gamma_c(D_su)
\nonumber\\
&&+\frac{i}{8}D_s(\delta g_{\alpha\beta}g^{\alpha\beta}u)
-\frac{i}{8}\delta g_{\alpha\beta}g^{\alpha\beta}D_su\nonumber\\
&&+\frac{1}{4}\delta g^{\alpha\beta}e^a_{\alpha}e^b_{\beta}
\gamma_b\nabla_au
+\frac{1}{4}\nabla_b(\delta g^{\alpha\beta}e^a_{\alpha}e^b_{\beta}
\gamma_au).
\end{eqnarray}

Using the same notation as in proposition \ref{SEMTresult} we
find the result:
\begin{theorem}\label{thm_Dcommutator}
For a double test-spinor $f\in\mathcal{D}_0(M)$ with
$\mathrm{supp}\ f\subset N^+$ and for $x\in K$:
\begin{eqnarray}
\frac{\delta}{\delta g^{\alpha\beta}(x)}\left(\beta_gB_M(f)
\right)&=&B_M\left(\frac{\delta}{\delta g^{\alpha\beta}(x)} D_gSf
\right)\nonumber\\
&=&
\frac{i}{2}e^a_{\alpha}e^b_{\beta}\left[B_M(f),T^s_{ab}(x,x)\right].
\end{eqnarray}
\end{theorem}
\begin{proof*}
Using proposition \ref{varRCE} and equations (\ref{var10},\ref{var11})
we notice that the terms with $D_c$ and $D_s$ cancel out, because
$B_M$ and $Sf$ satisfy the (doubled) Dirac equation:
\begin{eqnarray}
\delta(\beta_{\epsilon}B_M(f))=
B_M(\delta D_{\epsilon}Sf)&=&\nonumber\\
\frac{-i}{4}B_M(\delta g^{\alpha\beta}e^a_{\alpha}e^b_{\beta}
\gamma_b\nabla_aSRf)
-\frac{i}{4}B_M(\nabla_b(\delta g^{\alpha\beta}e^a_{\alpha}
e^b_{\beta}\gamma_aSRf))&=&\nonumber\\
\frac{-i}{4}\delta g^{\alpha\beta}e^a_{\alpha}e^b_{\beta}
\left(B_M(\gamma_{(b}\nabla_{a)}SRf)-\nabla_{(b}
B_M(\gamma_{a)}SRf)\right).
\end{eqnarray}
We now compare with proposition \ref{SEMTresult} to obtain the result.
\end{proof*}

This result compares well with the scalar field case, theorem 4.3 in
\cite{Brunetti+}. As particular cases we obtain for $\psi_M$ and
$\psi^+_M$:
\begin{eqnarray}
\frac{\delta}{\delta g^{\alpha\beta}(x)}(\beta_g\psi_M(v))&=&
\frac{i}{2}e^a_{\alpha}e^b_{\beta}
\left[\psi_M(v),T^s_{ab}(x,x)\right],\nonumber\\
\frac{\delta}{\delta g^{\alpha\beta}(x)}(\beta_g\psi^+_M(u))&=&
\frac{i}{2}e^a_{\alpha}e^b_{\beta}
\left[\psi^+_M(u),T^s_{ab}(x,x)\right].\nonumber
\end{eqnarray}

\begin{corollary}
Let $X\in\Alg{F}^0_M$ and $x\in K$, then
\[
\frac{\delta}{\delta g^{\alpha\beta}(x)}\left(\beta_g X\right)=
\frac{i}{2}e^a_{\alpha}e^b_{\beta}\left[X,T^s_{ab}(x,x)\right].
\]
\end{corollary}
\begin{proof*}
Theorem \ref{thm_Dcommutator} tells us that the equation is true if
$X=B_M(f)$ for any double test-spinor $f\in\mathcal{D}_0(M)$ with
$\mathrm{supp}\ f\subset N^+$. The same is then true for any monomial
of such terms, because for $X_1,X_2\in\Alg{F}^0_M$ we have
$\beta_g(X_1X_2)=\beta_g(X_1)\beta_g(X_2)$ and hence
\[
\frac{\delta}{\delta g^{\alpha\beta}(x)}\left(\beta_g(X_1X_2)\right)=
\frac{\delta}{\delta g^{\alpha\beta}(x)}\left(\beta_g(X_1)\right)X_2
+X_1\frac{\delta}{\delta g^{\alpha\beta}(x)}\left(\beta_g(X_2)\right)
\]
and $[X_1X_2,T]=[X_1,T]X_2+X_1[X_2,T]$ for any operator $T$. Finally,
because the equation is linear in $X$ it holds for any polynomial of
terms $B_M(f)$ with $f$ supported in $N^+$. Any
$X\in\Alg{F}^0_M$ is of this form, by lemma \ref{lem_Dts1}. This
completes the proof.
\end{proof*}

\chapter{The Reeh-Schlieder property in curved spacetime}\label{ch_RS}

\begin{quote}
Er st\"urzte sich allerdings in das Unerme\ss liche, das die
astro\-physische Wissenschaft zu messen sucht, nur um dabei zu Ma\ss en,
Zahlen, Gr\"o\ss enordnungen zu gelangen, zu denen der Menschen\-geist gar
kein Verh\"altnis mehr hat, und die sich im Theoretischen und
Abstrakten, im v\"ollig Unsinnlichen, um nicht zu sagen: Unsinnigen
verlieren.
\end{quote}
\begin{flushright}
Thomas Mann, Doktor Faustus, Ch. 27
\end{flushright}

The Reeh-Schlieder theorem \cite{Reeh+} is a result in axiomatic
quantum field theory which states that for a scalar Wightman field in
Minkowski spacetime any state in the Hilbert space can be approximated
arbitrarily well by acting on the vacuum with operations performed in
any prescribed open region. The physical meaning of this is that the
vacuum state has very many non-local correlations and an experimenter
in any given region can exploit the vacuum fluctuations by performing
a suitable measurement in order to produce any desired state up to
arbitrary accuracy.

The original proof uses analytic continuation arguments, an approach
which was extended to analytic spacetimes in \cite{Strohmaier+} by
replacing the spectrum condition of the Wightman axioms in Minkowski
spacetime by an analytic
microlocal spectrum condition. For spacetimes which are not analytic a
result by Strohmaier \cite{Strohmaier} (see also \cite{Verch3})
shows that in a stationary
spacetime all ground and thermal (KMS-)states of several types of free
fields (including the Klein-Gordon, Dirac and Proca field) also have the
Reeh-Schlieder property. To prove the existence of such states directly
one may need to make further assumptions, depending on the type of
field (see \cite{Strohmaier}).

In this chapter we will investigate whether Reeh-Schlieder states exist
in general globally hyperbolic spacetimes, which may be neither analytic
nor stationary. First we will define and discuss the Reeh-Schlieder
property in the context of locally covariant quantum field theory in
section \ref{sec_RSintro} and discuss the relevant fact that not all
states on an algebra need to be in the physical state space. Next we
will prove some general results in section \ref{sec_RSgen}, namely that
the Reeh-Schlieder property is
local and stable under purifications. We then proceed to discuss the
possibility of deforming a Reeh-Schlieder state on one spacetime into a
Reeh-Schlieder state on a diffeomorphic (but not isometric) spacetime in
section \ref{sec_RSdeform}. For this we use the time-slice axiom and the
technique of spacetime
deformation as pioneered in \cite{Fulling+} and as applied successfully
to prove a spin-statistics theorem in curved spacetime in \cite{Verch1}.
We will prove that, given a Reeh-Schlieder state on the initial
globally hyperbolic
spacetime, we can find for every region in the deformed spacetime a state
in the physical state space that has the Reeh-Schlieder property for that
particular region (but maybe not for all regions). After these general
results we specialise in section \ref{sec_WFRS} to the
Borchers-Uhlmann functor $\Func{U}$
of chapter \ref{ch_sf} and give a smoothly covariant condition on states
that guarantees that
a state has the Reeh-Schlieder property and satisfies the $\mu$SC.
Next we specialise even further to the real free scalar field in
Minkowski spacetime and we prove the existence of many Hadamard
Reeh-Schlieder states in section \ref{sec_RSM0}. Finally, we draw some
conclusions concerning the Reeh-Schlieder property in locally
covariant quantum field theory in section \ref{sc_RSconclusion}.

\section[The Reeh-Schlieder property in LCQFT]{The Reeh-Schlieder
property in a locally covariant quantum field theory}\label{sec_RSintro}

In locally covariant quantum field theory we define the Reeh-Schlieder
property as follows:
\begin{definition}\label{def_LCQFTRS}
Consider a locally covariant quantum field theory $\Func{A}$ with a
state space $\Func{S}$. A state $\omega\in S_M$ has the
\emph{Reeh-Schlieder property} for a cc-region $O\subset M$ iff
\[
\overline{\pi_{\omega}(\mathcal{A}_O)\Omega_{\omega}}=
\mathcal{H}_{\omega}.
\]
We then say that $\omega$ is a \emph{Reeh-Schlieder state for $O$}. We
say that $\omega$ is a \emph{(full) Reeh-Schlieder state}, or that
$\omega$ has the \emph{(full) Reeh-Schlieder property}, iff it is a
Reeh-Schlieder state for all cc-regions in $M$.
\end{definition}

The original result of \cite{Reeh+} then states that the vacuum state
$\omega$ of a scalar Wightman field theory in Minkowski spacetime $M_0$
has the full Reeh-Schlieder property. It implies that the vacuum is an
entangled state even over causally disjoint regions of spacetime
\cite{Landau,Clifton+2}. (Even if a state has the Reeh-Schlieder
property only for a certain cc-region $O$ and the theory is nowhere
classical, there exist non-local correlations between $O$ and any
cc-region $V$ space-like to it \cite{Redhead}). Furthermore, the
entanglement can be improved using a
distillation procedure (see \cite{Verch+}) to approximate a
maximal violation of the Bell inequalities (for appropriate observables
in the two disjoint regions). Moreover, it is argued in \cite{Clifton+}
that these non-local correlations cannot easily be avoided. Indeed, if
there is one vector in a Hilbert space which defines a state with the
Reeh-Schlieder property for a theory defined by $C^*$-algebras, and this
is the case for example for the Minkowski vacuum of the real free scalar
field in terms of the local Weyl algebras (i.e. using the functor
$\Func{A}^0$ of chapter \ref{ch_sf}),
then the same is true for a generic vector in that Hilbert space. (We will
make this statement more precise below in definition \ref{def_generic}.)
Therefore, by Fell's theorem (see \cite{Haag} theorem 3.2.2.13), we cannot
distinguish a Reeh-Schlieder vector-state from a vector-state that does not
have the Reeh-Schlieder property.\footnote{\cite{Clifton+} contrasts
this with the entanglement that can occur between two systems in quantum
mechanics and that can be undone by performing a measurement on one of
the systems. However, \cite{Clifton+} also argues that scientific
methodology is not in danger due to another property commonly found in
quantum field theories, namely the split property (see e.g.
\cite{Haag}). This allows one to isolate systems ``for all practical
purposes''.} Another consequence of the Reeh-Schlieder theorem is that
every nontrivial positive local operator has a strictly positive vacuum
expectation value (see e.g. proposition \ref{separability1} ahead). The
Reeh-Schlieder theorem therefore poses a problem for a notion
of localised particles, because it is impossible to create, annihilate
or even just count particles using local operations and again these
problems cannot easily be avoided \cite{Halvorson}.

Apparently the conclusion must be that entanglement is the rule rather
than the exception. On the other hand, it can be argued that the energy
required to approximate a given state increases with the desired
accuracy (see \cite{Haag} p.254, \cite{Clifton+}). In other words, it
may take an increasingly large ensemble in order for a selective
measurement to give a positive result within the desired range of
accuracy. It is also known that the strength of any non-classical
correlations decay exponentially with the separation between the regions
of interest (see \cite{Summers+}). The Reeh-Schlieder theorem also has
many theoretical implications, both of a mathematical and of a physical
nature. On the mathematical side it can be used to determine the type of
local von Neumann algebras \cite{Kadison,Araki2} and it allows the
application of Tomita-Takesaki modular theory, which has led to a field
of research in its own right (see \cite{Borchers} for a review).

An extension of the Reeh-Schlieder theorem to curved spacetimes would
have physical implications as well. One of its consequences would be
the existence of correlations between observables localised in regions
which are causally disjoint, or even in regions which have always been
causally disjoint. This has led to speculations about its importance for
the physical understanding of e.g. black holes (\cite{Banach}) and the
early universe. As Wald puts it in the beautiful little essay
\cite{Wald3}:
\begin{quote}
The point I \emph{do} wish to make is that -- at the very least -- it is
far from obvious, \emph{a priori}, that the relevant correlations beyond
the horizon in a quantum field theory model will be small, and the
neglect of such correlations in analyzing any phenomenon must be
justified by quantitative estimates rather than by a simple appeal to a
lack of causal communication. Indeed, because of the fundamental and
universal nature of these correlations, it would be surprising if they
did not play some important role in our understanding of the nature of
the early universe.
\end{quote}
Wald then goes on to state that
\begin{quote}
\ldots the strength and generality of the Reeh-Schleider [sic] theorem
in flat spacetime is such that it seems inconceivable that similar
correlations could fail to be present for essentially all states and
over essentially all regions in any curved spacetime, including
cosmological spacetimes with horizons.
\end{quote}

As we already mentioned, the Reeh-Schlieder property has indeed been
shown to hold in curved spacetimes too, under certain conditions.
However, there are also some arguments that suggest it may not be
quite as general as Wald suggests. Indeed, even in Minkowski spacetime
it can be argued that the property may not be as omnipresent as the
references above make us believe. To be specific, let us examine the
case of the Minkowski vacuum state $\omega_0$ of the real free scalar
field, described by a net of algebras \cite{Haag,Brunetti+}. If this
field is described
by the functor $\Alg{A}^0$ (see chapter \ref{ch_sf}), then the results
of \cite{Clifton+,Dixmier+} do show that a generic vector in
$\mathcal{H}_{\omega_0}$ defines a state in the state space
$\Stat{S}^0_{M_0}$ (for the state space defined in
chapter \ref{ch_sf}) which has the Reeh-Schlieder property. But now
suppose that we describe the real free scalar field by the free field
Borchers-Uhlmann functor $\Func{U}^0$. Although both functors are
meant to describe the same field, and despite the correspondence
between the descriptions (see the discussion above and below
definition \ref{def_regular}), the mathematical situation is very
different. One can check in this concrete example that a
Reeh-Schlieder state for $\Alg{U}^0_{M_0}$ is also a Reeh-Schlieder
state for
$\Alg{A}^0_{M_0}$ (see e.g. \cite{Araki3} theorem 4.16, \cite{Buchholz}),
however, how do we know that such a Reeh-Schlieder state, defined by a
generic vector in the Hilbert space $\mathcal{H}_{\omega_0}$, is
actually in the state space $\Stat{Q}^0_{M_0}$? In other words, how do
we know if such a Reeh-Schlieder state is of any physical interest?

The next question is therefore: how big is the difference between the
state spaces $\Stat{Q}^0_{M_0}$ for $\Alg{U}^0_{M_0}$ and
$\Stat{S}^0_{M_0}$ for $\Alg{A}^0_{M_0}$? Note that both of these state
spaces contain all quasi-free Hadamard states, which are certainly of
physical interest. However, to make $\Stat{S}^0_{M_0}$ closed under
operations from $\Alg{A}^0_{M_0}$, which is required by our definition of
a state space (see definition \ref{def_states}), we included all states
that are locally quasi-equivalent to a quasi-free Hadamard state. This
makes the state space $\Stat{S}^0_{M_0}$ much larger than
$\Stat{Q}^0_{M_0}$, at least in the sense that a generic vector in
$\mathcal{H}_{\omega}$ does not define a Hadamard state on
$\Alg{U}^0_{M_0}$. To close this subsection we will make this
statement precise, which requires some more terminology (see
e.g. \cite{Csaszar}).

\begin{definition}\label{def_generic}
A \emph{$G_{\delta}$ set} in a topological space $T$ is a countable
intersection of open sets. An \emph{$F_{\sigma}$ set} is the
complement of a $G_{\delta}$ set.

A \emph{Baire space} is a topological space in which any countable
intersection of dense open sets is a dense set.

We say that a property is \emph{generic} in a Baire space $T$ iff
there is a dense $G_{\delta}$ in $T$ of elements which have this
property.
\end{definition}
It is known that every complete pseudo-metric space, and in particular
every Hilbert space, is a Baire space by Baire's theorem (see \cite{Csaszar}
section 9.2b). Note that for any countable sequence
of generic properties $P_n$ the property $P$ of having all $P_n$ is
still generic. It follows from \cite{Dixmier+} (see also
\cite{Clifton+}) that the property that a vector in
$\mathcal{H}_{\omega_0}$ defines a Reeh-Schlieder state is a generic
property. We will now argue that the property of not defining a
Hadamard state is also generic. Indeed, a vector $\psi$ which defines a
Hadamard state must be in the domain of the unbounded field operator
$\Phi^{(\omega_0)}(f)$ for every $f\in\Test_0(M_0)$. Using
$\Phi^{(\omega_0)}(\bar{f})=\Phi^{(\omega_0)}(f)^*|_{\mathscr{D}_{\omega_0}}$
we see that $\psi$ must also be in the domain of
$T:=\Phi^{(\omega_0)}(f)^{**}\Phi^{(\omega_0)}(f)^*$, which is a
self-adjoint operator (see \cite{Kadison+} theorem 2.7.8v). The domain of
$T$, although dense, is the complement of a dense $G_{\delta}$:
\begin{lemma}\label{meagredomain}
The domain of a self-adjoint operator $T$ on a Hilbert space
$\mathcal{H}$ is a meagre $F_{\sigma}$, i.e. it is the
complement of a dense $G_{\delta}$.
\end{lemma}
\begin{proof*}
For every $n\in\N$ we define
$V_n:=\left\{\psi\in\mathrm{dom}(T)|\ \|T\psi\|\le n\|\psi\|\right\}$,
where $\mathrm{dom}(T)$ denotes the domain of $T$. Note that
$\mathrm{dom}(T)=\cup_n V_n$. To show that
$V_n$ is closed we choose a Cauchy sequence $\psi_i\in V_n$
such that $\psi_i\rightarrow\psi\in\mathcal{H}$. We let $E_{[-r,r]}$
denote the spectral projection of $T$ on the interval $[-r,r]$ and
compute: $\|TE_{[-r,r]}\psi\|\le\|TE_{[-r,r]}(\psi-\psi_i)\|+
\|TE_{[-r,r]}\psi_i\|\le r\|\psi-\psi_i\|+n\|\psi_i\|$. Taking
$i\rightarrow\infty$ shows that $\|TE_{[-r,r]}\psi\|\le n\|\psi\|$ for
all $r$ and hence $\|T\psi\|\le n\|\psi\|$, i.e. $\psi\in V_n$.
Finally the sets $V_n$ are nowhere dense, because $T$ is
unbounded. This completes the proof.
\end{proof*}

\section{Some general results on the Reeh-Schlieder
property}\label{sec_RSgen}

As a prelude to our study of the Reeh-Schlieder property in curved
spacetimes we will now prove some relatively easy statements which
hold under very general assumptions. In this subsection we will first
consider a fixed globally hyperbolic spacetime $M$ and a locally
covariant quantum field theory $\map{\Func{A}}{\CatSMan}{\CatAlg}$
with a state space $\Func{S}$.

We first prove a well-known consequence of the Reeh-Schlieder
theorem (see e.g. \cite{Haag}).
\begin{definition}
A vector $v$ in a Hilbert space $\mathcal{H}$ is a
\emph{separating vector} for a $C^*$-algebra $\Alg{A}$ of operators
on $\mathcal{H}$ iff $Av=0$ with $A\in\Alg{A}$ implies $A=0$.
\end{definition}
\begin{proposition}\label{separability1}
Let $\omega\in\Stat{S}_M$ be a state on the $C^*$-algebra $\Alg{A}_M$
which has the Reeh-Schlieder property for a cc-region $O\subset M$.
Assume that $\Func{A}$ is causal and let $V\subset O^{\perp}$ be a
cc-region, then $\Omega_{\omega}$ is a separating vector for
$\Alg{R}^{(\omega)}_V$.
\end{proposition}
\begin{proof*}
Suppose that $A\Omega_{\omega}=0$ for some $A\in\Alg{R}^{(\omega)}_V$,
then $A\pi_{\omega}(B)\Omega_{\omega}=\pi_{\omega}(B)A\Omega_{\omega}=0$
for all $B\in\Alg{A}_O$. By the Reeh-Schlieder property the set
$\pi_{\omega}(B)\Omega_{\omega}$ is dense, so by continuity of $A$ we
find $Av=0$ for all $v\in\mathcal{H}$ and hence $A=0$.
\end{proof*}
This result implies that every non-zero positive operator $A^*A$ in
$\Alg{R}^{(\omega)}_V$ has a strictly positive expectation value in the
state $\omega$, because if we have
$\omega(A^*A)=\|A\Omega_{\omega}\|^2=0$ then $A=0$.

Next we prove that the Reeh-Schlieder property is stable under
purification, which appears to be a hitherto unknown result:
\begin{proposition}
Let $\omega\in\Stat{S}_M$ be a state which has the Reeh-Schlieder
property for a cc-region $O\subset M$ and suppose that
$\omega$ is a mixture of $\omega_1,\omega_2\in\Stat{S}_M$, i.e.
$\omega=\lambda\omega_1+(1-\lambda)\omega_2$ with $0<\lambda\le 1$.
Then $\omega_1$ also has the Reeh-Schlieder property for $O$.
\end{proposition}
\begin{proof*}
We fix arbitrary $\psi\in\mathcal{H}_{\omega_1}$ and $\epsilon>0$
and use theorem \ref{GNSrep} to find an $A\in\Alg{A}_M$ such that
$\|\psi-\pi_{\omega_1}(A)\Omega_{\omega_1}\|<\frac{\epsilon}{2}$.
We can then  find $B\in\Alg{A}_O$ such that
$\omega((A-B)^*(A-B))=\|\pi_{\omega}(A-B)\Omega_{\omega}\|^2<
\frac{\lambda\epsilon^2}{4}$, by the assumed Reeh-Schlieder
property. Then
\[
\|\pi_{\omega_1}(A-B)\Omega_{\omega_1}\|^2=\omega_1((A-B)^*(A-B))\le
\frac{1}{\lambda}\omega((A-B)^*(A-B))<\frac{\epsilon^2}{4}
\]
and hence $\|\psi-\pi_{\omega_1}(B)\Omega_{\omega_1}\|\le
\|\psi-\pi_{\omega_1}(A)\Omega_{\omega_1}\|+
\|\pi_{\omega_1}(A-B)\Omega_{\omega_1}\|<\epsilon$.
\end{proof*}
It is interesting to note that the same purification argument works
for the Hadamard condition on states of the real free scalar field:
\begin{proposition}
If $\omega$ is a Hadamard state on $\Alg{U}^0_M$ and
$\omega=\lambda\omega_1+(1-\lambda)\omega_2$ with $0<\lambda\le 1$
for any states $\omega_i$ on $\Alg{U}^0_M$, then $\omega_1$ is a
Hadamard sate.
\end{proposition}
\begin{proof*}
By positivity of $\omega_1$ we have the Cauchy-Schwarz inequality
for all $A,B\in\Alg{U}^0_M$:
$|\omega_1(B^*A)|^2\le\omega_1(B^*B)\omega_1(A^*A)$. In particular,
$|\omega_1(A)|^2\le\omega_1(A^*A)\le\frac{1}{\lambda}\omega(A^*A)$. If
$A_n\rightarrow A$ in $\Alg{U}^0_M$ then $(A-A_n)^*(A-A_n)\rightarrow 0$
by definition \ref{def_BUalg} and the continuity of the canonical
projection $\map{p}{\Alg{U}_M}{\Alg{U}^0_M}$. Hence
$|\omega_1(A_n-A)|\rightarrow 0$, which proves that $\omega_1$ is a
continuous state on $\Alg{U}^0_M$. For the two-point distribution we have
$(\omega_1)_2(\overline{f},f)\le\frac{1}{\lambda}
\omega_2(\overline{f},f)$ which can be rewritten in terms of
Hilbert-space-valued distributions as
$\|\phi^{(\omega_1)}_1(f)\|\le\|\phi^{(\omega)}_1(f)\|$. (Note that
these distributions may take values in different Hilbert spaces.)
This implies $WF(\phi^{(\omega_1)}_1)\subset WF(\phi^{(\omega)}_1)$ and
hence by theorem \ref{WFHvalued}
$WF((\omega_1)_2)\subset\mathcal{N}^-\times\mathcal{N}^+$.
Using the commutation relations and propagation of singularities as in
the proof of proposition \ref{lem_ts2} it now follows that $\omega_1$ is
Hadamard.
\end{proof*}

To conclude this section we prove that for an additive
$C^*$-algebraic theory the Reeh-Schlieder property is a local property.
This is a new and interesting result, but we will not need it elsewhere.
\begin{theorem}\label{RSlocal}
Consider an additive locally covariant quantum field theory
$\map{\Func{A}}{\CatMan}{\CatAlg}$, a globally hyperbolic spacetime
$M$ and a state $\omega$ on $\Alg{A}_M$. Assume that every point
$p\in\mathcal{M}$ is contained in a cc-region $O$ such that
$\omega|_{\Alg{A}_O}$ has the Reeh-Schlieder property, i.e. such that
for every cc-region $V\subset O$ we have
$\overline{\pi_{\omega}(\Alg{A}_V)\Omega_{\omega}}=
\overline{\pi_{\omega}(\Alg{A}_O)\Omega_{\omega}}
=\mathcal{H}_{\omega|_{\Alg{A}_O}}$. Then $\omega$ is a
Reeh-Schlieder state.
\end{theorem}
As a matter of terminology we will say that $\omega$ has the
Reeh-Schlieder property \emph{on} $O$ iff $\omega|_{\Alg{A}_O}$ has
the Reeh-Schlieder property. (Cf. definition \ref{def_LCQFTRS}).
\begin{proof*}
Let $U_i$, $i\in I$ be an open covering of $M$ by cc-regions such that
$\omega|_{\Alg{A}_{U_i}}$ has the Reeh-Schlieder property on $U_i$ and
let $O\subset M$ be an arbitrary cc-region. We wish
to show that $\omega$ has the Reeh-Schlieder property for $O$. This is
certainly the case if $\omega$ has the Reeh-Schlieder property for a
subset of $O$, so without loss of generality we may shrink $O$ and
assume that
$O\subset U_a$ for some index $a\in I$. Now suppose for the moment that
$\omega$ has the Reeh-Schlieder property \emph{for} $U_a$ (and not just
\emph{on} $U_a$). Given arbitrary
$\psi\in\mathcal{H}_{\omega}$ and $\epsilon>0$ we can then find
$A\in\Alg{A}_{U_a}$ such that
$\|\psi-\pi_{\omega}(A)\Omega_{\omega}\|<\frac{\epsilon}{2}$. Moreover,
because the
restriction $\omega'$ of $\omega$ to $\Alg{A}_{U_a}$ is a
Reeh-Schlieder state we can find a $B\in\Alg{A}_O$ such that
\begin{eqnarray}\label{RSrepestimate}
\|\pi_{\omega}(A-B)\Omega_{\omega}\|^2&=&\omega((A-B)^*(A-B))\\
&=&\omega'((A-B)^*(A-B))
=\|\pi_{\omega'}(A-B)\Omega_{\omega'}\|^2<\frac{\epsilon^2}{4}.
\nonumber
\end{eqnarray}
Together this implies that
$\|\psi-\pi_{\omega}(B)\Omega_{\omega}\|<\epsilon$, so $\omega$
then has the Reeh-Schlieder property for $O$. It remains to prove that
$\omega$ has the Reeh-Schlieder property for $U_a$.
For this we prove the following lemma.

\begin{figure}
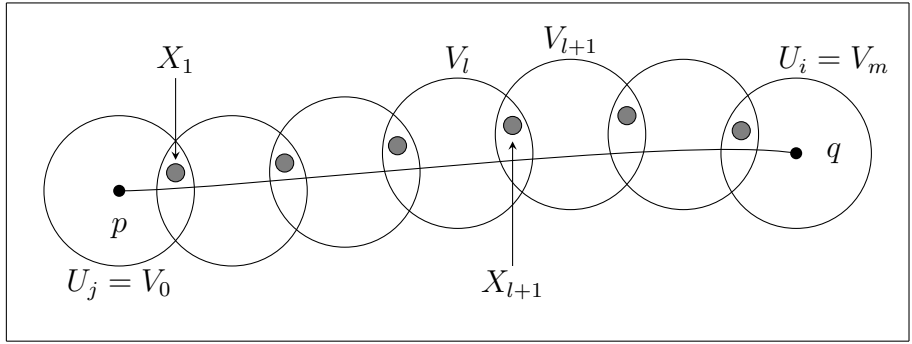

\begin{center}
\begin{pgfpicture}{0cm}{0cm}{12cm}{4.5cm}
\pgfline{\pgfxy(0,0)}{\pgfxy(12,0)}
\pgfline{\pgfxy(12,0)}{\pgfxy(12,4.5)}
\pgfline{\pgfxy(12,4.5)}{\pgfxy(0,4.5)}
\pgfline{\pgfxy(0,4.5)}{\pgfxy(0,0)}

\pgfxycurve(1.5,2)(3,2)(9,2.75)(10.5,2.5)
\pgfellipse[fill]{\pgfxy(1.5,2)}{\pgfxy(0.08,0)}{\pgfxy(0,0.08)}
\pgfellipse[fill]{\pgfxy(10.5,2.5)}{\pgfxy(0.08,0)}{\pgfxy(0,0.08)}

\pgfellipse[stroke]{\pgfxy(1.5,2)}{\pgfxy(1,0)}{\pgfxy(0,1)}
\pgfellipse[stroke]{\pgfxy(3,2)}{\pgfxy(1,0)}{\pgfxy(0,1)}
\pgfellipse[stroke]{\pgfxy(4.5,2.25)}{\pgfxy(1,0)}{\pgfxy(0,1)}
\pgfellipse[stroke]{\pgfxy(6,2.5)}{\pgfxy(1,0)}{\pgfxy(0,1)}
\pgfellipse[stroke]{\pgfxy(7.5,2.75)}{\pgfxy(1,0)}{\pgfxy(0,1)}
\pgfellipse[stroke]{\pgfxy(9,2.75)}{\pgfxy(1,0)}{\pgfxy(0,1)}
\pgfellipse[stroke]{\pgfxy(10.5,2.5)}{\pgfxy(1,0)}{\pgfxy(0,1)}

\color[gray]{0.5}
\pgfellipse[fill]{\pgfxy(2.25,2.24)}{\pgfxy(0.12,0)}{\pgfxy(0,0.12)}
\pgfellipse[fill]{\pgfxy(3.7,2.37)}{\pgfxy(0.12,0)}{\pgfxy(0,0.12)}
\pgfellipse[fill]{\pgfxy(5.2,2.6)}{\pgfxy(0.12,0)}{\pgfxy(0,0.12)}
\pgfellipse[fill]{\pgfxy(6.73,2.87)}{\pgfxy(0.12,0)}{\pgfxy(0,0.12)}
\pgfellipse[fill]{\pgfxy(8.25,3)}{\pgfxy(0.12,0)}{\pgfxy(0,0.12)}
\pgfellipse[fill]{\pgfxy(9.77,2.8)}{\pgfxy(0.12,0)}{\pgfxy(0,0.12)}
\color[gray]{0}
\pgfellipse[stroke]{\pgfxy(2.25,2.24)}{\pgfxy(0.12,0)}{\pgfxy(0,0.12)}
\pgfellipse[stroke]{\pgfxy(3.7,2.37)}{\pgfxy(0.12,0)}{\pgfxy(0,0.12)}
\pgfellipse[stroke]{\pgfxy(5.2,2.6)}{\pgfxy(0.12,0)}{\pgfxy(0,0.12)}
\pgfellipse[stroke]{\pgfxy(6.73,2.87)}{\pgfxy(0.12,0)}{\pgfxy(0,0.12)}
\pgfellipse[stroke]{\pgfxy(8.25,3)}{\pgfxy(0.12,0)}{\pgfxy(0,0.12)}
\pgfellipse[stroke]{\pgfxy(9.77,2.8)}{\pgfxy(0.12,0)}{\pgfxy(0,0.12)}

\pgfsetendarrow{\pgfarrowsingle}
\pgfline{\pgfxy(2.25,3.5)}{\pgfxy(2.25,2.44)}
\pgfline{\pgfxy(6.73,1)}{\pgfxy(6.73,2.63)}

\pgfputat{\pgfxy(1.5,1.5)}{\pgfbox[center,center]{$p$}}
\pgfputat{\pgfxy(11,2.5)}{\pgfbox[center,center]{$q$}}
\pgfputat{\pgfxy(1.5,0.75)}{\pgfbox[center,center]{$U_j=V_0$}}
\pgfputat{\pgfxy(6,3.75)}{\pgfbox[center,center]{$V_l$}}
\pgfputat{\pgfxy(7.5,4)}{\pgfbox[center,center]{$V_{l+1}$}}
\pgfputat{\pgfxy(11,3.75)}{\pgfbox[center,center]{$U_i=V_m$}}
\pgfputat{\pgfxy(2.25,3.75)}{\pgfbox[center,center]{$X_1$}}
\pgfputat{\pgfxy(6.73,0.75)}{\pgfbox[center,center]{$X_{l+1}$}}
\end{pgfpicture}
\end{center}
\caption{Sketch depicting the proof of lemma
\ref{RShopping}.}\label{fig5}
\end{figure}

\begin{lemma}\label{RShopping}
For all $\epsilon>0$, $i,j\in I$ and $A\in\Alg{A}_{U_i}$ there is a
$B\in\Alg{A}_{U_j}$ such that
$\|\pi_{\omega}(A-B)\Omega_{\omega}\|<\epsilon$.
\end{lemma}
\begin{proof*}
We refer to figure \ref{fig5} for a depiction of the geometry of this
proof.

Let $\epsilon>0$, $i,j\in I$ and $A\in\Alg{A}_{U_i}$ be given. Because
$M$ is connected we can find a continuous curve
$\map{\gamma}{[0,1]}{M}$ between points $p:=\gamma(0)\in U_j$ and
$q:=\gamma(1)\in U_i$. For each point $t\in [0,1]$ we choose $k(t)\in I$
such that $U_{k(t)}$ contains the point $\gamma(t)$. The image of
$\gamma$ is a compact subset of $M$, because it is the continuous image
of a compact set. Therefore, we can find a finite number of points
$t_1,\ldots,t_n\in[0,1]$ such that the cc-regions $V_n:=U_{k(t_n)}$
cover the image of $\gamma$. We now choose a subcover inductively as
follows. First we take $p\in V_0:=U_j$. We let
$t_1:=\min\left\{s\in[0,1]|\ \gamma(s)\not\in V_0 \right\}$ and choose
$V_1$ such that $t_1\in V_1$. We then set
$t_2:=\min\left\{s\in[t_1,1]|\ \gamma(s)\not\in V_1\right\}$ and choose
$V_2$ such that $t_2\in V_2$. We proceed in this way until we find
an index $m$ with $q\in V_m$. Extending the sequence $V_0,\ldots,V_m$
by one set if necessary we may assume that $V_m:=U_i$. Notice that
by construction $V_{l-1}\cap V_l\not=\emptyset$ for $l=1,\ldots,m$, so
we can choose cc-regions $X_l\subset V_{l-1}\cap V_l$ for
$l=1,\ldots,m$. Using the same calculation as in equation
(\ref{RSrepestimate}) and the hypothesis that $\omega$ is a
Reeh-Schlieder state on each $V_l$ we can
find $B_m\in\Alg{A}_{X_m}$ such that
\[
\|\pi_{\omega}(A-B_m)\Omega_{\omega}\|<\frac{\epsilon}{m+1},
\]
and proceed inductively to find $B_l\in\Alg{A}_{V_l}$ for
$l=m-1,\ldots,0$ such that
$\|\pi_{\omega}(B_{l+1}-B_l)\Omega_{\omega}\|<\frac{\epsilon}{m+1}$.
This provides us with a $B:=B_0\in\Alg{A}_{U_j}$ such that
\[
\|\pi_{\omega}(A-B)\Omega_{\omega}\|\le
\|\pi_{\omega}(A-B_m)\Omega_{\omega}\|+
\sum_{l=1}^m\|\pi_{\omega}(B_m-B_{m-1})\Omega_{\omega}\|
<\epsilon.
\]
\end{proof*}

We resume the proof of theorem \ref{RSlocal} and prove that
$\omega$ has the Reeh-Schlieder property for $U_a$ for a fixed
but arbitrary index $a\in I$. We consider a monomial
$A_1\cdots A_n$ where $A_i\in\Alg{A}_{U_{k(i)}}$ for some indices
$k(i)\in I$ and set $k(0)=a$. We assume
that all $A_i$ are non-zero and we define $r>0$ by
$r:=2\max_{i=1,\ldots,n}\|\pi_{\omega}(A_i)\|$. Given
$\epsilon>0$ we can then find elements $B_i\in\Alg{A}_{U_{k(i)}}$ for
$i=0,\ldots,n-1$ such that
\begin{equation}\label{hopestimate}
\|\pi_{\omega}(A_{i+1}\cdots A_n-B_i)\Omega_{\omega}\|<
\frac{\epsilon}{r^i}.
\end{equation}
Indeed, for $i=n-1$ this follows directly from the lemma. We can then
proceed inductively to find $B_{n-2},\ldots,B_1,B_0$ as follows.
We notice that $A_iB_i\in\Alg{A}_{U_{k(i)}}$ for $i\ge 1$ and apply
the lemma to choose $B_{i-1}\in\Alg{A}_{U_{k(i-1)}}$ such that
$\|\pi_{\omega}(A_iB_i-B_{i-1})\Omega_{\omega}\|<\frac{\epsilon}{2r^{i-1}}$.
Then we use the estimate
\begin{eqnarray}
&&\|\pi_{\omega}(A_i\cdots A_n-B_{i-1})\Omega_{\omega}\|\nonumber\\
&\le&\|\pi_{\omega}(A_i)\|\cdot
\|\pi_{\omega}(A_{i+1}\cdots A_n-B_i)\Omega_{\omega}\|+
\|\pi_{\omega}(A_iB_i-B_{i-1})\Omega_{\omega}\|\nonumber\\
&<&\frac{r}{2}\cdot\frac{\epsilon}{r^i}+
\|\pi_{\omega}(A_iB_i-B_{i-1})\Omega_{\omega}\|<\frac{\epsilon}{r^{i-1}}
,\nonumber
\end{eqnarray}
which is (\ref{hopestimate}) for the index $i-1$. This provides us with
$B_0\in\Alg{A}_{U_a}$ such that
$\|\pi_{\omega}(A_1\cdots A_n-B_0)\Omega_{\omega}\|<\epsilon$.

Now let $\Alg{P}$ be the $^*$-algebra of all (finite) polynomials of
elements in $\cup_{i\in I}\Alg{A}_{U_i}$. Given $P\in\mathcal{P}$ and
$\epsilon>0$ we can apply the result of the previous paragraph to each
monomial in $P$ and find a $B\in\Alg{A}_{U_a}$ such that
$\|\pi_{\omega}(P-B)\Omega_{\omega}\|<\epsilon$. In other words,
$\pi_{\omega}(\Alg{A}_{U_a})\Omega_{\omega}$ is dense in
$\pi_{\omega}(\Alg{P})\Omega_{\omega}$. Notice that $\Alg{P}$
is the smallest $^*$-algebra that contains all algebras $\Alg{A}_{U_i}$
and that we have $\Alg{A}_M=\overline{\Alg{P}}$ by additivity, where we
take the norm closure. It follows that
$\pi_{\omega}(\Alg{P})\Omega_{\omega}$ is dense in
$\mathcal{H}_{\omega}$. This completes the proof.
\end{proof*}



\section[Spacetime deformation of the Reeh-Schlieder
property]{The Reeh-Schlieder property under spacetime
deformation}\label{sec_RSdeform}

The existence of Hadamard states of the free scalar field in certain
curved spacetimes was proved in \cite{Fulling+} by deforming Minkowski
spacetime into another globally hyperbolic spacetime. Using a similar
but slightly more technical spacetime deformation argument \cite{Verch1}
proved a spin-statistics theorem for locally covariant quantum field
theories with a spin structure, given that such a theorem holds in
Minkowski spacetime. In subsection \ref{subs_RSdeform} we will assume
the existence of a Reeh-Schlieder state in one globally hyperbolic
spacetime and try to
deduce the existence of such states on a deformed spacetime along
the same lines. As a geometric prerequisite we will state and prove in
subsection \ref{subs_deformation} a spacetime deformation result
employing similar methods to the references mentioned above.

\subsection{Spacetime deformation}\label{subs_deformation}

First we recall the spacetime deformation result due to \cite{Fulling+}:
\begin{proposition}\label{deformation1}
Consider two globally hyperbolic spacetimes $M_i$, $i=1,2$, with
space-like Cauchy surfaces $C_i$ both diffeomorphic to $C$. Then there
exists a globally hyperbolic spacetime $M'=(\R\times C,g')$ with
space-like Cauchy surfaces $C'_i$, $i=1,2$, such that $C'_i$ is
isometrically diffeomorphic to $C_i$ and an open neighbourhood of $C'_i$
is isometrically diffeomorphic to an open neighbourhood of $C_i$.
\end{proposition}
We omit the proof of this result, because we will prove the stronger
proposition \ref{deformation2} later on. Note, however, the following
interesting corollary (cf. \cite{Brunetti+} section 4):
\begin{corollary}\label{isomorphism1}
Two globally hyperbolic spacetimes $M_i$ with diffeomorphic Cauchy
surfaces are mapped to isomorphic $^*$-algebras $\Alg{A}_{M_i}$ by
any locally covariant quantum field theory $\Func{A}$ satisfying the
time-slice axiom (with some state space $\Func{S}$).
\end{corollary}
\begin{proof*}
Consider two diffeomorphic globally hyperbolic spacetimes $M_i$ for
$i=1,2$, let $M'$ be the deforming spacetime of proposition
\ref{deformation1} and let $W_i\subset\mathcal{M}_i$ be open
neighbourhoods of the Cauchy surfaces $C_i\subset\mathcal{M}_i$ which
are isometrically diffeomorphic under $\psi_i$ to the open
neighbourhoods $W'_i\subset\mathcal{M}'$ of the Cauchy surfaces
$C'_i\subset\mathcal{M}'$. We may take the $W_i$ and $W'_i$ to be
cc-regions (as will be shown in proposition \ref{deformation2}), so
that the maps $\map{\psi_i}{W_i}{W'_i}$ determine isomorphisms
$\Psi_i$ in $\CatMan$. It then follows from lemma \ref{Cauchydevelopment}
that
\begin{eqnarray}
\Alg{A}_{M_1}&\simeq&\Alg{A}_{W_1}\simeq\Alg{A}_{\psi_1^{-1}(W_1')}
\simeq\alpha_{\Psi_1}^{-1}(\Alg{A}_{W_1'})\simeq\alpha_{\Psi_1}^{-1}
(\Alg{A}_{M'})\nonumber\\
&\simeq&\alpha_{\Psi_1}^{-1}\circ\alpha_{\Psi_2}(\Alg{A}_{M_2}),\nonumber
\end{eqnarray}
where the $\alpha_{\Psi_i}$ are $^*$-isomorphisms. This proves the
assertion.
\end{proof*}

At this point a warning seems in place. When $g_1,g_2$ are two
Lorentzian metrics on a manifold $\mathcal{M}$ such that both
$M_i:=(\mathcal{M},g_i)$ are objects in $\CatMan$, corollary
\ref{isomorphism1} gives a $^*$-isomorphism $\alpha$ between the algebras
$\Alg{A}_{M_i}$. Hence, if $O\subset\mathcal{M}$ is a cc-region for
$g_1$ then $\alpha$ is a $^*$-isomorphism from $\Alg{A}_{(O,g_1)}$
into $\Alg{A}_{M_2}$. However, the image cannot always be identified with
$\Alg{A}_{(O,g_2)}$, because $O$ need not be causally convex for $g_2$,
in which case the object is not defined.

\begin{figure}
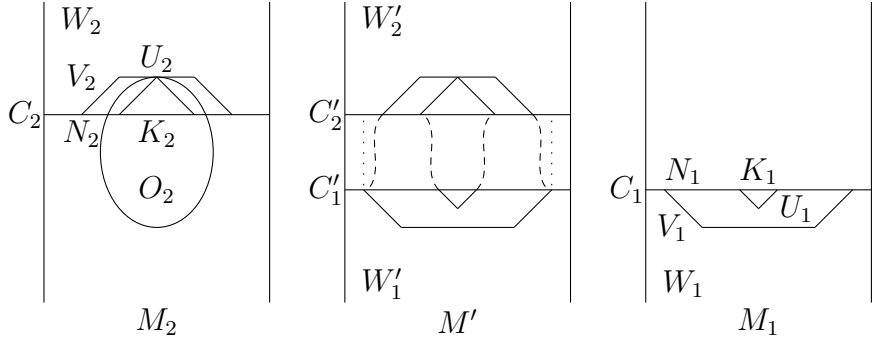

\begin{center}
\begin{pgfpicture}{0cm}{0cm}{13cm}{4.5cm}

\pgfline{\pgfxy(1,0.5)}{\pgfxy(1,4.5)}
\pgfline{\pgfxy(4,0.5)}{\pgfxy(4,4.5)}
\pgfline{\pgfxy(5,0.5)}{\pgfxy(5,4.5)}
\pgfline{\pgfxy(8,0.5)}{\pgfxy(8,4.5)}
\pgfline{\pgfxy(9,0.5)}{\pgfxy(9,4.5)}
\pgfline{\pgfxy(12,0.5)}{\pgfxy(12,4.5)}

\pgfline{\pgfxy(1,3)}{\pgfxy(4,3)}
\pgfline{\pgfxy(5,3)}{\pgfxy(8,3)}
\pgfline{\pgfxy(5,2)}{\pgfxy(8,2)}
\pgfline{\pgfxy(9,2)}{\pgfxy(12,2)}

\pgfline{\pgfxy(1.5,3)}{\pgfxy(2,3.5)}
\pgfline{\pgfxy(2,3.5)}{\pgfxy(3,3.5)}
\pgfline{\pgfxy(3,3.5)}{\pgfxy(3.5,3)}
\pgfline{\pgfxy(5.5,3)}{\pgfxy(6,3.5)}
\pgfline{\pgfxy(6,3.5)}{\pgfxy(7,3.5)}
\pgfline{\pgfxy(7,3.5)}{\pgfxy(7.5,3)}

\pgfline{\pgfxy(6,3)}{\pgfxy(6.5,3.5)}
\pgfline{\pgfxy(6.5,3.5)}{\pgfxy(7,3)}
\pgfline{\pgfxy(2,3)}{\pgfxy(2.5,3.5)}
\pgfline{\pgfxy(2.5,3.5)}{\pgfxy(3,3)}

\pgfline{\pgfxy(5.25,2)}{\pgfxy(5.75,1.5)}
\pgfline{\pgfxy(5.75,1.5)}{\pgfxy(7.25,1.5)}
\pgfline{\pgfxy(7.25,1.5)}{\pgfxy(7.75,2)}
\pgfline{\pgfxy(9.25,2)}{\pgfxy(9.75,1.5)}
\pgfline{\pgfxy(9.75,1.5)}{\pgfxy(11.25,1.5)}
\pgfline{\pgfxy(11.25,1.5)}{\pgfxy(11.75,2)}

\pgfline{\pgfxy(6.25,2)}{\pgfxy(6.5,1.75)}
\pgfline{\pgfxy(6.5,1.75)}{\pgfxy(6.75,2)}
\pgfline{\pgfxy(10.25,2)}{\pgfxy(10.5,1.75)}
\pgfline{\pgfxy(10.5,1.75)}{\pgfxy(10.75,2)}

\pgfellipse[stroke]{\pgfxy(2.5,2.5)}{\pgfxy(0.75,0)}{\pgfxy(0,1)}
\pgfsetdash{{0.2mm}{1.2mm}}{0.8mm}
\pgfline{\pgfxy(5.25,2)}{\pgfxy(5.25,3)}
\pgfline{\pgfxy(7.75,2)}{\pgfxy(7.75,3)}
\pgfsetdash{{1mm}{0.8mm}}{0mm}
\pgfxycurve(5.5,3)(5.25,2.75)(5.55,2.25)(5.3,2)
\pgfxycurve(7.5,3)(7.75,2.75)(7.45,2.25)(7.7,2)
\pgfxycurve(6.25,2)(6,2.25)(6.3,2.75)(6.05,3)
\pgfxycurve(6.75,2)(7,2.25)(6.7,2.75)(6.95,3)

\pgfputat{\pgfxy(2.5,0.25)}{\pgfbox[center,center]{$M_2$}}
\pgfputat{\pgfxy(6.5,0.25)}{\pgfbox[center,center]{$M'$}}
\pgfputat{\pgfxy(10.5,0.25)}{\pgfbox[center,center]{$M_1$}}
\pgfputat{\pgfxy(0.75,3)}{\pgfbox[center,center]{$C_2$}}
\pgfputat{\pgfxy(4.75,3)}{\pgfbox[center,center]{$C'_2$}}
\pgfputat{\pgfxy(1.5,4.25)}{\pgfbox[center,center]{$W_2$}}
\pgfputat{\pgfxy(5.5,4.25)}{\pgfbox[center,center]{$W'_2$}}
\pgfputat{\pgfxy(5.5,0.75)}{\pgfbox[center,center]{$W'_1$}}
\pgfputat{\pgfxy(9.5,0.75)}{\pgfbox[center,center]{$W_1$}}
\pgfputat{\pgfxy(2.5,2)}{\pgfbox[center,center]{$O_2$}}
\pgfputat{\pgfxy(1.5,3.5)}{\pgfbox[center,center]{$V_2$}}
\pgfputat{\pgfxy(2.5,3.75)}{\pgfbox[center,center]{$U_2$}}
\pgfputat{\pgfxy(2.5,2.75)}{\pgfbox[center,center]{$K_2$}}
\pgfputat{\pgfxy(1.5,2.75)}{\pgfbox[center,center]{$N_2$}}
\pgfputat{\pgfxy(4.75,2)}{\pgfbox[center,center]{$C'_1$}}
\pgfputat{\pgfxy(8.75,2)}{\pgfbox[center,center]{$C_1$}}
\pgfputat{\pgfxy(10.5,2.25)}{\pgfbox[center,center]{$K_1$}}
\pgfputat{\pgfxy(9.5,2.25)}{\pgfbox[center,center]{$N_1$}}
\pgfputat{\pgfxy(9.35,1.5)}{\pgfbox[center,center]{$V_1$}}
\pgfputat{\pgfxy(11,1.75)}{\pgfbox[center,center]{$U_1$}}
\end{pgfpicture}
\end{center}
\caption{Sketch of the geometry of proposition
\ref{deformation2}.}\label{fig1}
\end{figure}

We now formulate and prove our spacetime deformation result. The
geometric situation is schematically depicted in figure \ref{fig1}.
\begin{proposition}\label{deformation2}
Consider two globally hyperbolic spacetimes $M_i$, $i=1,2$, with
diffeomorphic Cauchy surfaces and a bounded cc-region
$O_2\subset M_2$ with non-empty causal complement,
$O_2^{\perp}\not=\emptyset$. Then there are a globally hyperbolic
spacetime $M'=(\mathcal{M}',g')$, space-like Cauchy surfaces
$C_i\subset M_i$ and $C_1',C_2'\subset M'$ and
bounded cc-regions $U_2,V_2\subset M_2$ and
$U_1,V_1\subset M_1$ such that the following hold:
\begin{enumerate}
\item there are isometric diffeomorphisms $\map{\psi_i}{W_i}{W'_i}$
where $W_1:=I^-(C_1)$, $W'_1:=I^-(C'_1)$, $W_2:=I^+(C_2)$ and
$W'_2:=I^+(C'_2)$,
\item $U_2,V_2\subset W_2$, $U_2\subset D(O_2)$, $O_2\subset D(V_2)$,
\item $U_1,V_1\subset W_1$, $U_1\not=\emptyset$,
$V_1^{\perp}\not=\emptyset$, $\psi_1(U_1)\subset D(\psi_2(U_2))$
and $\psi_2(V_2)\subset D(\psi_1(V_1))$.
\end{enumerate}
\end{proposition}

\begin{proof*}
First we recall the result of \cite{Bernal+1} that for any globally
hyperbolic spacetime $(\mathcal{M},g)$ there is a diffeomorphism
$\map{F}{\mathcal{M}}{\R\times C}$ for some smooth three
dimensional manifold $C$ in such a way that for each $t\in\R$
the surface $F^{-1}(\left\{t\right\}\times C)$ is a space-like Cauchy
surface. The pushed-forward metric $g':=F_*g$ makes
$(\R\times C,g')$ a globally hyperbolic manifold, where $g'$
is given by
\begin{equation}\label{metric}
g'_{\mu\nu}=\beta dt_{\mu}dt_{\nu}-h_{\mu\nu}.
\end{equation}
Here $dt$ is the differential of the canonical projection on the
first coordinate $\map{t}{\R\times C}{\R}$, which is
a smooth time function, $\beta$ is a strictly positive smooth function
and $h_{\mu\nu}$ is a (space and time dependent) Riemannian metric on
$C$. The orientation and time-orientation of $M$ induce an
orientation and time-orientation on $\R\times C$ via $F$. (If
necessary we may compose $F$ with the time-reversal diffeomorphism
$(t,x)\mapsto(-t,x)$ of $\R\times C$ to ensure that the function
$t$ increases in the positive time direction.) Applying the above to
the $M_i$ gives us two diffeomorphisms
$\map{F_i}{\mathcal{M}_i}{\mathcal{M}'}$, where
$\mathcal{M}'=\R\times C$ as a manifold. Note that we can take
the same $C$ for both $i=1,2$ by the assumption that the $M_i$
have diffeomorphic Cauchy surfaces.

Define $O'_2:=F_2(O_2)$ and let $t_{\min}$ and $t_{\max}$ be the minimum
and maximum value that the function $t$ attains on the compact set
$\overline{O'_2}$. We now prove that
$F_2^{-1}((t_{\min},t_{\max})\times C)\cap O_2^{\perp}\not=\emptyset$.
Indeed, if this were empty, then we see that $\overline{J(O_2)}$
contains $F_2^{-1}(\left[t_{\min},t_{\max}\right]\times C)$ and hence
also $C_{\max}:=F_2^{-1}(\left\{t_{\max}\right\}\times C)$ and
$C_{\min}:=F_2^{-1}(\left\{t_{\min}\right\}\times C)$. In fact, we have
$C_{\min}\subset\overline{J^-(O_2)}$. Indeed, if
$p:=F_2^{-1}(t_{\min},x)$ is in $\overline{J^+(O_2)}$ then we can
consider a basis of neighbourhoods of $p$ of the form
$I^-(F_2^{-1}(t_{\min}+1/n,x))\cap
I^+(F_2^{-1}(\left\{t_{\min}-1/n\right\}\times C))$. Now, if
$q_n\in J^+(O_2)$ is in such a basic neighbourhood, then the same
neighbourhood also contains a point $p_n\in O_2$. Hence, given a
sequence $q_n$ in $J^+(O_2)$ converging to $p$ we find a sequence $p_n$
in $O_2$ converging to $p$ and we conclude that
$p\in\overline{O_2}\subset\overline{J^-(O_2)}$. Similarly we can show
that $C_{\max}\subset\overline{J^+(O_2)}$. It then follows that
$I^+(C_{\max})\subset\overline{J^+(O_2)}$ and
$I^-(C_{\min})\subset\overline{J^-(O_2)}$, so together with
$F_2^{-1}([t_{\min},t_{\max}]\times C)\subset\overline{J(O_2)}$ we find
that $\overline{J(O_2)}=M$ and $O^{\perp}=\emptyset$. This
contradicts our assumption on $O_2$, so we must have
$F_2^{-1}((t_{\min},t_{\max})\times C)\cap O_2^{\perp}\not=\emptyset$.
Then we may choose $t_2\in (t_{\min},t_{\max})$ such that
$C_2:=F_2^{-1}(\left\{t_2\right\}\times C)$ intersects both
$O_2$ and $O_2^{\perp}$. We define $C'_2:=F_2(C_2)$, $W_2:=I^+(C_2)$
and $W'_2:=(t_2,\infty)\times C$.

Note that $C_2\cap J(\overline{O_2})$ is compact (see \cite{Baer+}
corollary A.5.4).
This means that we can find relatively compact open sets
$K,N\subset C$ such that $K'_2:=\left\{t_2\right\}\times K$, $K_2:=F_2^{-1}(K'_2)$,
$N'_2:=\left\{t_2\right\}\times N$ and $N_2:=F_2^{-1}(N'_2)$ satisfy
$K\not=\emptyset$, $\overline{N}\not= C$, $\overline{K_2}\subset O_2$
and $C_2\cap J(\overline{O_2})\subset N_2$. We let
$C_{\max}:=F_2^{-1}(\left\{t_{\max}\right\}\times C)$ and define
$U_2:=D(K_2)\cap I^+(K_2)\cap I^-(C_{\max})$ and
$V_2:=D(N_2)\cap I^+(N_2)\cap I^-(C_{\max})$. It follows from lemma
\ref{subspacetimes} that $U_2,V_2$ are bounded cc-regions in $M_2$.
Clearly $U_2,V_2\subset W_2$, $U_2\subset D(O_2)$, $O_2\subset D(V_2)$
and $V_2^{\perp}\not=\emptyset$.

Next we choose $t_1\in (t_{\min},t_2)$ and define
$C'_1:=\left\{t_1\right\}\times C$, $C_1:=F_1^{-1}(C'_1)$,
$W_1:=I^-(C_1)$ and $W'_1:=(-\infty,t_1)\times C$. Let
$N',K'\subset C$ be relatively compact connected open sets such that
$K'\not=\emptyset$, $\overline{N'}\not=C$, $\overline{K'}\subset K$
and $\overline{N}\subset N'$. We define
$N'_1:=\left\{t_1\right\}\times N'$,
$K'_1:=\left\{t_1\right\}\times K'$, $N_1:=F_1^{-1}(N'_1)$,
$K_1:=F_1^{-1}(K'_1)$ and
$C_{\min}:=F_1^{-1}(\left\{t_{\min}\right\}\times C)$. Let
$U_1:=D(K_1)\cap I^-(K_1)\cap I^+(C_{\min})$ and
$V_1:=D(N_1)\cap I^-(N_1)\cap I^+(C_{\min})$. Again by lemma
\ref{subspacetimes} these are bounded cc-regions in $M_1$.
Note that $U_1,V_1\subset W_1$ and $V_1^{\perp}\not=\emptyset$.

The metric $g'$ of $\mathcal{M}'$ is now chosen to be of the form
\[
g'_{\mu\nu}:=\beta dt_{\mu}dt_{\nu}-f\cdot (h_1)_{\mu\nu}-
(1-f)\cdot(h_2)_{\mu\nu}
\]
where we have written
$((F_i)_*g_i)_{\mu\nu}=\beta_i dt_{\mu}dt_{\nu}-(h_i)_{\mu\nu}$,
$f$ is a smooth function on $\mathcal{M}'$ which is identically $1$ on
$W'_1$, identically $0$ on $W'_2$ and $0<f<1$ on the intermediate region
$(t_1,t_2)\times C$ and $\beta$ is a strictly positive smooth function
which is
identically $\beta_i$ on $W'_i$. It is then clear that the maps $F_i$
restrict to isometric diffeomorphisms $\map{\psi_i}{W_i}{W'_i}$.

The function $\beta$ may be chosen small enough on the region
$(t_1,t_2)\times C$ to make $(\mathcal{M},g')$ globally hyperbolic. (As
pointed out in \cite{Fulling+} in their proof of proposition
\ref{deformation1}, choosing $\beta$ small ``closes up'' the light cones and
prevents causal curves from ``running off to spatial infinity'' in the
intermediate region.) Furthermore, using the compactness of
$(t_1,t_2)\times N'$ and the continuity of $(h_i)_{\mu\nu}$ we see
that we may choose $\beta$ small enough on this set to ensure that any
causal curve through $\overline{K'_1}$ must also intersect $K'_2$ and
any causal curve through $\overline{N'_2}$ must also intersect $N'_1$.
This means that $\overline{K'_1}\subset D(K'_2)$ and
$\overline{N'_2}\subset D(N'_2)$ and hence
$\psi_1(U_1)\subset D(\psi_2(U_2))$ and
$\psi_2(V_2)\subset D(\psi_1(V_1))$. This completes the proof.
\end{proof*}

\begin{figure}
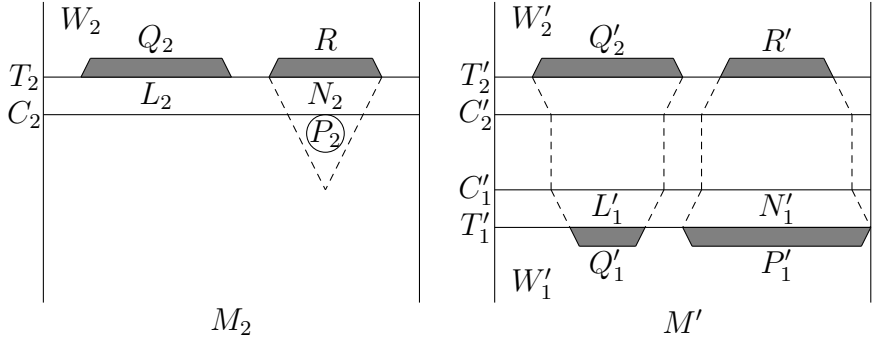

\begin{center}
\begin{pgfpicture}{0cm}{0cm}{13cm}{4.5cm}


\pgfline{\pgfxy(0.5,0.5)}{\pgfxy(0.5,4.5)}
\pgfline{\pgfxy(5.5,0.5)}{\pgfxy(5.5,4.5)}
\pgfline{\pgfxy(6.5,0.5)}{\pgfxy(6.5,4.5)}
\pgfline{\pgfxy(11.5,0.5)}{\pgfxy(11.5,4.5)}

\pgfline{\pgfxy(0.5,3)}{\pgfxy(5.5,3)}
\pgfline{\pgfxy(0.5,3.5)}{\pgfxy(5.5,3.5)}
\pgfline{\pgfxy(6.5,3)}{\pgfxy(11.5,3)}
\pgfline{\pgfxy(6.5,3.5)}{\pgfxy(11.5,3.5)}
\pgfline{\pgfxy(6.5,2)}{\pgfxy(11.5,2)}
\pgfline{\pgfxy(6.5,1.5)}{\pgfxy(11.5,1.5)}

\pgfellipse[stroke]{\pgfxy(4.25,2.75)}{\pgfxy(0.25,0)}{\pgfxy(0,0.25)}

\color[gray]{0.5}
\pgfmoveto{\pgfxy(1,3.5)}
\pgflineto{\pgfxy(1.125,3.75)}
\pgflineto{\pgfxy(2.875,3.75)}
\pgflineto{\pgfxy(3,3.5)}
\pgffill
\color[gray]{0}
\pgfline{\pgfxy(1,3.5)}{\pgfxy(1.125,3.75)}
\pgfline{\pgfxy(1.125,3.75)}{\pgfxy(2.875,3.75)}
\pgfline{\pgfxy(2.875,3.75)}{\pgfxy(3,3.5)}
\pgfline{\pgfxy(3,3.5)}{\pgfxy(1,3.5)}

\color[gray]{0.5}
\pgfmoveto{\pgfxy(7,3.5)}
\pgflineto{\pgfxy(7.125,3.75)}
\pgflineto{\pgfxy(8.875,3.75)}
\pgflineto{\pgfxy(9,3.5)}
\pgffill
\color[gray]{0}
\pgfline{\pgfxy(7,3.5)}{\pgfxy(7.125,3.75)}
\pgfline{\pgfxy(7.125,3.75)}{\pgfxy(8.875,3.75)}
\pgfline{\pgfxy(8.875,3.75)}{\pgfxy(9,3.5)}
\pgfline{\pgfxy(9,3.5)}{\pgfxy(7,3.5)}
\color[gray]{0.5}
\pgfmoveto{\pgfxy(3.5,3.5)}
\pgflineto{\pgfxy(3.625,3.75)}
\pgflineto{\pgfxy(4.875,3.75)}
\pgflineto{\pgfxy(5,3.5)}
\pgffill
\color[gray]{0}
\pgfline{\pgfxy(3.5,3.5)}{\pgfxy(3.625,3.75)}
\pgfline{\pgfxy(3.625,3.75)}{\pgfxy(4.875,3.75)}
\pgfline{\pgfxy(4.875,3.75)}{\pgfxy(5,3.5)}
\pgfline{\pgfxy(5,3.5)}{\pgfxy(3.5,3.5)}

\color[gray]{0.5}
\pgfmoveto{\pgfxy(9.5,3.5)}
\pgflineto{\pgfxy(9.625,3.75)}
\pgflineto{\pgfxy(10.875,3.75)}
\pgflineto{\pgfxy(11,3.5)}
\pgffill
\color[gray]{0}
\pgfline{\pgfxy(9.5,3.5)}{\pgfxy(9.625,3.75)}
\pgfline{\pgfxy(9.625,3.75)}{\pgfxy(10.875,3.75)}
\pgfline{\pgfxy(10.875,3.75)}{\pgfxy(11,3.5)}
\pgfline{\pgfxy(11,3.5)}{\pgfxy(9.5,3.5)}

\color[gray]{0.5}
\pgfmoveto{\pgfxy(7.5,1.5)}
\pgflineto{\pgfxy(7.625,1.25)}
\pgflineto{\pgfxy(8.375,1.25)}
\pgflineto{\pgfxy(8.5,1.5)}
\pgffill
\color[gray]{0}
\pgfline{\pgfxy(7.5,1.5)}{\pgfxy(7.625,1.25)}
\pgfline{\pgfxy(7.625,1.25)}{\pgfxy(8.375,1.25)}
\pgfline{\pgfxy(8.375,1.25)}{\pgfxy(8.5,1.5)}
\pgfline{\pgfxy(8.5,1.5)}{\pgfxy(7.5,1.5)}

\color[gray]{0.5}
\pgfmoveto{\pgfxy(9,1.5)}
\pgflineto{\pgfxy(9.125,1.25)}
\pgflineto{\pgfxy(11.375,1.25)}
\pgflineto{\pgfxy(11.5,1.5)}
\pgffill
\color[gray]{0}
\pgfline{\pgfxy(9,1.5)}{\pgfxy(9.125,1.25)}
\pgfline{\pgfxy(9.125,1.25)}{\pgfxy(11.375,1.25)}
\pgfline{\pgfxy(11.375,1.25)}{\pgfxy(11.5,1.5)}
\pgfline{\pgfxy(11.5,1.5)}{\pgfxy(9,1.5)}

\pgfsetdash{{1mm}{0.8mm}}{0mm}
\pgfline{\pgfxy(5,3.5)}{\pgfxy(4.25,2)}
\pgfline{\pgfxy(3.5,3.5)}{\pgfxy(4.25,2)}

\pgfline{\pgfxy(9.5,3.5)}{\pgfxy(9.25,3)}
\pgfline{\pgfxy(9.25,3)}{\pgfxy(9.25,2)}
\pgfline{\pgfxy(9.25,2)}{\pgfxy(9,1.5)}

\pgfline{\pgfxy(11,3.5)}{\pgfxy(11.25,3)}
\pgfline{\pgfxy(11.25,3)}{\pgfxy(11.25,2)}
\pgfline{\pgfxy(11.25,2)}{\pgfxy(11.5,1.5)}

\pgfline{\pgfxy(7,3.5)}{\pgfxy(7.25,3)}
\pgfline{\pgfxy(7.25,3)}{\pgfxy(7.25,2)}
\pgfline{\pgfxy(7.25,2)}{\pgfxy(7.5,1.5)}

\pgfline{\pgfxy(9,3.5)}{\pgfxy(8.75,3)}
\pgfline{\pgfxy(8.75,3)}{\pgfxy(8.75,2)}
\pgfline{\pgfxy(8.75,2)}{\pgfxy(8.5,1.5)}

\pgfputat{\pgfxy(3,0.25)}{\pgfbox[center,center]{$M_2$}}
\pgfputat{\pgfxy(9,0.25)}{\pgfbox[center,center]{$M'$}}
\pgfputat{\pgfxy(0.25,3)}{\pgfbox[center,center]{$C_2$}}
\pgfputat{\pgfxy(0.25,3.5)}{\pgfbox[center,center]{$T_2$}}
\pgfputat{\pgfxy(6.25,3)}{\pgfbox[center,center]{$C'_2$}}
\pgfputat{\pgfxy(6.25,3.5)}{\pgfbox[center,center]{$T'_2$}}
\pgfputat{\pgfxy(6.25,2)}{\pgfbox[center,center]{$C'_1$}}
\pgfputat{\pgfxy(6.25,1.5)}{\pgfbox[center,center]{$T'_1$}}
\pgfputat{\pgfxy(1,4.25)}{\pgfbox[center,center]{$W_2$}}
\pgfputat{\pgfxy(7,4.25)}{\pgfbox[center,center]{$W'_2$}}
\pgfputat{\pgfxy(7,0.75)}{\pgfbox[center,center]{$W'_1$}}
\pgfputat{\pgfxy(4.25,2.75)}{\pgfbox[center,center]{$P_2$}}
\pgfputat{\pgfxy(4.25,4)}{\pgfbox[center,center]{$R$}}
\pgfputat{\pgfxy(10.25,4)}{\pgfbox[center,center]{$R'$}}
\pgfputat{\pgfxy(10.25,1)}{\pgfbox[center,center]{$P'_1$}}
\pgfputat{\pgfxy(8,1)}{\pgfbox[center,center]{$Q'_1$}}
\pgfputat{\pgfxy(8,4)}{\pgfbox[center,center]{$Q'_2$}}
\pgfputat{\pgfxy(2,4)}{\pgfbox[center,center]{$Q_2$}}
\pgfputat{\pgfxy(4.25,3.25)}{\pgfbox[center,center]{$N_2$}}
\pgfputat{\pgfxy(10.25,1.75)}{\pgfbox[center,center]{$N'_1$}}
\pgfputat{\pgfxy(8,1.75)}{\pgfbox[center,center]{$L'_1$}}
\pgfputat{\pgfxy(2,3.25)}{\pgfbox[center,center]{$L_2$}}
\end{pgfpicture}
\end{center}
\caption{Sketch of the proof of the second part of
proposition \ref{isomorphism2}.}\label{fig2}
\end{figure}

The analogue of corollary \ref{isomorphism1} for the situation of
proposition \ref{deformation2} is:
\begin{proposition}\label{isomorphism2}
Consider a locally covariant quantum field theory $\Func{A}$ with a
state space $\Func{S}$ satisfying the time-slice axiom and let
$M_i$, $i=1,2$, be two
globally hyperbolic spacetimes with diffeomorphic Cauchy
surfaces. For any bounded cc-region $O_2\subset M_2$ with
non-empty causal complement there are bounded cc-regions
$U_1,V_1\subset M_1$ and a $^*$-isomorphism
$\map{\alpha}{\Alg{A}_{M_2}}{\Alg{A}_{M_1}}$ such that
$V_1^{\perp}\not=\emptyset$ and
\begin{equation}\label{estimate1}
\Alg{A}_{U_1}\subset\alpha(\Alg{A}_{O_2})\subset
\Alg{A}_{V_1}.
\end{equation}

Moreover, if the space-like Cauchy surfaces of the $M_i$ are non-compact
and $P_2\subset M_2$ is any bounded cc-region, then there are
bounded cc-regions $Q_2\subset M_2$ and
$P_1,Q_1\subset M_1$ such that
$Q_i\subset P_i^{\perp}$ for $i=1,2$ and
\begin{equation}\label{estimate2}
\alpha(\Alg{A}_{P_2})\subset\Alg{A}_{P_1},\quad
\Alg{A}_{Q_1}\subset\alpha(\Alg{A}_{Q_2}),
\end{equation}
where $\alpha$ is the same $^*$-isomorphism as in the first part of this
proposition.
\end{proposition}
\begin{proof*}
We apply proposition \ref{deformation2} to obtain sets $U_i,V_i$ with
and isomorphisms $\Psi_i:W_i\rightarrow W'_i$ associated to the
isometric diffeomorphisms $\psi_i$. As in the proof of corollary
\ref{isomorphism1} the $\Psi_i=(\psi_i)$ give rise to $^*$-isomorphisms
$\alpha_{\Psi_i}$ and
$\alpha:=\alpha_{\Psi_1}^{-1}\circ\alpha_{\Psi_2}$ is a $^*$-isomorphism
from $\Alg{A}_{M_2}$ to $\Alg{A}_{M_1}$. Using the properties of
$U_i,V_i$ stated in proposition \ref{deformation2} we deduce:
\begin{eqnarray}
\Alg{A}_{U_1}&=&\alpha_{\Psi_1}^{-1}(\Alg{A}_{U'_1})\subset
\alpha_{\Psi_1}^{-1}(\Alg{A}_{D(U'_2)})=
\alpha_{\Psi_1}^{-1}(\Alg{A}_{U'_2})
=\alpha(\Alg{A}_{U_2})\subset\alpha(\Alg{A}_{O_2})\nonumber\\
&\subset&\alpha(\Alg{A}_{V_2})=
\alpha_{\psi_1}^{-1}(\Alg{A}_{V'_2})\subset
\alpha_{\psi_1}^{-1}(\Alg{A}_{D(V'_1)})=
\alpha_{\psi_1}^{-1}(\Alg{A}_{V'_1})=\Alg{A}_{V_1}.\nonumber
\end{eqnarray}
Here we repeatedly used equation (\ref{embedding}) and lemma
\ref{Cauchydevelopment} (the time-slice axiom). This proves the first
part of the proposition.

Now suppose that the Cauchy-surfaces are non-compact and let $P_2$ be any
bounded cc-region. We refer to figure \ref{fig2} for a depiction of this
part of the proof.

First choose Cauchy surfaces $T_2,T_+\subset W_2$ such that
$T_+\subset I^+(T_2)$. Note that
$J(\overline{P_2})\cap T_2$ is compact, so it has a relatively compact
connected open neighbourhood $N_2\subset T_2$. Choosing $T_+$
appropriately we see that $R:=D(N_2)\cap I^+(N_2)\cap I^-(T_+)$ is a
bounded cc-region in $M_2$ by lemma \ref{subspacetimes} and
as usual we set $R':=\psi_2(R)$.

Now let $T'_-,T'_1\subset W_1'$ be Cauchy surfaces such that
$T'_-\subset I^-(T'_1)$ and note that $J(\overline{R'})\cap T'_1$ is
again compact, so we can find a relatively compact connected open
neighbourhood $N_1'\subset T'_1$ and use lemma \ref{subspacetimes} to
define the bounded cc-region
$P'_1:=D(N'_1)\cap I^-(N'_1)\cap I^+(T'_-)$ and
$P_1:=\psi_1^{-1}(P'_1)$.

Next we let $L'_1\subset T'_1$ be a connected relatively compact set
such that $L'_1\cap N'_1=\emptyset$. Such an $L'_1$ exists because
$T'_1$ is non-compact. We then define
$Q'_1:=D(L'_1)\cap I^-(L'_1)\cap I^+(T'_-)$ and
$Q_1:=\psi_1^{-1}(Q'_1)$. We see that $Q_1\subset P_1^{\perp}$ is
a bounded cc-region and $Q'_1\subset D(\psi_2(L_2))$ where
$L_2\subset T_2\setminus N$ is a relatively compact open set. In fact,
we can choose $L_2$ to be connected because $Q'_1$ lies in a connected
component $C$ of $D(\psi_2(T_2\setminus N))$. We now define the
bounded cc-region $Q_2:=D(L_2)\cap I^+(L_2)\cap I^-(T_+)$ and
$Q'_2:=\psi_2(Q_2)$, so that $Q_1\subset P_1^{\perp}$ and
$Q'_1\subset D(Q'_2)$.

So far the geometry of the proof. Now note that
$\Alg{A}_{P_2}\subset\Alg{A}_R$ by lemma
\ref{Cauchydevelopment} on $D(N_2)\cap I^+(N_2)$ and that
$\Alg{A}_{R'}=\alpha_{\Psi_2}(\Alg{A}_R)$. Applying
lemma \ref{Cauchydevelopment} in $D(N'_1)\cap I^-(N'_1)$ we see that
$\Alg{A}_{R'}\subset\Alg{A}_{P'_1}$ and we have
$\Alg{A}_{P_1}=\alpha_{\Psi_1}^{-1}(\Alg{A}_{P'_1})$.
Putting this together yields the inclusion:
\[
\alpha(\Alg{A}_{P_2})\subset\alpha(\Alg{A}_R)=
\alpha_{\Psi_1}^{-1}(\Alg{A}_{R'})\subset
\alpha_{\Psi_1}^{-1}(\Alg{A}_{P'_1})=\Alg{A}_{P_1}.
\]
Similarly we have
$\Alg{A}_{Q_1}=\alpha_{\Psi_1}^{-1}(\Alg{A}_{Q'_1})$,
$\Alg{A}_{Q'_2}=\alpha_{\Psi_2}(\Alg{A}_{Q_2})$ and
$\Alg{A}_{Q'_1}\subset\Alg{A}_{Q'_2}$ by lemma
\ref{Cauchydevelopment}. This yields the inclusion:
\[
\alpha(\Alg{A}_{Q_2})=\alpha_{\Psi_1}^{-1}(\Alg{A}_{Q'_2})
\supset\alpha_{\Psi_1}^{-1}(\Alg{A}_{Q'_1})=\Alg{A}_{Q_1}.
\]
\end{proof*}

\subsection{Deformation of the Reeh-Schlieder property}\label{subs_RSdeform}

We will now describe some of the consequences of the spacetime deformation
argument of the previous subsection for the Reeh-Schlieder property.
Unfortunately it is not clear that we can deform a Reeh-Schlieder state into
another (full) Reeh-Schlieder state, but we do have the following more limited
result:
\begin{theorem}\label{RSdeformation}
Consider a locally covariant quantum field theory $\Func{A}$ with state space
$\Func{S}$ which satisfies the time-slice axiom. Let $M_i$, $i=1,2$,
be two globally hyperbolic spacetimes with diffeomorphic Cauchy surfaces
and suppose that $\omega_1\in\Stat{S}_{M_1}$ is a Reeh-Schlieder state. Then
given any bounded cc-region $O_2\subset M_2$ with non-empty
causal complement, $O_2^{\perp}\not=\emptyset$, there is a
$^*$-isomorphism $\map{\alpha}{\Alg{A}_{M_2}}{\Alg{A}_{M_1}}$
such that $\omega_2:=\alpha^*(\omega_1)$ has the Reeh-Schlieder property
for $O_2$.

Moreover, if the Cauchy surfaces of the $M_i$ are non-compact and
$P_2\subset M_2$ is a bounded cc-region, then there is a
bounded cc-region $Q_2\subset P_2^{\perp}$ for which $\omega_2$ has the
Reeh-Schlieder property.
\end{theorem}
\begin{proof*}
For the first statement let $\alpha$ and $U_1$ be as in the first part
of proposition \ref{isomorphism2} and note that $\alpha$ gives rise to a
unitary map
$\map{\mathtt{U}_{\alpha}}{\mathcal{H}_{\omega_2}}{\mathcal{H}_{\omega_1}}$.
This map is the expression of the essential uniqueness of the
GNS-representation, so that
$\mathtt{U}_{\alpha}\Omega_{\omega_2}=\Omega_{\omega_1}$ and
$\mathtt{U}_{\alpha}\pi_{\omega_2}\mathtt{U}_{\alpha}^*=\pi_{\omega_1}
\circ\alpha$.
The Reeh-Schlieder property for $O_2$ then follows from the observation
that $\mathtt{U}_{\alpha}\pi_{\omega_2}(\Alg{A}_{O_2})
\mathtt{U}_{\alpha}^*\supset\pi_{\omega_1}(\Alg{A}_{U_1})$:
\[
\overline{\pi_{\omega_2}(\Alg{A}_{O_2})\Omega_{\omega_2}}\supset
\overline{\mathtt{U}_{\alpha}^*\pi_{\omega_1}(\Alg{A}_{U_1})\Omega_{\omega_1}}
=\mathtt{U}_{\alpha}^*\mathcal{H}_{\omega_1}=
\mathcal{H}_{\omega_2}.
\]

Similarly for the second statement, given a bounded cc-region $P_2$ and
choosing $Q_1,Q_2$ as in the second statement of proposition
\ref{isomorphism2} we see that
$\mathtt{U}_{\alpha}\pi_{\omega_2}(\Alg{A}_{Q_2})
\mathtt{U}_{\alpha}^*\supset\pi_{\omega_1}(\Alg{A}_{Q_1})$.
\end{proof*}
The second part of theorem \ref{RSdeformation} means that $\omega_2$
is a Reeh-Schlieder state for all cc-regions that are big enough.
Indeed, if $V_2$ is a sufficiently small cc-region then $V_2^{\perp}$ is
connected (recall that we work with four-dimensional spacetimes) and
therefore $\omega_2$ has the Reeh-Schlieder property for some
cc-region in $V_2^{\perp}$ and hence also for $V_2^{\perp}$ itself.

In the remainder of this subsection we consider only
$C^*$-algebraic theories, because they allow us to draw stronger
conclusions than for general topological $^*$-algebras. We begin with
the following consequence of theorem \ref{RSdeformation}:
\begin{corollary}\label{separability}
In the situation of theorem \ref{RSdeformation}, if
$\map{\Func{A}}{\CatMan}{\CatAlg}$ is a
causal locally covariant quantum field theory, then $\Omega_{\omega_2}$
is a cyclic and separating vector for the local von Neumann algebra
$\Alg{R}^{\omega_2}_{O_2}$. If the Cauchy surfaces are non-compact
$\Omega_{\omega_2}$ is a separating vector for all
$\Alg{R}^{\omega_2}_{P_2}$ where $P_2$ is a bounded cc-region.
\end{corollary}
\begin{proof*}
Recall that a vector is a separating vector for a von Neumann algebra
$\Alg{R}$ iff it is a cyclic vector for the commutant $\Alg{R}'$
(\cite{Kadison+} proposition 5.5.11, see also our proof of
proposition \ref{separability1}). Choosing $V_1$ as in the
first part of proposition \ref{isomorphism2} we have
$\mathtt{U}_{\alpha}\pi_{\omega_2}(\Alg{A}_{O_2})
\mathtt{U}_{\alpha}^*\subset\pi_{\omega_1}(\Alg{A}_{V_1})$ by the
inclusion (\ref{estimate1}). Therefore the commutant of
$\mathtt{U}_{\alpha}\Alg{R}^{\omega_2}_{O_2}\mathtt{U}_{\alpha}^*$
contains $(\Alg{R}^{\omega_1}_{V_1})'$. As
$V_1^{\perp}\not=\emptyset$ this commutant contains the local algebra of
some cc-region for which $\Omega_{\omega_1}$ is cyclic. Hence
$\Omega_{\omega_1}$ is a separating vector for
$\Alg{R}^{\omega_1}_{V_1}$ and $\Omega_{\omega_2}$ for
$\Alg{R}^{\omega_2}_{O_2}$.

If the Cauchy surfaces are non-compact, $P_2$ is a bounded region
and $Q_2$ is as in theorem \ref{RSdeformation}, then
$(\Alg{R}^{\omega_2}_{P_2})'$ contains
$\pi_{\omega_2}(\Alg{A}_{Q_2})$, for which $\Omega_{\omega_2}$ is
cyclic. It follows that $\Omega_2$ is separating for
$\Alg{R}^{\omega_2}_{P_2}$.
\end{proof*}
If the theory is nowhere classical then this corollary implies that
there exist non-local correlations between $O_2$ and any cc-region $V_2$
space-like to it, just as in the Minkowski spacetime case (see e.g.
\cite{Redhead}). Also, if the Cauchy surfaces are non-compact, any
localised non-trivial positive observable has a strictly positive
expectation value.

If the state space is locally quasi-equivalent and large enough it is
possible to show the existence of full Reeh-Schlieder states. The proof
uses abstract existence arguments, as opposed to the proof of theorem
\ref{RSdeformation} which is constructive, at least in principle.
\begin{theorem}\label{RSdense}
Let $\map{\Func{A}}{\CatMan}{\CatAlg}$ be a locally covariant quantum
field theory with a
locally quasi-equivalent state space $\Func{S}$ which is causal and
satisfies the time-slice axiom. Assume that $\Func{S}$ is maximal in
the sense that for any state $\omega$ on some $\Alg{A}_{M}$ which is
locally quasi-equivalent to a state in $\Stat{S}_M$ we have
$\omega\in\Stat{S}_M$.

Let $M_i$, $i=1,2$, be two globally hyperbolic spacetimes with
diffeomorphic non-compact Cauchy surfaces and assume that
$\omega_1\in\Stat{S}_{M_1}$ is a Reeh-Schlieder state. Then
$\Stat{S}_{M_2}$ contains a (full) Reeh-Schlieder state.
\end{theorem}
\begin{proof*}
Let $\left\{O_n\right\}_{n\in\mathbb{N}}$ be a countable basis for the
topology of $M_2$ consisting of bounded cc-regions with non-empty causal
complement. (That every open set contains a cc-region can be seen by
using a convex normal neighbourhood and choosing a sufficiently small
region of the form $I^+(p)\cap I^-(q)$, cf. \cite{Wald} theorem 8.1.2
and our lemma \ref{subspacetimes}).
We then apply theorem \ref{RSdeformation} to each $O_n$ to
obtain a sequence of states $\omega^{n}_2\in\Stat{S}_{M_2}$ which have
the Reeh-Schlieder property for $O_n$. We write $\omega:=\omega^{1}_2$
and let $(\mathcal{H},\pi,\Omega)$ denote its GNS-triple.

For all $n\ge 2$ we now find a bounded cc-region
$V_n\subset M_2$ such that $V_n\supset O_1\cup O_n$. For this
purpose we first choose a Cauchy surface $C\subset M_2$ and
note that $K_n:=C\cap J(\overline{O_n})$ is compact. Letting
$L_n\subset C$ be a compact connected set containing $K_1\cup K_n$ in
its interior it suffices to choose
$V_n:=\mathrm{int}(D(L_n))\cap I^-(C_+)\cap I^+(C_-)$ for Cauchy
surfaces $C_{\pm}$ to the future, respectively to the past, of $O_1$,
$O_n$ and $C$.
Note that $\Omega$ and $\Omega_{\omega^{n}_2}$ are cyclic and separating
vectors for $\Alg{R}^{\omega}_{V_n}$ and $\Alg{R}^{\omega^{n}_2}_{V_n}$
respectively, by
$O_1\cup O_n\subset V_n$ and by corollary \ref{separability}.
Because $\omega$ and $\omega^n_2$ are locally quasi-equivalent there
is a $^*$-isomorphism $\map{\phi}{\Alg{R}^{\omega^n_2}_{V_n}}
{\Alg{R}^{\omega}_{V_n}}$. In the presence of the cyclic and
separating vectors $\Omega$ and $\Omega_{\omega_2^n}$ the $^*$-isomorphism
$\phi$ is implemented by a unitary map
$\map{\mathtt{U}_n}{\mathcal{H}_{\omega^n_2}}{\mathcal{H}}$ (see
\cite{Kadison+} theorem 7.2.9). We claim that
$\psi_n:=\mathtt{U}_n\Omega_{\omega^n_2}$ is cyclic for
$\Alg{R}^{\omega}_{O_n}$. Indeed, by the definition of
quasi-equivalence we have $\phi\circ\pi_{\omega^n_2}=\pi_{\omega}$ on
$\mathcal{A}_{V_n}$, so
\[
\overline{\pi_{\omega}(\Alg{A}_{O_n})\psi_n}=
\overline{\mathtt{U}_n\pi_{\omega^n_2}(\Alg{A}_{O_n})
\Omega_{\omega^n_2}}=\mathtt{U}_n\mathcal{H}_{\omega^n_2}=
\mathcal{H}_{\omega}.
\]

We now apply the results of \cite{Dixmier+} to conclude that
$\mathcal{H}$ contains a dense set of vectors $\psi$ which are cyclic and
separating for all $\Alg{R}^{\omega}_{O_n}$ simultaneously. Because
each cc--region $O\subset M_2$ contains some $O_n$ we see that
$\omega_{\psi}:A\mapsto\frac{\langle\psi,\pi_{\omega}(A)\psi\rangle}
{\|\psi\|^2}$ defines a full Reeh-Schlieder state. Finally, because the
GNS-triple of $\omega_{\psi}$ is just $(\mathcal{H},\pi,\psi)$
we see that it is locally quasi-equivalent to $\omega$ and hence
$\omega_{\psi}\in\Stat{S}_{M_2}$.
\end{proof*}

Although the state space may in general not be big enough to contain
full Reeh-Schlieder states, theorem \ref{RSdeformation} is already
enough for some useful applications. As an example we present the
following conclusion concerning the type of local von Neumann algebras:
\begin{corollary}\label{not_finite}
Consider a nowhere classical causal locally covariant quantum field
theory $\map{\Func{A}}{\CatMan}{\CatAlg}$ with a locally
quasi-equivalent state space $\Func{S}$ which satisfies the time-slice
axiom. Let $M_i$, $i=1,2$, be two
globally hyperbolic spacetimes with diffeomorphic Cauchy surfaces and
let $\omega_1\in\Stat{S}_{M_1}$ be a Reeh-Schlieder state. Then for
any state $\omega\in\Stat{S}_{M_i}$ and any cc-region
$O\subset M_i$ the local
von Neumann algebra $\Alg{R}^{\omega}_O$ is not finite.
\end{corollary}
\begin{proof*}
We will use proposition 5.5.3 in \cite{Baumgartel+}, which says that
$\Alg{R}^{\omega}_O$ is not finite if the GNS-vector $\Omega$ is a
cyclic and separating vector for $\Alg{R}^{\omega}_O$ and for a
proper sub-algebra $\Alg{R}^{\omega}_V$. Note that we can drop
the superscript $\omega$ if $O$ and $V$ are bounded, by local
quasi-equivalence.

First we consider $M_1$. For any bounded cc-region
$O_1\subset M_1$ such that $O_1^{\perp}\not=\emptyset$ we
can find bounded cc-regions $O'\subset O_1^{\perp}$ and
$U,V\subset O_1$ such that $U\subset V^{\perp}$. By the Reeh-Schlieder
property the GNS-vector $\Omega_{\omega_1}$ is cyclic for
$\Alg{R}_V$ and hence also for $\Alg{R}_{O_1}$. Moreover it is
cyclic for $\Alg{R}_{O_1}'\supset\Alg{R}_{O'}$ and therefore it is
separating for $\Alg{R}_{O_1}$ and $\Alg{R}_V$. Now suppose that
$\Alg{R}_{O_1}=\Alg{R}_V$. Then, by causality:
\[
\pi_{\omega}(\Alg{A}_U)\subset \pi_{\omega}(\Alg{A}_V)'=
\pi_{\omega}(\Alg{A}_{O_1})'\subset
\pi_{\omega}(\Alg{A}_U)'.
\]
It follows that $\Alg{R}_U\subset\Alg{R}_U'$, which
contradicts the nowhere-classicality. Therefore, the inclusion
$\Alg{R}_V\subset\Alg{R}_{O_1}$ must be proper and the cited
theorem applies. Of course, if
$O\subset M_1$ is a cc-region that is not bounded, then it
contains a bounded cc-region $O_1$ as above and
$\Alg{R}^{\omega}_O\supset\Alg{R}^{\omega}_{O_1}
\simeq\Alg{R}_{O_1}$ isn't finite either for any $\omega\in
\Stat{S}_{M_1}$. (If $V$ is a partial isometry in the smaller algebra
such that $I=V^*V$ and $E:=VV^*<I$ then the same $V$ shows that $I$ is
not finite in the larger algebra.)

Next we consider $M_2$ and let $O\subset M_2$ be any
cc-region. It contains a cc-region $O_2$ with
$O_2^{\perp}\not=\emptyset$,
so we can apply theorem \ref{RSdeformation}. Using the unitary map
$\map{\mathtt{U}_{\alpha}}{\mathcal{H}_{\omega_2}}{\mathcal{H}_{\omega_1}}$
we see that $\Alg{R}_{O_2}\simeq\Alg{R}^{\omega_2}_{O_2}$
contains $\alpha^{-1}(\Alg{R}^{\omega_1}_{O_1})$, which is not
finite by the first paragraph. Hence $\Alg{R}_{O_2}$ is not finite
and the statement for $O$ then follows again by inclusion.
\end{proof*}
Instead of the nowhere-classicality we could have assumed that the local
von Neumann algebras in $M_1$ are infinite, which allows us to derive
the same conclusion for $M_2$. Unfortunately it is in general impossible
to completely derive the type
of the local algebras using this kind of argument. Even if we know the
types of the algebras $\Alg{A}_{U_1}$ and $\Alg{A}_{V_1}$ in the
inclusions (\ref{estimate1}), we can't deduce the type of
$\Alg{A}_{O_2}$.

Another important consequence of proposition \ref{RSdeformation} in the
$C^*$-algebraic case is that
corollary \ref{separability} enables us to apply the Tomita-Takesaki
modular theory to $\Alg{R}^{\omega_2}_{O_2}$ (or to the von Neumann
algebra of any bounded cc-region $V_2$ which contains $O_2$, if the
Cauchy surfaces are non-compact). More precisely, let
$O_2\subset M_2$ be given and let $U_1,V_1\subset M_1$
be the bounded cc-regions and
$\map{\alpha}{\Alg{A}_{M_1}}{\Alg{A}_{M_1}}$ the $^*$-isomorphism of
proposition \ref{isomorphism2}, so that $\Alg{A}_{O_1}\subset
\alpha(\Alg{A}_{O_2})\subset\Alg{A}_{V_1}$. We can then define
$\Alg{R}:=\mathtt{U}_{\alpha}\Alg{R}^{\omega_2}_{O_2}
\mathtt{U}_{\alpha}^*$ and obtain $\Alg{R}^{\omega_1}_{U_1}\subset
\Alg{R}\subset\Alg{R}^{\omega_1}_{V_1}$. It then follows that
the respective Tomita-operators are extensions of each other,
$S_{U_1}\subset S_{\Alg{R}}\subset S_{V_1}$ (see e.g.
\cite{Kadison+}).

\section[$WF_{qA}$ and the Reeh-Schlieder property]{The quasi-analytic
wave front set and the Reeh-Schlieder property for scalar
fields}\label{sec_WFRS}

After the general results on the Reeh-Schlieder property presented in
sections \ref{sec_RSgen} and \ref{subs_deformation} we now specialise
to the real scalar field, described by the Borchers-Uhlmann functor
$\Func{U}$. The main result of this section is a smoothly covariant
condition on the continuous states of this algebra that guarantees the
Reeh-Schlieder property as well as the fulfillment of the microlocal
spectrum condition. This condition, which we call
the quasi-analytic microlocal spectrum condition, is analogous to the
analytic microlocal spectrum condition of \cite{Strohmaier+} and
the microlocal spectrum condition. As a preparation to the formulation
of our condition we need to study analytic wave front sets in more
detail in subsection \ref{subs_supp}, in particular their relation to
the boundary of the support of a distribution. We refer to appendix
\ref{ch_ma} for the definition of analytic wave front sets and their
properties.

\subsection{Wave front sets and the support}\label{subs_supp}

We begin with a definition concerning the boundary of a closed set
(see e.g. \cite{Hoermander}):
\begin{definition}
Let $\mathcal{O}$ be a closed subset of a smooth manifold $\mathcal{M}$.
The \emph{exterior normal set} $N_e(\mathcal{O})$ of $\mathcal{O}$
consists of all $(x,k)\in T^*\mathcal{M}$ such that
$x\in\mathcal{O}$ and there is a real-valued function $f\in C^2(\mathcal{M})$
with $f(y)\le f(x)$ for all $y\in\mathcal{O}$ and $df(x)=k\not= 0$.

The \emph{normal set} $N(\mathcal{O})$ of $\mathcal{O}$ consists of
$(x,k)\in T^*\mathcal{M}$ such
that either $(x,k)\in N_e(\mathcal{O})$ or $(x,-k)\in N_e(\mathcal{O})$.
\end{definition}
If $(x,k)\in N_e(\mathcal{O})$ then $x$ cannot be in the interior of
$\mathcal{O}$, because
an extremum of a $C^2$ function $f$ can only be attained in the interior
of $\mathcal{O}$ if $df=0$, as is well-known. Conversely,
\begin{lemma}\label{Ndense}
For a closed subset $\mathcal{O}$ of a smooth manifold $\mathcal{M}$ the
projection of
$N_e(\mathcal{O})$ on $\mathcal{M}$ is dense in the boundary of $\mathcal{O}$.
\end{lemma}
\begin{proof*}
See \cite{Hoermander} proposition 8.5.8.
\end{proof*}
Thus the normal set characterises the boundary of the closed set $\mathcal{O}$
very well. Moreover, it is by definition a subset of
$T^*\mathcal{M}\setminus\mathcal{Z}$ and it is seen to be conic by multiplying
the function $f$ in the definition by a positive real number. This
means that it can be compared with the wave front set:
\begin{proposition}\label{boundary1}
Let $u$ be a scalar distribution on an open set $X\subset\R^n$, then
$\overline{N(\mathrm{supp}\ u)}\setminus\mathcal{Z}\subset WF_A(u)$.
\end{proposition}
\begin{proof*}
This is \cite{Hoermander} theorem 8.5.6'.
\end{proof*}

\begin{remark}
As an illustration we consider the case where the analytic wave front
set of the distribution $u$ is empty, $WF_A(u)=\emptyset$. Proposition
\ref{boundary1} then tells us that $N(\mathrm{supp}\ u)=\emptyset$ and
by lemma \ref{Ndense} the boundary of $\mathrm{supp}\ u$ must be empty.
Another way to reach the same conclusion is to notice that $u$ is an
analytic function, so if there is an $x\in X$ which is not in the
support of $u$, then $u\equiv 0$ on an open subset of $X$ and hence
$u\equiv 0$ by analyticity. The support of $u$ is either all of $X$
or empty and in any case the boundary of the support is empty.
It would be unreasonable to expect a similar result for the smooth
wave front set, because one can easily construct smooth compactly
supported functions.

Another instructive example, which shows that the support of a
distribution cannot be characterised entirely in terms of the analytic
wave front set, is the following. Let $\delta_0(x)$ be the Dirac
measure on $\R$ at the point $x=0$ and consider the distribution
$u(x):=1+\delta_0(x)$ on $\R$. The support of $u$ is all of $\R$, so
its normal set is empty. On the other hand,
$WF_A(u)=\left\{0\right\}\times(\R\setminus\left\{0\right\})$. Indeed,
$u$ fails to be analytic only at $x=0$, so there must be some $k\not=0$
with $(x,k)\in WF_A(u)$ and because $u$ is real-valued we then also have
$(x,-k)\in WF_A(u)$. The conclusion then follows because $WF_A(u)$ is
conic.
\end{remark}

It follows from proposition \ref{boundary1} and lemma \ref{Ndense}
that the analytic wave front set gives only an upper bound on the
boundary of the support of a distribution. Moreover, because this
result requires the use of analytic wave front sets it seems that
it can only be formulated on an analytic manifold. Of course every
$C^1$ manifold allows an analytic structure compatible with the
$C^1$ structure (\cite{Hirsch} section 2.5), so this is not really a
restriction. However, this analytic structure is highly non-unique
(unlike the $C^k$ and $C^{\infty}$ structures, which are unique,
\cite{Hirsch} theorem 2.2.9 and 2.3.4) and there does not
seem to be a natural choice. The following definition helps us to
avoid making a choice and allows us to sharpen the result of proposition
\ref{boundary1} considerably:
\begin{definition}\label{defWFqA}
Let $u$ be a distribution on a smooth manifold $\mathcal{M}$ with values
in a Banach space $\mathcal{B}$. The
\emph{quasi-analytic wave front set} $WF_{qA}(u)$ of $u$ is defined to
be the conic subset of $T^*\mathcal{M}\setminus\mathcal{Z}$ consisting
of all $(x,k)\in T^*\mathcal{M}$ such that for every smooth coordinate map
$\map{\kappa}{O\subset\mathcal{M}}{\R^n}$ near $x$ we have
\[
(x,k)\in\kappa^*(WF_A(\kappa_*u)),
\]
where $\kappa^*(y,l):=(\kappa^{-1}(y),d\kappa^{-1}l)$.
\end{definition}

The quasi-analytic wave front set is a closed conic subset of
$T^*\mathcal{M}\setminus\mathcal{Z}$, because it is locally the intersection
of the sets $\kappa^*(WF_A(\kappa_*u))$ which are closed in
$T^*\mathcal{M}\setminus\mathcal{Z}$. It is worth noting that
\[
\overline{\cup_{\phi\in\mathcal{B}'}WF_{qA}(\phi(u))}\setminus
\mathcal{Z}\subset WF_{qA}(u),
\]
because of theorem \ref{WFAclosure} and the closedness of
$WF_{qA}(u)$, but it is not clear that equality
holds in general, because the union over $\phi$ does not commute with
the intersection over the choices of coordinates. Also some of the
results in theorem \ref{WFAresults2}, such as the estimate for
the wave front set of a sum, will fail in general for the quasi-analytic
wave front set. Its usefulness is entirely based on the fact that it
is by definition covariant under smooth diffeomorphisms, but at the
same time contains some of the information of the analytic wave front
set (in any given analytic structure on the manifold $\mathcal{M}$). More
precisely:
\begin{lemma}
Let $u$ be a distribution on an analytic manifold $\mathcal{M}$ with
values in a Banach space $\mathcal{B}$. Then
$WF(u)\subset WF_{qA}(u)\subset WF_A(u)$.
\end{lemma}
\begin{proof*}
In any choice of local coordinates $\kappa$ we have
$WF(\kappa_*u)\subset WF_A(\kappa_*u)$, from which the first inclusion
follows. For the second we only need to choose $\kappa$ to be analytic
and use the definition.
\end{proof*}
In general neither of these inclusion is an equality. For the second
inclusion we can see this by considering $u=f\circ\phi$, where $f$ is an
analytic function on $\mathcal{M}$ and $\phi$ is a smooth diffeomorphism
of $\mathcal{M}$
which is not analytic. In this case $u$ cannot be expected to be
analytic. For the first inclusion we can choose $u$ to be a compactly
supported smooth function and use the following sharpening of
proposition \ref{boundary1}:
\begin{proposition}\label{boundary2}
For a distribution $u$ on a smooth manifold $\mathcal{M}$ with values in a
Banach space $\mathcal{B}$ we have
$\overline{N(\mathrm{supp}\ \phi(u))}\setminus\mathcal{Z}\subset
WF_{qA}(u)$ for all $\phi\in\mathcal{B}'$.
\end{proposition}
\begin{proof*}
If $(x,k)\in\overline{N(\mathrm{supp}\ \phi(u))}\setminus\mathcal{Z}$
and $\kappa$ is a smooth choice of coordinates near $x$ then
$d\kappa^T(x,k)\in WF_A(\kappa_*\phi(u))\subset WF_A(\kappa_*u)$ by
proposition \ref{boundary1} and theorem \ref{WFAclosure}.
\end{proof*}

\begin{remark}
To see that proposition \ref{boundary2} is indeed a sharpening of
proposition \ref{boundary1} one can consider the example of a smooth
real-valued function $f\in\Test(\R,\R)$.
We endow $\R$ with the usual analytic structure. It is known that a
generic real-valued function in $\Test(\R)$ is nowhere analytic
\cite{Cater,Darst} in which case we have
$WF_A(f)=T^*\R\setminus\mathcal{Z}$, i.e. the analytic wave front set
could not be larger. The same is presumably true on an analytic
manifold. On the other hand, a generic smooth function is a
Morse function (\cite{Hirsch} theorem 6.1.2), which can be expressed
locally as a polynomial of degree $\le 2$ in suitable coordinates. In
these coordinates the function is certainly analytic, so in the generic
case we have $WF_{qA}(f)=\emptyset$. In the two statements above the
notion ``generic'' actually refers to two distinct topologies on the
set of smooth functions, namely the weak and the strong topology
respectively (see \cite{Hirsch} section 2.1). However, for a compact
manifold these topologies coincide, which would imply that at least
on a compact
manifold we generically have that $WF_A(f)$ is maximal, whereas
$WF_{qA}(f)=\emptyset$. The conclusion is that the many possible choices
of local coordinates allow us to get a much stricter upper bound of the
normal set $N(\mathrm{supp}\ f)$.
\end{remark}

The following proposition is a microlocal analogue of (a corollary of)
the edge-of-the-wedge theorem (see e.g. \cite{Streater+} theorem 2.16
and 2.17 or \cite{Hoermander} theorem 9.3.5):
\begin{proposition}\label{EotW}
Let $u$ be a distribution on a connected smooth manifold $\mathcal{M}$
with values in a Banach space $\mathcal{B}$ such that
\[
WF_{qA}(u)\cap -WF_{qA}(u)=\emptyset.
\]
If $\mathcal{O}\subset \mathcal{M}$ is a non-empty open region,
$\phi\in\mathcal{B}'$ and $\phi(u)|_{\mathcal{O}}=0$, then
$\phi(u)\equiv 0$.
\end{proposition}
\begin{proof*}
If $(x,k)\in N(\mathrm{supp}\ \phi(u))$, then
$(x,-k)\in N(\mathrm{supp}\ \phi(u))$ so by lemma \ref{boundary2} both
$(x,k)$ and $(x,-k)$ are in $WF_{qA}(u)$, which contradicts the
assumption. This means that $N(\mathrm{supp}\ \phi(u))$ is empty and
hence the boundary of $\mathrm{supp}\ \phi(u)$ is empty too by lemma
\ref{Ndense}. As $\mathcal{M}$ is connected and $\mathrm{supp}\ \phi(u)$
is not all of $\mathcal{M}$, we must have
$\mathrm{supp}\ \phi(u)=\emptyset$, i.e. $\phi(u)\equiv 0$.
\end{proof*}


\subsection{The quasi-analytic microlocal spectrum condition}

We are now in a position to prove the Reeh-Schlieder property for states
that satisfy an appropriate microlocal condition. We will first
reproduce the result of \cite{Strohmaier+}, using an analytic
microlocal spectrum condition. Then we will generalise this to all
states that satisfy a certain quasi-analytic microlocal spectrum
condition and we will discuss the implications and usefulness of this
condition.

The analytic microlocal spectrum condition of \cite{Strohmaier+} is a
direct generalisation of the (smooth) microlocal spectrum condition:
\begin{definition}\label{AmSC}
A spacetime $M$ is an \emph{analytic spacetime} if it is endowed with
an analytic structure in which the metric is analytic. (Equivalently,
all component functions $g_{\mu\nu}$ of the metric are analytic in any
choice of coordinates on the analytic manifold $\mathcal{M}$.)

A state $\omega$ on the Borchers-Uhlmann algebra $\Alg{U}_M$ of an
analytic spacetime $M$ satisfies
the \emph{analytic microlocal spectrum condition} (A$\mu$SC) if and only
if $WF_A(\omega_n)\subset\Gamma_n$ for all $n\in\N$.
\end{definition}
Note that the A$\mu$SC implies the $\mu$SC, because
$WF(\omega_n)\subset WF_A(\omega_n)$. It also implies the Reeh-Schlieder
property as follows:
\begin{theorem}\label{RSAmSC}
Let $\omega$ be a state on the Borchers-Uhlmann algebra $\Alg{U}_M$ of
an analytic globally hyperbolic spacetime $M$ that satisfies the A$\mu$SC.
Then $\omega$ has the Reeh-Schlieder property.
\end{theorem}
\begin{proof*}
Our proof follows that of \cite{Strohmaier+}, which is a generalisation
of the proof in \cite{Streater+} for the case of a Wightman field in
Minkowski spacetime, using the spectrum condition of the Wightman
axioms.

Let $O\subset M$ be any cc-region and set $\mathcal{D}_O:=
\left\{\pi_{\omega}(A)\Omega_{\omega}|\ A\in\Alg{U}_O\right\}$. Notice
that $\mathcal{D}_O\subset\mathcal{H}_{\omega}$ is dense if and only if
$\mathcal{D}_O^{\perp}=\left\{0\right\}$. We now suppose that
$\psi\in\mathcal{D}_O^{\perp}$, which means that the distribution
\[
w_n(x_n,\ldots,x_1):=\langle\psi,\phi^{(\omega)}_n(x_n,\ldots,x_1)\rangle
\]
is identically zero on the open neighbourhood $O^{\times n}$ in
$M^{\times n}$ for all $n\in\N$. Now suppose that
$(x,\pm k)\in WF_A(\phi_n^{(\omega)})$ for both choices of the sign. By
theorem \ref{WFAresults2} we then have
$(x,\mp k;x,\pm k)\in WF_A(\omega_{2n})\subset\Gamma_{2n}$, or
$(x,-k;x,k)\in\Gamma_{2n}\cap -\Gamma_{2n}=\emptyset$ by proposition
\ref{propGamma}. This is a contradiction, which proves that
$WF_A(\phi_n^{(\omega)})\cap -WF_A(\phi_n^{(\omega)})=\emptyset$
and hence also $WF_A(w_n)\cap-WF_A(w_n)=\emptyset$.
We may therefore apply proposition \ref{EotW} to conclude that $w_n=0$
on all of $M^{\times n}$. This in turn implies that
$\langle\psi,\pi_{\omega}(A)\Omega_{\omega}\rangle=0$ for all
$A\in\Alg{U}_M$. The vectors $\pi_{\omega}(A)\Omega_{\omega}$ with
$A\in\Alg{U}_M$ form a dense subspace of $\mathcal{H}_{\omega}$, so we
conclude $\psi\in\mathcal{H}_{\omega}^{\perp}=\left\{0\right\}$, which
completes the proof.
\end{proof*}

The proof of theorem \ref{RSAmSC} can be generalised considerably.
In fact, the A$\mu$SC allows us to derive
$WF_A(\phi_n^{(\omega)})\cap-WF_A(\phi_n^{(\omega)})=\emptyset$, but
we only need
$WF_{qA}(\phi_n^{(\omega)})\cap-WF_{qA}(\phi_n^{(\omega)})=\emptyset$
to arrive at the conclusion of theorem \ref{RSAmSC}. We will now show
that the A$\mu$SC
can be weakened to a quasi-analytic microlocal spectrum condition which
is still strong enough to make the proof above work. There is a subtlety
involved, however, because theorem \ref{WFAresults2} does not hold for
quasi-analytic wave front sets without modification. We therefore define
the following:
\begin{definition}
Let $u_n$ be a distribution on the $n$-fold product
$\mathcal{M}^{\times n}$ of a
smooth manifold $\mathcal{M}$ with values in a Banach space
$\mathcal{B}$. We define the wave front set $WF^{(n)}_{qA}(u_n)$ to be
the conic subset of $T^*\mathcal{M}^{\times n}\setminus\mathcal{Z}$
consisting of
all points $(x_1,k_1;\ldots;x_n,k_n)$ such that
\[
(x_1,k_1;\ldots;x_n,k_n)\in(\kappa^{\times n})^*
(WF_A((\kappa^{\times n})_*u)),
\]
where $\kappa$ is a smooth coordinate map on a neighbourhood of all
$x_i$, $1\le i\le n$.
\end{definition}
Given a finite number of points $x_1,\ldots,x_n$ we can always find a
coordinate map which contains these points in its domain by theorem
16.26.9 of \cite{Dieudonne}. Moreover, when these points are distinct
we can choose the analytic structure near every point arbitrarily by
theorem 8.3.1 of \cite{Hirsch}.

Notice that $WF_{qA}(u_n)\subset WF_{qA}^{(n)}(u_n)$, but the converse
is not true. The point of this definition is the following result,
which presumably fails for $WF_{qA}$:
\begin{theorem}\label{WFqAHvalued}
Let $\mathcal{H}$ be a Hilbert space, $\mathcal{M}$ a smooth manifold
and $u_i$, $i=1,2$, two $\mathcal{H}$-valued distributions on
$\mathcal{M}^{\times n_i}$ for some $n_i\in\N$. We define the
distributions $w_{ij}$ on $\mathcal{M}^{\times(n_i+n_j)}$ by
$w_{ij}(f_1,f_2):=\langle u_i(\overline{f}_1),u_j(f_2)\rangle$. Then
\[
(x,k)\in WF_{qA}^{(n_1)}(u_1)\quad \Leftrightarrow
(x,-k;x,k)\in WF_{qA}^{(2n_1)}(w_{11})
\]
and
\[
WF_{qA}^{(n_i+n_j)}(w_{ij})\subset
\left(-WF_{qA}^{(n_i)}(u_i)\cup\mathcal{Z}\right)\times
\left(WF_{qA}^{(n_j)}(u_j)\cup\mathcal{Z}\right).
\]
\end{theorem}
\begin{proof*}
If $(x,-k;x,k)\not\in WF_{qA}^{(2n_1)}(w_{11})$ then there is a choice
of coordinates $\kappa$ near all $x_i$ such
that $(x,-k;x,k)\not\in(\kappa^{\times 2n_1})^*
WF_A((\kappa^{\times 2n_1})_*w_{11})$. Theorem \ref{WFAHvalued} therefore
implies $(x,k)\not\in(\kappa^{\times n_1})^*WF_A((\kappa^{\times n_1})_*u_1)$
and hence $(x,k)\not\in WF_{qA}^{(n_1)}(u_1)$. This proves one direction of
the first statement. On the other hand, if
$(x,k)\not\in WF_{qA}^{(n_i)}(u_i)$ with $k\not=0$ then there exists a
choice of coordinates $\kappa$ such that $(x,k)\not\in
(\kappa^{\times n_i})^*WF_A((\kappa^{\times n_i})_*u_i)$. Now consider
any point $(x',k')\in T^*\mathcal{M}^{\times n_j}$. We can find a choice
of coordinates $\lambda$ such that $(x,x')$ and $(x',x)$ are in the
domain of $(\lambda)^{\times(n_i+n_j)}$ (\cite{Dieudonne} loc.~cit.).
By composing $\lambda$ with a suitable diffeomorphism we can ensure that
the analytic structure determined by $\lambda$ near $x$ coincides with
that of $\kappa$ (\cite{Hirsch} loc.~cit.). It then follows from
theorem \ref{WFAHvalued} that
$(x',k';x,k)\not\in (\lambda^{\times(n_i+n_j)})^*
WF_{A}(\lambda^{\times(n_i+n_j)})(w_{ij})$ and
$(x,-k;x',k')\not\in (\lambda^{\times(n_i+n_j)})^*
WF_{A}(\lambda^{\times(n_i+n_j)})(w_{ji})$, which completes the proof.
\end{proof*}

\begin{definition}
We say that a state $\omega$ on the Borchers-Uhlmann algebra $\Alg{U}_M$
of a globally hyperbolic spacetime $M$ satisfies the
\emph{quasi-analytic microlocal spectrum condition} (qA$\mu$SC) iff
$WF_{qA}^{(n)}(\omega_n)\subset\Gamma_n$ for all $n$.
\end{definition}
Note that this condition is well-defined on a smooth spacetime $M$
and that it is independent of a choice of coordinates. Morever we have
\begin{corollary}
Let $\omega$ be a state on the Borchers-Uhlmann algebra $\Alg{U}_M$ of
a globally hyperbolic spacetime $M$ that satisfies the qA$\mu$SC. Then
$\omega$ has the Reeh-Schlieder property.
\end{corollary}
\begin{proof*}
The proof is essentially the same as that of theorem \ref{RSAmSC}.
Note in particular that for any $\psi\in\mathcal{H}_{\omega}$ we have
\[
WF_{qA}(\langle\psi,\phi_n^{(\omega)}\rangle)\subset
WF_{qA}^{(n)}(\langle\psi,\phi_n^{(\omega)}\rangle)\subset
WF_{qA}^{(n)}(\phi_n^{(\omega)})
\]
and $WF_{qA}^{(n)}(\phi_n^{(\omega)})\cap-WF_{qA}^{(n)}
(\phi_n^{(\omega)})=\emptyset$ by theorem \ref{WFqAHvalued} and the
qA$\mu$SC.
\end{proof*}

The qA$\mu$SC is a condition that implies the $\mu$SC as well as the
Reeh-Schlieder property, but its practical use is limited unless we
can find states that satisfy this condition. Indeed, on a generic
smooth spacetime the class that is singled out by this condition could
be empty. Unfortunately the condition is very hard to work with in this
respect. It is clear that any state that satisfies the A$\mu$SC of
\cite{Strohmaier+} also satisfies the qA$\mu$SC, but in order to find
more examples one would need to have a better understanding of how the
analytic wave front set changes under a smooth diffeomorphism. Even for
ground and KMS-states on a stationary spacetime, states that are known
to have the Reeh-Schlieder property \cite{Strohmaier}, it is not clear
whether they satisfy the qA$\mu$SC. One approach would be to consider
spacetimes whose metric is analytic in a time-coordinate. For such
metrics some unique continuation results are known \cite{Robbiano+} and
these can possibly be extended to our situation, although we were unable
to obtain a result in this direction. Of course this idea may be
criticised, because a generic metric is not analytic in a time
coordinate. On the other hand it can be argued that we are not
interested in generic metrics, but only in solutions to Einstein's
equation, which makes the situation less clear. In fact, it may well
be easier to use another approach to find Reeh-Schlieder states with the
$\mu$SC, for example using the spacetime deformation argument of
section \ref{sec_RSdeform}. This means that the importance of the
qA$\mu$SC is largely academic, but it does prove the existence of a
smoothly
covariant condition that ensures the $\mu$SC and the full Reeh-Schlieder
property.

\section[The real free scalar field on $M_0$]{The Reeh-Schlieder property
for the real free scalar field on Minkowksi spacetime}\label{sec_RSM0}

It is well-known that the Minkowski vacuum $\omega_0$ has the
Reeh-Schlieder property (both as a state on $\Alg{A}^0_{M_0}$ and on
$\Alg{U}^0_{M_0}$). It is also known \cite{Clifton+,Dixmier+} that
$\mathcal{H}_{\omega_0}$ contains a dense $G_{\delta}$ of vectors which
define Reeh-Schlieder states on $\Alg{A}^0_{M_0}$, and these include at
least all states of bounded energy \cite{Haag}. We now turn to the
question whether we can find many vectors in $\mathcal{H}_{\omega_0}$
that define Reeh-Schlieder states that are also Hadamard states.

For a first result we consider the space $\mathcal{S}(\R^n)$ of
Schwartz-functions, which is a Fr\'echet space (\cite{Hoermander}
definition 7.1.2), and we define the following algebra, in analogy with
the Borchers-Uhlmann algebra\footnote{Actually, Borchers
\cite{Borchers2} works
in Minkowski spacetime and defines the algebra $\Alg{U}'_{M_0}$, so
by right this algebra can also be called ``Borchers-Uhlmann algebra''.
Note however that the notion of Schwartz-functions cannot be generalised
to general manifolds.}:
\begin{definition}
We define the algebra
$\Alg{U}'_{M_0}:=\oplus_{n=0}^{\infty}\mathcal{S}(M_0^{\times n})$,
(in the algebraic sense), equipped with:
\begin{enumerate}
\item the product $f(x_1,\ldots,x_n)g(x_{n+1},\ldots,x_{n+m}):=
(f\otimes g)(x_1,\ldots,x_{n+m})$, extended linearly,
\item the $^*$-operation
$f(x_1,\ldots,x_n)^*:=\overline{f}(x_n,\ldots,x_1)$,
extended anti-linearly,
\item a topology such that $f_j=\oplus_n f_j^{(n)}$ converges to
$f=\oplus_n f^{(n)}$
if and only if for all $n$ we have $f_j^{(n)}\rightarrow f^{(n)}$ in
$\mathcal{S}(M_0^{\times n})$ and for some $N>0$ we have $f_j^{(n)}=0$
for all $j$ and $n\ge N$.
\end{enumerate}
\end{definition}
This is a topological $^*$-algebra in the same way as $\Alg{U}_{M_0}$.
In fact, it contains $\Alg{U}_{M_0}$ as a dense linear subspace and the
canonical embedding is a continuous linear map (see \cite{Hoermander}
lemma 7.1.8). Note that the multiplication in $\Alg{U}'_{M_0}$ is
jointly continuous, because the map
$(f^{(i)},h^{(j)})\mapsto f^{(i)}\otimes h^{(j)}$ is jointly continuous.
A state $\omega$ on $\Alg{U}'_{M_0}$ consists of a sequence of $n$-point
distributions $\omega_n$, which are tempered distributions.

The state $\omega_0$ on $\Alg{U}^0_{M_0}$ has $n$-point
distributions which are tempered, so if
$\map{p}{\Alg{U}_{M_0}}{\Alg{U}^0_{M_0}}$ is the canonical projection
map, then the state $\omega_0\circ p$ on $\Alg{U}_{M_0}$ can be extended
in a unique way to $\Alg{U}'_{M_0}$. We will denote this extension by
$\omega_0$ too.
\begin{theorem}
Consider the Minkowski vacuum state $\omega_0$ on $\Alg{U}'_{M_0}$ for
a positive mass
$m>0$ and an $A\in\Alg{U}'_{M_0}$ such that
$v:=\pi_{\omega_0}(A)\Omega_{\omega_0}\not=0$. Then there exists a
sequence of elements
$A_n\in\Alg{U}'_{M_0}$ such that $A_n\rightarrow A$ in $\Alg{U}'_{M_0}$
as $n\rightarrow\infty$ and such that the vectors
$v_n:=\pi_{\omega_0}(A_n)\Omega_{\omega_0}$ define states on
$\Alg{U}'_{M_0}$ which restrict to Hadamard Reeh-Schlieder states on
$\Alg{U}_{M_0}$.
\end{theorem}
\begin{proof*}
For all $n\in\N$ we set $h_n(x):=\frac{n^4}{\pi^2}e^{-n^2\|x\|^2}$,
where $\|x\|^2$ denotes the Euclidean norm on $\R^4$. Notice that
$h_n\in\mathcal{S}(M_0)$ is analytic for all $n$ and that
$h_n\rightarrow\delta_0$ in the space $\mathcal{S}'(M_0)$ of tempered
distributions.

Every element $A'$ in $\Alg{U}'_{M_0}$ can be approximated by an
elements $A\in\Alg{U}'_{M_0}$ of the form
$A=\oplus_{i=0}^Nf_1^{(i)}\otimes\ldots\otimes f^{(i)}_i$ with
$f_j^{(i)}\in\mathcal{S}(M_0)$ and the result for $A'$ follows from
that for $A$. To prove it for an $A$ of this form we define the $A_n$ by
$A_n:=\oplus_{i=0}^N(h_n*f_1^{(i)})\otimes\ldots\otimes(h_n*f_i^{(i)})$,
where $*$ denotes the convolution,
\[
h_n*f_j^{(i)}(x_j)=\int h_n(x_j-y_j)f_j^{(i)}(y_j) dy_j.
\]
We then have $h_n*f_j^{(i)}\rightarrow f_j^{(i)}$ in $\mathcal{S}(M_0)$
and hence $A_n\rightarrow A$ as $n\rightarrow\infty$.

Now we let $u_j\in\Test_0(M_0^{\times 2})$ be a sequence such that
$u_j\rightarrow (\omega_0)_2$ as $j\rightarrow\infty$ (see
\cite{Hoermander} theorem 4.1.5 for the existence of such a sequence).
For every pair of functions $\phi_1,\phi_2\in\Test_0(M_0)$ we then
have (cf. Parseval's formula, \cite{Hoermander} theorem 7.1.6)
\[
\int u_j(x,y)\phi_1(x)\phi_2(y)dx\ dy=(2\pi)^{-4}\int
e^{ix\cdot\xi}\hat{u}_j(\xi,\eta)\phi_1(x)
\hat{\phi}_2(-\eta)dx\ d\xi\ d\eta.
\]
Using the fact that $\widehat{(\omega_0)_2}(\xi,\eta)=
(2\pi)^5\delta(\xi+\eta)\delta(\eta^2-m^2)\theta(\eta_0)$
(see the notations and conventions in the preface) and taking the
limit $j\rightarrow\infty$ yields:
\begin{eqnarray}\label{partialParseval}
(\omega_0)_2(\phi_1,\phi_2)&=&(2\pi)^{-4}\int e^{ix\cdot\xi}
\widehat{(\omega_0)_2}(\xi,\eta)\phi_1(x)
\hat{\phi}_2(-\eta)dx\ d\xi\ d\eta\nonumber\\
&=&2\pi\int e^{-ix\cdot\eta}\delta(\eta^2-m^2)\theta(\eta_0)
\phi_1(x)\hat{\phi}_2(-\eta)dx\ d\eta.
\end{eqnarray}
This uses the fact that a distribution on $M_0^{\times 2}$
is uniquely determined by its action on functions of the form
$\phi_1\otimes\phi_2$ (see \cite{Hoermander} theorem 5.1.1).
Notice that
$\int e^{-ix\cdot\eta}\delta(\eta^2-m^2)\theta(\eta_0)
\phi_1(x)dx=\hat{\phi}_1(\eta)\delta(\eta^2-m^2)\theta(\eta_0)$
is a tempered distribution in $\eta$, so using the fact that
$\Test_0(M_0)\subset\mathcal{S}(M_0)$ is dense (\cite{Hoermander}
lemma 7.1.8) we may extend equation (\ref{partialParseval}) to all
$\phi_2\in\mathcal{S}(M_0)$. We can then write for all
$\phi_2\in\mathcal{S}(M_0)$:
\[
(\omega_0)_2(x,\phi_2)= 2\pi\int e^{-ix\cdot\eta}\delta(\eta^2-m^2)
\theta(\eta_0)\hat{\phi}_2(-\eta)d\eta,
\]
where the expression on the right hand side is well-defined for
each point $x$. We now substitute $\phi_2=h_n*f$ with
$f\in\mathcal{S}(M_0)$ so that
\begin{eqnarray}\label{analyticity}
(\omega_0)_2(x,h_n*f)&=&2\pi\int e^{-ix\cdot\eta}\delta(\eta^2-m^2)
\theta(\eta_0)e^{-\|\eta\|^2/(4n^2)}\hat{f}(-\eta)d\eta\nonumber\\
&=&2\pi\int e^{-ix\cdot\eta}\delta(\eta_0-\omega_{\eta}^2)
e^{-\|\eta\|^2/(4n^2)}\hat{f}(-\eta)\frac{d\eta}{2\eta_0}
\end{eqnarray}
where $\omega_{\eta}:=\sqrt{\|(\eta_1,\eta_2,\eta_3)\|^2+m^2}$ and
$\|\eta\|$ again denotes the Euclidean norm on $\R^4$. The
Gaussian on the right-hand side ensures that this expression is
well-defined for every $x\in\C$, so $(\omega_0)_2(x,h_n*f)$ can be
extended to a function on $\C$. Moreover, for all $z\in\C^4$ we
can substitute $e^{-i(x-z)\cdot\eta}$ for $e^{-ix\cdot\eta}$ in
equation (\ref{analyticity}) and use $e^{-i(x-z)\cdot\eta}=
e^{-ix\cdot\eta}(1-iz\cdot\eta+R(z,\eta))$, where the remainder term
$R(z,\eta)=-(z\cdot\eta)^2\int_0^1(1-s)e^{-isz\cdot\eta}ds$ satisfies
$|R(z,\eta)|\le |z\cdot\eta|^2(1+|e^{-iz\cdot\eta}|)$. Using the
Gaussian for convergence it follows that $(\omega_0)_2(x,h_n*f)$ is
complex differentiable and hence analytic. The same is then true for
$(\omega_0)_2(h_n*f,y)=\overline{(\omega_0)_2(y,h_n*\overline{f})}$.

Notice that $v_n\rightarrow v$, by $\|v_n-v\|^2=\omega_0((A_n-A)^*(A_n-A))$
and the joint continuity of the multiplication in $\Alg{U}'_{M_0}$.
So if $v\not=0$ then $v_n\not=0$ for
all sufficiently large $n$. By dropping a finite number of indices at
the start of the sequence we may assume that
this holds for all $n$. We now claim that each $v_n$ defines a state
that satisfies the A$\mu$SC, from which the result follows. To prove
this claim we note the fact that $\omega_0$ is quasi-free, that
$(\omega_0)_2(.,h_n*f)$ and $(\omega_0)_2(h_n*f,.)$ are analytic for
all $f=f^{(i)}_j$ and that $WF_A((\omega_0)_2)\subset\Gamma_2$ (see
\cite{Strohmaier+} theorem 6.3 and \cite{Araki} section 4.2). The
result then follows from definition \ref{def_qf} and
proposition \ref{propGamma}.
\end{proof*}

To conclude this section we prove that the Minkowski vacuum has a
property which is stronger than the Reeh-Schlieder property.
\begin{theorem}\label{strongRS}
Consider the Minkowski vacuum state $\omega_0$ on $\Alg{U}^0_{M_0}$
with its GNS-quadruple $(\mathcal{H}_{\omega_0},\pi_{\omega_0},
\Omega_{\omega_0},\mathscr{D}_{\omega_0})$. If $O\subset M_0$ is any
non-empty cc-region, then $\pi_{\omega_0}(\Alg{U}^0_O)\Omega_{\omega_0}$
is dense in $\mathscr{D}_{\omega_0}$ in the graph topology
(see definition \ref{def_graphtop}).
\end{theorem}
\begin{proof*}
Recall that we need to find for each $\phi\in\mathscr{D}_{\omega_0}$ a
sequence of vectors $\phi_n\in\pi_{\omega_0}(\Alg{U}^0_O)\Omega_{\omega_0}$
such that $\pi_{\omega_0}(A)\phi_n\rightarrow\pi_{\omega_0}(A)\phi$ for all
$A\in\Alg{U}^0_{M_0}$. We will use the fact that $\omega_0$ is a quasi-free
state that satisfies the A$\mu$SC.
First we consider the real Hilbert space
$\mathcal{H}^1:=\overline{\left\{\phi_1^{(\omega_0)}(f)|\ f\in
\Test_0(M_0,\R)\right\}}$ and the subspace
$\mathcal{H}^1_O:=\overline{\left\{\phi_1^{(\omega_0)}(f)|\ f\in
\Test_0(O,\R)\right\}}$ with
$(\psi,\chi):=\mathrm{Re}\langle\psi,\chi\rangle$ as inner product. (See
\cite{Kay2} appendix A1 for a similar one-particle Reeh-Schlieder result.)
If $\psi\in\mathcal{H}^1$ is in $(\mathcal{H}^1_O)^{\perp}$, where $\perp$
refers to the inner product $(,)$ on $\mathcal{H}^1_O$, then the
$\mathcal{H}_{\omega_0}$-valued distribution
$w(x):=\langle\psi,\phi_1^{(\omega_0)}(x)\rangle$ is identically $0$
on $O$ by complex linearity.
By the A$\mu$SC we find that $WF_A(w)\subset\mathcal{N}^+$, so
$WF_A(w)\cap-WF_A(w)=\emptyset$ and $w\equiv 0$ everywhere by
proposition \ref{EotW}. Hence, for every $f\in\Test_0(M_0,\R)$ we can
find a sequence $f_n$ of elements in $\Test_0(O,\R)$ such that
$\phi_1^{(\omega_0)}(f_n)\rightarrow\phi_1^{(\omega_0)}(f)$ in
$\mathcal{H}^1$ as $n\rightarrow\infty$. This also means that:
\[
\|\phi_1^{(\omega_0)}(f-f_n)\|^2_{\mathcal{H}_{\omega_0}}=
(\phi_1^{(\omega_0)}(f-f_n),\phi_1^{(\omega_0)}(f-f_n))\rightarrow 0.
\]
In other
words, $\phi_1^{(\omega_0)}(f_n)\rightarrow\phi_1^{(\omega_0)}(f)$ in
$\mathcal{H}_{\omega_0}$. If we decompose $f\in\Test_0(M_0)$ as
$f=u+iv$ where $u,v\in\Test_0(M_0,\R)$, then we may apply the previous
reasoning to find sequences $u_n$ and $v_n$ in $\Test_0(O,\R)$ such that
$\phi_1^{(\omega_0)}(u-u_n)\rightarrow 0$ and
$\phi_1^{(\omega_0)}(v-v_n)\rightarrow 0$ as $n\rightarrow\infty$. This
implies that for $f_n:=u_n+iv_n$ we have
$\phi_1^{(\omega_0)}(f-f_n)\rightarrow 0$ and
$\phi_1^{(\omega_0)}(\overline{f}-\overline{f}_n)\rightarrow 0$ as
$n\rightarrow\infty$.

Now consider the two homogeneous elements $A:=f^i\otimes\ldots\otimes f^1$
and $B:=h^{(r)}$ of $\Alg{U}^0_{M_0}$ with $f^j\in\Test_0(M_0)$ and
$h^{(r)}\in\Test_0(M_0^{\times r})$. By the previous paragraph we
can find sequences $f^j_n$ in $\Test_0(O)$ such that
$\phi_1^{(\omega_0)}(f^j_n-f^j)\rightarrow 0$ and
$\phi_1^{(\omega_0)}(\overline{f}^j_n-\overline{f}^j)\rightarrow 0$
as $n\rightarrow\infty$. We set $A_n:=f^1_n\otimes\ldots\otimes f^i_n$
and notice that
\begin{eqnarray}\label{telescope}
B(A-A_n)&=&h^{(r)}\otimes f^i\otimes\ldots\otimes f^2\otimes
(f^1-f^1_n)\nonumber\\
&&+
h^{(r)}\otimes f^i\otimes\ldots\otimes f^3\otimes(f^2-f^2_n)\otimes
f^1_n\nonumber\\
&&+\ldots+h^{(r)}\otimes (f^i-f^i_n)\otimes f^{i-1}_n\otimes\ldots\otimes
f^1_n.
\end{eqnarray}
We wish to show that
$\pi_{\omega_0}(B(A-A_n))\Omega_{\omega_0}\rightarrow 0$ as
$n\rightarrow\infty$. For this it is sufficient to show that each term
in equation (\ref{telescope}) converges to $0$, so we consider a fixed
term containing $f^j-f^j_n$. Without loss of generality
we may absorb the factor $f^i\otimes\ldots\otimes f^{j+1}$ into $h^{(r)}$,
so that the term looks like
\[
h^{(r)}\otimes(f^j-f^j_n)\otimes f^{j-1}_n\otimes
\ldots\otimes f^1_n.
\]
The norm squared of $\pi_{\omega_0}(.)\Omega_{\omega_0}$ of this term is of
the form:
\begin{eqnarray}\label{normsquared}
\|\pi_{\omega_0}(h^{(r)}\otimes (f^j-f^j_n)\otimes f^{j-1}_n\otimes\ldots
\otimes f^1_n)\Omega_{\omega_0}\|^2&=&\nonumber\\
\omega_0(\overline{f}^1_n\otimes\ldots\otimes\overline{f}^{j-1}_n\otimes
(\overline{f}^j-\overline{f}^j_n)\otimes(h^{(r)})^*\otimes&&\\
h^{(r)}\otimes (f^j-f^j_n)\otimes f^{j-1}_n\otimes\ldots\otimes f^1_n).
&&\nonumber
\end{eqnarray}
Using the fact that $\omega_0$ is quasi-free we write this as a
sum of terms and show that each term converges to $0$. To see this we
first note that the sequences $\|\phi_1^{(\omega_0)}(f^j_n)\|$ and
$\|\phi_1^{(\omega_0)}(\overline{f}^j_n)\|$ remain bounded, so all
factors of the form $(\omega_0)_2(f_n,f'_n)$, where $f_n$ and $f'_n$ are
either $f^j_n$ or $\overline{f}^j_n$, remain bounded by the
Cauchy-Schwarz inequality. Next we notice that
\begin{eqnarray}
\left|\int (\omega_0)_2(x,y)\chi(x)h(y,y')dx\ dy\right|^2
&\le&\nonumber\\
(\omega_0)_2(\chi,\overline{\chi})\cdot
\int(\omega_0)_2(x,y)\overline{h}(x,y')h(y,y')dx\ dy\nonumber
\end{eqnarray}
is a compactly supported smooth function of $y'$. If we substitute
for $\chi$ either
$f^j_n$ or $\overline{f}^j_n$, the right-hand side is
estimated by a bounded constant times a compactly supported smooth
function of $y'$. Using these facts we can integrate out all but a few
variables in a summand of equation (\ref{normsquared}) to obtain:
\begin{eqnarray}\label{integrated}
&&\left|\int \overline{f}^1_n(x_1)\cdots\overline{f}^{j-1}_n(x_{j-1})
(\overline{f}^j-\overline{f}^j_n)(x_j)
\overline{h}(x_{j+r},\ldots,x_{j+1})\right.\nonumber\\
&&h(x_{j+r+1},\ldots,x_{j+2r})
(f^j-f^j_n)(x_{j+2r+1})f^{j-1}_n(x_{j+2r+2})\cdots f^1_n(x_{2r+2j})
\nonumber\\
&&\biggl.(\omega_0)_2(x_{\pi^{-1}(1)},x_{\pi^{-1}(2)})\cdots
(\omega_0)_2(x_{\pi^{-1}(2j+2r-1)},x_{\pi^{-1}(2j+2r)})
dx_1\cdots dx_{2j+2r}\biggr|\nonumber\\
&\le&
C\sum_{\pi}\left|\int (f^j-f^j_n)(x_1)H(x_2)
(\omega_0)_2(x_{\pi(1)},x_{\pi(2)})dx_1\ dx_2\right|\\
&&+
C\sum_{\pi}\left|\int (f^j-f^j_n)(x_1)\chi(x_2)
(\omega_0)_2(x_{\pi(1)},x_{\pi(2)})dx\ dy\right|\nonumber
\end{eqnarray}
where $C>0$ is a constant, $H\in\Test_0(M_0)$, $\chi$ is either
$f^i_n$ or $\overline{f}^i_n$ for some index $i$ and we sum over both
permutations $\pi$ of the set $\left\{1,2\right\}$. The first term
appears whenever the variable $y$ occurs in
$\overline{f}^j\otimes h^*\otimes h$ rather than in some $f_n^i$ or
$\overline{f}^i_n$.
Both terms in equation (\ref{integrated}) can be estimated using the
Cauchy-Schwarz inequality and are then seen to converge to $0$, because
$\|\phi_1^{(\omega_0)}(f^j-f^j_n)\|$ and
$\|\phi_1^{(\omega_0)}(\overline{f}^j-\overline{f}^j_n)\|$
converge to $0$ as $n\rightarrow\infty$. This proves that the
norm-squared in equation (\ref{normsquared}) converges to $0$. The same
conclusion remains true when we replace $B$ by a finite sum of
homogenous terms and hence $\pi_{\omega_0}(A-A_n)\Omega_{\omega_0}$
converges to $0$ in the graph topology.

By definition of $\pi_{\omega_0}$ the linear space
$\pi_{\omega_0}(\Alg{U}^0_{M_0})\Omega_{\omega_0}$ is dense in
$\mathscr{D}_{\omega_0}$ in the graph topology (see definition
\ref{def_graphtop} and theorem \ref{GNSrep}). We can approximate every
$A\in\Alg{U}^0_{M_0}$ by a sequence $A_n$ of elements of the form
$A_n=\oplus_{i=0}^Nf^{(i)}_{1,n}\otimes\ldots\otimes f^{(i)}_{i,n}$ with
$f^{(i)}_{j,n}\in\Test_0(M_0)$. By joint continuity of the multiplication we
find that $(A-A_n)^*B^*B(A-A_n)$ converges to $0$ for every
$B\in\Alg{U}^0_{M_0}$ and hence that $\pi_{\omega_0}(B(A-A_n))\Omega_{\omega_0}$
converges to $0$ for all $B$. Hence, the elements of the form
$\sum_{i=1}^n\phi_i^{(\omega_0)}(f^{(i)}_1,\ldots,f^{(i)}_i)$, where
$f^{(i)}_j\in\Test_0(M_0)$ and
$\phi_i^{(\omega_0)}$ are the Hilbert space-valued distributions, are dense in
$\mathscr{D}_{\omega_0}$ in the graph topology. By the previous paragraph every
term in this sum can be approximated in the
graph topology by a term of the same form but with $f^{(i)}_j\in\Test_0(O)$.
\end{proof*}

\section[The RS-property for the real free scalar field]{The
Reeh-Schlieder property for the real free scalar field}\label{sc_RSconclusion}

To conclude this chapter we apply the results of the previous sections
to the free scalar and Dirac field as presented in chapters \ref{ch_sf}
and \ref{ch_Df} and consider what conclusions we may draw.
We first consider the free field Borchers-Uhlmann functor $\Func{U}^0$ with
the state space $\Func{Q}^0$ of Hadamard states defined in
chapter \ref{ch_sf}:
\begin{proposition}\label{freefield1}
Let $M$ be a globally hyperbolic spacetime, let $O\subset M$ a
bounded cc-region with non-empty causal complement and assume
that the mass $m>0$ is strictly positive. Then there is a state
$\omega\in\Stat{Q}^0_M$ on $\Alg{U}^0_M$ which has the Reeh-Schlieder
property for $O$.
\end{proposition}
\begin{proof*}
We can find an ultrastatic (and hence
stationary) spacetime $M'$ diffeomorphic to $M$. Because $m>0$ we may
apply the results of \cite{Kay}, which imply the existence of a
regular quasi-free ground state $\omega'$ on $\Alg{A}^0_{M'}$. This
state is a Reeh-Schlieder state (see \cite{Strohmaier}) and is
Hadamard because it satisfies the microlocal spectrum condition (see
\cite{Strohmaier+,Radzikowski}). It follows that $\omega'$ also defines
a Hadamard state $\tilde{\omega}$ on $\Alg{U}^0_{M'}$. To see that
$\tilde{\omega}$ is a Reeh-Schlieder state we choose a non-empty
cc-region $O\subset M'$ and compare the GNS-representations of
$\rho':=\omega'|_{\Alg{A}^0_O}$ and
$\tilde{\rho}:=\tilde{\omega}|_{\Alg{U}^0_O}$. Notice that we may take
$\mathcal{H}_{\tilde{\rho}}\subset\mathcal{H}_{\rho'}$ with
$\Omega_{\tilde{\rho}}=\Omega_{\rho'}$, by the essential
uniqueness of the GNS-representation (theorem \ref{GNSrep}). Because
we can identify
$\pi_{\omega'}(W(f))=\mathrm{exp}(i\Phi^{(\tilde{\omega})}(f))$
(see \cite{Bratteli+} proposition 5.2.4) we see that $\Omega_{\rho'}$
must be cyclic for $\pi_{\omega'}(\Alg{U}^0_O)$, otherwise it would not
be cyclic for $\pi_{\omega'}(\Alg{A}^0_O)$. The fact that
$\tilde{\omega}$ has the Reeh-Schlieder property therefore follows from
the fact that $\omega'$ has it. Now recall that the locally covariant
quantum field theory $\Func{U}^0$ and state space $\Func{Q}^0$
satisfy the time-slice axiom (see proposition \ref{U0Q0}). We can
therefore apply theorem \ref{RSdeformation} with the state
$\tilde{\omega}$, from which the result follows immediately.
\end{proof*}

As we noticed in subsection \ref{subs_RSdeform} we can draw
stronger conclusions when the theory is $C^*$-algebraic:
\begin{proposition}\label{freefield2}
Let $M$ be a globally hyperbolic spacetime with a non-compact Cauchy
surface and assume that the mass
$m>0$ is strictly positive. Then there is a state
$\omega\in\Stat{S}^0_M$ on $\Alg{A}^0_M$ such that $\mathcal{H}_{\omega}$
contains a dense $G_{\delta}$ $\mathcal{G}$ of vectors which define (full)
Reeh-Schlieder states. For all bounded cc-regions $V\subset M$
the local von Neumann algebra $\Alg{R}_V$ is not finite and if $V$ has
non-zero causal complement then each vector $\psi\in\mathcal{G}$ is
cyclic and separating for $\Alg{R}_V$.
\end{proposition}
\begin{proof*}
The theory is causal, locally quasi-equivalent, satisfies the time-slice
axiom and is nowhere classical (see proposition \ref{prop_A0prop}). Note
that $\Alg{R}_V$ is well-defined, independent of $\omega\in\Stat{S}^0_M$
by local quasi-equivalence. As
in the proof of proposition \ref{freefield1} we can find a Reeh-Schlieder
state $\omega'$ on $\Alg{A}^0_{M'}$, where $M'$ is a spacetime
diffeomorphic (but not isometric) to $M$. Now theorem \ref{RSdense} and
the definition of $\Stat{S}^0_M$ prove the existence of a full Reeh-Schlieder
state $\omega\in\Stat{S}^0_M$ and the results of \cite{Dixmier+} (see also
the proof of theorem \ref{RSdense}) provide the dense $G_{\delta}$ set
$\mathcal{G}$
in $\mathcal{H}_{\omega}$. The other conclusions then follow from
proposition \ref{separability1} and corollary \ref{not_finite}.
\end{proof*}
Note that stronger results on the type of the local algebras are known
\cite{Verch2}, but we have used a different and interesting method of
proof.

It seems likely that our deformation results can be extended from
spacetimes to spin spacetimes, so that similar results can be obtained
for the Dirac. In the case of the functors $\Func{F}^0$ and $\Func{R}^0$
we formulate:
\begin{conjecture}\label{Diracfield1}
Let $M$ be a globally hyperbolic spin spacetime and let $O\subset\mathcal{M}$
be a bounded cc-region with non-empty causal complement. Then there is a
state $\omega\in\Stat{R}^0$ on $\Alg{F}^0_M$ which has the Reeh-Schlieder
property for $O$.
\end{conjecture}
\begin{proofsketch*}
We can find an ultrastatic (and hence stationary) spacetime $M'$ diffeomorphic
to $M$. There then exists a quasi-free KMS state $\omega'$ on
$\overline{\Alg{F}}^0_{M'}$, which has the Reeh-Schlieder property (see
\cite{Strohmaier}). By \cite{Sahlmann+2} this state is Hadamard. Because
$\Alg{F}^0_{M'}$ is dense in $\overline{\Alg{F}}^0_{M'}$ we see that
$\omega'$ defines a Reeh-Schlieder state on $\Alg{F}^0_{M'}$, which
is Hadamard by definition \ref{def_DmSC} and hence satisfies the $\mu$SC
by proposition \ref{lem_Dts2}.
The locally covariant quantum field theory $\Func{F}^0$ and the state
space $\Func{R}^0$ satisfy the time-slice axiom (see proposition
\ref{F0R0}). The proof then comes down to a generalisation of
theorem \ref{RSdeformation} with the state $\omega'$, from which the result
would follow immediately.
\end{proofsketch*}

To find full Reeh-Schlieder states for the free Dirac field we could again
use the $C^*$-algebraic approach and a generalisation of theorem \ref{RSdense}.
However, theorem \ref{RSdense} requires the theory to be causal, which
means that we would have to use $\Func{B}$ and not $\overline{\Func{F}}^0$.
\begin{conjecture}\label{Diracfield2}
Let $M$ be a globally hyperbolic spin spacetime with a non-compact
Cauchy surface. Then there is a state
$\omega\in\Stat{T}_M$ on $\Alg{B}_M$ such that $\mathcal{H}_{\omega}$
contains a dense $G_{\delta}$ $\mathcal{G}$ of vectors which define (full)
Reeh-Schlieder states. For all bounded cc-regions $V\subset M$
with non-zero causal complement each vector $\psi\in\mathcal{G}$ is
cyclic and separating for $\Alg{R}_V$.
\end{conjecture}
\begin{proofsketch*}
As in the proof of proposition \ref{Diracfield1} we can find an
ultrastatic (and hence stationary) spacetime $M'$ diffeomorphic to $M$
and a quasi-free Reeh-Schlieder state on $\overline{\Alg{F}}^0_{M'}$
which is Hadamard. The restriction of $\omega'$ to $\Alg{F}^0_{M'}$
satisfies the $\mu$SC by proposition \ref{lem_Dts2} and hence
$\omega'$ as a state on $\overline{\Alg{F}}^0_{M'}$ satisfies the
$\mu$SC, because this condition depends only the $n$-point distributions
(see definition \ref{def_DmSC}).

In the Hilbert space $\mathcal{H}_{\omega'}$ we can define the closed
subspaces $\mathcal{H}^0$ respectively $\mathcal{H}^1$, generated by
the even respectively odd polynomials of elements $B_{M'}(f)$,
$f\in\mathcal{D}_0(M')$. Because $\omega'$ is quasi-free we see that
these spaces are orthogonal and hence $\mathcal{H}_{\omega'}$ is the
direct sum $\mathcal{H}_{\omega'}=\mathcal{H}^0\oplus\mathcal{H}^1$.
The restriction of $\omega'$ to $\Alg{B}_{M'}$ has the GNS-triple
$(\mathcal{H}^0,\pi_{\omega'}|_{\Alg{B}_{M'}},\Omega_{\omega'})$,
by the essential uniqueness of the GNS-representation (theorem
\ref{GNSrep}). Because $\omega'$ has the Reeh-Schlieder property we
see that for a non-empty cc-region $O$ the linear space
$\pi_{\omega'}(\Alg{B}_O)\Omega_{\omega'}$ is dense in
$\mathcal{H}^0$ and the space spanned by the odd polynomials of
$B_{M'}(f)$ with $f\in\mathcal{D}_0(O)$ is dense in $\mathcal{H}^1$.
The first of these two statements  implies that the restriction of
$\omega'$ to $\Alg{B}_{M'}$ is a Reeh-Schlieder state.

The locally covariant quantum field theory $\Func{B}$ with the state
space functor $\Func{T}$ is causal, locally quasi-equivalent and
satisfies the time-slice axiom by proposition \ref{BTprop}.
A generalisation of theorem \ref{RSdense} and the definition of
$\Stat{T}_M$ (definition \ref{def_qfD}) would then prove the existence
of a full Reeh-Schlieder
state $\omega\in\Stat{T}_M$ and the results of \cite{Dixmier+} (see also
the proof of theorem \ref{RSdense}) provide the dense $G_{\delta}$ set
$\mathcal{G}$ in $\mathcal{H}_{\omega}$. The final conclusion
follows from proposition \ref{separability1}.
\end{proofsketch*}

To conclude this chapter we return to the question whether Hadamard states
with the (full) Reeh-Schlieder property exist for the free scalar field in
any globally hyperbolic spacetime. Whereas proposition \ref{freefield1}
provides us with Hadamard states that have the Reeh-Schlieder property only
for a fixed but arbitrary region, proposition \ref{freefield2}
provides us with full
Reeh-Schlieder states that are possibly not Hadamard (recall lemma
\ref{meagredomain}). The main problem is that theorem \ref{RSdense} is
formulated in the Hilbert space topology, a topology which is not
suitable to obtain results on Hadamard states. We believe that the
invariant dense domain $\mathscr{D}_{\omega}$ in the graph topology, where
$\omega$ is any Hadamard state, might be more suited for this purpose. A first
question of interest is whether this space can be shown to be a Baire space.
The result of theorem \ref{strongRS} may also be of interest for
investigations along these lines.

\chapter{Conclusions}\label{ch_concl}

In this thesis we have presented and discussed results on several aspects of
locally covariant quantum field theory \cite{Brunetti+}.

First of all we have tried to put
the theory in a philosophical context in chapter \ref{ch_phil} and
described how its morphisms can be interpreted as a subsystem relation,
which makes the framework a model for modal logic.

In chapter \ref{ch_LCQFT}
we gave a precise mathematical formulation of locally covariant quantum field
theory, following closely the existing literature except for the sharpened
definition of the time-slice axiom and the introduction of
nowhere-classicality.

Chapter \ref{ch_sf} and \ref{ch_Df} describe two
examples of locally covariant quantum fields, namely the real free scalar
field and the free Dirac field. The scalar field is described in two
well-known approaches in chapter \ref{ch_sf}, namely the distributional
approach based on the Borchers-Uhlmann algebra and the $C^*$-algebraic
approach which uses the CCR-algebra (or Weyl-algebra). This chapter also
contains the elegant new results that the Hadamard condition on the
two-point distribution of a state automatically implies the $\mu$SC and that
all its truncated $n$-point distributions are smooth for $n\not=2$, due to
the commutation relations. Chapter \ref{ch_Df} describes the free Dirac
field as a locally covariant quantum field and shows that this can be
done in a representation independent way, so that the physics is determined
entirely by the relations between the adjoint map, charge conjugation and
the Dirac operation. This chapter also contains the proof of a relation
between the stress-energy-momentum tensor and the relative Cauchy evolution,
similar to a result that was already known for the scalar field
\cite{Brunetti+}.

In chapter \ref{ch_RS} we considered the Reeh-Schlieder property
in locally covariant quantum field theories. We discussed the
meaning and importance of this property and proved several general
results and their application to the real free scalar field. The main
issue in finding full Reeh-Schlieder states was the size of the state space.
If the state space is sufficiently large, we can find many such states.
However, if we restrict our attention to Hadamard states, we have only
proved the existence of Hadamard states with the Reeh-Schlieder property
for an arbitrarily given region. The question whether Hadamard states with
the full
Reeh-Schlieder property exist in general curved spacetimes is still
open, although we have given a smoothly covariant sufficient condition
in terms of the new notion of quasi-analytic wave front sets.
We also suggested that the use of the graph topology could be useful to
answer it, if it can be shown that $\mathscr{D}_{\omega}$ is a Baire space.
As a first result in this direction we proved that the Minkowski vacuum state
has a strong form of the Reeh-Schlieder property.

Finally, the appendix explains the notion of smooth and analytic wave
front set and gives a systematic and elegant treatment of these notions
for distributions with values in a Banach space, including some new (but
expected) results.

\appendix
\renewcommand{\thesection}{A.\arabic{section}}

\chapter{Some results on wave front sets}\label{ch_ma}

\begin{quote}
`\ldots [A]s Aristotle expressly declares on page 633 of the
Louvre edition:
\begin{quote}
\greektext Enteleq'eia tis >'esti ka'i l'ogos to\~u dun'amin >'eqontos\\
toio\~udi >'eitai.\latintext '
\end{quote}
`I am not very well versed in Greek,' said the giant.\\
`Nor I either,' said the philosophical mite.\\
`Why then do you quote that same Aristotle in Greek?' resumed the
Sirian.\\
`Because,' answered the other, `it is but reasonable we should quote
what we do not comprehend in a language we do not understand.'
\end{quote}
\begin{flushright}
Voltaire, Micromegas: a philosophical tale, Ch. 7
\end{flushright}

In this appendix we will explain the language of wave front sets, which
is used to formulate some of the results in this thesis. We will define
smooth and analytic wave front sets for Banach space-valued
distributions on complex vector bundles and derive a number of useful
results in an elegant way that directly generalises the scalar-valued
cases. For a detailed introduction to scalar distributions we refer
to \cite{Hoermander}. More information on Hilbert and Banach
space-valued distributions can be found in \cite{Strohmaier+,Fewster+2}
and for distributions on vector bundles we refer to
\cite{Sahlmann+} and also \cite{Dencker}.

\section{The smooth wave front set}

Let $\mathcal{B}$ be a Banach space with continuous dual space
$\mathcal{B}'$ and let $u$ be a $\mathcal{B}$-valued distribution on
an open set $X\subset\R^n$, i.e.
$\map{u}{\Test_0(X)}{\mathcal{B}}$ is a continuous linear map, where
$\Test_0(X)$ is the space of test-functions on $X$ in the
test-function topology. This
means that for every compact subset $K\subset X$ there are constants
$C>0$ and $m\in\N$ such that
\begin{eqnarray}\label{def_distr}
\|u(f)\|\le C\sum_{|\alpha|\le m}\sup_{x\in K}|D^{\alpha}f(x)|
\end{eqnarray}
for all $f\in\Test_0(K)$. The following lemma will
come in useful:
\begin{lemma}\label{Banachlemma}
Let $u$ be a $\mathcal{B}$-valued distribution on an open set
$X\subset\R^n$, where $\mathcal{B}$ is a Banach space. If
$O\subset X$ is an open subset, then $u$ is smooth on $O$ if and only
if $\phi\circ u$ is smooth on $O$ for all $\phi\in\mathcal{B}'$.
\end{lemma}
\begin{proof*}
If $u$ is smooth on $O$ then $\phi\circ u$ is smooth on $O$ for each
$\phi\in\mathcal{B}'$, because $\map{\phi}{\mathcal{B}}{\C}$ is smooth.
Notice that we can identify every continuous function
$\map{u}{O}{\mathcal{B}}$, and hence also every smooth function, with
a distribution using Bochner integrals (see e.g. \cite{Hille} section
7.5 for the definition of Bochner integrals). This works as follows.
For each $f\in\Test_0(O)$ the product $fu$ is Bochner-integrable and
$u(f):=\int_O fu$ is the unique element in $\mathcal{B}$ such that
$\phi(u(f))=(\phi\circ u)(f)=\int_O (\phi\circ u)f$ for all
$\phi\in\mathcal{B}'$. Clearly $f\mapsto u(f)$ is linear and
\begin{equation}\label{BDest}
\|u(f)\|\le\int_O \|u(x)\|\cdot|f(x)|dx.
\end{equation}
To prove that $u$ is a distribution we note that $\|u(x)\|$ attains a
maximum $C\ge 0$ on any given compact set $K$, so equation (\ref{BDest})
implies for $f\in\Test_0(K)$ that
$\|u(f)\|\le C\int_O |f(x)|dx$, which implies equation (\ref{def_distr}).

For the converse we suppose that $u$ is a distribution on $X$ such that
$\phi\circ u$ is smooth on $O$ for all $\phi\in\mathcal{B}'$.  For any
compact subset $K\subset O$ we consider the space $C^0(K)$ of continuous
functions on $K$, which is a Banach space in the supremum norm
$\|f\|_{C^0}:=\sup_K|f|$. The Banach space dual of $C^0(K)$ is
$\mathcal{E}^0(K)$, the space of distributions of order $0$ with support
in $K$, which has the norm
$\|v\|_{\mathcal{E}^0}:={\displaystyle\sup_{f\not=0}}\frac{|v(f)|}{\|f\|_{C^0}}$.
For each $\phi\in\mathcal{B}'$ and $f\in \Test_0(K)\subset\mathcal{E}^0(K)$
we then have
\[
|\phi\circ u(f)|=\left|\int_O(\phi\circ u)f\right|\le
C_{\phi}\|f\|_{\mathcal{E}^0}
\]
for some constant $C_{\phi}\ge 0$. For each $f\not=0$ the map
$\phi\mapsto\frac{1}{\|f\|_{\mathcal{E}^0}}\phi\circ u(f)$ is a bounded
linear map on $\mathcal{B}'$, so we can apply the uniform boundedness principle
(\cite{Kadison+} theorem
1.8.10) to find $\|u(f)\|\le C\|f\|_{\mathcal{E}^0}$ for all $f\in\Test_0(K)$.
(Here we also use
the fact that the canonical map $\mathcal{B}\subset\mathcal{B}''$ is
isometric, \cite{Hille} theorem 7.2.2, so the norm $\|u(f)\|$ can be taken to
be the norm in $\mathcal{B}$.) Moreover, because
$\Test_0(K)\subset\mathcal{E}^0(K)$ is dense we can extend $u$ to a bounded
linear map from $\mathcal{E}^0(K)$ to $\mathcal{B}$. Because we can do
this for all compact subsets $K\subset O$ we can obtain a continuous
linear map $\map{u}{\mathcal{E}^0(O)}{\mathcal{B}}$, where $\mathcal{E}^0(O)$
is the space of compactly supported distributions on $O$ of order $0$.
This space contains the Dirac delta distribution $\delta_x$ at each point
$x\in O$, so we can define a function $\map{L}{O}{\mathcal{B}}$ by
$L(x):=u(\delta_x)$. We wish to show that $L$ is smooth and gives rise to
the original distribution $u$.

For each convex compact subset $K\subset O$ and each $\phi\in\mathcal{B}'$
we can find a constant $C_{\phi}$ such that
$|\phi\circ L(x)-\phi\circ L(y)|=|\phi\circ u(x)-\phi\circ u(y)|\le
C_{\phi}\|x-y\|$ for all $x,y\in K$, because the first order derivatives of
the smooth function $\phi\circ u$ remain bounded on $K$. Applying the
uniform boundedness principle (and the isometry
$\mathcal{B}\subset\mathcal{B}''$) again we find a constant $C$ such that
$\|L(x)-L(y)\|\le C\|x-y\|$ for $x,y\in K$, showing that $L$ is continuous.
For all $f\in\Test_0(O)$ and $\phi\in\mathcal{B}'$ the Bochner integral $L(f)$
satisfies $\phi\circ L(f)=\int (\phi\circ L)(x)f(x)\ dx
=\int (\phi\circ u)(x)f(x)\ dx=\phi\circ u(f)$, i.e. $L(f)=u(f)$ and we may
identify $u$ with the continuous function $L$ on $O$.

Applying the argument of the previous paragraphs to the distributions
$\partial^{\alpha}u$ for each multi-index $\alpha$ gives rise to continuous
functions $\map{L^{\alpha}}{O}{\mathcal{B}}$. To see that the $L^{\alpha}$
really are the derivatives of $L$ we argue as follows. For each $x\in O$,
$i\in\left\{1,\ldots,n\right\}$, multi-index $\alpha$ and
$\phi\in\mathcal{B}'$ there is a constant $C_{\alpha,\phi}$ such that for
all sufficiently small $h\in\R$, $h\not=0$ we have:
\[
\left|\frac{\phi\circ L^{\alpha}(x+he_i)-\phi\circ L^{\alpha}(x)}{h}-
\phi\circ L^{\alpha'}(x)\right|\le C_{\alpha,\phi}|h|
\]
by Taylor's theorem. Here $e_i$ is a basis vector of $\R^n$ and
$\alpha'$ is the multi-index obtained from $\alpha$ by increasing
$\alpha_i$ by one. The maps
$\frac{1}{h}(L^{\alpha}(x+he_i)-L^{\alpha}(x))-L^{\alpha'}(x)$
are continuous linear maps on $\mathcal{B}'$, so by the uniform
boundedness principle we obtain
\[
\left\|\frac{L^{\alpha}(x+he_i)-L^{\alpha}(x)}{h}-L^{\alpha'}(x)\right\|
\le C_{\alpha}|h|
\]
for some constant $C_{\alpha}$. Hence, $L^{\alpha'}$ is the derivative of
$L^{\alpha}$ in the direction $e_i$. It follows that all derivatives of $L$
exist and are continuous, so $L$ is smooth.
\end{proof*}

\begin{definition}
A smooth \emph{regular direction} for a Banach space-valued
distribution $u$ is a point
$(x,k)\in X\times(\R^n\setminus\left\{0\right\})$ for
which there exist an $f\in\Test_0(X)$ with
$f(x)\not=0$, a conic open neighbourhood
$V\subset(\R^n\setminus\left\{0\right\})$ of $k$ (i.e. an open
neighbourhood such that $\xi\in V$ and $r>0$ imply $r\xi\in V$) and a
sequence of constants $C_N$, $N\in\N$, such that
$\|u(e^{-i\xi\cdot}f)\|\le\frac{C_N}{1+\|\xi\|^N}$ for all $\xi\in V$,
where $\|\xi\|$ denotes the Euclidean norm.

The \emph{wave front set} $WF(u)$ of $u$ is defined as
\begin{eqnarray}
WF(u)&:=&\left\{(x,k)\in X\times(\R^n\setminus\left\{0\right\})|\
(x,k)\mathrm{\ is\ not\ a\ smooth\ regular}\right.\nonumber\\
&&\left.\mathrm{direction\ for\ } u\right\}.\nonumber
\end{eqnarray}
\end{definition}

It is clear from the definition that the wave front set is a closed
conic subsets of $X\times(\R^n\setminus\left\{0\right\})$.
The case that $\mathcal{B}=\C$ and the general case are
related by the following new theorem, which also gives an alternative
way of defining the wave front set for Banach space-valued distributions.
\begin{theorem}\label{WFsum}
$WF(u)=\overline{\cup_{\phi\in\mathcal{B}'} WF(\phi\circ u)}
\setminus\mathcal{Z}$.
\end{theorem}
\begin{proof*}
We let $R_u$ and $R_{\phi}$ denote the set of regular directions for $u$
and $\phi\circ u(.)$ respectively, where $\phi\in\mathcal{B}'$. If
$(x,k)\in R_u$ then there are an open neighbourhood $O$ of $x$ and an
open conic neighbourhood $V$ of $k$ such that
$O\times V\subset R_u$. For any point $(x',k')\in O\times V$ and any
$\phi\in\mathcal{B}'$ we then have $(x',k')\in R_{\phi}$, because
$\|\phi\circ u(e^{-ik'\cdot}f)\|\le \|\phi\|\cdot\|u(e^{-ik'\cdot}f)\|$
where $\|\phi\|<\infty$. Therefore,
\[
R_u\subset\mathrm{int}(\cap_{\phi\in\mathcal{B}'}R_{\phi}).
\]
To prove the converse of this inclusion we let
$(x,k)\in\mathrm{int}(\cap_{\phi\in\mathcal{B}'}R_{\phi})$. It follows
that there are an open neighbourhood $O$ of $x$ and a conic open
neighbourhood $V$ of $k$ such that
$O\times V\subset\mathrm{int}(\cap_{\phi\in\mathcal{B}'}R_{\phi})$.
Now choose a function $f\in\Test_0(O)$ such
that $f(x)\not=0$ and a conic open neighbourhood $V'$ of $k$ such that
$\overline{V'}\setminus\left\{0\right\}\subset V$. For each
$\phi\in\mathcal{B}'$ we can then find constants $C_{N,\phi}$ such that
\begin{equation}\label{uniformbound}
|\phi\circ u(e^{-ik'\cdot}f)|\le\frac{C_{N,\phi}}{1+\|k'\|^N}
\end{equation}
for all $k'\in V'$ and $N\in\N$ by \cite{Hoermander} lemma 8.2.1.
We now consider the family $(1+\|k'\|^N)u(e^{-ik'\cdot}f)$ for all
$k'\in V'$ and for fixed (but arbitrary) $N\in\N$ as a family of
bounded linear operators on the Banach space $\mathcal{B}'$. By the
estimate
(\ref{uniformbound}) these linear operators are bounded pointwise on
each $\phi\in\mathcal{B}'$. The uniform boundedness principle
(\cite{Kadison+} theorem 1.8.10) implies that we can choose constants
$C_N$ independently of $\phi$ such that
\[
\|u(e^{-ik'\cdot}f)\|\le\frac{C_N}{1+\|k'\|^N}
\]
for all $k'\in V'$ and $N\in\N$. Hence, $(x,k)\in R_u$ and
$R_u=\mathrm{int}(\cap_{\phi\in\mathcal{B}'}R_{\phi})$ and therefore:
\begin{eqnarray}\label{WFclosure}
WF(u)\cup\mathcal{Z}&=&R_u^c=(\mathrm{int}(\cap_{\phi}R_{\phi}))^c
=\overline{(\cap_{\phi}R_{\phi})^c}=\overline{\cup_{\phi}R_{\phi}^c}
\nonumber\\
&=&\overline{\cup_{\phi}WF(\phi\circ u(.))\cup\mathcal{Z}}=
\overline{\cup_{\phi}WF(\phi\circ u(.))}\cup\mathcal{Z}.\nonumber\\
WF(u)&=&\overline{\cup_{\phi}WF(\phi\circ u(.))}\setminus\mathcal{Z}.
\end{eqnarray}
\end{proof*}

Theorem \ref{WFsum} allows some standard results on scalar distributions
(see \cite{Hoermander}) to be generalised as follows:
\begin{theorem}\label{WFresults}
If $u,v$ are $\mathcal{B}$-valued distributions on an open set
$X\subset\R^n$ and $\mathcal{B}$ is a Banach space, then
\begin{enumerate}
\item $\mathrm{sing\ supp}(u)$ is the projection of
$WF(u)$ on the first variable,
\item $u\in C^{\infty}(X,\mathcal{B})$ if and only if
$WF(u)=\emptyset$,
\item $WF(u+v)\subset WF(u)\cup WF(v)$,
\item if $P$ is a linear partial differential operator on $X$
with smooth coefficients and principal symbol\footnote{We refer
to \cite{Baer+} definition A.4.2 for the definition of the
principal symbol.} $p(x,\xi)$, then
\[
WF(Pu)\subset WF(u)\subset WF(Pu)\cup\mathrm{Char}(P),
\]
where $\mathrm{Char}(P):=\left\{(x,\xi)\in X\times(\R^n\setminus
\left\{0\right\})|\ \xi\not=0,
p(x,\xi)=0\right\}$,
\item if $\map{f}{Y}{X}$ is a diffeomorphism between open sets
$X,Y\subset\R^n$ and $\mathrm{supp}\ u\subset X$, then
$WF(f^*u)=f^*(WF(u))$, where the wave front set is pulled back as a
subset of the cotangent bundle $T^*X$.
\end{enumerate}
\end{theorem}
\begin{proof*}
The last three statements follow directly from equation
(\ref{WFclosure}) and the statements for scalar distributions, which
are proved in \cite{Hoermander}. The second statement follows from the
first, so it remains to prove the first statement.

The distribution $u$ is smooth on the open set $O$ if and only if
$\phi\circ u$ is smooth on $O$ for all $\phi\in\mathcal{B}'$ by lemma
\ref{Banachlemma}. This is true if and only if
$WF(\phi\circ u)\cap(O\times(\R^n\setminus\left\{0\right\}))$ for all
$\phi$, by \cite{Hoermander} section 8.1. In view of theorem
\ref{WFsum} this is true if and only if
$WF(u)\cap(O\times(\R^n\setminus\left\{0\right\})=\emptyset$.
\end{proof*}

The last item of theorem \ref{WFresults} allows us to define the wave
front set of a distribution $u$ on a manifold $\mathcal{M}$ as a subset
of the cotangent bundle which is closed in
$T^*\mathcal{M}\setminus\mathcal{Z}$ and which coincides in each
coordinate chart $\kappa$ with $\kappa^*WF(u\circ\kappa^{-1})$.

If $\mathcal{X}$ is an $m$-dimensional (complex) vector bundle
on an $n$-dimensional manifold $\mathcal{M}$ then the space of compactly
supported smooth sections of $\mathcal{X}$ can be given a test-function
topology. We can define the wave front set of a $\mathcal{B}$-valued
distribution on such test-functions in a local trivialisation. Let
$\left\{e_i\right\}_{i=1,\ldots,m}$ be a local frame for $\mathcal{X}$
and define the $\mathcal{B}$-valued distributions $u_i$ by
$u_i(h):=u(he_i)$. Then $u$ is determined completely by
$u(\sum_if_ie_i)=\sum_iu_i(f_i)$. We define
\[
WF(u):=\cup_{i=1}^m WF(u_i).
\]
If $e'_i$ is a different local frame, then $e'_i=e_jM^j_{\ i}$ for a
local $\mathrm{Aut}(\C^m)$-valued function $M$. Using theorem
\ref{WFresults} it follows that $WF(u)$ is independent of the choice
of local frame and transforms as a subset of the cotangent
bundle.\footnote{Note that $u$ is locally equivalent to a distribution
$\tilde{u}$ with values in the Banach space $\mathcal{B}\otimes(\C^m)^*$
and defined by: $\tilde{u}(h):=\sum_iu(fe_i)\otimes d^i$, where $d^i$ is
a basis of $(\C^m)^*$. We can recover $u$ as
$u(\sum_if_ie_i)=\sum_i\langle\tilde{u}(f_i),d_i\rangle$, where $d_i$ is
a basis of $\C^m$ dual to $d^i$ and the brackets denote the action of
the second factor of $\mathcal{B}\otimes(\C^m)'$ on $\C^m$. In this case
we have $WF(u)=WF(\tilde{u})$ by theorem \ref{WFsum}.}

\begin{theorem}\label{WFresults2}
If $u,v$ are $\mathcal{B}$-valued distributions on smooth sections of
a complex vector bundle $\mathcal{X}$ over a smooth manifold
$\mathcal{M}$ and $\mathcal{B}$ is a Banach space, then
\begin{enumerate}
\item $\mathrm{sing\ supp}(u)$ is the projection of
$WF(u)$ on the first variable,
\item $u\in C^{\infty}(\mathcal{X}^*,\mathcal{B})$ if and only if
$WF(u)=\emptyset$,
\item $WF(u+v)\subset WF(u)\cup WF(v)$,
\item if $P$ is a linear partial differential operator on $\mathcal{X}$
with smooth coefficients and (matrix-valued) principal symbol\footnote{See
\cite{Baer+} definition A.4.2 for the definition of the principal symbol.}
$p(x,\xi)$, then $WF(Pu)\subset WF(u)\subset WF(Pu)\cup\Omega_P$, where\\
$\Omega_P:=\left\{(x,\xi)\in T^*\mathcal{M}|\ \xi\not=0,
\det p(x,\xi)=0\right\}$.
\end{enumerate}
\end{theorem}
\begin{proof*}
These results follow directly from theorem \ref{WFresults} and the
definition of the wave front set for a distribution on
vector-bundle-valued sections on a manifold, except the second inclusion
of the
last statement. For this result we refer to \cite{Dencker}.
\end{proof*}

We now follow \cite{Strohmaier+} and prove a useful result in the case
where $\mathcal{B}$ is a Hilbert space. We refer to definition
\ref{boxtimes} for the exterior tensor product $\boxtimes$ of two vector
bundles.
\begin{theorem}\label{WFHvalued}
Let $\mathcal{H}$ be a Hilbert space and $\mathcal{X}_i$, $i=1,2$, two
finite dimensional (complex) vector bundles over smooth $n_i$-dimensional
manifolds $\mathcal{M}_i$ with complex conjugations $J_i$, i.e. the $J_i$
are anti-linear, base-point preserving bundle isomorphisms
$\map{J_i}{\mathcal{X}_i}{\mathcal{X}_i}$ such that $J_i^2=\id$. Let
$u_i$, $i=1,2$, be two $\mathcal{H}$-valued distributions on the
test-sections of $\mathcal{X}_i$ and let $w_{ij}$ be the distributions
on sections of the vector bundle $\mathcal{X}_i\boxtimes\mathcal{X}_j$
over $\mathcal{M}_i\times\mathcal{M}_j$ determined by
$w_{ij}(f_1\boxtimes f_2):=\langle u_i(Jf_1),u_j(f_2)\rangle$. Then
\[
(x,k)\in WF(u_1)\quad \Leftrightarrow (x,-k;x,k)\in WF(w_{11})
\]
and
\[
WF(w_{ij})\subset \left(-WF(u_i)\cup\mathcal{Z}\right)\times
\left(WF(u_j)\cup\mathcal{Z}\right).
\]
\end{theorem}
Note that $w_{ij}$ does indeed uniquely define a distribution on sections
of $\mathcal{X}_i\boxtimes\mathcal{X}_j$, essentially by the Schwartz
kernel theorem (\cite{Hoermander} theorem 5.2.1).

\begin{proof*}
The proof is a straightforward generalisation of the proof of
proposition 3.2 part (iii) in \cite{Fewster+2}, where we notice the
following.
We may work in local coordinates on $\mathcal{M}_i$ and choose a local
frame $\left\{e^{(i)}_r\right\}$ that is real w.r.t $J_i$, i.e. such that
$J_ie^{(i)}_r=e^{(i)}_r$. Notice that $e^{(i)}_r\times e^{(j)}_s$ is a
local frame for $\mathcal{X}_i\boxtimes\mathcal{X}_j$ and for $w_{ii}$
there is no loss of generality in using the same frame in both entries,
because any two points in $\mathcal{M}_i$ can be contained in a single
local trivialisation (using \cite{Dieudonne} theorem 16.26.9). This,
together with the complex conjugation and theorem \ref{WFresults} part 3),
essentially reduces the problem to distributions on
test-functions rather than test-sections. In \cite{Fewster+2} one takes
the inner product of two distributions on the same manifold, but the key
ingredient of the proof, the Cauchy-Schwarz inequality, still works if
we allow the manifolds to be different.
\end{proof*}

Finally we collect the wave front sets of some useful distributions,
which may be found in \cite{Radzikowski}:
\begin{proposition}\label{WFE}
Let $E^{\pm}$ be the advanced $(-)$ and retarded $(+)$ fundamental
solutions of the Klein-Gordon operator $K$ or of the operator $\tilde{D}D$
of section \ref{quantisation} on a globally hyperbolic spin spacetime
$M$, then
\begin{eqnarray}
WF(E^{\pm})&=&\left\{(x,\xi;y,\xi')\in T^*(M\times M)|\
(x,-\xi)\sim (y,\xi'), x\in J^{\pm}(y)\right\}
\setminus\mathcal{Z},\nonumber
\end{eqnarray}
where $(x,-\xi)\sim (y,\xi')$ if and only if $(x,-\xi)=(y,\xi')$ or
there is an affinely parameterised light-like geodesic between $x$ and $y$
to which $-\xi,\xi'$ are cotangent (and hence $-\xi$ and $\xi'$ are
parallel transports of each other along the geodesic).
\end{proposition}
Strictly speaking, \cite{Radzikowski} only states this proposition for
advanced and retarded fundamental solutions of the scalar Klein Gordon
operator, not for the Lichnerowicz wave operator $\tilde{D}D$. The latter
acts on sections of a vector bundle, which complicates the situation
somewhat. Nevertheless, the principal part is diagonal and this is what
determines the bicharacteristic strips and allows the construction of the
advanced and retarded fundamental solutions. Therefore we believe the
result should still hold, although we could not produce a reference
for this fact.

\section{The analytic wave front set}

Results for analytic wave front sets are mostly analogous to those for
smooth wave front sets, except that they are more involved to
formulate. The difficulty is that we cannot localise singularities
at a point $x$ by multiplying with a compactly supported analytic
function $f$ with $f(x)\not=0$.

Consider again a $\mathcal{B}$-valued distribution $u$ on an open
set $X\subset\R^n$.
\begin{definition}
An \emph{analytic regular direction} for $u$ is a point
$(x,k)\in\R^n\times(\R^n\setminus\left\{0\right\})$ for which there
exist an open neighbourhood $O$ of $x$, a conic open neighbourhood
$V$ of $k$, a bounded sequence of compactly supported distributions
$u_N$, $N\in\N$, which equal $u$ on $O$ and a constant $C>0$ such that
$\|u_N(e^{-i\xi\cdot})\|\le C\left(\frac{C(N+1)}{\|\xi\|}\right)^N$ for
all $\xi\in V$ and $N\in\N$, where $\|\xi\|$ denotes the Euclidean norm.

The \emph{analytic wave front set} $WF_A(u)$ of $u$ is defined as
\begin{eqnarray}
WF_A(u)&:=&\left\{(x,k)\in\R^n\times(\R^n\setminus\left\{0\right\})|\
(x,k)\mathrm{\ is\ not\ an\ analytic\ regular}\right.\nonumber\\
&&\left.\mathrm{direction\ for\ }u\right\}.\nonumber
\end{eqnarray}
\end{definition}

Like the smooth wave front set the analytic wave front set is closed
and we have $WF(u)\subset WF_A(u)$. Analogous to theorem \ref{WFsum}
we have the following equivalent characterisation of the analytic wave
front set of a Banach space-valued distribution in terms of the
analytic wave front sets of scalar distributions:
\begin{theorem}\label{WFAclosure}
$WF_A(u)=\overline{\cup_{\phi\in\mathcal{B}'}WF_A(\phi\circ u)}
\setminus\mathcal{Z}$.
\end{theorem}
\begin{proof*}
If $(x,k)$ is an analytic regular direction for $u$ then there are an
open neighbourhood $O$ of $x$ and an open conic neighbourhood $V$ of $k$
such that $O\times V\cap WF_A(U)=\emptyset$. For any
$\phi\in\mathcal{B}'$ any point $(x',k')\in O\times V$ is then an
analytic regular direction, because
$\|\phi\circ u_N(e^{-ik'\cdot})\|\le \|\phi\|\cdot\|u_N(e^{-ik'\cdot})\|$
where $\|\phi\|<\infty$.

For the converse we suppose that $(x,k)\not\in\overline{
\cup_{\phi\in\mathcal{B}'}WF_A(\phi\circ u)}$ for $k\not=0$ and we choose
an open neighbourhood $O$ of $x$ and a closed conic neighbourhood $V$ of
$k$ such that $O\times (V\setminus\left\{0\right\})\cap
\overline{\cup_{\phi\in\mathcal{B}'}WF_A(\phi\circ u)}=\emptyset$. If
$K\subset O$ is a compact neighbourhood
of $x$ then we may find a sequence $\chi_N\in\Test_0(O)$ such that
$\chi_N\equiv 1$ on $K$ and $\sup_O|D^{\alpha+\beta}\chi_N|\le
C_{\alpha}^{1+|\beta|}(N+1)^{|\beta|}$ for $|\beta|\le N$
(see \cite{Hoermander} theorem 1.4.2, cf. the proof of proposition
8.4.2 and lemma 8.4.4). By
\cite{Hoermander} lemma 8.4.4 we then have for some constants
$C_{\phi}>0$ and all $\xi\in V$:
\begin{equation}\label{WFAestimate}
\left(\frac{\|\xi\|}{N+1}\right)^N|\phi\circ u(\chi_Ne^{-i\xi\cdot})|
\le C_{\phi}^{N+1}.
\end{equation}

Now define for each $p\in\N$ the Banach space
\[
l^{\infty}_p:=\left\{x=\left\{x_i\right\}_{i\in\N}\in\mathcal{B}^{\times\N}|\
\sup_{i\in\N}\|x_i\|p^{-i}<\infty\right\}
\]
and the inductive limit $k^{\infty}:=\cup_{p\in\N}l^{\infty}_p$, which is a
locally convex space (cf. \cite{Bonet+} section 3). The estimate
(\ref{WFAestimate}) now means that for a fixed $\phi\in\mathcal{B}'$ the set
\[
\left\{\left\{\left(\frac{\|\xi\|}{N+1}\right)^N|\phi\circ u(\chi_Ne^{-i\xi\cdot .})|
\right\}_{N\in\N}|\ \xi\in V\right\}\subset k^{\infty}
\]
is bounded. By the (generalised) uniform boundedness principle, theorem 3.4.2
in \cite{Schaefer}, the set
$X:=\left\{\left\{\left(\frac{\|\xi\|}{N+1}\right)^N\|u(\chi_Ne^{-i\xi\cdot})\|
\right\}_{N\in\N}|\ \xi\in V\right\}\subset k^{\infty}$ is bounded. This means that
$X\subset l^{\infty}_p$ is a bounded subset for some $p\in\N$ (\cite{Schaefer} 2.6.5)
and hence we have for some $p\in\N$ and all $\xi\in V$:
\begin{eqnarray}\label{WFAestimate2}
\left(\frac{\|\xi\|}{N+1}\right)^N\|u(\chi_Ne^{-i\xi\cdot .})\|\le Cp^N
\end{eqnarray}
We conclude that $(x,k)\not\in WF_A(u)$.
\end{proof*}

Analogously to theorem \ref{WFresults} we can now generalise some
results for scalar distributions:
\begin{theorem}\label{WFAresults}
If $u,v$ are $\mathcal{B}$-valued distributions on an open set
$X\subset\R^n$ and $\mathcal{B}$ is a Banach space, then
\begin{enumerate}
\item $\mathrm{sing\ supp}_A(u)$ is the projection of
$WF_A(u)$ on the first variable,
\item $u\in C^{\omega}(X,\mathcal{B})$ if and only if
$WF_A(u)=\emptyset$,
\item $WF_A(u+v)\subset WF_A(u)\cup WF_A(v)$,
\item if $P$ is a linear partial differential operator on $X$
with real-analytic coefficients and principal symbol $p(x,\xi)$, then
\[
WF_A(Pu)\subset WF_A(u)\subset WF_A(Pu)\cup\mathrm{Char}(P),
\]
where $\mathrm{Char}(P):=\left\{(x,\xi)\in X\times(\R^n\setminus
\left\{0\right\})|\ \xi\not=0,p(x,\xi)=0\right\}$,
\item if $\map{f}{Y}{X}$ is an analytic diffeomorphism between open sets
$X,Y\subset\R^n$ and $\mathrm{supp}\ u\subset X$, then
$WF_A(f^*u)=f^*(WF_A(u))$, where the wave front set is pulled back as a
subset of the cotangent bundle $T^*X$.
\end{enumerate}
\end{theorem}
\begin{proof*}
The last three statements follow directly from theorem \ref{WFAclosure}
and the corresponding statements for scalar distributions, which
are proved in \cite{Hoermander}. The second statement follows from the
first, so it remains to prove the first statement.

If the distribution $u$ is analytic on the open set $O$ then
$\phi\circ u$ is analytic on $O$ for all $\phi\in\mathcal{B}'$ and hence
$WF_A(u)\cap(O\times(\R^n\setminus\left\{0\right\}))=\emptyset$ by
theorem \ref{WFAclosure}. Conversely, if
$WF_A(u)\cap(O\times(\R^n\setminus\left\{0\right\}))=\emptyset$
then $u$ is a smooth function by theorem \ref{WFresults} and the fact
that $WF(u)\subset WF_A(u)$. It remains to prove that $u$ is
analytic on $O$. In the case $O=\R$ this is \cite{Bonet+} proposition 9
(see also the references there). For completeness we prove the required
generalisation. Given $x\in O$
we can choose a compact neighbourhood $K\subset O$ of $x$ and functions
$\chi_N$ as in the proof of theorem \ref{WFAclosure}. For each
$\phi\in\mathcal{B}'$ and $x\in K$ we then have
\begin{eqnarray}\label{WFintegral}
\phi(\partial^{\alpha}u)(x)&=&\partial^{\alpha}(\phi(\chi_Nu))(x)
\nonumber\\
&=&(2\pi)^{-n}\int_{\R^n} e^{i\xi\cdot x}(i\xi)^{\alpha}
\widehat{\phi(\chi_Nu)}(\xi)d\xi.
\end{eqnarray}
Because $u$ is a distribution we have
$|\widehat{\phi(\chi_Nu)}(\xi)|\le C\|\phi\|(1+\|\xi\|)^M$ for some order
$M$, which we use to estimate the integral over $\|\xi\|\le 1$. For
$\|\xi\|\ge 1$ we use the estimate (\ref{WFAestimate2}) with
$N=|\alpha|+n$ to find:
\[
|\phi(\partial^{\alpha}u)(x)|\le C^{N+1}(N+1)^N\|\phi\|
\]
for some $C>0$ and hence $\|\partial^{\alpha}u(x)\|\le C^{N+1}(N+1)^N$,
which implies
\[
\|\partial^{\alpha}u(x)\|\le C^{|\alpha|+1}(|\alpha|+1)^{|\alpha|}
\]
for some $C>0$, using $(|\alpha|+1)^n\le ce^{|\alpha|}$ and
$(|\alpha|+n+1)\le(|\alpha|+1)(n+1)$. Now let $r>0$ be such that the
disc around $x_0$ with radius $r$ is contained in $K$. A general term
in the Taylor series of $u$ can then be estimated by
\[
\left\|\frac{(x-x_0)^{\alpha}}{\alpha!}\partial^{\alpha}u(x_0)\right\|
\le C(nrC)^{|\alpha|}\frac{(N+1)^N}{N!}.
\]
Here we used $n^N=(1+\ldots+1)^N=\sum_{|\alpha|\le N}\frac{N!}{\alpha!}$
(by Newton's binomial theorem) to obtain
$\frac{1}{\alpha!}\le\frac{n^N}{N!}$. The Taylor series contains no
more than $n^N$ terms with $|\alpha|=N$, so
\[
\sum_{\alpha}\left\|\frac{(x-x_0)^{\alpha}}{\alpha!}
\partial^{\alpha}u(x_0)\right\|\le \sum_{N=0}^{\infty}
C(n^2rC)^{|\alpha|}\frac{(N+1)^N}{N!}.
\]
Because $\frac{(N+1)^N}{N!}\le c^N$ for some constant $c>0$ we can
choose $r$ small enough to ensure that the series is absolutely
convergent. For all $\phi\in\mathcal{B}'$ and $\|x-x_0\|$ within the radius
of convergence we then have
\[
\phi\left(\sum_{\alpha}\frac{(x-x_0)^{\alpha}}{\alpha!}
\partial^{\alpha}u(x_0)\right)=
\sum_{\alpha}\frac{(x-x_0)^{\alpha}}{\alpha!}
\partial^{\alpha}(\phi\circ u)(x_0)
=\phi\circ u(x).
\]
This shows that the limit of the series is the function $u(x)$ itself.
\end{proof*}

The last statement of theorem \ref{WFAresults} implies that we can
define analytic wave front sets on analytic manifolds as a subset of
the cotangent bundle in a similar way as for the smooth wave front set
on smooth manifolds.

As in the smooth case we can consider an $m$-dimensional (complex)
real-analytic vector bundle $\mathcal{X}$ on an $n$-dimensional analytic
manifold $\mathcal{M}$ and endow the space of compactly supported smooth
sections of $\mathcal{X}$ with a test-function topology. Given an analytic
local frame $\left\{e_i\right\}_{i=1,\ldots,m}$ for $\mathcal{X}$, the
analytic wave front
set of a $\mathcal{B}$-valued distribution $u$ on test-sections of
$\mathcal{X}$ can be defined as
\[
WF_A(u):=\cup_{i=1}^m WF_A(u_i),
\]
where $u_i(h):=u(he_i)$ are $\mathcal{B}$-valued distributions as before.
If $e'_i$ is a different analytic local frame, then $e'_i=e_jM^j_{\ i}$ for
an analytic local $\mathrm{Aut}(\C^m)$-valued function $M$. Using theorem
\ref{WFAresults} it follows that $WF_A(u)$ is independent of the choice
of local frame and transforms as a subset of the cotangent
bundle.

\begin{theorem}\label{WFAresults2}
If $u,v$ are $\mathcal{B}$-valued distributions on smooth sections of
a complex, real-analytic vector bundle $\mathcal{X}$ over an analytic
manifold $\mathcal{M}$ and $\mathcal{B}$ is a Banach space, then
\begin{enumerate}
\item $\mathrm{sing\ supp}_A(u)$ is the projection of
$WF_A(u)$ on the first variable,
\item $u\in C^{\omega}(\mathcal{X}^*,\mathcal{B})$ if and only if
$WF_A(u)=\emptyset$,
\item $WF_A(u+v)\subset WF_A(u)\cup WF_A(v)$,
\item if $P$ is a linear partial differential operator on $\mathcal{X}$
with real-analytic coefficients and (matrix-valued) principal symbol
$p(x,\xi)$, then $WF_A(Pu)\subset WF_A(u)$.
\end{enumerate}
\end{theorem}
\begin{proof*}
These results follow directly from theorem \ref{WFAresults} and the
definition of the wave front set for a distribution on
vector-bundle-valued sections on a manifold.
\end{proof*}

Again we can follow \cite{Strohmaier+} and prove a useful result in the
case where $\mathcal{B}$ is a Hilbert space (and again we refer to
definition \ref{boxtimes} for the exterior tensor product $\boxtimes$
of two vector bundles):
\begin{theorem}\label{WFAHvalued}
Let $\mathcal{H}$ be a Hilbert space and $\mathcal{X}_i$, $i=1,2$, two
finite dimensional (complex) real-analytic vector bundles over analytic
$n_i$-dimensional manifolds $\mathcal{M}_i$ with real-analytic complex
conjugations $\map{J_i}{\mathcal{X}_i}{\mathcal{X}_i}$. Let $u_i$,
$i=1,2$, be two $\mathcal{H}$-valued distribution on the
test-sections of $\mathcal{X}_i$ and let $w_{ij}$ be the distributions
on sections of the vector bundle $\mathcal{X}_i\boxtimes\mathcal{X}_j$
over $\mathcal{M}_i\times\mathcal{M}_j$ determined by
$w_{ij}(f_1,f_2):=\langle u_i(Jf_1),u_j(f_2)\rangle$. Then
\[
(x,k)\in WF_A(u_1)\quad \Leftrightarrow (x,-k;x,k)\in WF_A(w_{11})
\]
and
\[
WF_A(w_{ij})\subset \left(-WF_A(u_i)\cup\mathcal{Z}\right)\times
\left(WF_A(u_j)\cup\mathcal{Z}\right).
\]
\end{theorem}
\begin{proof*}
The proof is a straightforward generalisation of the proof of
proposition 2.6 part 2) in \cite{Strohmaier+}, where we notice the
following.
We may work in local coordinates on $\mathcal{M}_i$ and choose a local
frame $\left\{e^{(i)}_r\right\}$ that is real w.r.t $J_i$, i.e. such that
$J_ie^{(i)}_r=e^{(i)}_r$. Notice that $e^{(i)}_r\times e^{(j)}_s$ is a
local frame for $\mathcal{X}_i\boxtimes\mathcal{X}_j$ and for $w_{ii}$
there is no loss of generality in using the same frame in both entries,
because any two points in $\mathcal{M}_i$ can be contained in a single
local trivialisation (using \cite{Dieudonne} theorem 16.26.9 and
\cite{Hirsch} theorem 8.3.1 to guarantee that we have the right analytic
structure near the given points). In \cite{Strohmaier+} one takes the
inner product of a distribution with itself, but the key ingredient of
the proof, the Cauchy-Schwarz inequality, still works if we take the
inner product of two different distributions.
\end{proof*}


\backmatter

\end{document}